\documentstyle[12pt,dina4p,epsfig,rotating]{article} 

\begin{document}

\newcommand{\lwig}{\mbox{\,\raisebox{.3ex}
{$<$}$\!\!\!\!\!$\raisebox{-.9ex}{$\sim$}\,}}
\newcommand{\gwig}{\mbox{\,\raisebox{.3ex}
{$>$}$\!\!\!\!\!$\raisebox{-.9ex}{$\sim$}}\,}
\newcommand{\gev}{\mbox{$\,{{\rm GeV}}$}}
\newcommand{\gee}{\mbox{$g_1$}}
\newcommand{\gz}{\mbox{$g_2$}}
\newcommand{\gd}{\mbox{$g_3$}}
\newcommand{\ges}{\mbox{$g_1^2$}}
\newcommand{\gzs}{\mbox{$g_2^2$}}
\newcommand{\gds}{\mbox{$g_3^2$}}
\newcommand{\gesz}{\mbox{$g_{10}^2$}}
\newcommand{\gzsz}{\mbox{$g_{20}^2$}}
\newcommand{\gdsz}{\mbox{$g_{30}^2$}}
\newcommand{\la}{\mbox{$\lambda$}}
\newcommand{\gt}{\mbox{$g_t$}}
\newcommand{\gts}{\mbox{$g_t^2$}}
\newcommand{\gtsz}{\mbox{$g_{t0}^2$}}
\newcommand{\gb}{\mbox{$g_b$}}
\newcommand{\gbs}{\mbox{$g_b^2$}}
\newcommand{\gbsz}{\mbox{$g_{b0}^2$}}
\newcommand{\gta}{\mbox{$g_{\tau}$}}
\newcommand{\gtas}{\mbox{$g_{\tau}^2$}}
\newcommand{\gtasz}{\mbox{$g_{\tau 0}^2$}}
\newcommand{\htt}{\mbox{$h_t$}}
\newcommand{\hts}{\mbox{$h_t^2$}}
\newcommand{\htsz}{\mbox{$h_{t0}^2$}}
\newcommand{\hb}{\mbox{$h_b$}}
\newcommand{\hbs}{\mbox{$h_b^2$}}
\newcommand{\hbsz}{\mbox{$h_{b0}^2$}}
\newcommand{\hta}{\mbox{$h_{\tau}$}}
\newcommand{\htas}{\mbox{$h_{\tau}^2$}}
\newcommand{\htasz}{\mbox{$h_{\tau 0}^2$}}
\newcommand{\mt}{\mbox{$m_t$}}
\newcommand{\mb}{\mbox{$m_b$}}
\newcommand{\mta}{\mbox{$m_{\tau}$}}
\newcommand{\mh}{\mbox{$m_H$}}
\newcommand{\mtp}{\mbox{$m_t^{{\rm pole}}$}}
\newcommand{\mbp}{\mbox{$m_b^{{\rm pole}}$}}
\newcommand{\mfp}{\mbox{$m_f^{{\rm pole}}$}}
\newcommand{\mhp}{\mbox{$m_H^{{\rm pole}}$}}
\newcommand{\mgut}{\mbox{$M_{{\rm GUT}}$}}
\newcommand{\msusy}{\mbox{$M_{{\rm SUSY}}$}}
\newcommand{\rt}{\mbox{$\rho_t$}}
\newcommand{\rb}{\mbox{$\rho_b$}}
\newcommand{\rta}{\mbox{$\rho_{\tau}$}}
\newcommand{\rtz}{\mbox{$\rho_{t0}$}}
\newcommand{\rbz}{\mbox{$\rho_{b0}$}}
\newcommand{\rtaz}{\mbox{$\rho_{\tau 0}$}}
\newcommand{\lam}{\mbox{$\Lambda$}}
\newcommand{\rh}{\mbox{$\rho_H$}}
\newcommand{\re}{\mbox{$\rho_1$}}
\newcommand{\rz}{\mbox{$\rho_2$}}
\newcommand{\rez}{\mbox{$\rho_{10}$}}
\newcommand{\rzz}{\mbox{$\rho_{20}$}}
\newcommand{\tanb}{\mbox{$\tan\beta$}}

\begin{center}
{\Large \bf Top Quark and Higgs Boson Masses:\\[2mm] 
Interplay between Infrared and Ultraviolet Physics}\,\footnote{\normalsize to be published in {\it Progress in Particle and Nuclear Physics}, Vol. 37, 1996,\\\hspace*{6mm} copyright Elsevier Science Ltd.}\\[10mm]
{\large Barbara Schrempp}$\, ^{\dag\,\ddag}$ \ \ and \ \ {\large Michael
Wimmer}$\, ^{\dag}$\,\footnote{\normalsize supported by Deutsche
  Forschungsgemeinschaft}\\[12mm] $ ^{\dag}$ {\it Institut
  f\"ur Theoretische Physik, Universit\"at Kiel, D-24118 Kiel}\\ 
$ ^{\ddag}$ {\it Deutsches Elektronen-Synchrotron DESY,
  D-22603 Hamburg}\\[3mm]
May 1996\\[10mm]
{\bf Abstract}
\end{center} 

We review recent efforts to explore the information on masses of heavy
matter particles, notably of the top quark and the Higgs boson, as
encoded at the quantum level in the renormalization group equations.
The Standard Model (SM) and the Minimal Supersymmetric Standard Model
(MSSM) are considered in parallel throughout.\\[1mm] First, the
question is addressed to which extent the infrared physics of the
``top-down'' renormalization group flow is independent of the
ultraviolet physics. The central issues are i) infrared attractive
fixed point values for the top and the Higgs mass, the most
outstanding one being $\mt=O(190\gev)\sin\beta$ in the MSSM, ii)
infrared attractive relations between parameters, the most prominent
ones being an infrared fixed top-Higgs mass relation in the SM,
leading to \mh=O(156\gev) for the experimental top mass, and an
infrared fixed relation between the top mass and the parameter \tanb\ 
in the MSSM, and iii) a systematical analytical assessment of their
respective strengths of attraction. The triviality and vacuum
stability bounds on the Higgs and top masses in the SM as well as the
upper bound on the mass of the lightest Higgs boson in the MSSM are
reviewed. The mathematical backbone for all these features, the rich
structure of infrared attractive fixed points, lines, surfaces,...  in
the corresponding multiparameter space, is made transparent.
Interesting hierarchies emerge: i) infrared attraction in the MSSM is
systematically stronger than in the SM, ii) generically, nontrivial
higher dimensional fixed manifolds are more strongly infrared
attractive than the lower dimensional ones.\\[1mm] Tau-bottom-(top)
Yukawa coupling unification as an ultraviolett symmetry property of
supersymmetric grand unified theories and its power to focus the
``top-down'' renormalization group flow into the IR top mass fixed
point and, more generally, onto the infrared fixed line in the
\mt-\tanb-plane is reviewed.
\\[1mm] The program of reduction of parameters, a systematic search for
renormalization group invariant relations between couplings, guided by the
requirement of asymptotically free couplings in the complementary
``bottom-up'' renormalization group evolution, is summarized; its
interrelations with the search for IR attractive fixed manifolds are pointed
out.



{\Large\bf Table of Content}

\vspace*{.2cm}

{\bf 1. Introduction}\\[0.5cm] {\bf 2. Theoretical framework}\\[3mm]
\hspace*{7mm}2.1 Standard Model\\[1mm] \hspace*{7mm}2.2 Minimal
Supersymmetric Standard Model\\[1mm] \hspace*{7mm}2.3 Grand
Unification\\[1mm] \hspace*{7mm}2.4 Renormalization Group Equations\\[1mm]
\hspace*{7mm}2.5 Relations between Pole Masses and $\overline{{\rm MS}}$
Couplings\\[1mm]\hspace*{7mm}2.6 Effective Potential and Vacuum Stability\\[0.5cm] {\bf 3. Preview of Infrared Fixed Manifolds and Bounds
  in the SM and MSSM}\\[0.5cm] {\bf 4. Infrared Fixed Points, Lines,
  Surfaces and Mass Bounds}\\ \phantom{{\bf 4. }}{\bf in Absence of
  Electroweak Gauge Couplings}\\[3mm] \hspace*{7mm}4.1 The Pure Higgs
Sector of the SM -- Triviality and an Upper Bound on the Higgs Mass\\ 
\hspace*{7mm}4.2 The Higgs-Top Sector of the SM - a First IR Fixed Line and
a First Vacuum Stability \hspace*{7mm}\phantom{4.5} Bound\\[1mm] \hspace*{7mm}4.3 The Top-\gd\ Sector
of the SM and MSSM -- a Non-Trivial IR Fixed Point\\[1mm] \hspace*{7mm}4.4
The Higgs-Top-\gd\ Sector of the SM -- a First Non-Trivial
Approximation\\[1mm] \hspace*{7mm}4.5 The Top-Bottom-\gd\ Sector of the SM
and MSSM -- Top-Bottom Yukawa Unification as \\ 
\hspace*{7mm}\phantom{4.5} an IR fixed Property\\[1mm] \hspace*{7mm}4.6 The
Higgs-Top-Bottom-\gd\ Sector of the SM -- a First IR Fixed Surface\\[0.5cm]
{\bf 5. Infrared Fixed Points, Lines, Surfaces}\\ 
\phantom{{\bf 5. }}{\bf in Presence of All Gauge Couplings}\\[3mm]
\hspace*{7mm}5.1 The Top Sector of the SM and MSSM\\[1mm] \hspace*{7mm}5.2
The Top-Bottom-Sector of the SM and MSSM\\[1mm] \hspace*{7mm}5.3 The
Higgs-Top-Bottom Sector of the SM\\[0.5cm] {\bf 6. Infrared Attractive Top
  and Higgs Masses, Mass Relations and Mass Bounds}\\[3mm] \hspace*{7mm}6.1
Top Mass and \tanb\ in the MSSM\\[1mm]\hspace*{7mm}6.2 Top and Higgs Masses
and Top-Higgs Mass Relation in the SM\\[1mm]\hspace*{7mm}6.3 Lower Bound on
the Higgs Mass in the SM\\[1mm]\hspace*{7mm}6.4 Upper Bound on the Lightest Higgs Mass in the MSSM\\[5mm]{\bf 7.
  Supersymmetric Grand Unification Including Yukawa Unification}\\[5mm]
{\bf 8. Program of Reduction of Parameters}\\[5mm]{\bf 9. Conclusions}

\newpage

\section{Introduction\label{intro}}

The Standard Model (SM) is highly successful at describing the
electromagnetic, weak and strong gauge interactions among the elementary
particles up to presently accessible energies. It has, however, a
conceptual weakness: the masses of the matter particles, i.e. of the
quarks, leptons and the theoretically predicted Higgs boson, enter as free
parameters. This deficiency largely persists in prominent extensions of the
SM, i) its embedding into an underlying grand unified theory (GUT),
unifying the three gauge interactions in a single one, ii) its extension by
the fermion-boson supersymmetry which is considered to be instrumental for
an additional implementation of gravity.

Starting point for this review are investigations of the {\it quantum
  effects} in the framework of this wide class of theories, which have been
performed over the last decade or so, with peak activities during the last
few years. The main interest focuses on the inherent potential of the
quantum effects for i) relating ultraviolet (UV) physics issues to infrared
(IR) physics and vice versa and in particular for ii) providing
informations on (at least the heavy) particle masses. In their mildest form
these informations imply upper and lower bounds for (heavy) particle
masses.  Ultimatively, however, there even appears to open up the
fascinating possibility that one does not have to go beyond the SM and its
extensions in search for the dynamical origin of (heavy) particle masses
but that this dynamical origin is provided on the level of the quantum
effects -- in a sense to be specified below.

The top quark and the Higgs boson are the heaviest matter particles of the
Standard Model. Only recently the top quark has been observed directly at
the proton-antiproton collider at FERMILAB
\begin{eqnarray}
  {\rm CDF\ collaboration \cite{cdf}:}\;\;\;\;\;\mt&=&176\;\pm 8\;\;({\rm
    stat.})\pm 10\; ({\rm syst.})\;\gev,\\ {\rm D0\ collaboration
    \cite{d0}:}\;\;\;\;\;\mt&=&199\begin{array}{l}+19\\-21\\ 
\end{array}({\rm stat.})\pm 22\;({\rm syst.})\;\gev.
\label{CDF}
\end{eqnarray}
This mass value is in good agreement with the present indirect evidence
from the electron-positron collider LEP at CERN \cite{lep}
\begin{equation}
  {\rm LEP\ collaborations,\ 
    combined:}\;\;\;\;\;\mt\;=\;178\begin{array}{l}+11+18\\-11-19\\ 
\end{array}\gev;
\label{LEP}
\end{equation}
the central value and the first errors quoted refer to a Higgs mass of 300
\gev, the second errors correspond to a variation of the central value when
varying the Higgs mass between 60 \gev\ and 1000 \gev.  The top quark is
much heavier than all the other quarks and leptons, even substantially
heavier than its partners in the heaviest fermion generation, the bottom
quark with mass 
\begin{equation}
  \mb=4.25\pm 0.15\gev\,
\label{massb}
\end{equation}
(the $\overline{\rm MS}$ mass at $\mu=\mb$) as determined \cite{gas} from
QCD sum rules, and the tau lepton with mass \cite{par}
\begin{equation}
  \mta=1.7771\begin{array}{l}+0.0004\\-0.0005\\\end{array}\gev,
\label{masstau}
\end{equation}
which will also play a role in this review.

For the Higgs mass there exits only an experimental lower bound from LEP
\cite{par}
\begin{equation}
  \mh>58.4\gev
\end{equation}
at $95\%$ confidence level. From the upgrade LEP200 of LEP and the future
collider LHC one expects soon an extended experimental reach for the Higgs
boson. In expectation of these future Higgs searches the activities for a
precise determination of theoretical bounds on the Higgs mass have
increased in the recent literature \cite{cab}-\cite{wagne}, where
Ref. \cite{mai} has played the role of a primer in the field. This applies in
particular to a lower (vacuum stability) bound within the SM and to an
upper bound for the lightest Higgs boson within the minimal supersymmetric
extension of the SM, the MSSM. The
upshot of these developments will be included in this review.

Altogether, one may expect the top and Higgs masses to be very roughly of
the order of the weak interaction scale
\begin{equation}
  v/\sqrt{2}\simeq 174\gev,
\label{vv}
\end{equation}
given in terms of the vacuum expectation value $v$ of the Higgs field.
Clearly, it is a great challenge to understand the dynamical origin for why
the top and Higgs masses should be of $O(v)$ and also for the mass
disparity with respect to the other quarks and leptons. As mentioned
already, important clues for an answer lie in the wealth of informations on
the top quark and Higgs boson masses, which have emerged from analyses of
the quantum effects over the last decade.

In all the above mentioned frameworks of the SM, possibly embedded in a
grand unified theory and possibly endorsed with supersymmetry, the Higgs
mass and the quark and lepton masses are related to couplings; these
couplings are a measure for the strength of the Higgs self interaction and
the Higgs-fermion-antifermion Yukawa interactions, respectively. A
characteristic signature of the quantum effects is that these couplings are
not constant but ``run'' as functions of a momentum scale $\mu$. The
running is encoded in the renormalization group equations (RGE), a set of
nonlinear coupled differential equations. They describe the response of all
couplings of the theory, the Higgs and Yukawa couplings as well as the
electromagnetic, weak and strong gauge couplings, to a differential change
in the momentum scale $\mu$. They have been calculated in two-loop order in
the framework of perturbation theory. The RGE allow to relate physics at
different momentum scales $\mu$. Of interest in this review are scales
generically between an infrared (IR) scale of the order of the presently
accessible weak interaction scale (\ref{vv}) and some ultraviolet (UV)
scale \lam. Let us emphasize that throughout this review the notions IR
scale, IR region or IR behaviour refer to the scale $\mu\simeq v$ and {\it
  not to the limit} $\mu\rightarrow 0$. The UV scale may be as large as
$\lam=\mgut\simeq O(10^{16}\gev)$ in the framework of (supersymmetric)
grand unification, in which the theory is supposed to continue to hold up
to the scale \mgut\ where the three gauge couplings unify, or ultimatively
as large as the Planck scale $\lam=M_{{\rm Planck}}\simeq 10^{19}\gev$,
where graviational interactions become important.



Let us anticipate and emphasize already here that whichever the framework,
SM or its minimal supersymmetric extension (MSSM), and whichever the size
of \lam, the interplay between IR and UV physics and the implications for
particle masses turn out to be similar {\it in principle}, even though
different on the quantitative level. This makes it a challenging task to
consider all these cases simultaneously and treat them in parallel, as
intended in this review.

In solving the RGE, which are a set of first order differential equations,
in the first instance one faces the same inherent deficiency as on the
classical level: the initial values of the couplings and thus the particle
masses are still free parameters.

There are, however, strong physical motivations from different sources to
be spelt out below, which in essence tend to single out {\it special
  solutions} of the RGE. From the mathematical point of view these special
solutions are distinguished by being determined by suitable {\it boundary
  conditions} in contradistinction to initial value conditions.

The theoretical motivations in the literature pointing towards such special
solutions of the RGE are the following.
\begin{itemize}
\item Consider the so-called ``top-down'' RG evolution, from the UV scale
  \lam\ to the IR scale. Determine the corresponding RG flow, which
  comprises all solutions of the RGE for {\it any} UV initial values for
  the Higgs self coupling and the Yukawa couplings which are admitted
  within the framework of perturbation theory. An important issue
  \cite{frogatt}, \cite{cab}, \cite{mai}, \cite{pen}-\cite{ross1}, \cite{mond},
  \cite{zoup} has recently been to determine {\it the extent to which the
    IR physics is independent of the UV physics}, i.e. independent of the
  UV initial values. This happens if the IR behaviour is dominated by
  special solutions of the RGE
\begin{itemize}
\item which correspond to fixed points, fixed lines, fixed surfaces,..., in
  general to fixed manifolds in the space of ratios of couplings,
\item which are IR attractive for the whole RG flow.
\end{itemize}
Indeed a rich structure of such IR attractive fixed points, lines,
surfaces,... exists, singling out IR attractive top and Higgs mass values
and IR attractive relations between masses. The most conspicuous IR fixed
point and line with the widest coverage in the literature
\cite{alv}-\cite{schwi}, \cite{taubuni}-\cite{ross1}, \cite{mond}, \cite{zoup} leads within the MSSM to a
top mass value of
\begin{equation}
  \mt\simeq {\rm O}(190-200\gev)\sin\beta
\label{val}
\end{equation}
and to a fixed line in the \tanb-\mt-plane, where \tanb\ is a ratio of two
vacuum expectation values characteristic for the Higgs sector of the MSSM.
The top mass value (\ref{val}) is well compatible with the experimental
value (\ref{CDF}).  Within the SM the corresponding IR fixed point leads,
though less conspicuously, to a mass value O(215\gev), not too far from the
experimental value, and the IR fixed point Higgs mass value also of
O(210\gev). Furtherreaching results are IR attractive relations between the
top and Higgs mass or, more generally, between the top, Higgs and bottom
masses within the SM and an IR attractive relation between the top mass and
\tanb. Within the SM e.g. an IR attractive top-Higgs mass relation
leads to 
\begin{equation}
\mh=O(156\gev)\ \ \ \  {\rm for\ the\ experimental\ value}\ \ \ \mt=176\gev.
\end{equation}
These mass values are clearly interesting and make the research into
IR fixed manifolds of the RG equations a very appealing subject. 

Let us also anticipate two interesting (phenomenological) hierarchies in
the IR attraction of the RG flow which emerge from the presentation of the
material in this review.: i) Typically the RG flow is roughly first
attracted towards a fixed surface, say, then within this surface along a
fixed line and finally along this line towards the fixed point. Now, it
turns out that the higher dimensional manifolds imply always highly
non-trivial relations between the involved couplings. This pattern of
course enhances the importance of higher dimensional IR fixed manifolds. A
further interesting hierarchy emerges in the comparison of the SM with the
MSSM: the IR fixed manifolds in the MSSM seem to be systematically more
strongly attractive than in the corresponding SM ones.

IR fixed manifolds may be considered to be interesting in their own right,
since they correspond to RG invariant values for couplings or to RG
invariant relations between couplings and thus between particle masses.
Their significance is certainly enhanced if the IR attraction is
sufficiently strong as to attract the RG flow into their close vicinity. Of
course the RG flow comes the closer to the IR fixed manifold the longer is
the evolution path from the UV to the IR. This explains an interest into
{\it high} UV scales in this context, even though the IR fixed manifolds
themselves are independent of the UV scale.

Next, let us place the well-studied upper (triviality) and lower (vacuum
stability) bounds for the Higgs mass and the top mass in context with IR
attractive fixed manifolds and the RGE ``top-bottom'' flow towards them.
The existence of these bounds may be traced back to the most strongly IR
attractive fixed manifold and the shape of these bounds in the
multiparameter space strongly reflects the position of this IR attractive
manifold. In fact, since the evolution path from the UV to the IR is
finite, there are IR images of UV initial values which fail to reach the
most strongly IR attractive manifold; it is their boundaries which
constitute the bounds. Clearly the bounds will be the tighter the longer is
the evolution path, i.e. the higher is the UV scale.  The bounds are thus
straight consequences of the perturbatively calculated quantum effects. In
certain approximations they are supported by non-perturbative lattice
calculations which will also be included in this review. The bounds to be
discussed are the so-called triviality bound, a top-mass dependent upper
Higgs mass bound, and the vacuum stability bound a top-mass dependent lower
Higgs mass bound or, conversely, a Higgs-mass dependent upper top mass
bound.  As has already been mentioned earlier, the bounds on the Higgs mass
in the SM, resp. on the lightest Higgs mass in the MSSM, are of high
actuality in view of the search for the Higgs at LEP200 and LHC in the near
future and the ongoing efforts to pin down the top mass at FERMILAB.

\item A second issue \cite{taubuni}-\cite{ross1} of high recent interest in the context of the
  ``top-down'' RG evolution runs under the headline {\it Yukawa
    coupling unification};
\begin{itemize}
\item One access is within supersymmetric grand unification. It starts from
  the very appealing and economical symmetry property of some grand unified
  models to provide a unification of the tau-bottom or even furtherreaching
  of the tau-bottom-top Yukawa couplings at the UV grand unification scale
  \lam=\mgut\ accompanying the unification of the gauge couplings. In the
  first instance it does not single out a special solution of the RGE in
  the sense proclaimed above.  It rather furnishes symmetry relations
  between UV initial values, thus reducing the number of free parameters in
  the ``top-down'' RG flow. As it turns out i) this constrained RG flow
  \cite{taubuni}-\cite{ross1} focuses in the IR region much more closely
  onto the IR attractive fixed point and line than the unconstrained one;
  ii) it appears to be the very existence of the IR attractive fixed point
  and line which allows the implementation of tau-bottom Yukawa
  unification.
\item Another access is the observation that there exists an IR attractive
  fixed manifold which implements approximate top-bottom Yukawa coupling
  unification {\it at all scales $\mu$}, which again is only of interest in
  the supersymmetric theory, the MSSM.
\end{itemize}  


\item The complementary approach is the so-called ``bottom-up'' evolution
  of the RGE, from the IR scale up to some UV scale \lam\ (and
  mathematically also to $\mu\rightarrow\infty$), the direction of
  evolution advocated by the interesting so-called program of reduction of
  parameters. The central issues are to establish {\it renormalization
    group invariant} relations (in principle to all orders in perturbation
  theory) between as many couplings as possible such that ii) they become
  simultaneously {\it asymptotically free}. The program as applied to the
  SM amounts to a systematic search for special solutions of the RGE which
  link the top Yukawa coupling and the Higgs self coupling in such a way to
  the strong gauge coupling that they decrease simultaneously towards zero,
  i.e. become simultaneously asymtotically free. It should be emphasized
  that this approach has been the first to concentrate on a {\it systematic
    search for special solutions} of the RGE, subject to the implementation
  of asymptotic freedom; and it certainly has influenced later developments
  in the systematic search for special solutions of the RGE, subject to
  being IR attractive.
\end{itemize}

There are interesting interrelations between the results from the different
approaches:
\begin{itemize}
\item among the special solutions of the RGE which are singled out as IR
  attractive (at the one-loop level) in the ``top-down'' approach are
  solutions which implement asymptotic freedom within the ``bottom-up''
  approach and vice versa;
\item supersymmetric grand unified theories with additional features like
  the tau-bottom Yukawa unification or parameter reduction beyond the grand
  unification scale \mgut\ drive the RG flow into the IR fixed manifolds;
\item top-bottom Yukawa coupling unification may be viewed as an UV
  symmetry input, as motivated from grand unified theories, it also appears
  to be encoded in an IR attractive fixed manifold which implies
  approximate top-bottom Yukawa unification {\it at all scales $\mu$}.
\end{itemize}
Thus, different aspects pointing towards special solutions of the RGE may
be viewed as different facets of some global regularities in the interplay
between IR and UV physics. This is a strong incentive to review all these
issues under the same headline as intended in this review.

Altogether, it is clearly a fascinating task to trace to which extent the
quantum effects encoded in the RGE for couplings yield information about
the Higgs, top and further quark and lepton masses and to which extent IR
and UV physics issues are interlocked.

Let us next further specify the scope within which these questions are
addressed.

First, a physical interpretation of the UV scale \lam\ is required. In the
framework of the SM it is the scale at which new physics beyond the SM is
encountered. Generically it is envisaged that the SM is embedded in a more
complete underlying theory at a higher momentum scale \lam; accordingly the
SM can be viewed as an effective theory with \lam\ acting as the UV cutoff.
The UV scale will presumably be smaller than the Planck mass $\lam\lwig
M_{{\rm Planck}}\simeq 10^{19}\gev$, the scale at which gravity becomes
important. In case of the SM there are physical motivations to consider the
large possible range $O(10^3\gev)\lwig\lam\lwig M_{{\rm Planck}}\simeq
O(10^{19}\gev)$. An UV scale as small as $\lam=O(10^3\gev)$ is realized
e.g. in a technicolor scenario, where the Higgs boson is a composite
particle. Intermediate scales \lam\ are possible, e.g. accounting for
compositeness of leptons and quarks or for embedding the SM into a
left-right symmetric gauge theory. A scale as large as $\lam=\mgut\simeq
O(10^{15}\gev)$ is appropriate in a grand unification scenario.  Though
there are strong recent doubts whether this unification can work out on the
quantitative level, we shall continue to include a large UV scale
O($10^{15}\gev$) into the discussion.

Here is where the Minimal Supersymmetric Extension of the Standard Model
(MSSM), which implements in a minimal way the appealing boson-fermion
supersymmetry into the SM, has its merits. First of all its improved
renormalizability properties allow naturally two vastly different scales in
the theory, the weak interaction scale (\ref{v}) and a grand unification
scale \mgut. Furthermore, the MSSM has recently experienced a strong
revival, since unification of the gauge couplings at a unification scale
$\mgut\simeq O(2\cdot 10^{16}\gev)$ appears to work out very well
quantitatively in supersymmetric grand unification. The MSSM is reasonably
considered {\it only} in the grand unification framework with a high UV
scale $\lam=\mgut$.


The material in this review is organized as follows. Sect. \ref{thfr}
summarizes the theoretical basis for the perturbative RGE evolution within
the SM as well as in the MSSM. Some minimal background for grand
unification with emphasis on unification of Yukawa couplings is provided. A
collection of all radiative corrections relating the running masses in the
$\overline{\rm MS}$ scheme to the physical pole masses is also included.
Sect.  \ref{preview} serves as a preview and logical guideline for the
material developped in detail in Sects. \ref{noelweak} and \ref{allgauge} :
it contains a summary of all IR attractive fixed manifolds in form of a
table, in which the coupling parameter space is enlarged entry by entry. In
Sects.  \ref{noelweak} and \ref{allgauge} much effort is spent in
developping a comprehensive insight into the highly non-trivial IR
attractive fixed points, fixed lines, fixed surfaces,... in a space of
ratios of couplings which is enlarged from a one-parameter space (for the
Higgs selfcoupling) step by step to a five parameter space. This allows to
develop the material pedagogically, to include the large body of pioneering
publications which have considered reduced parameter sets analytically and
to finally culminate with the latest developments in the literature. This
procedure also allows a comparison with non-perturbative results from
lattice calculations relevant for the pure Higgs and the Higgs-fermion
sector of the SM.  Sect.  \ref{noelweak} provides the detailed derivation
in absence of the electroweak gauge couplings, Sect. \ref{allgauge} treats
the non-trivial inclusion of the electroweak couplings. In both sections
the SM and the MSSM are treated strictly in parallel; they include also a
(largely) analytical assessment of the respective strengths of IR
attraction of the IR fixed manifolds. Sect.  \ref{masses} then summarizes
the resulting IR attractive fixed point masses for the top and Higgs, the
IR attractive top-Higgs, top-bottom and Higgs-top-bottom mass relations,
the IR attractive relation between the top mass and \tanb\ in the MSSM on
the level of the present state of the art. The dynamical origin for the
triviality (upper) bounds and vacuum stability (lower) bounds in the
Higgs-top mass plane of the SM is developed step by step in Sects.
\ref{higgs}-\ref{Htgd}; Sect. \ref{higgs} also contains an estimate of an
absolute upper bound on the SM Higgs mass from various sources (including
lattice calculations). The most
recent determinations of the SM bounds as well as an upper bound for the
lightest Higgs boson mass in the MSSM are presented in Sects.  \ref{lowsm}
and \ref{upmssm}. Sect. \ref{yukuni} is devoted to the interrelated issues
of implementing tau-bottom(-top) Yukawa unification into supersymmetric
unification and the IR attractive top fixed point mass which has received
so much attention in the literature. Finally, Sect.  \ref{parred}
summarizes the program of reduction of parameters in its application to the
SM and its interrelation with a search for IR attractive manifolds.

\section{Theoretical Framework\label{thfr}}

In order to render the review selfcontained on the one hand and to avoid
repetition of too much text book material on the other hand, we shall
introduce in detail only those elements pertinent to the physics issues
addressed in this review.

\subsection{Standard Model\label{sm}}


The SM of elementary particle theory comprises the Glashow-Weinberg-Salam
model of electroweak interactions \cite{gla} and quantum chromodynamics,
the theory of strong interactions \cite{fri}. These fundamental
interactions among elementary particles derive from a local gauge principle
with gauge group
\begin{equation}
  SU(3)\times SU(2)\times U(1)
\end{equation}
which is broken spontaneously to $SU(3)\times U(1)_{em}$ by means of the
Higgs mechanism .

The field content of the theory is given in terms of the gauge fields,
which mediate the gauge interactions, the fermionic quark and lepton matter
fields and the Higgs field responsible for the spontaneous symmetry
breakdown.  For the purpose of this review we confine the discussion to the
third, heaviest generation of quarks and leptons consisting of the
left-handed $SU(2)$ top-bottom and tau-neutrino-tau doublets
\begin{eqnarray}
  q_{{\rm L}}&=&\big(\begin{array}{l}t_{{\rm L}}\\b_{{\rm L}}\\ 
\end{array}\big), \nonumber\\ l_{{\rm L}}&=&\big(\begin{array}{l}\nu_{
    \tau_{\rm L}}\\\tau_{{\rm L}}\\ 
\end{array}\big)
\end{eqnarray} 
and the corresponding right-handed $SU(2)$ singlets $t_{{\rm R}},\,b_{{\rm
    R}},\,\tau_{{\rm R}}$. The complex $SU(2)$ doublet Higgs field
$\Phi(x)$ with $U(1)$ hypercharge $Y=1$ is
\begin{equation}
  \Phi=\bigg(\begin{array}{l}\phi^{+}\\\phi^0\\ 
\end{array}\bigg),
\label{phi}
\end{equation}
where the suffixes +,0 characterize the electric charge +1, 0 of the
components.

The most general gauge invariant and renormalizable interaction Lagrangian
is
\begin{equation}
  {\cal L}={\cal L}_{{\rm gauge}}+{\cal L}_{{\rm Yukawa}}-V(\Phi).
\end{equation}
${\cal L}_{{\rm gauge}}$ contains the gauge interactions in terms of the
respective $SU(3)\times SU(2)\times U(1)$ gauge couplings
$g_3,\,g_2,\,g_1$, with $g_1$ normalized as motivated by grand unification,
$g_1=\sqrt{5/3}g'$. The potential
\begin{equation}
  V(\Phi)=-m^2\Phi^{\dag}\Phi+\lambda(\Phi^{\dag}\Phi)^2
\label{pot}
\end{equation}
contains the Higgs field self interaction in terms of the a
priori unknown Higgs self coupling $\lambda$. It is parametrized such that
the Higgs field acquires a vacuum expectation value responsible for the
spontaneous electroweak symmetry breakdown
\begin{equation}
  <\Phi>=\frac{1}{\sqrt{2}}\bigg(\begin{array}{l}0\\v\\ 
\end{array}\bigg)
\end{equation}
with
\begin{equation}
  v=\frac{m}{\sqrt{\lambda}}.
\end{equation}
The numerical value of $v$ is given in terms of the Fermi constant
\begin{equation}
  G_F=1.16639(2)\,10^{-5}\,\gev^{-2}
\label{gf}
\end{equation}
to be
\begin{equation}
  v=(\sqrt{2}G_F)^{-1/2}=246.218(2)\,\gev.
\label{v}
\end{equation}

Of the four Higgs degrees of freedom three are Goldstone degrees of
freedom, furnishing the longitudinal degrees of freedom for the massive
weak gauge bosons, thus providing the W boson mass $m_W=vg_2/2$. The
remaining one corresponds to the physical Higgs boson field
\begin{equation}
  h=\sqrt{2}\big(Re\phi^0-v/\sqrt{2}).
\end{equation} 

${\cal L}_{{\rm Yukawa}}$ describes the interactions of the doublet Higgs
field with the fermion matter fields
\begin{equation}
  {\cal L}_{{\rm Yukawa}}=-\gt\overline{q}_{{\rm L}}\Phi^{c}
  t_{R}-\gb\overline{q}_{{\rm L}}\Phi b_{R}-\gta\overline{l}_{{\rm
      L}}\Phi\tau_{R} +{\rm h.c.}
\end{equation}
$\Phi^{c}=i\tau_2\Phi^{*}$ is the charge conjugate of $\Phi$, $\tau_2$ the
second Pauli matrix, $\gt,\,\gb,\,\gta$ are the a priori unknown top,
bottom and tau Yukawa couplings.

All masses in the SM are induced by the spontaneous symmetry breakdown and
are proportional to $v$. The weak gauge boson masses allow to determine the
size of $v$ which was already introduced in Eq. (\ref{v}). The tree level
top, bottom, tau and Higgs masses are given in terms of the vacuum
expectation value $v$ and their respective couplings
\begin{equation}
  \mt=\gt\frac{v}{\sqrt{2}},\;\;\;\mb=\gb\frac{v}{\sqrt{2}},\;\;\;
  \mta=\gta\frac{v}{\sqrt{2}}
\label{mf}
\end{equation}
and
\begin{equation}
  \mh=\sqrt{2\lambda}v.
\label{mh}
\end{equation}
Since \gt, \gb, \gta\ and \la\ are free parameters,
the tree level masses \mt, \mb, \mta\ and \mh\ are a priori undetermined.



\subsection{Minimal Supersymmetric Standard Model\label{mssm}}

A strong reason to implement supersymmetry into the SM is the improvement
in renormalizability properties, which allows to retain the Higgs boson as
elementary particle up to a high UV scale \lam\ without running into the
(interrelated) problems of naturalness, fine tuning and hierarchy.

Within the SM higher order corrections to the Higgs mass are quadratically
divergent, i.e. the ``natural'' size of the Higgs mass is the high UV
cut-off \lam. Renormalization brings down this mass to $O(v)$ only by means
of an unnatural finetuning of parameters order by order in perturbation
theory. Thus, the theory does not supply any dynamical mechanism which
allows naturally the coexistence of two vastly different scales, the weak
interaction scale and a very high UV scale, as is e.g. necessary in grand
unified theories. This is the hierarchy problem. A dynamical mechanism
could be supplied by an appropriate additional symmetry. This is indeed the
case for supersymmetry.

In a supersymmetric theory (see Ref. \cite{wess} for an excellent textbook) particles are classified in
supermultiplets containing bosons and fermions. In a supersymmetric
extension of the SM the quadratic divergence is naturally cancelled by the
related loop diagrams involving the fermionic supersymmetric partners of the SM
particles contributing to the divergent loops.  Supersymmetry has, however, to be broken in order
to account for the fact that so far no supersymmetric partners for the SM
particles have been found experimentally. A soft supersymmetry breaking at
a scale \msusy\ close to the weak interaction scale can be arranged; this
way the naturalness and hierarchy problems remain resolved. The masses of the
supersymmetric partners will be of the order of this scale \msusy. We shall
elaborate more on this scale at the end of this subsection.

In order to understand the implications of the Minimal Supersymmetric
extension of the Standard Model (MSSM) for the renormalization group
equations relevant for the SM particle masses only a few important
ingredients have to be introduced. For excellent reviews see e.g. Refs.
\cite{nilles}, \cite{hab}.

Following the text book \cite{hab} the interactions of Higgs bosons and
third generation fermions is obtained from the supersymmetric
superpotential given by
\begin{equation}
  W=\epsilon_{ij}(\htt\hat{Q}^{i}\hat{H}_2^{j}\hat{T}+\hb\hat{Q}^{j}
  \hat{H}_1^{i}\hat{B}+\hta
  \hat{L}^{j}\hat{H}_1^{i}\hat{\tau})+\mu\epsilon_{ij}\hat{H}_1^{i}
  \hat{H}_2^{j}
\label{supo}
\end{equation}
in terms of the unknown Yukawa couplings \htt, \hb, \hta\ and the parameter
$\mu$; $\epsilon_{ij}$ is the antisymmetric tensor in two dimensions.

Here $\hat{H}_1$ and $\hat{H}_2$ are the two Higgs $SU(2)$ doublet
superfields, containing besides their scalar components $H_{1,2}$ the
respective chiral supersymmetric partners. $\hat{Q}$, $\hat{L}$ are the
$SU(2)$ weak doublet top-bottom and tau-neutrino-tau superfields,
respectively, and $\hat{T},\,\hat{B},\,\hat{\tau}$ are the $SU(2)$ singlet
top, bottom, tau superfields, respectively. They contain besides the SM quark
and lepton fields their respective supersymmetric partners, the scalar
squark and slepton fields. The $SU(2)$ indices are contracted in a gauge
invariant way. The two Higgs doublet superfields are necessary in order to
i) provide masses for the up type top quark as well as for the down type
bottom quark (since the appearance of $\hat{H}_1^{*}$ and $\hat{H}_2^{*}$
is forbidden in Eq. (\ref{supo}) on account of their fermionic components)
and ii) provide mutual cancellation of the anomalies introduced by the
fermionic components of $\hat{H}_1$ and $\hat{H}_2$.

The scalar field potential at tree level for the Higgs sector, arising in
this supersymmetric and gauge invariant theory, is given in terms of the
scalar field components $H_{1,2}$ of the superfields $\hat{H}_{1,2}$, as
follows
\begin{equation}
  V(H_1,H_2)=\frac{1}{8}(\frac{3}{5}\ges+\gzs)\big(H_1^{i*}H_1^{i}-
  H_2^{i*}H_2^{i}\big)^2+\frac{1}{2}\gzs|H_1^{i*}H_2^{i}|^2+\mu^2
  (H_1^{i*}H_1^{i}+H_2^{i*}H_2^{i})
\end{equation}
with the Higgs field notation in analogy to Eq. (\ref{phi})
\begin{eqnarray}
  H_1\,=\,\bigg(\begin{array}{l}H_1^1\\H_1^2\\\end{array}\bigg) & =
  &\bigg(\begin{array}{l}\phi_1^{0\,*}\\-\phi_1^{-}\\\end{array}\bigg)
  =\,\Phi_1^{c}\nonumber\\ 
  H_2\,=\,\bigg(\begin{array}{l}H_2^1\\H_2^2\\\end{array}\bigg) & =
  &\bigg(\begin{array}{l}\phi_2^{+}\\\phi_2^{0}\\\end{array}\bigg)=\Phi_2.
\end{eqnarray}
The dimension four terms involve the electroweak gauge couplings $g_1$ and
$g_2$ exclusively, a characteristic feature of the supersymmetric theory.
The necessity for quark masses requires both Higgs fields to have
nonvanishing vacuum expectation values $v_1,\,v_2$ which may be different
\begin{equation}
  <H_1>=\frac{1}{\sqrt{2}}\bigg(\begin{array}{l}v_1\\0\\ 
\end{array}\bigg),\hspace{2cm}<H_2>=\frac{1}{\sqrt{2}}
\bigg(\begin{array}{l}0\\v_2\\\end{array}\bigg)
\end{equation}
with $v_1$, $v_2$ positive and
\begin{equation}
  v_1^2+v_2^2=v^2
\end{equation}
with $v$ given in Eq. (\ref{v}). This leads to the sensible introduction of
an angle $\beta$ as an additional key parameter in the MSSM defined by
\begin{equation}
  \tan\beta=v_2/v_1
\end{equation}
with $0\leq\beta\leq\pi/2$.  In terms of $\beta$ the tree level fermion
masses become naively (disregarding for the moment the effect of the soft
SUSY breaking to be discussed below)
\begin{equation}
  \mt=\frac{v}{\sqrt{2}}\htt\sin\beta,\;\;\;
  \mb=\frac{v}{\sqrt{2}}\hb\cos\beta,\;\;\;
  \mta=\frac{v}{\sqrt{2}}\hta\cos\beta.
\label{susymasses}
\end{equation}

Of the eight Higgs degrees of freedom, three are Goldstone degrees of
freedom, serving to give mass to the weak gauge bosons as in the SM, the
remaining five correspond to physical Higgs particles. The lightest one of
them is the SUSY analogon of the physical SM Higgs boson, the heavier ones
comprise two charged and a CP-even and a CP-odd neutral Higgs boson.

A soft supersymmetry breaking has to be introduced since the supersymmetric
partners of the SM particles must have masses beyond their experimental
limits. An appropriate soft supersymmetry breaking is achieved by
introducing additional dimension two terms into the Higgs potential (higher
dimensional terms would destroy the naturalness achieved by supersymmetry),
soft mass terms for the gauginos, the fermionic superpartners of the gauge
bosons, soft mass terms for the squarks and sleptons, the scalar
superpartners of the quarks and leptons, and trilinear
Higgs-squark-antisquark and Higgs slepton-antislepton couplings. None of
these parameters spoil the cancellation of quadratic divergencies. The new
free parameters are monitored by the two conditions, that the
supersymmetric partners and the heavy Higgs bosons have masses large enough
not to be in conflict with lower experimental bounds and small enough
(smaller than 1-10 TeV), in order to keep the theory natural. The resulting
spectrum will be spread over a whole range of masses, see e.g. Refs.
\cite{bert},\cite{ross}.

In this review we are concerned with the effect of supersymmetry on
the RGE responsible for the top, bottom, tau masses and the (lightest)
Higgs mass.  Generically, at energies high with respect to all masses,
the MSSM RGE (in the $\overline{{\rm MS}}$ renormalization scheme) are
valid, at energies well below the masses of the superpartners and the
heavy Higgs bosons the SM RGE hold.  The intermediate region is
characterized by a change in the RGE each time a superparticle or
heavy Higgs boson mass threshold is passed. It has been argued
\cite{msusy}-\cite{lang2} however, that effectively this transition
region can be approximately lumped into one scale, the supersymmetry
scale $\msusy$, absorbing the effect of the soft supersymmetry
breaking parameters.  This idealization of the complex situation has
been widely used in the literature and will be also adhered to in the
following. In crude approximation, the size of \msusy\ is expected to
be of $O(v)-O(1\,{\rm TeV})$, in practical applications it is treated
as a free parameter, allowed to vary typically from $\mt$ to several
TeV, at most 10 TeV. In Ref. \cite{lang1} it has been pointed out that
even values for \msusy\ below $m_{Z}$ may be appropriate.

The transition of MSSM running couplings to SM running couplings at
$\mu=\msusy$ is approximated as usual by the continuous (but not
differentiable) matching conditions
\begin{eqnarray}
  g_i(\msusy^{-}) & = & g_i(\msusy^{+})\hspace{2cm}{\rm for}\;i = 1,2,3,\\ 
  \gt(\msusy^{-}) & = & \htt(\msusy^{+})\sin\beta,\\ \gb(\msusy^{-}) & = &
  \hb(\msusy^{+})\cos\beta,\\ \gta(\msusy^{-}) & = &
  \hta(\msusy^{+})\cos\beta.
\label{match}
\end{eqnarray} 

In the frequently considered not unlikely case that the heavier Higgs
bosons are sufficiently much heavier than the lightest one, integrating out
the heavy Higgs field combinations at the scale $\mu=\msusy$ leaves the
combination
\begin{equation}
  h_0=\sqrt{2}\big((Re\phi_1^0-v_1/\sqrt{2})\cos\beta+
  (Re\phi_2^0-v_2/\sqrt{2})\sin\beta\big),
\end{equation}
the light Higgs boson field $h_0$ to be identified with the SM Higgs field
h below \msusy. There is, however, a crucial difference to the SM Higgs.
While the SM Higgs selfcoupling \la\ is undetermined at the tree level the
MSSM Higgs selfcoupling \la\ is subject to the tree level condition at
$\mu=\msusy$
\begin{equation}
  \la(\msusy^{-})=
  \frac{1}{8}\big(\frac{3}{5}\ges(\msusy)+\gzs(\msusy))\cos^2 2\beta,
\label{lowmh}
\end{equation}
which leads to a tree level Higgs mass $\mh^2\leq m_Z^2\cos^2\beta\leq
m_Z^2$. The Higgs mass is lifted by radiative corrections as a function of
the top mass and the size of the scale \msusy, which will be
summarized in Sect. 
\ref{vac}. The relation (\ref{lowmh}) is the origin for
the rather low upper mass bound for the lightest Higgs particle in the
MSSM. Since the Higgs selfcoupling is fixed at \msusy\ in terms of the
electroweak gauge couplings to be rather small, it has only the SM RG
evolution from $\mu=\msusy$ down to $\mu=\mh$ available to increase its
value and correspondingly the value of the Higgs mass. In contradistinction
to the SM, where the upper Higgs mass bound depends on the UV scale \lam,
the upper Higgs mass bound in the MSSM depends on \msusy. How this works
out in detail in professional analyses will be reviewed in Sect.
\ref{upmssm}.

This idealized MSSM framework, appropriate to describe the RG evolution of
the gauge couplings and the couplings relevant for the Higgs, top, bottom
and tau masses, may be viewed -- for practical applications -- to involve
the free parameters \htt, \hb, \hta, \la\ (the latter one below \msusy),
the new SUSY key parameter $\tan\beta$ and the effective parameter \msusy\ 
(varying between bounds).

\subsection{Grand Unification\label{gut}}


Grand unification is a magnificent theoretical framework in itself.  From
the point of view of this review, it is an appealing scheme which allows to
single out solutions of the RGE by providing symmetry relations between UV
initial values for couplings or in short, which constrains the ``top-down''
RG flow considerably.

In a grand unified theory \cite{gut} the SM is embedded into an underlying
gauge theory with a gauge group containing the SM gauge group $SU(3)\times
SU(2)\times U(1)$. The minimal grand unifying gauge group is $SU(5)$;
further groups of interest are e.g. $SO(10)$ and $E_6$. The different
scenarios are defined mathematically by the grand unifying gauge group, the
classification of the quark and lepton fields with respect to (irreducible)
representations of the unifying group and the specific Higgs sector of the
theory, responsible for the spontaneous breakdown of the grand unifying
gauge symmetry to the SM gauge symmetry at the grand unification scale
\mgut. The grand unification framework, endorsed \cite{sugut} with global
supersymmetry, is the natural starting point for supergravity and
superstring theories.  Excellent textbooks on grand unification and
supersymmetric grand unification are e.g. Refs.  \cite{revgut}.

For the purpose of this review, it suffices to discuss a kind of minimal
framework popular in the literature. The grand unification gauge symmetry
is assumed to establish at the \mgut\ scale a symmetry relation between the
three gauge couplings of the SM as provided in a minimal $SU(5)$ theory
\begin{equation}
  \gee\,(\mu=\mgut)=\gz\,(\mu=\mgut)=\gd\,(\mu=\mgut).
\label{gegzgd}
\end{equation}
At $\mu=\mgut$ the spontaneous breakdown of the grand unifying gauge group
to the SM gauge group becomes effective; thus, for values of the scale
$\mu\lwig\mgut$ below \mgut\ the running gauge couplings
$\gee(\mu),\;\gz(\mu),\;\gd(\mu)$ are subject to the ``top-down'' (two-loop)
RG evolution of the SM; in case of minimal supersymmetric grand unification
it is subject to the RG evolution of the MSSM down to $\mu=\msusy$, and
below again to the RG evolution of the SM; the correponding RGE in their
two-loop form will be given explicitely in the next Sect. \ref{ren}. 

This scheme is successful essentially if i) the initial value condition
(\ref{gegzgd}) combined with the high precision data for $\alpha$ and
$\sin\theta_W$ at $\mu=m_Z$ leads to a (two-loop) value for $\gd(\mu=m_Z)$
compatible with data and ii) if \mgut\ turns out to be sufficiently large,
in order not to run into conflict with the experimental limits on proton
decay which is mediated by the exchange of heavy gauge bosons of the grand
unifying gauge theory. The MSSM has the advantage over the SM of an
additional parameter, the effective scale \msusy, which, however, for
consistency reasons is strongly constrained as has been detailed in Sect.
\ref{mssm}. (Threshold corrections and non-renormalizable operator
corrections at the high scale as well as at the low scale are usually
neglected; see Refs. \cite{msusy}-\cite{lang2} for an estimate of these
effects).

Applying these criteria, recent reevaluations of gauge coupling unification
\cite{ssgut},\cite{bar},\cite{lang1},\cite{langa},\cite{lang2} have singled out supersymmetric grand
unification as successful with a grand unification scale $\mgut\simeq$
2-3$\,10^{16}\gev$, and the strong gauge coupling \cite{lang2}
$\gds\,(\mu=m_Z)/(4\pi)\simeq 0.129\pm 0.010$ being a bit on the high
side, but within the errors compatible with the experimental value. The
grand $SU(5)$ unification without supersymmetry is strongly disfavoured on
the quantitative level.

This revival of interest into supersymmetric grand unification has also
renewed the interest into what is called in the literature Yukawa coupling
unification
\cite{taubuni}-\cite{ross1}.
It is a well-known feature of the minimal $SU(5)$ grand unified theory that
it implies the symmetry property of tau-bottom Yukawa coupling unification
\cite{taubuni}-\cite{dimop} at $\mu=\mgut$
\begin{equation}
  \hta\,(\mu=\mgut)=\hb\,(\mu=\mgut).
\label{bta}
\end{equation}
Yukawa couplings involve fermion (quark and lepton) as well as Higgs
fields; correspondingly Yukawa coupling unification is a symmetry property
not only dependent on the classification of the fermions but also of the
Higgs fields with respect to the gauge group. Thus, tau-bottom Yukawa
unification holds more generally in grand unified theories such as $SU(5)$,
$SO(10)$ and $E_6$ theories for Yukawa couplings which involve Higgs fields
in the fundamental {\bf 5},{\bf 10} and {\bf 27} representations,
repectively.  Even more appealing and economic is the option of
tau-bottom-top Yukawa unification at the scale $\mu=\mgut$,
\begin{equation}
  \hta\,(\mu=\mgut)=\hb\,(\mu=\mgut)=\htt\,(\mu=\mgut),
\label{tbta}
\end{equation}
a symmetry property provided e.g. in some $SO(10)$ models involving a
single complex Higgs 10-plet.

Implementing the UV symmetry properties of tau-bottom or even
tau-bottom-top Yukawa unification into minimal supersymmetric grand
unification strongly constrains the IR parameters such as the top mass and
the MSSM parameter \tanb. These fascinating
investigations will be reviewed in Sect. \ref{yukuni}.

\subsection{Renormalization Group Equations}{\label{ren}}

Quantization and renormalization of an interacting field theory introduces
a scale $\mu$ with the dimension of a momentum (dimensional
transmutation); it is a hidden parameter which has to be introduced in
order to define the parameters of the theory. The important property of RG
invariance ensures, however, that within any given order of perturbation
theory {\it measurable quantities are $\mu$ independent}, i.e. that their
explicit $\mu$ dependence is 
cancelled by implicit $\mu$ dependences, introduced through the $\mu$
dependences of the renormalized parameters of the theory, i.e.  couplings
and masses, and of renormalized
wave functions.

This implies in particular that the couplings vary with the momentum
transfer with which they are probed; vacuum polarization effects, for
example, screen the electric charge, resulting in an effective electric
coupling which grows with $\mu$. Quite generally, the response of the set
of renormalized couplings of the SM to a differential scale change from
$\mu$ to $\mu+{\rm d}\,\mu$ reflects according to the uncertainty principle
the differentially increased resolution: it allows to ``see'' higher order
radiative corrections resulting from all those virtual particle emissions
and reabsorptions due to allowed interactions in the considered order of
perturbation theory.  This response is summarized in the RGE, a system of
nonlinear coupled differential equations.

The renormalization group equations up to two loops in perturbation theory
were calculated in the mass independent $\overline{{\rm MS}}$
renormalization scheme for the couplings \ges, \gzs, \gds, \gts, \gbs,
\gtas\ and \la\ of the SM in Refs. \cite{rgsm} and for the couplings \ges,
\gzs, \gds, \hts, \hbs, \htas\ of the MSSM in Refs. \cite{rgmssm}; a
compact summary may be found in Ref. \cite{bar}. The Yukawa couplings of
the first and second generation quarks and leptons are so small that they
lead to negligible contributions to all quantities relevant in this review.
They are assumed to vanish identically in the following.  This assumption
also precludes generation mixing. Then the two-loop renormalization group
equations, valid well above $\mu=\mt,\ \mh$, take the following form in
terms of the common independent variable
\begin{equation}
  t=ln\frac{\mu}{\lam}.
\end{equation}
\newpage

{\bf In the SM:}

\begin{eqnarray}
  \frac{{\rm d}\,\ges}{{\rm d}\, t}&=&\frac{g_1^4}{8\pi^2}\,
  \bigg(\;\;\;\,\frac{41}{10}\;\;+\frac{1}{16\pi^2}\Big(\frac{199}{50}\ges
  +\frac{27}{10}\gzs+\frac{44}{5}\gds-\frac{17}{10}\gts-\frac{1}{2}
  \gbs-\frac{3}{2}\gtas\Big)\bigg)
\label{RG1}\\ \nonumber\\
\frac{{\rm d}\,\gzs}{{\rm d}\,t}&=&\frac{g_2^4}{8\pi^2}\,\bigg(
-\frac{19}{6}\;\;+\frac{1}{16\pi^2}\Big(\frac{9}{10}\ges
+\frac{35}{6}\gzs+12\gds-\frac{3}{2}\gts-\frac{3}{2}\gbs
-\frac{1}{2}\gtas\Big)\bigg)
\label{RG2}\\ \nonumber\\
\frac{{\rm d}\,\gds}{{\rm d}\,t} & = & \frac{g_3^4}{8\pi^2}\,
\bigg(-\,\,7\,\;\;+\;\frac{1}{16\pi^2}\Big(\frac{11}{10}\ges
+\frac{9}{2}\gzs-26\gds-2\gts-2\gbs\Big)\bigg)
\label{RG3}\\ \nonumber\\
\frac{{\rm d}\,\gts}{{\rm d}\,t}&=&\frac{\gts}{8\pi^2}\,\bigg(
-\frac{17}{20}\ges-\frac{9}{4}\gzs-8\gds+\frac{9}{2}\gts
+\frac{3}{2}\gbs+\gtas\nonumber\\ \nonumber\\ & &
+\frac{1}{16\pi^2}\Big(\frac{1187}{600}g_1^4-\frac{9}{20}\ges\gzs
-\frac{23}{4}g_2^4+\frac{19}{15}\ges\gds+9\gzs\gds-108 g_3^4
+\frac{393}{80}\ges\gts\nonumber\\ \nonumber\\ & &
+\frac{7}{80}\ges\gbs+\frac{15}{8}\ges\gtas+\frac{225}{16}
\gzs\gts+\frac{99}{16}\gzs\gbs+\frac{15}{8}\gzs\gtas
+36\gds\gts+4\gds\gbs\nonumber\\ \nonumber\\ & & -12g_t^4
-\frac{11}{4}\gts\gbs-\frac{1}{4}g_b^4-\frac{9}{4}\gts\gtas+
\frac{5}{4}\gbs\gtas-\frac{9}{4}g_{\tau}^4-12\gts\la-4\gbs\la
+6\la^2\Big)\bigg)
\label{RGt}\\ \nonumber\\
\frac{{\rm d}\,\gbs}{{\rm d}\,t}&=&\frac{\gbs}{8\pi^2}\,\bigg(
-\frac{1}{4}\ges-\frac{9}{4}\gzs-8\gds+\frac{3}{2}\gts+\frac{9}{2}
\gbs+\gtas\nonumber\\ \nonumber\\ & & +\frac{1}{16\pi^2}\Big(-
\frac{127}{600}g_1^4-\frac{27}{20}\ges\gzs-\frac{23}{4}g_2^4
+\frac{31}{15}\ges\gds+9\gzs\gds-108 g_3^4+\frac{91}{80}\ges\gts
\nonumber\\ \nonumber\\ & & +\frac{237}{80}\ges\gbs+\frac{15}{8}
\ges\gtas+\frac{99}{16}\gzs\gts+\frac{225}{16}\gzs\gbs+\frac{15}{8}
\gzs\gtas+4\gds\gts+36\gds\gbs\nonumber\\ \nonumber
  \\ & & -\frac{1}{4}g_t^4-\frac{11}{4}\gts\gbs-12 g_b^4
  +\frac{5}{4}\gts\gtas-\frac{9}{4}\gbs\gtas-\frac{9}{4}g_{\tau}^4
  -4\gts\la-12\gbs\la+6\la^2\Big)\bigg)
\label{RGb}\\ \nonumber\\
\frac{{\rm d}\,\gtas}{{\rm d}\,t}&=&\frac{\gtas}{8\pi^2}\,\bigg(
-\frac{9}{4}\ges-\frac{9}{4}\gzs+3\gts+3\gbs+\frac{5}{2}\gtas \nonumber\\ 
\nonumber\\ & & +\frac{1}{16\pi^2}\Big(\frac{1371}{200}
g_1^4+\frac{27}{20}\ges\gzs-\frac{23}{4}g_2^4+\frac{17}{8}\ges\gts
+\frac{5}{8}\ges\gbs+\frac{537}{80}\ges\gtas+\frac{45}{8}\gzs\gts
\nonumber\\ \nonumber\\ & & +\frac{45}{8}\gzs\gbs+\frac{165}{16}
\gzs\gtas+20\gds\gts+20\gds\gbs\nonumber\\ \nonumber\\ & &
-\frac{27}{4}g_t^4+\frac{3}{2}\gts\gbs-\frac{27}{4}g_b^4-\frac{27}
{4}\gts\gtas-\frac{27}{4}\gbs\gtas-3 g_{\tau}^4-12\gtas\la+6\la^2
\Big)\bigg)
\label{RGta}
\end{eqnarray}
\newpage
\begin{eqnarray}
\frac{{\rm d}\,\la}{{\rm d}\,t}&=&\frac{1}{16\pi^2}\,\bigg(
\frac{27}{200}g_1^4+\frac{9}{20}\ges\gzs+\frac{9}{8}g_2^4-\frac{9}{5}
\ges\la-9\gzs\la-6\gt^4-6\gb^4-2\gta^4\nonumber\\ \nonumber\\ & &
\hspace{1.4cm}+12\gts\la+12\gbs\la+4\gtas\la+24\la^2\nonumber\\ \nonumber\\ 
& &+\frac{1}{16\pi^2}\Big(-\frac{3411}{2000}g_1^6-\frac
{1677}{400}g_1^4\gzs-\frac{289}{80}\ges g_2^4+\frac{305}{16}g_2^6
-\frac{171}{100}g_1^4\gts+\frac{9}{20}g_1^4\gbs\nonumber\\ \nonumber
  \\& &-\frac{9}{4}g_1^4\gtas+\frac{63}{10}\ges\gzs\gts+\frac{27}{10}
  \ges\gzs\gbs+\frac{33}{10}\ges\gzs\gtas-\frac{9}{4}g_2^4\gts
  -\frac{9}{4}g_2^4\gbs-\frac{3}{4}g_2^4\gtas\nonumber\\ \nonumber\\&
  &+\frac{1887}{200}g_1^4\la+\frac{117}{20}\ges\gzs\la
  -\frac{73}{8}g_2^4\la-\frac{8}{5}\ges\gt^4+\frac{4}{5}\ges\gb^4
  -\frac{12}{5}\ges\gta^4-32\gds\gt^4\nonumber\\ \nonumber\\& &
  -32\gds\gb^4+\frac{17}{2}\ges\gts\la+\frac{5}{2}\ges\gbs\la
  +\frac{15}{2}\ges\gtas\la+\frac{45}{2}\gzs\gts\la+\frac{45}{2}
  \gzs\gbs\la+\frac{15}{2}\gzs\gtas\la\nonumber\\ \nonumber\\& &
  +80\gds\gts\la+80\gds\gbs\la+\frac{108}{5}\ges\la^2+108\gzs\la^2
  \nonumber\\ \nonumber\\& &+30\gt^6-6\gt^4\gbs-6\gts\gb^4+30\gb^6
  +10\gta^6-3\gt^4\la+6\gts\gbs\la-3\gb^4\la\nonumber\\ \nonumber\\ &
  &-\gta^4\la-144\gts\la^2-144\gbs\la^2-48\gtas\la^2-312\la^3 \Big)\bigg)
\label{RGla}
\end{eqnarray}\\

{\bf In the MSSM:}

\begin{eqnarray}
  \frac{{\rm d}\,\ges}{{\rm d}\, t}&=&\frac{g_1^4}{8\pi^2}\,
  \bigg(\,\;\frac{33}{5}\;\;+\frac{1}{16\pi^2}\Big(\frac{199}{25}
  \ges+\frac{27}{5}\gzs+\frac{88}{5}\gds-\frac{26}{5}\hts-\frac{14}{5}
  \hbs-\frac{18}{5}\htas\Big)\bigg)
\label{RG1s}\\ \nonumber\\ 
\frac{{\rm d}\,\gzs}{{\rm d}\,t}&=&\frac{g_2^4}{8\pi^2}\,\bigg(
\;\;\;1\,\;\;+\frac{1}{16\pi^2}\Big(\frac{9}{5}\ges+25\gzs+24\gds-6
\hts-6\hbs-2\htas\Big)\bigg)
\label{RG2s}\\ \nonumber\\ 
\frac{{\rm d}\,\gds}{{\rm d}\,t} & = & \frac{g_3^4}{8\pi^2}\,
\bigg(-3\;+\;\frac{1}{16\pi^2}\Big(\frac{11}{5}\ges+9\gzs+14\gds
-4\hts-4\hbs\Big)\bigg)
\label{RG3s}\\ \nonumber\\
\frac{{\rm d}\,\hts}{{\rm d}\,t}&=&\frac{\hts}{8\pi^2}\,\bigg(
-\frac{13}{15}\ges-3\gzs-\frac{16}{3}\gds+6\hts+\hbs\nonumber\\ \nonumber\\ 
& & +\frac{1}{16\pi^2}\Big(\frac{2743}{450}g_1^4
+\ges\gzs+\frac{15}{2}g_2^4+\frac{136}{45}\ges\gds+8\gzs\gds
-\frac{16}{9}g_3^4+\frac{6}{5}\ges\hts\nonumber\\ \nonumber\\ & &
+\frac{2}{5}\ges\hbs+6\gzs\hts+16\gds\hts-22 h_{t}^4-5 \hts\hbs-5
h_{b}^4-\hbs\htas\Big)\bigg)
\label{RGts}
\end{eqnarray}
\newpage 
\begin{eqnarray}
\frac{{\rm d}\,\hbs}{{\rm d}\,t}&=&\frac{\hbs}{8\pi^2}\,\bigg(
-\frac{7}{15}\ges-3\gzs-\frac{16}{3}\gds+\hts+6 \hbs+\htas\nonumber\\ 
\nonumber\\ & & +\frac{1}{16\pi^2}\Big(
\frac{287}{90}g_1^4+\ges\gzs+\frac{15}{2}g_2^4+\frac{8}{9}\ges\gds
+8\gzs\gds-\frac{16}{9}g_3^4+\frac{4}{5}\ges\hts+\frac{2}{5}
\ges\hbs\nonumber\\ \nonumber\\ & &+\frac{6}{5}\ges\htas
+6\gzs\hbs+16\gds\hbs-5h_{t}^4-5\hts\hbs-22 h_{b}^4-3\hbs \htas-3
h_{\tau}^4\Big)\bigg)
\label{RGbs}\\ \nonumber\\ 
\frac{{\rm d}\,\htas}{{\rm d}\,t}&=&\frac{\htas}{8\pi^2}\,\bigg(
-\frac{9}{5}\ges-3\gzs+3\hbs+4\htas\nonumber\\ \nonumber\\ & &
+\frac{1}{16\pi^2}\Big(\frac{27}{2}g_1^4+\frac{9}{5}\ges\gzs
+\frac{15}{2}g_2^4-\frac{2}{5}\ges\hbs+\frac{6}{5}\ges\htas
+6\gzs\htas+16\gds\hbs-3\hts\hbs\nonumber\\ \nonumber\\ & &
-9h_{b}^4-9\hbs\htas -10h_{\tau}^4\Big)\bigg),
\label{RGtas}
\end{eqnarray}\\

The running of the parameter $\tan\beta$ is negligible and ignored as
usual.

In considering the perturbative ``top-down'' RG flow in future sections one
has first to make sure that one does not leave the region of validity of
pertubation theory. This requires that all involved couplings, in the
generic forms $g^2/(4\pi)$, $h^2/(4\pi)$ or $\la/(4\pi)$, have to be
sufficiently small as compared to 1. This is well guaranteed for all gauge
couplings between a grand unification UV scale and the weak interaction IR
scale $O(v)$: the RGE (\ref{RG1})-(\ref{RGtas}) are valid in the given form
only above the top and Higgs mass thresholds anyway; but even if one were
to use the appropriate variants below this threshold, one would eventually
run into the region where \gd\ increases towards its Landau pole and leaves
the perturbative region; going well above the UV scale \lam, besides being
physically unmotivated, leads into the region where the non-asymptotically
free coupling \gee\ increases
toward its Landau pole and thus leaves the perturbative region. The UV
initial values for all the other couplings
$\la/(4\pi),\;g^2_{t,b,\tau}/(4\pi)$ resp. $h^2_{t,b,\tau}/(4\pi)$ have to
be chosen sufficiently much smaller than 1, then their ``top-down'' RG flow
down to an IR scale O($v$) automatically remains within the perturbative
region.

\subsection{Relations between Pole Masses and $\overline{{\rm MS}}$ 
  Couplings}{\label{rad}}

The SM tree level relations between the top (bottom, tau) and Higgs masses
and their respective couplings, Eqs. (\ref{mf}) and (\ref{mh}), have to be
adapted to the order in perturbation theory under discussion. The physical
mass, which is gauge invariant, infrared finite and renormalization scheme
independent, is the so-called pole mass \cite{wil}. It is defined as the
real part of the complex pole position of the propagator in the considered
order of perturbation theory and will henceforth be denoted by $m^{{\rm
    pole}}$. Since the RGE (\ref{RGt})-(\ref{RGla}) furnish the running
couplings in the modified minimal subtraction ($\overline{\rm MS}$) scheme, we
need the relation between the $\overline{\rm MS}$ running couplings and the
pole masses which are as follows
\begin{eqnarray}
  m_f(\mu)&=&\mfp(1+\delta_f(\mu)) \;\;\;{\rm with}\label{delf}\\ 
  m_f(\mu)&=&\frac{v}{\sqrt{2}}
  g_f(\mu)=\frac{1}{\sqrt{2\sqrt{2}G_F}}g_f(\mu);\;\;{\rm for}\; f=t,b,\tau
\label{mfr}
\end{eqnarray} 
and
\begin{eqnarray}
  \mh(\mu)&=&\mhp(1+\delta_H(\mu)) \;\;\;{\rm with}\label{delh}\\ 
  \mh(\mu)&=&\sqrt{2\lambda(\mu)}v=\sqrt{\frac{\sqrt{2}}{G_F}\la(\mu)},
\label{mhr}
\end{eqnarray}
respectively, where $m(\mu)$ is the running mass in terms of the running
coupling and - as a reminder -
\begin{equation}
  v=(\sqrt{2}G_F)^{-1/2},
\end{equation}
given in terms of the Fermi constant
\begin{equation}
  G_F=1.16639(2)\,10^{-5}\,\gev^{-2}.
\end{equation}
The radiative corrections $\delta_f(\mu)$ and $\delta_H(\mu)$ have to be
taken into account to order \ges, \gzs, \gds, \gts, \gbs, \gtas\ and \la,
if running couplings and correspondingly running masses resulting from the
{\it two-loop} RGE (\ref{RGt})-(\ref{RGla}) are used. It seems worthwhile
to collect here all relevant formulae for $\delta_f(\mu)$ and
$\delta_H(\mu)$ since their contributions are scattered over a number of
publications and in particular since this is an occasion to correct various
typographical errors in the literature.\footnote{We are grateful to authors
  of Refs. \cite{kni},\cite{isi} for agreeing with us on the following
  typographical errors: i) in Ref. \cite{barb} the sign in front of the
  bracket in the last formula of Eq. (29) has to changed from - to + and
  the formula (28) applies only in the limit $\mh\gg\mt$, ii) in Ref.
  \cite{isi} the first term, $\frac{1}{24}\la$, in the bracket of Eq. (12)
  has to be replaced by $\frac{1}{12}\la$ and the whole Eq.  (12) holds
  only in the limit $\mh\gg\mt$, iii) the entry for the quantity $a_t$ in
  the table of Ref. \cite{kni} has to replaced by its negative as
  implemented in Table \ref{kniehl} in this review.}
 
The correction terms $\delta_f(\mu)$ and $\delta_H(\mu)$ may be taken from
the partially very recent literature, \cite{tar}-\cite{kni}, \cite{isi} and
\cite{sir2}, \cite{cas}, respectively. Following Ref. \cite{kni}
$\delta_f(\mu)$ is decomposed into a weak, electromagnetic and QCD
contribution which are separately finite and gauge independent
\begin{equation}
  \delta_f(\mu)=\delta_f^{{\rm w}}(\mu)+\delta_f^{{\rm QED}}(\mu)
  +\delta_f^{{\rm QCD}}(\mu).
\label{delt}
\end{equation}
The QCD correction $\delta_f^{{\rm QCD}}(\mu)$, applicable only for quarks,
is numerically the largest one. It has been calculated to $O(\gds)$ in
Refs. \cite{tar},\cite{gas} and to $O(g_3^4)$ in Refs. \cite{mas}, \cite{schi}. It seems
worthwhile to collect the relevant formulae defining $\delta_f^{{\rm
    QCD}}(\mu)$ implicitely (disregarding for the moment the electroweak contributions)
\begin{eqnarray}
  \frac{\mfp}{m_f(\mu=\mfp)}&=&1+\frac{4}{3}\frac{\alpha_3(\mfp)}
  {\pi}\nonumber\\ & &+\left[16.11-1.04\sum_{i=1}^{n_f-1}(1-\frac{m_i^{{\rm
        pole}}}
  {\mfp})\right]\left(\frac{\alpha_3(\mfp)}{\pi}\right)^2+O\left(
  \alpha_3^3(\mfp)\right)
\end{eqnarray}
with $\alpha_3=\gds/(4\pi)$ in the $\overline{\rm MS}$ scheme. The second term
is an accurate approximation for $n_f$-$1$ light quarks with pole masses
$m_i^{{\rm pole}}<\mfp$. From Refs. \cite{mas}
\begin{eqnarray}
  \frac{m_t(\mu=\mtp)}{m_t(\mu)}&=&\left(\frac{\alpha_3(\mtp)}
  {\alpha_3(\mu)}\right)^{{\scriptstyle \frac{12}{21}}}\frac{\left[1+1.398
    {\displaystyle \frac{\alpha_3(\mtp)}{\pi}}+1.794\left({\displaystyle
      \frac{\alpha_3(\mtp)}{\pi}}\right)^2\right]}{\left[1+1.398
    {\displaystyle \frac{\alpha_3(\mu)}{\pi}}+1.794\left({\displaystyle
      \frac{\alpha_3(\mu)}{\pi}}\right)^2\right]},\\ 
  \frac{m_b(\mu=\mbp)}{m_b(\mu)}&=&\left(\frac{\alpha_3(\mbp)}
  {\alpha_3(\mu)}\right)^{{\scriptstyle \frac{12}{23}}}\frac{\left[1+1.176
    {\displaystyle \frac{\alpha_3(\mbp)}{\pi}}+1.50\left({\displaystyle
      \frac{\alpha_3(\mbp)}{\pi}}\right)^2\right]}{\left[1+1.176
    {\displaystyle \frac{\alpha_3(\mu)}{\pi}}+1.50\left({\displaystyle
      \frac{\alpha_3(\mu)}{\pi}}\right)^2\right]}.
\end{eqnarray}
From these formulae $\delta_f^{{\rm QCD}}(\mu)$ to $O(\alpha_3)$ becomes
\begin{equation}
  \delta_f^{{\rm QCD}}(\mu)=\frac{4}{3}\frac{\alpha_3(\mu)}{\pi}(-1
  +\frac{3}{4}\ln\frac{m_f^2}{\mu^2})+O(\left(
  \frac{\alpha_3(\mu)}{\pi}\right)^2).
\end{equation}
$\delta_f^{{\rm QED}}$ is obtained from the leading order $\delta_f^{{\rm
    QCD}}$ by substituting $\frac{4}{3}\alpha_3$ by $Q_f^2\alpha$, where
$Q_f$ is the electric charge of the fermion $f=t,\ b,\ \tau$.

$\delta_f^{{\rm w}}(\mu)$ was determined in Ref. \cite{kni} in the limit
where \mh\ and/or \mt\ are large as compared to all the other particles of
the SM. Setting the latter masses equal to zero and omitting the subleading
terms the authors obtain
\begin{eqnarray}
  \delta_t^{{\rm w}}(\mu)&=&\frac{G_F
    \mt^2}{8\pi^2\sqrt{2}}\left[-\frac{9}{2}
  \ln\frac{\mt^2}{\mu^2}+\frac{11}{2}-r+2r(2r-3)\ln(4r)\right.\nonumber\\
& &\hspace*{1.7cm}\left. -8r^2\left(
  1-\frac{1}{r}\right)^{3/2}{\rm arcosh}\sqrt{r}\right],
\label{tgen}\\
\delta_b^{{\rm w}}(\mu)&=&\frac{G_F}{8\pi^2\sqrt{2}}\left[\mt^2\left(
-\frac{3}{2}\ln\frac{\mt^2}{\mu^2}+\frac{1}{4}\right)+\frac{\mh^2}{4}
\right],\label{bgen}\\ \delta_{f\neq t,b}^{{\rm
    w}}(\mu)&=&\frac{G_F}{8\pi^2\sqrt{2}}\left[
3\mt^2\left(-\ln\frac{\mt^2}{\mu^2}+\frac{1}{2}\right)+\frac{\mh^2}{4}\right]
\label{tagen}
\end{eqnarray}
with
\begin{equation}
  r=\frac{\mh^2}{4\mt^2}.
\end{equation}
Eq. (\ref{tgen}) is valid for $r\geq 1$. For $r<1$ one has to replace
$(1-1/r)^{3/2} {\rm arcosh}\sqrt{r}$ by $(1/r-1)^{3/2}{\rm
  arccos}\sqrt{r}$. Following Ref. \cite{kni} one can expand Eq.
(\ref{tgen}) for $\mh\gg 2\mt$ ($r\gg 1$) and $\mh\ll 2\mt$ ($r\ll 1$),
which leads to
\begin{eqnarray}
  \delta_t^{{\rm w}}(\mu)&=&\frac{G_F}{8\pi^2\sqrt{2}}\left[\frac
  {\mh^2}{4}+\mt^2\left(-3\ln\frac{\mt^2}{\mu^2}+\frac{3}{2}-\frac{3}{2}
  \ln\frac{\mh^2}{\mu^2}+\frac{7}{4}\right.\right.\nonumber\\
& &\hspace*{1.7cm}\left.\left. +O\left(\frac{\mt^2}{\mh^2}\ln\frac{\mh^2}{\mt^2}\right)\right)\right],  
\label{rg1}\\
\delta_t^{{\rm w}}(\mu)&=&\frac{G_F \mt^2}{8\pi^2\sqrt{2}}\left[-\frac
{9}{2}\ln\frac{\mt^2}{\mu^2}+\frac{11}{2}-2\pi\frac{\mh}{\mt}
+O\left(\frac{\mh^2}{\mt^2}\ln\frac{\mt^2}{\mh^2}\right)\right],
\end{eqnarray}
respectively. In Ref. \cite{kni} also the subleading corrections have been
calculated. It is found that $\delta_f^{{\rm w}}(m_f)+\delta_f^{{\rm QED}}$
including these subleading corrections are very well approximated by Eqs.
(\ref{tgen})-(\ref{tagen}) if the terms
\begin{equation}
  d_f=a_f+b_f\ln\frac{\mh}{300\,\gev}+c_f\ln\frac{\mt}{175\,\gev}
\label{abc}
\end{equation} 
are added on their right hand sides. The coefficients for $f=t,\,b,\,\tau$
are listed in Table \ref{kniehl}.
\begin{table}[hbt]
\begin{center}
\begin{tabular}{|c||c|c|c|}\hline
  $f$ & $a_f$ & $b_f$ & $c_f$ \\ \hline\hline $t$ & $-6.90\times 10^{-3}$ &
  $1.73\times 10^{-3}$ & $-5.82 \times 10^{-3}$\\ \hline $b$ & $1.52\times
  10^{-2}$ & $1.73\times 10^{-3}$ & $0$\\ \hline $\tau$ & $1.59\times
  10^{-2}$ & $1.73\times 10^{-3}$ & $0$\\ \hline
\end{tabular} 
\end{center}
\caption[dum]{Coefficients $a_f$, $b_f$ and $c_f$ for $f=t,\,b,\,\tau$, to
  be inserted into Eq. (\ref{abc}).}
\label{kniehl}
\end{table}
The correction $\delta_H(\mu)$ has been calculated in Ref. \cite{sir2} and
more recently in Ref. \cite{cas}. The results turn out\footnote{We are
  grateful to J.R.  Espinosa for undertaking the effort to compare the
  results and for providing us with the conclusion.} to be identical apart from
a numerically negligible redefinition of v. The results are summarized as
follows \cite{sir2}

with $\xi = m_H^2/m_Z^2,\, s^2 = \sin^2 \theta_W,\,
  c^2 = \cos^2 \theta_W$ and $\theta_W$ being the Weinberg angle:
\begin{eqnarray}
 \delta_H(\mu) & = & \frac{G_F}{\sqrt{2}}
                     \frac{m_Z^2}{16 \pi^2}
                     \left\{\, \xi f_1(\xi, \mu)
                        + f_0(\xi, \mu)
                        + \xi^{-1} f_{-1}(\xi, \mu) \,\right\}, \\
 {\rm with}\ \ f_1(\xi, \mu) & = & 6\ln\frac{\mu^2}{m_H^2} +\frac{3}{2}\ln\xi
                     -\frac{1}{2} Z\!\left(\frac{1}{\xi}\right)
                     -Z\!\left(\frac{c^2}{\xi}\right)
                     -\ln c^2
                     +\frac{9}{2}
                        \left[ \frac{25}{9} - \sqrt{\frac{1}{3}}\:\pi\right],\\
 f_0(\xi, \mu) & = & -6\ln\frac{\mu^2}{m_Z^2}
                        \left[ 1 +2 c^2 -2\frac{m_t^2}{m_Z^2} \right]
                     +\frac{3 c^2 \xi}{\xi-c^2} \ln\frac{\xi}{c^2}
                     +2 Z\!\left( \frac{1}{\xi} \right) \nonumber\\
               &   & +\;4 c^2 Z\!\left( \frac{c^2}{\xi} \right)
                     +\frac{3 c^2 \ln c^2}{s^2} +12 c^2 \ln c^2
                     -\frac{15}{2} \left( 1 +2 c^2 \right) \nonumber\\
               &   & -\;3\frac{m_t^2}{m_Z^2} \left[
                        2 Z\!\left( \frac{m_t^2}{m_Z^2 \xi} \right)
                        +4 \ln\frac{m_t^2}{m_Z^2} -5 \right], \\
 f_{-1}(\xi,\mu) &=& 6\ln\frac{\mu^2}{m_Z^2}
                        \left[ 1 +2 c^4 -4\frac{m_t^4}{m_Z^4} \right]
                     -6 Z\!\left( \frac{1}{\xi} \right)
                     -12 c^4 Z\!\left( \frac{c^2}{\xi} \right)
                     -12 c^4 \ln c^2 \nonumber\\
               &   & +\;8\left( 1 +2 c^4 \right)
                     +24 \frac{m_t^4}{m_Z^4} \left[
                        \ln\frac{m_t^2}{m_Z^2} -2
                       + Z\!\left( \frac{m_t^2}{m_Z^2 \xi} \right) \right],
\end{eqnarray}

\hfill\hfill\parbox{9cm}{\begin{eqnarray*}
 Z(z) &=& \left\{ \begin{array}{l@{\quad}l}
                2 A \arctan(1/A) &
                        (z > {\textstyle\frac{1}{4}} ) \\
                A \ln \left[ (1+A)/(1-A) \right] &
                        (z < {\textstyle\frac{1}{4}} )\; ,
        \end{array} \right.
\end{eqnarray*}}\hfill
\parbox{6cm}{\begin{equation}
 A = \sqrt{ \left| 1 - 4 z \right| }\; .\hspace*{1cm}\mbox{}
\end{equation}}

\begin{figure}
\begin{center}
\epsfig{file=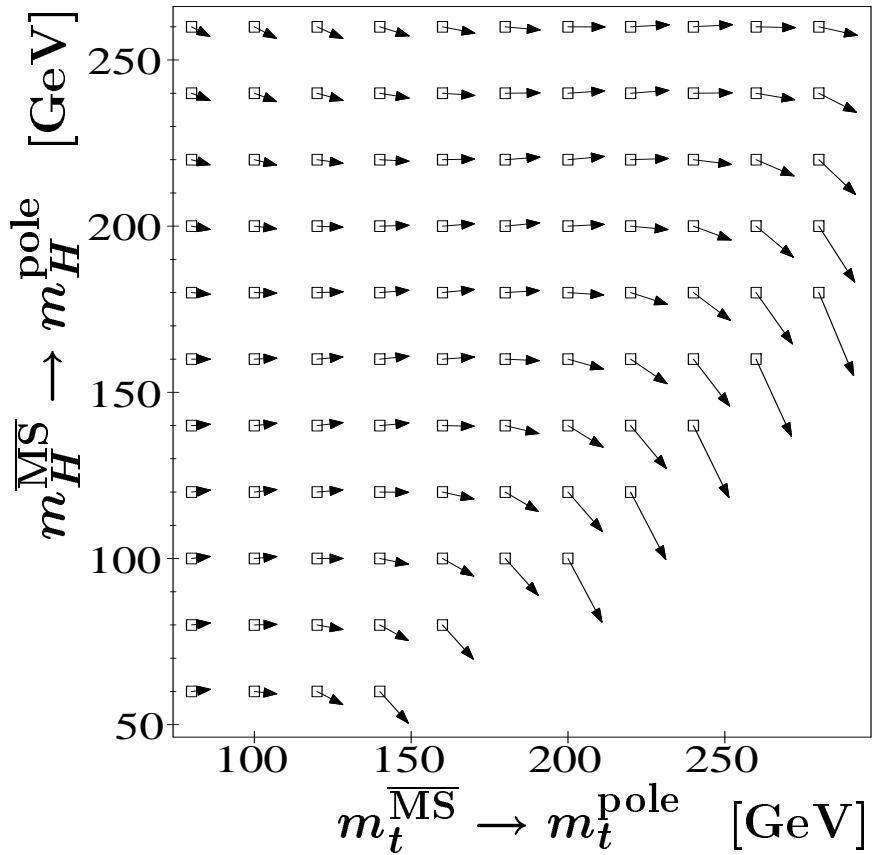,bbllx=114pt,bblly=249pt,bburx=548pt,
bbury=679pt,width=11.5cm}
\end{center}
\caption[dum]{Radiative corrections, applied to the 
  $\overline{\rm MS}$
  Higgs-top mass pairs, denoted by open squares, leading to the
  corresponding physical pole mass pairs at the tips of the
  arrows.}
\label{msbarmpole}
\end{figure}

The following asymptotic expressions hold \cite{sir2}:
\begin{eqnarray}
 \delta_H(\mu) \;=\; \frac{G_F}{\sqrt{2}} \frac{m_Z^2}{16 \pi^2}
        &\!\!\!\!\xi\!\!\!\!&
                \left[\, 6 \ln\frac{\mu^2}{m_H^2}
                +\frac{9}{2}
                        \left( \frac{25}{9} - \sqrt{\frac{1}{3}}\:\pi \right)
                \right], \qquad\qquad (\xi \gg 1) \\
 \delta_H(\mu) \;=\; \frac{G_F}{\sqrt{2}} \frac{m_Z^2}{16 \pi^2}
        &\!\!\!\!{\displaystyle\frac{1}{\xi}}\!\!\!\!&
                \left[\, 6 \ln\!\left(\frac{\mu^2}{m_Z^2}\right)
                        \left(1+2 c^4 -4\frac{m_t^4}{m_Z^4} \right)
                - 4 \left( 1 + 2 c^4 \right) \right. \nonumber\\
        &\!\!\!\!\!\!\!\!&
                \left.
                \;\; -\;12 c^4 \ln c^2 + 24 \frac{m_t^4}{m_Z^4}
                        \ln\frac{m_t^2}{m_Z^2}
                \;\right] \qquad\qquad\hspace{1mm} (\xi \ll 1).
\end{eqnarray}
All radiative corrections are collectively exposed in Fig.
\ref{msbarmpole}. More precisely, the squares represent representative
starting values for pairs of masses in the $\mh^{\overline{\rm
    MS}}$-$\mt^{\overline{\rm MS}}$-plane. The arrows attached to the
squares indicate where these pairs end up in the \mhp-\mtp-plane. We stop
the presentation in the region of the absolute lower bound (to be
discussed in Sect. \ref{lowsm}). This diagram allows to discern at a glance
where the radiative corrections become sizeable and in which directions they aim.

\subsection{Effective Potential and Vacuum Stability\label{vac}}

For small values of the Higgs selfcoupling \la\ radiative corrections to
the Higgs potential become important. For a very nice and comprehensive
review on the subject and relevant references we refer to Ref.
\cite{sherrev}, supplemented by Ref. \cite{for}. These radiative
corrections become relevant for two physics issues addressed in this review
\begin{itemize}
\item for the calculation of the vacuum stability bound
  \cite{mai},\cite{linvac}-\cite{esp} within the SM, a lower bound on the Higgs mass,
  which increases for increasing top mass, resp. an upper bound for the top
  mass which increases for increasing Higgs mass;
\item for the calculation of an upper bound \cite{kun}-\cite{wagne} for the mass of
the lightest Higgs boson in the MSSM as function of the SUSY breaking
  scale \msusy.
\end{itemize}
The radiative corrections to the Higgs potential imply the calculation of
the effect of virtual particle emission and reabsorption on the interaction
energy. The ``quantum corrections'' to the scalar potential of the SM,
$V(\phi)=-\frac{1}{2}m^2\phi^2+\frac{\la}{4}\phi^4$ emerge in form of a
loop expansion (with $\phi$ to be identified with $\sqrt{2} {\rm
  Re}\,\phi^0$ in Eq. (\ref{phi})).  The contribution to a given number of
loops results from summing all the one-particle irreducible graphs with any
number of external legs with the classical scalar field on the external
legs and with zero external momentum.  The loop expansion typically
contains terms
\begin{equation}
\alpha^{n+1}\left( \ln\left(\frac{\phi^2}{\mu_0^2}\right)\right)^n
\end{equation}
where $\alpha$ is a generic SM coupling, $\phi$ the classical scalar
field, $\mu_0$ the arbitrary momentum scale and $n$ the loop order. 

In order for the loop expansion to be reliable, it is not sufficient for
the coupling $\alpha$ to be small, but also the loop expansion parameter
$\alpha \ln\left(\phi^2/\mu_0^2\right)$ has to be smaller than one. Of
course the renormalization scale $\mu_0$ can be chosen to make
$\ln\left(\phi^2/\mu_0^2\right)$ as small as possible, but $\mu_0$ can
only take one value. 

In order to keep the effect of the logarithmic terms small over large
ranges of large $\phi$, as necessary for the investigation of the issue of
vacuum instability, one has to take recourse to the {\it RG improved}
effective potential. The relevant state of the art is to treat the one-loop
effective potential with RG improvement on the two-loop level; in this case
the leading and next-to-leading logarithms are summed to all-loop in the
effective potential.

Including the one-loop correction in the 't
Hoft Landau gauge  \cite{for} (keeping the top Yukawa coupling \gt\ as only
non-zero fermion coupling), the one-loop effective potential of the SM reads
\begin{eqnarray}
V(\phi)&=&-\frac{1}{2}m^2\phi^2+\frac{\la}{4}\phi^4 + \frac{1}{16\pi^2}\left[\frac{1}{4}H^2\left(\ln\frac{H}{\mu_0^2}-\frac{3}{2}\right)+\frac{3}{4}G^2\left(\ln\frac{G}{\mu_0^2}-\frac{3}{2}\right)\right.\nonumber\\
& & \left. +\frac{3}{2}W^2\left(\ln\frac{W}{\mu_0^2}-\frac{5}{6}\right)+\frac{3}{4}Z^2\left(\ln\frac{Z}{\mu_0^2}-\frac{5}{6}\right)-3T^2\left(\ln\frac{T}{\mu_0^2}-\frac{3}{2}\right)\right]\\
{\rm with}\ \ H & = & -m^2+3\la\phi^2,\nonumber\\
T & = & \frac{1}{2}\gts\phi^2,\nonumber\\
G & = & -m^2+\la\phi^2,\nonumber\\
W & = & \frac{1}{4}\gzs\phi^2,\nonumber\\
Z & = & \frac{1}{4}(\gzs+\frac{3}{5}\ges)\phi^2.
\end{eqnarray}
The renormalization group improvement consists in introducing a
variable rescaling 
\begin{equation}
\mu(t)=\mu_0 e^t.
\end{equation} 
Since the effective potential is independent of the renormalization
scale, the effect of the rescaling on the explicit scale dependence in
the effective potential has to be absorbed into changes of the
couplings and the field $\phi$. For the resulting couplings,
$\la(\mu)$, $\gt(\mu)$,..., the differential change at the two-loop
level is given in terms of the two-loop RGE (\ref{RGla}),
(\ref{RGt}),... in terms of the variable t or $\mu$.

Next, the physical requirement of vacuum stability, i.e. of the
stability of the (radiatively corrected) electroweak vacuum, has to be
fulfilled. This stability is only in danger at large values of the
field $\phi$. The key point is now the following. The rescaling allows
to choose the scaling parameter t such that $\mu(t)\sim \phi$ at large
$\phi$. It has been shown \cite{for} that for this choice the only
term which is of importance in the potential is the
$\frac{1}{4}\la\phi^4$ term in the tree level potential, but now with
the constant tree level coupling \la\ replaced by the two-loop running
coupling $\la(\mu)$. Correspondingly the question of the existence of
a false, deep minimum which could destabilize the electroweak minimum
at some scale is simply the question of whether the running coupling
$\la(\mu)$ goes negative as t increases. Even for very small negative
\la, the fact that this happens at $\phi/m_Z\gg 1$ means that the term
$\frac{1}{4}\la(\mu)\phi^4$ drives the potential well below the
electroweak minimum. The importance of the evolution of \la\ to the
stability of the vacuum was in fact already recognized in Ref.
\cite{mai}. Since the RGE for $\la(\mu)$ contains a large contibution
from the large top Yukawa coupling, the vacuum stability bound is
strongly top mass dependent. Of course no extrapolation to values of
the field larger than the physical cut-off \lam\ should be made: the
maximal scale $\mu$ at which a zero of the running coupling $\la(\mu)$
signals a destabilization of the vacuum is $\mu=\lam$. Thus, given the
UV scale \lam, the corresponding vacuum stability bound in the
$g_t$-$\la$-plane is to a good approximation given by the {\it lower
  boundary} of the IR end points $\la(\mu\approx m_Z)$ of the RG flow
in the $g_t$-$\la$-plane constrained to solutions $\la(\mu)$ which
{\it do not become negative} in the interval between $\mu=\lam$ and
$\mu\approx m_Z$.

In Refs. \cite{cas}, \cite{casa} the choice of scale was refined to the
effect that the scale dependence of the one-loop effective potential, RG improved at the two-loop
level, becomes minimal. This leads the authors to replace the role of \la\
played in the argument led above by that of the slightly shifted variable
\begin{equation}
\tilde{\la}=\la-\frac{1}{16\pi^2}\left[3\gt^4\left(\ln\frac{\gt^2}{2}-1\right)-\frac{3}{8}\gz^4\left(\ln\frac{\gzs}{4}-\frac{1}{3}\right)-\frac{3}{16}(\gzs+\frac{3}{5}\ges)^2\left(\ln\frac{(\gzs+\frac{3}{5}\ges)}{4}-\frac{1}{3}\right)\right].
\end{equation}

\section{Preview of Infrared Fixed Manifolds and Bounds 
  in the SM and MSSM\label{preview}}

Since IR fixed manifolds in the multiparameter space imply interesting
relations between parameters of the SM or of the MSSM with a likely bearing
on physical reality, it is well worthwhile to scrutinize them quasi with a
mathematical magnifying glass. This section is meant as a first guideline
only; it contains a brief {\it summary} of the exact one-loop IR fixed
manifolds as well as some one-loop bounds in the space of couplings.
Neither a thorough understanding nor precise predictions for particle
masses or mass relations or mass bounds may be inferred from it. In fact
all derivations of these manifolds and the strength of their IR
attraction, the latest state of the art including higher order
radiative corrections, the resulting IR attractive masses, relations
between masses and the resulting mass bounds are filled in in the following
Sects.  \ref{noelweak}-\ref{masses}. These sections also will contain the
large number of references to the literature, which are omitted here
altogether. The purpose of this summary is twofold: it allows to take in at
a glance the richness of the IR structures thus hopefully wetting the
appetite for more information; and it allows most transparently to see each
of the following chapters in perspective to the complete picture to emerge
only towards the end.

Starting point is to trade the independent variable t for the variable \gds, and
consider the {\it ratios} of couplings
\begin{equation}
  \la/\gds,\;\;\;g^2_{t,b,\tau}/\gds,\,\;{\rm
    resp.}\,\;h^2_{t,b,\tau}/\gds, \;\;\;{\rm and}\;\;\;g^2_{1,2}/\gds.
\label{rho}
\end{equation}
This leads to a set of RGE for the new dependent variables
(\ref{rho}) and the new independent variable \gds. The most interesting IR
fixed manifolds will appear in this set of RGE and will as a rule be exact
only in one-loop approximation. Therefore this exploratory chapter is
strictly based on the one-loop RGE results.

As has been mentioned in the introduction, bounds for masses are boundaries
of IR points of the RG flow which fail to be fully attracted onto the IR
fixed manifolds. They lie the closer to the IR manifolds the longer is the
evolution path, i.e. the larger is the value of the UV cut-off \lam.
Important bounds of this kind will be shown alongside the IR manifolds in
this exploratory section. The most well known are the bounds on the Higgs
selfcoupling \la\ responsible for the triviality (upper) bound and the
vacuum stability (lower) bounds on the Higgs mass, cutting out a wedge-like
allowed region in the \la-\gt-plane. 

It is most instructive to present the material here and in the
following sections by {\it gradually increasing the parameter space}
for two reasons.  One is pedagogical: it allows to proceed in small
steps from warm-up exercises with simple examples to the complex
structure, observing how with each step the physical implications
become less and less trivial. The second one is that a number of
crucial results in the literature as e.g. the non-perturbative lattice
results have been obtained within such reduced parameter spaces and
may be included most systematically this way. At each step in the
gradual increase of the subset of considered parameters their
evolution according to the RGE (either of the SM or of the MSSM) is
evaluated {\it by setting all the excluded parameters identical to
zero}.  An a posteriori justification for this procedure emerges in
the following sections.

With the exception of the RG evolution of the Higgs selfcoupling,
which is only treated in the SM, the MSSM results are displayed in
parallel to the SM results. For simplicity the scale \msusy\ for the
transition from REG of the MSSM to the RGE of the SM is chosen equal
to $\mt=176\gev$ in this exploratory section. 


\begin{table}
\parbox{171mm}{
\begin{sideways}
\begin{tabular}[b]{|c|c||c|c|c|c|}\hline 

  \raisebox{-6mm}{model} & \raisebox{-6mm}{\begin{tabular}{c}considered\\
      couplings\end{tabular}} & \raisebox{-6mm}{IR fixed point} & \raisebox{-6mm}{IR
    fixed line} & \raisebox{-6mm}{\begin{tabular}{c}figure of IR fixed line
      (fat),\\ bounds at $\mu=m_{t}$ (thin) for\\
      $\lam=10^4,\;10^6\;10^{10},\;10^{15}\gev$\end{tabular}} &
  \raisebox{-6mm}{enlargement} \\[8mm] \hline

 
  \raisebox{-3mm}{SM} & \raisebox{-3mm}{$\lambda$} &
  \raisebox{-3mm}{$\lambda=0$} & &
  \multicolumn{2}{c|}{\raisebox{-3mm}{$\lambda\leq 1.63,\;0.37,\;0.22\;$
      resp. for $\;\Lambda = 10^4,\;10^{10},\;10^{15}\gev$}}
   \\[6mm] \hline


   \raisebox{2.6cm}{SM} & \raisebox{2.6cm}{$\lambda,\,g_{t}$} &
   \raisebox{2.6cm}{\parbox{2cm}{\begin{eqnarray} \nonumber \lambda & =& 0,
       \\ \nonumber g_{t}^2 & = & 0\\ \nonumber \end{eqnarray}}} &
   \raisebox{2.6cm}{\parbox{3.1cm}{$\;\lambda = {\displaystyle
         \frac{\sqrt{65}-1}{16}}\;g_{t}^2,$ \newline \newline fat line in
       figure}} & \raisebox{-4mm}{\parbox{51mm}{\begin{picture}(50,50)
       \epsfig{file=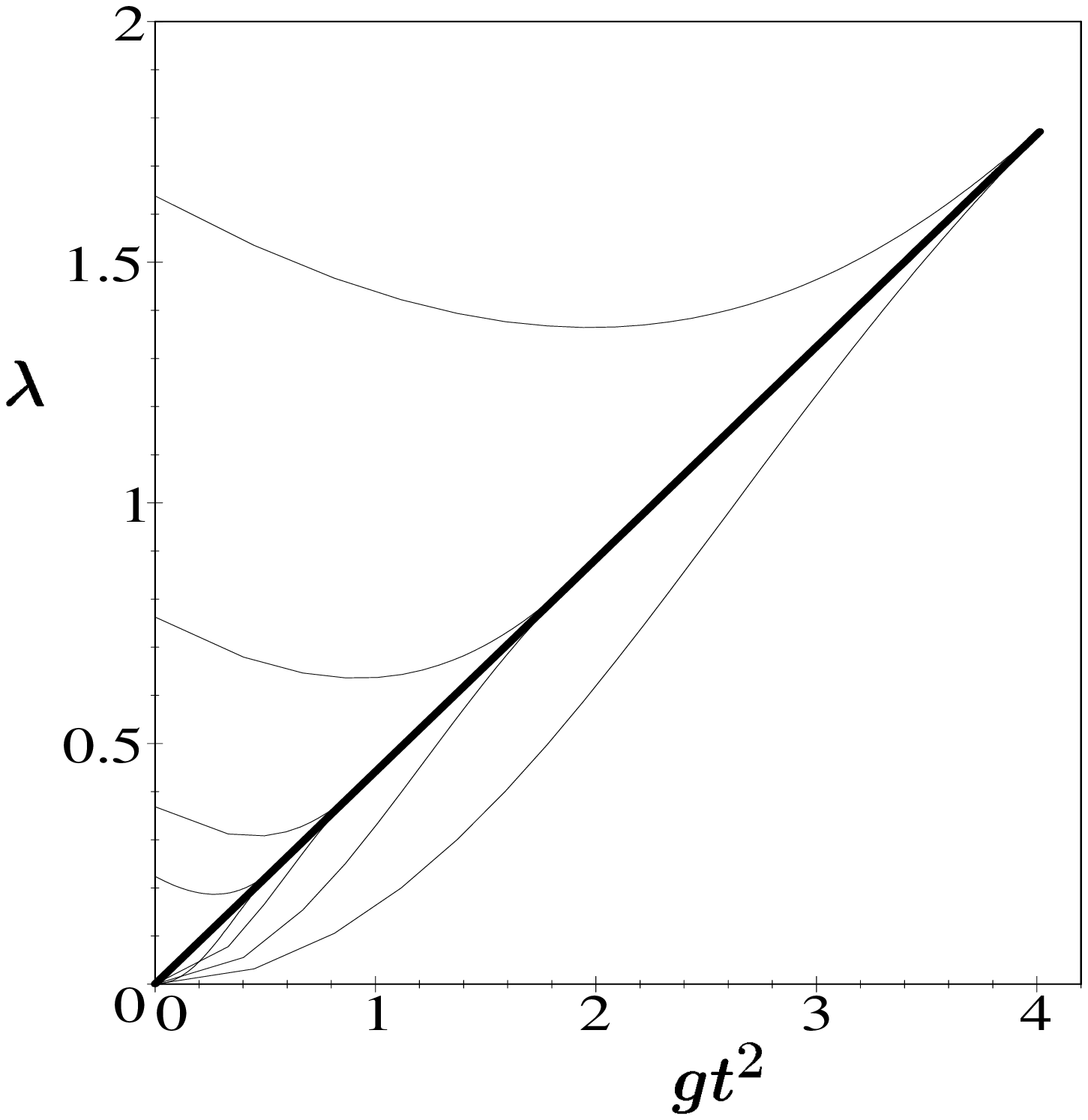,bbllx=52pt,bblly=338pt, bburx=471pt,
         bbury=770pt,height=5cm,width=5cm}\end{picture}}} &
   \raisebox{-4mm}{\parbox{51mm}{\begin{picture}(50,50)
       \epsfig{file=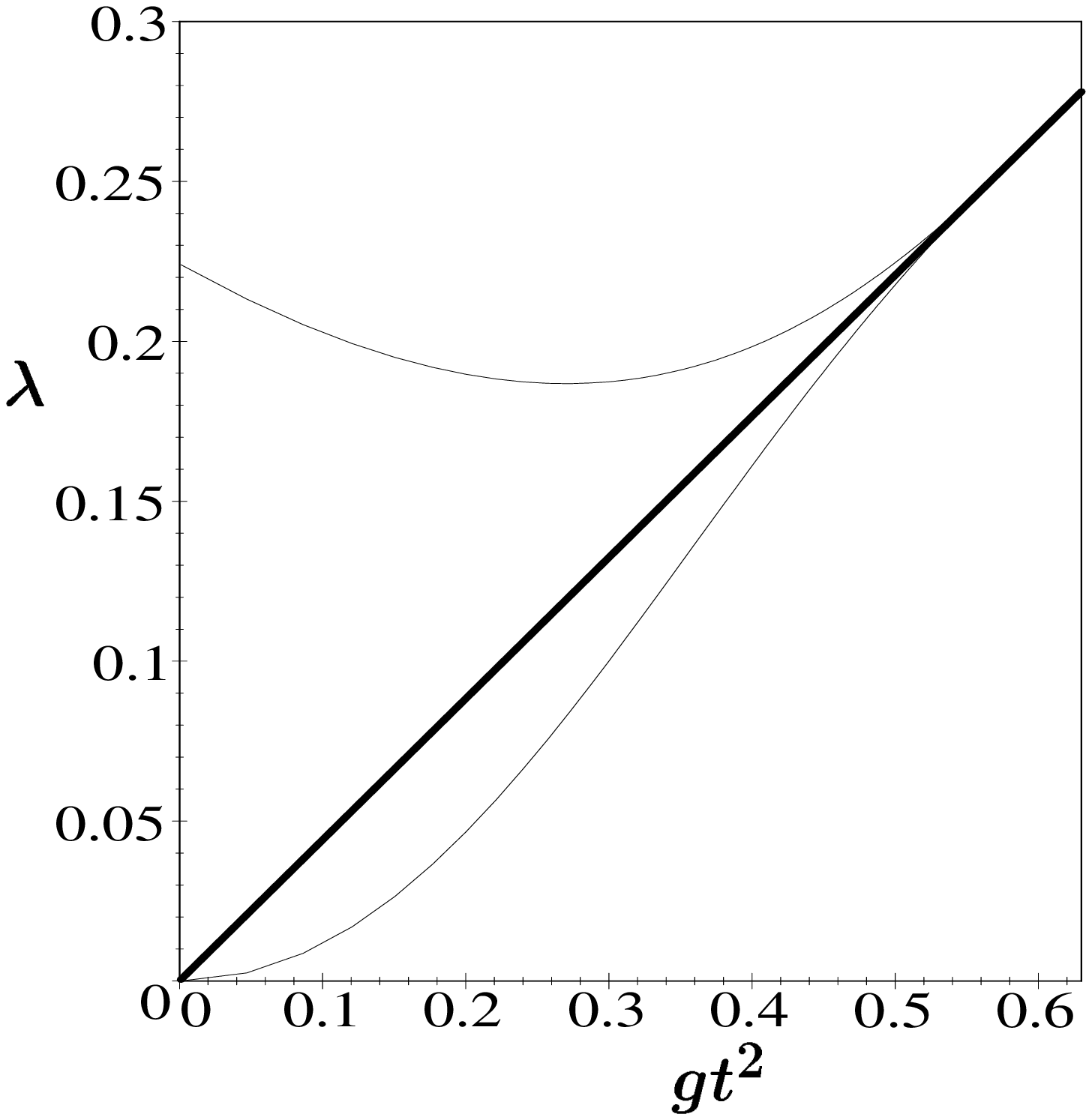,bbllx=54pt,bblly=339pt,
         bburx=471pt,bbury=770pt,height=5cm,width=5cm}\end{picture}}}
  \\[2mm] \hline

  
  \raisebox{-4mm}{SM} & \raisebox{-4mm}{$g_{t},\,g_3$} &
  \raisebox{-4mm}{${\displaystyle \frac{g_{t}^2}{g_3^2}}={\displaystyle
      \frac{2}{9}}$} & & & \\[8mm] \hline


  \raisebox{-4mm}{MSSM} & \raisebox{-4mm}{$h_{t},\,g_3$} &
  \raisebox{-4mm}{${\displaystyle \frac{h_{t}^2}{g_3^2}}={\displaystyle
      \frac{7}{18}}$} & & & \\[8mm] \hline


  \raisebox{2.6cm}{SM} & \raisebox{2.6cm}{$\lambda,\,g_{t},\,g_3$} &
  \raisebox{2.6cm}{\parbox{3.6cm}{\begin{eqnarray} \nonumber
      \frac{\lambda}{g_3^2} & = & \frac{\sqrt{689}-25}{72},\\ \nonumber
      \frac{g_{t}^2}{g_3^2} & = & \frac{2}{9}\\ \nonumber \end{eqnarray}
      symbol $\Diamond$ in figure}} & \raisebox{2.6cm}{fat line in figure}
  & \raisebox{2mm}{\parbox{51mm}{\begin{picture}(50,50)
      \epsfig{file=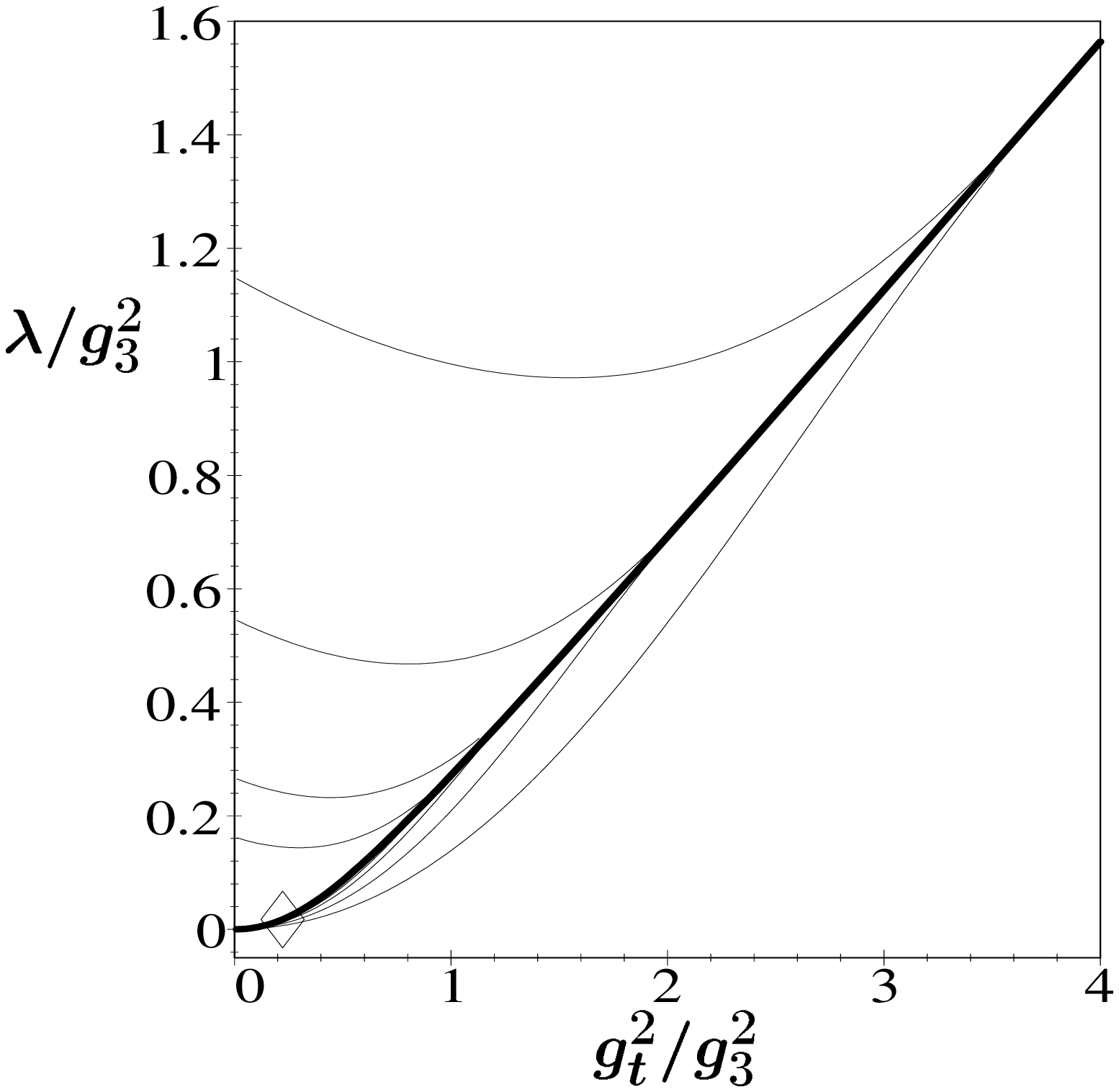,bbllx=16pt,bblly=339pt,
        bburx=464pt,bbury=776pt,height=5cm,width=5cm}\end{picture}}} &
  \raisebox{2mm}{\parbox{51mm}{\begin{picture}(50,50)
      \epsfig{file=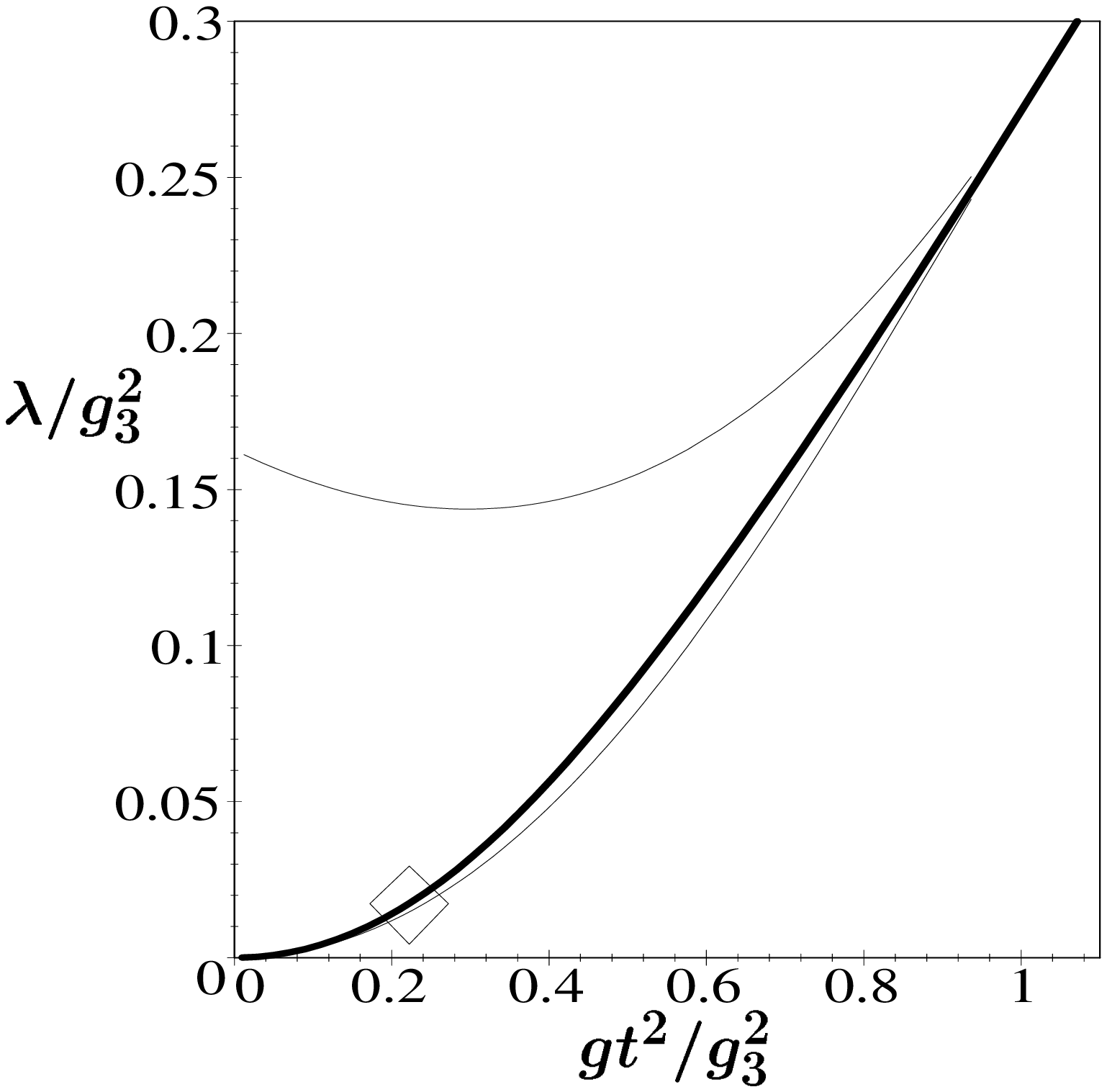,bbllx=16pt,bblly=338pt,
        bburx=458pt,bbury=776pt,height=5cm,width=5cm}\end{picture}}}\\ 
  \hline
\end{tabular}
\end{sideways}
}
\end{table}


\begin{table}
\parbox{171mm}{
\begin{sideways}
\begin{tabular}[b]{|c|c||c|c|c|c|} \hline 

  \raisebox{-6mm}{model} & \raisebox{-6mm}{\begin{tabular}{c}considered\\
      couplings\end{tabular}} & \raisebox{-6mm}{IR fixed point} & \raisebox{-6mm}{IR
    fixed lines} & \raisebox{-6mm}{\begin{tabular}{c}figure of IR fixed lines
      (fat),\\ bounds at $\mu=m_{t}$ (thin) for\\
      $\lam=10^4,\;10^6,\;10^{10},\; 10^{15}\gev$\end{tabular}} &
  \raisebox{-6mm}{\parbox{5cm}{enlargement}}
  \\[7mm] \hline 


  \raisebox{0cm}{SM} & \raisebox{0cm}{$g_{t},\,g_{b},\,g_3$} &
  \raisebox{0cm}{\parbox{3.6cm}{\begin{eqnarray} \nonumber
      \frac{g_{t}^2}{g_3^2} & = & \frac{1}{6},\\ \nonumber
      \frac{g_{b}^2}{g_3^2} & = & \frac{1}{6}\\ \nonumber \end{eqnarray}
      symbol $\Diamond$ in figure}} & \raisebox{0cm}{\begin{tabular}{c}fat
      line {\bf 1} in figure\\ (strongly IR\\
      attractive)\\[7mm] fat line {\bf 2} in figure\\
      (weakly IR\\ attractive)\end{tabular}} &
  \raisebox{-22mm}{\parbox{49mm}{\begin{picture}(49,49)\epsfig{file=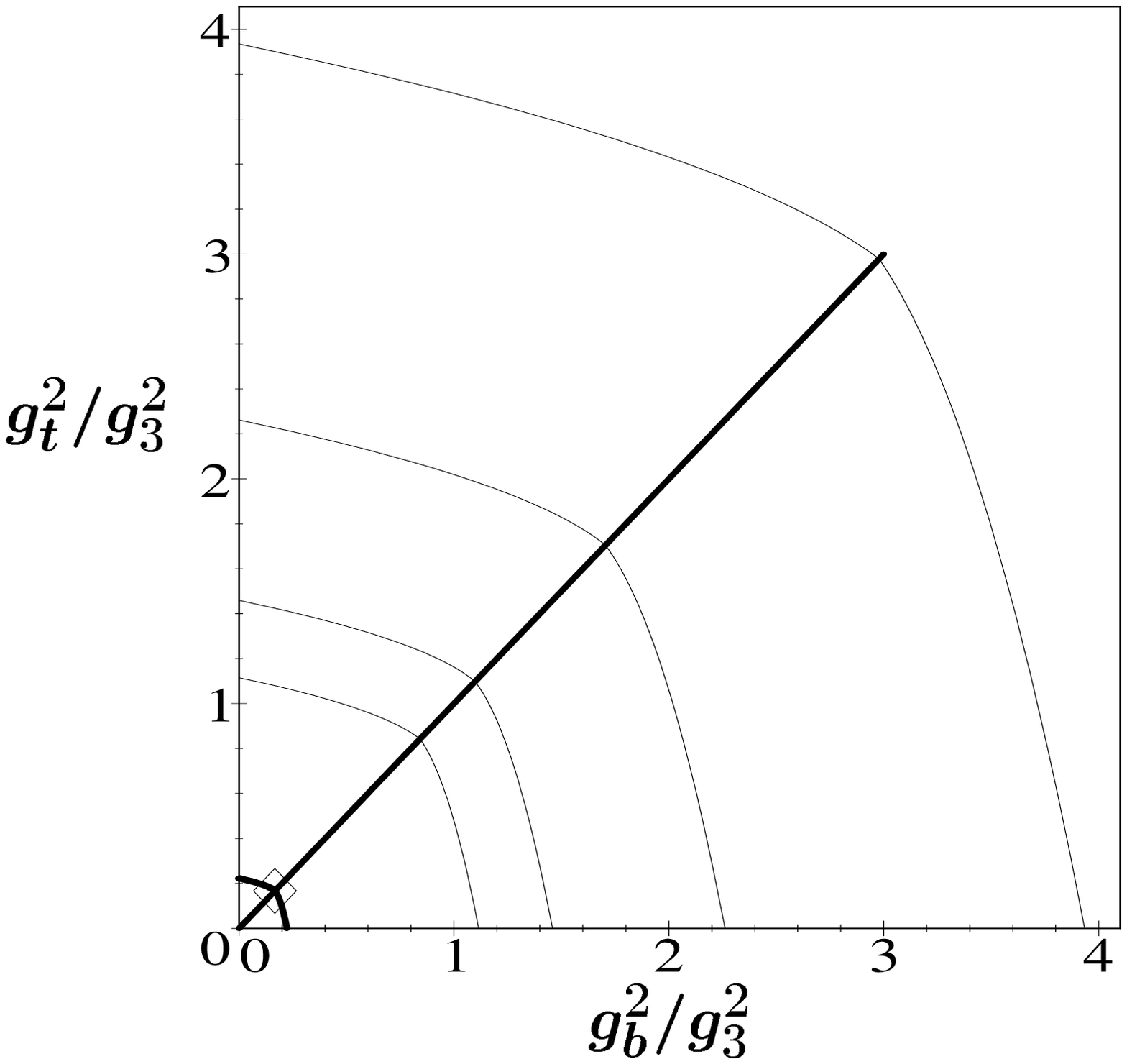,bbllx=14pt,bblly=338pt,
        bburx=472pt,bbury=771pt,height=4.9cm,width=4.9cm}\end{picture}}} &
\raisebox{-22mm}{\parbox{49mm}{\begin{picture}(49,49)\epsfig{file=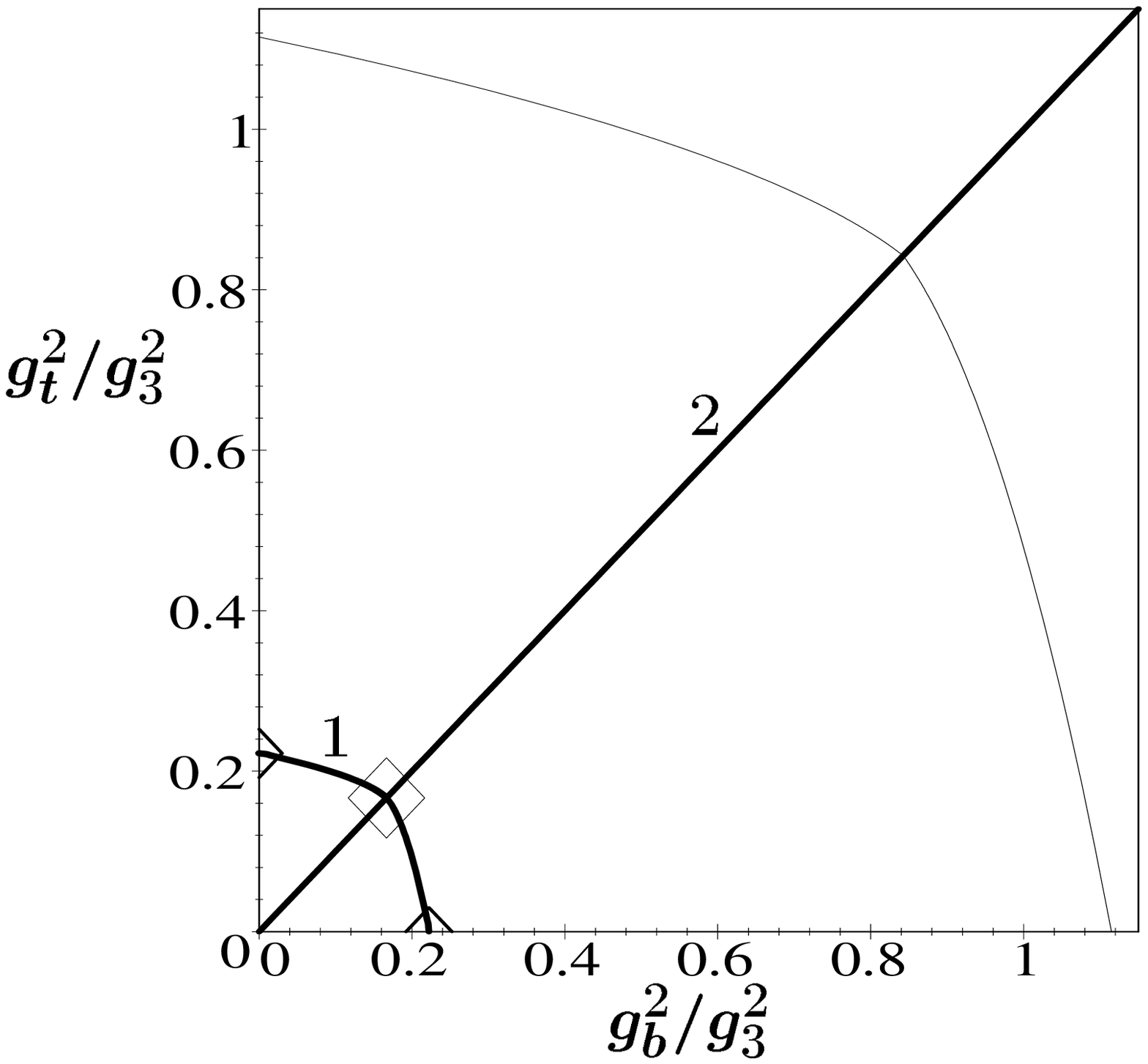,bbllx=6pt,bblly=339pt,
        bburx=472pt,bbury=771pt,height=4.9cm,width=4.9cm}\end{picture}}}
  \\ \hline


  \raisebox{0mm}{MSSM} 
& \raisebox{0mm}{$h_{t},\,h_{b},\,g_3$} 
& \raisebox{0mm}{\parbox{3.5cm}{\begin{eqnarray}
  \nonumber \frac{h_{t}^2}{g_3^2} & = &
  \frac{1}{3},\\ \nonumber \frac{h_{b}^2}{g_3^2} & = &
  \frac{1}{3}\\ \nonumber \end{eqnarray} symbol $\Diamond$
  in figure}} 
& \raisebox{0mm}{\begin{tabular}{c}fat line {\bf 1}  in
  figure\\ (strongly IR\\ attractive)\\[7mm] fat line {\bf 2} in
  figure\\ (weakly IR\\ attractive)\end{tabular}}
& 
& \raisebox{-21mm}{\parbox{49mm}{\begin{picture}(49,49)
  \epsfig{file=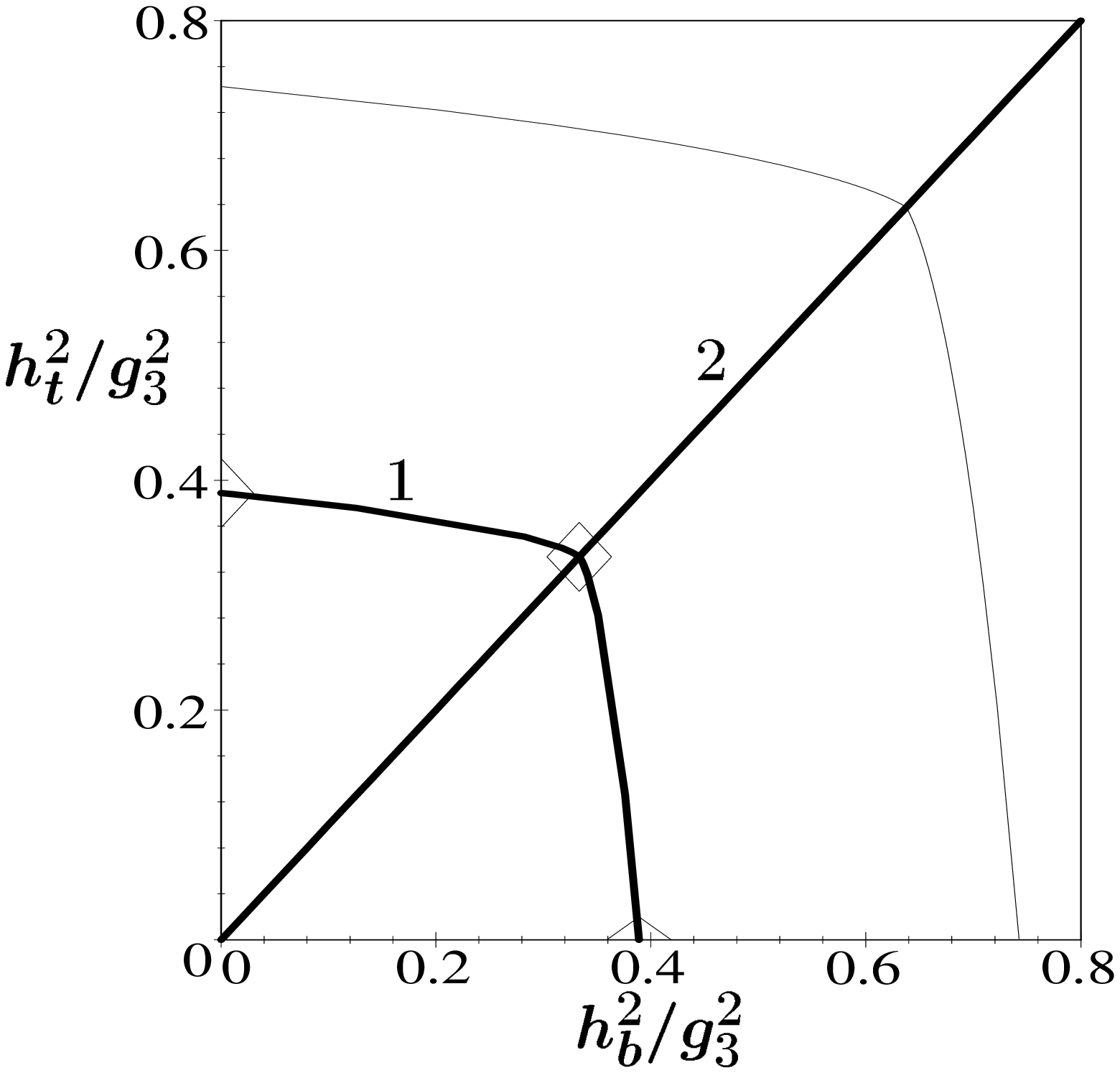,bbllx=21pt,bblly=338pt,
  bburx=471pt,bbury=770pt,height=4.9cm,width=4.9cm}\end{picture}}}
  \\ \hline

 & & & &figure of IR fixed surface& \\ \hline

  \raisebox{0cm}{SM}
& \raisebox{0cm}{$\lambda,\,g_{t},\,g_{b},\,g_3$}
& \raisebox{0mm}{\parbox{3.5cm}{\begin{eqnarray}
  \nonumber \frac{\la}{\gds} & = &
\frac{\sqrt{89}-9}{24}\\ \nonumber \frac{g_{t}^2}{g_3^2} & = &
  \frac{1}{6},\\ \nonumber \frac{g_{b}^2}{g_3^2} & = &
  \frac{1}{6}\nonumber \end{eqnarray} symbol $\Diamond$
  in figure}} 
& \raisebox{0cm}{\begin{tabular}{c}fat line {\bf 1}  in
  figure\\ (strongly IR\\ attractive)\\[7mm] fat line {\bf 2} in
  figure\\ (weakly IR\\ attractive)\end{tabular}}
& \raisebox{-17mm}{\parbox{49mm}{\begin{picture}(49,49)\epsfig{file=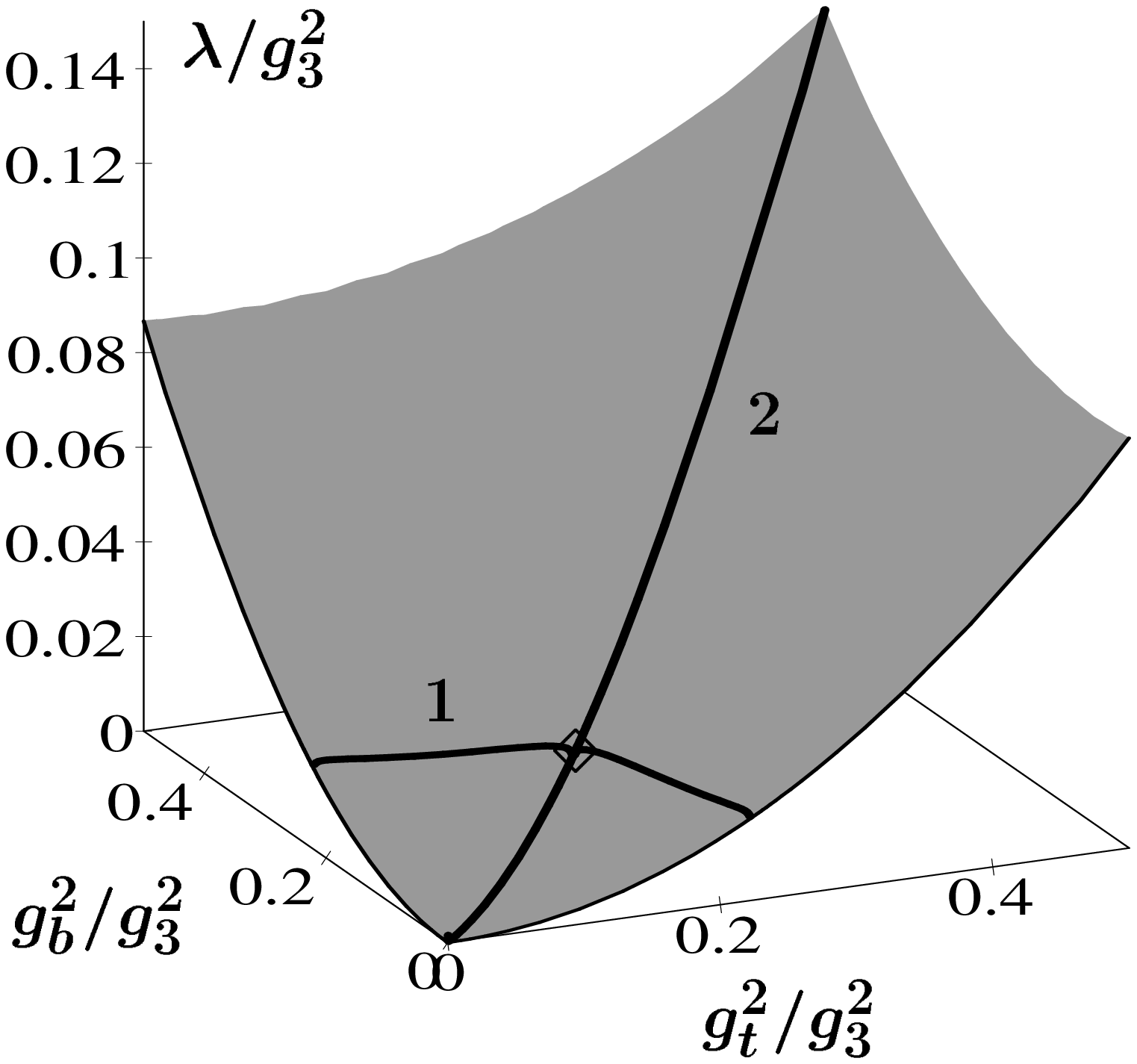,bbllx=118pt,bblly=253pt,
  bburx=558pt,bbury=663pt,height=4.9cm,width=4.9cm}\end{picture}}} 
& \\ \hline
\end{tabular}
\end{sideways}
}
\end{table}


\begin{table}
\parbox{171mm}{
\begin{sideways}
\begin{tabular}[b]{|c|c||c|c|c|c|} \hline

  \raisebox{-4mm}{model} & \raisebox{-4mm}{\begin{tabular}{c}considered\\
      couplings\end{tabular}} & \raisebox{-6mm}{IR
    fixed surface} & & \raisebox{-4mm}{\begin{tabular}{c}IR fixed line
    for given\\ initial values for \gee, \gz, \gd \end{tabular}} &
    \raisebox{-4mm}{\begin{tabular}{c}IR fixed point\\for $\mu=176\gev$
    \end{tabular}}
  \\ \hline


  \raisebox{-0mm}{SM} 
& \raisebox{0mm}{$g_{t},\,g_1,\,g_2,\,g_3$} 
& \raisebox{-21mm}{\parbox{48mm}{\begin{picture}(48,48)\epsfig{file=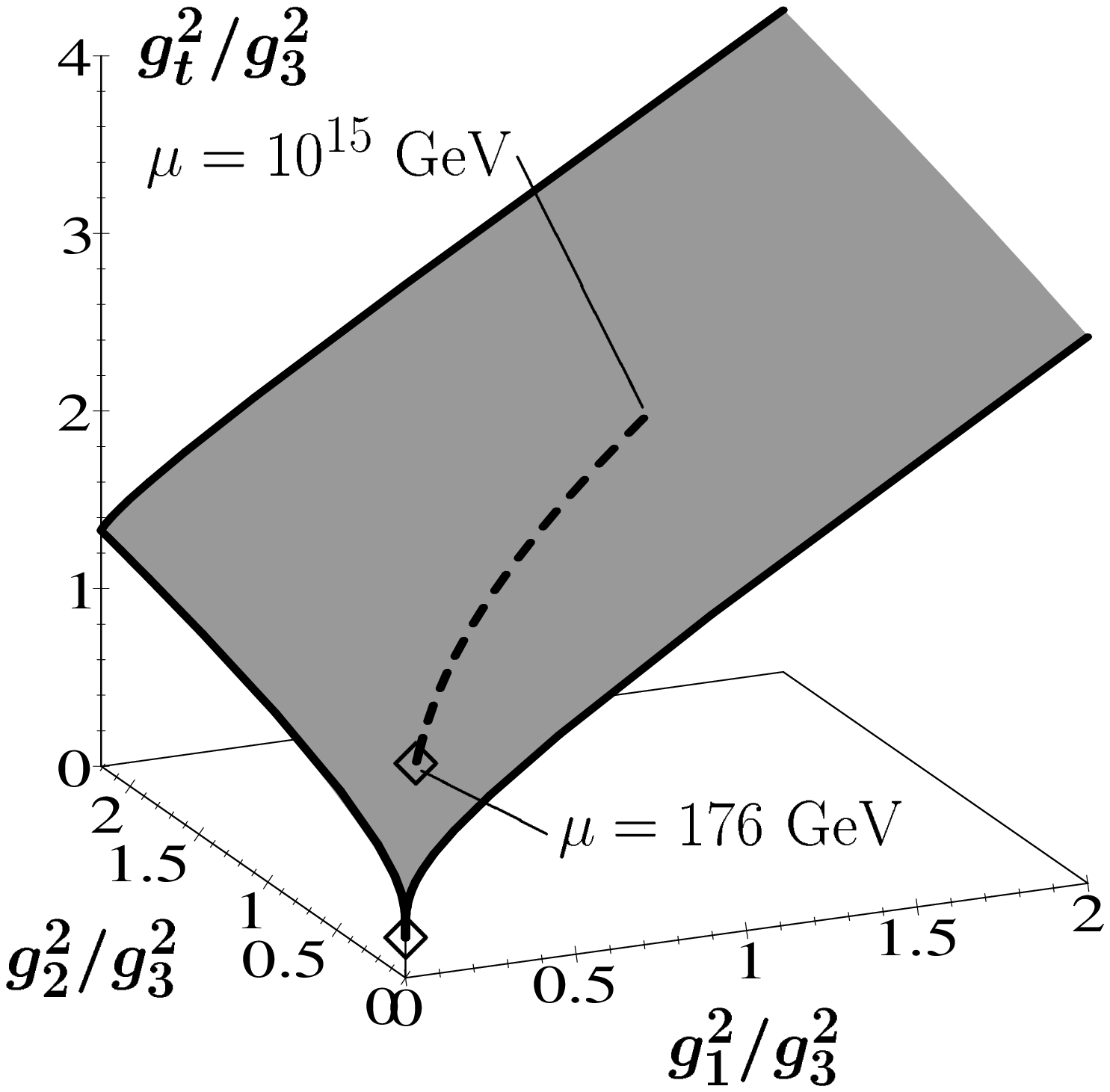,bbllx=127pt,
  bblly=254pt,bburx=548pt,
  bbury=675pt,height=4.8cm,width=4.8cm}\end{picture}}}
& \raisebox{0mm}{\begin{tabular}{c}given initial values\\ for
  $g_1,\,g_2,\,g_3$:\\[7mm] fat broken line in\\ the surface =\\
  fat line in plot\\ \gts/\gds\ versus 1/\gds\\\end{tabular}}
& \raisebox{-21mm}{\parbox{48mm}{\begin{picture}(48,48)
  \epsfig{file=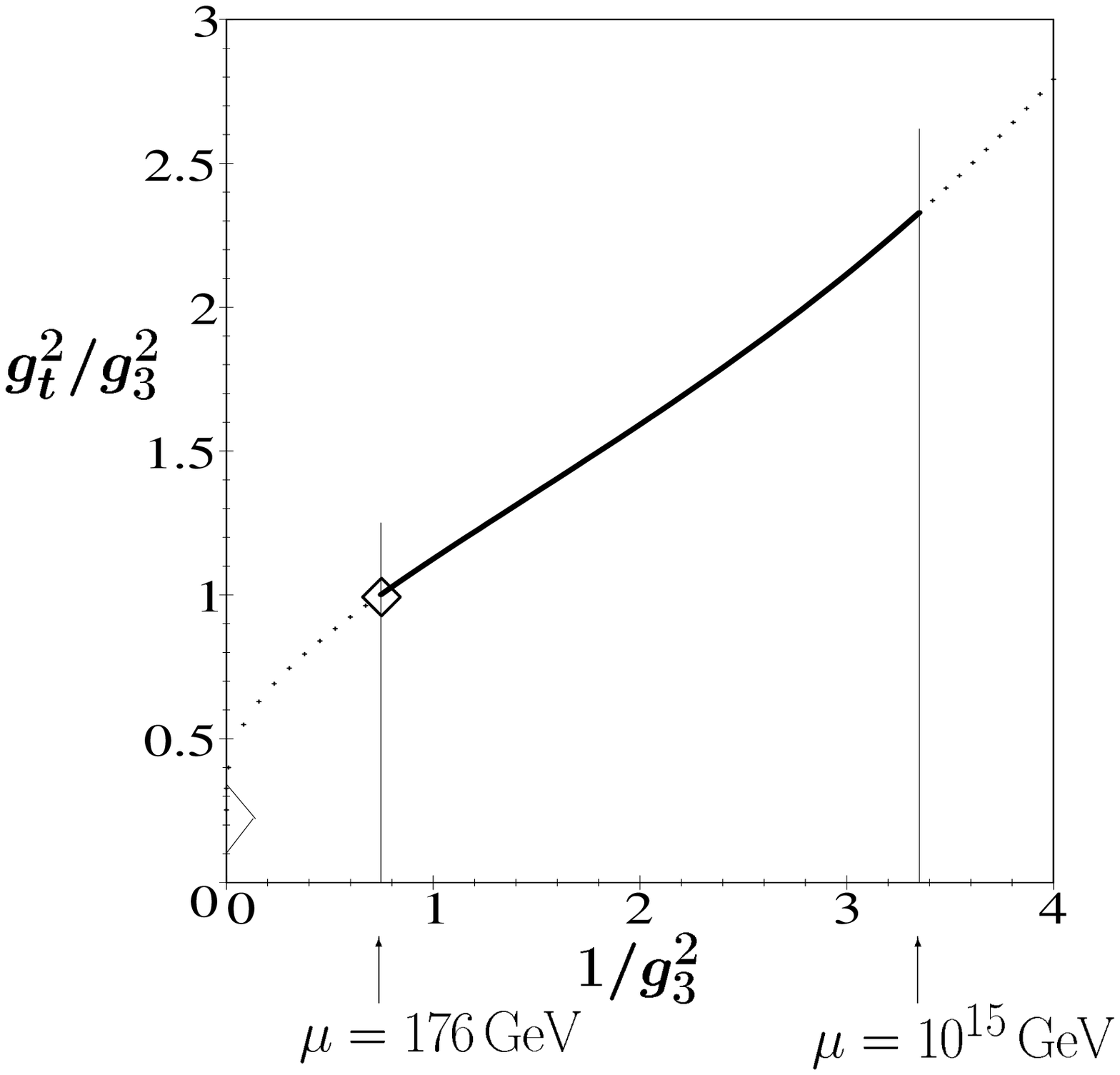,bbllx=14pt,bblly=313pt,bburx=490pt,
  bbury=768pt,height=4.8cm,width=4.8cm}\end{picture}}} 
& \raisebox{0mm}{symbol $\Diamond$ in figure}
\\ \hline


  \raisebox{0mm}{MSSM} 
& \raisebox{0mm}{$h_{t},\,g_1,\,g_2,\,g_3$}
& \raisebox{-23mm}{\parbox{48mm}{\begin{picture}(48,48)
  \epsfig{file=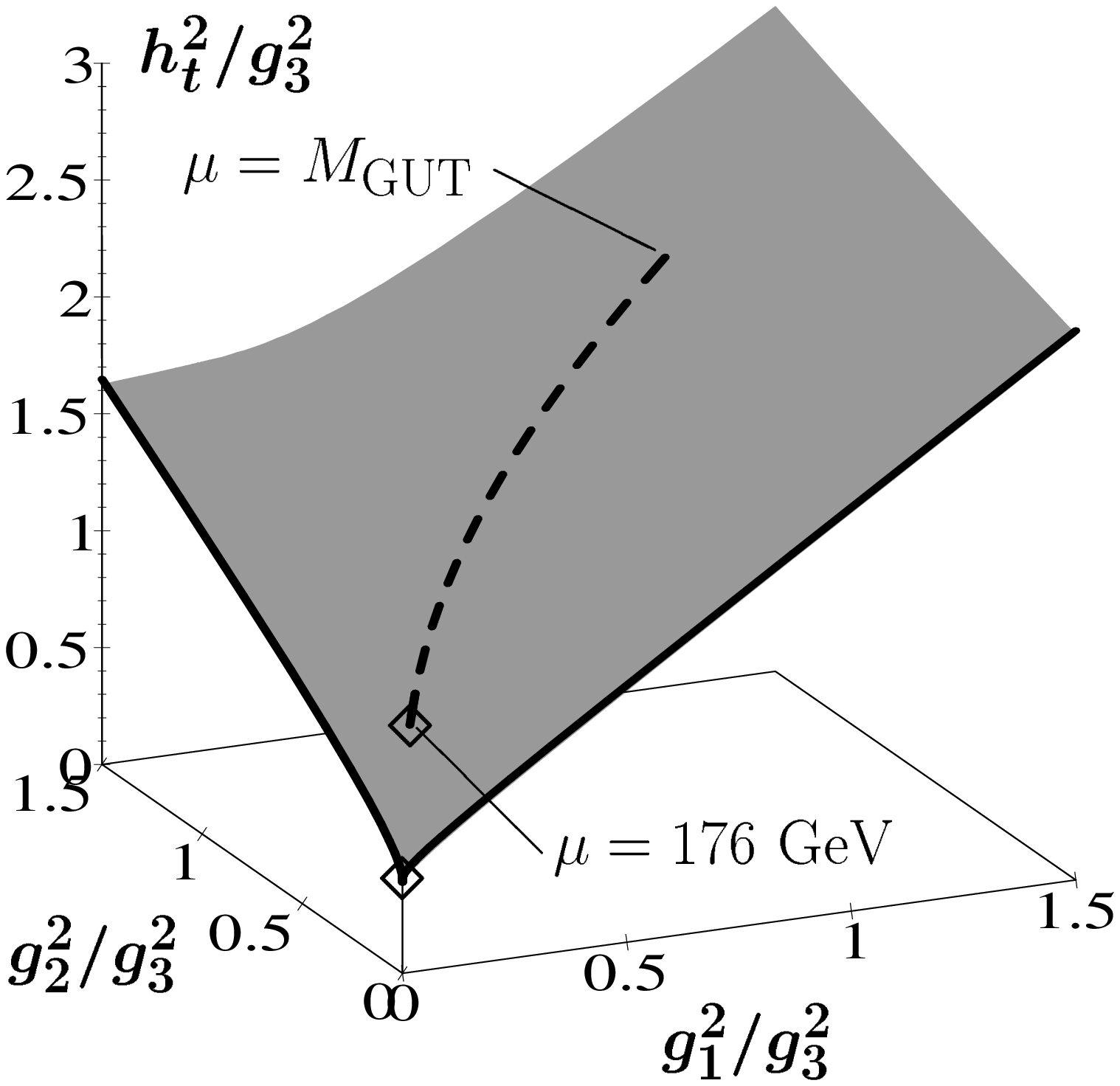,bbllx=125pt,bblly=254pt,bburx=562pt,
  bbury=679pt,height=4.8cm,width=4.8cm}\end{picture}}}
& \raisebox{0mm}{\begin{tabular}{c}given initial values\\ for
  $g_1,\,g_2,\,g_3$:\\[3mm] fat broken line in\\ the surface =\\
  fat line in plot\\ \hts/\gds\ versus 1/\gds\\ \end{tabular}}
& \raisebox{-23mm}{\parbox{48mm}{\begin{picture}(48,48)
  \epsfig{file=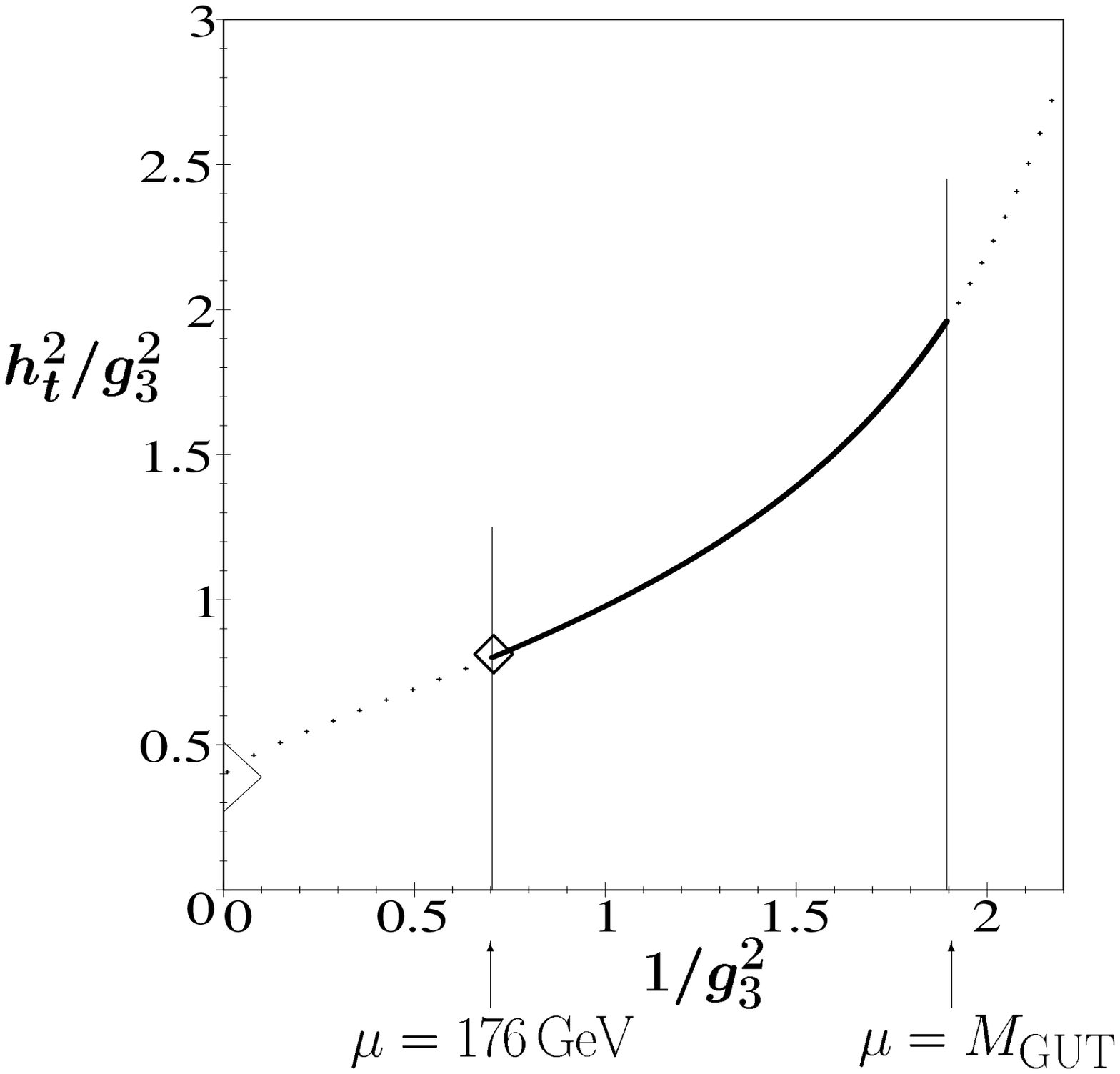,bbllx=16pt,bblly=315pt,bburx=492pt,
  bbury=769pt,height=4.8cm,width=4.8cm}\end{picture}}} 
& \raisebox{0mm}{symbol $\Diamond$ in figure}
\\ \hline

 & & \raisebox{-4mm}{\begin{tabular}{c} IR fixed surface for given\\
initial values of  \gee, \gz, \gd\end{tabular}} & \raisebox{-4mm}{\begin{tabular}{c}IR fixed line\\
    for $\mu=176\gev$ \end{tabular}} &
    \raisebox{-4mm}{\begin{tabular}{c}IR fixed point\\for $\mu=176\gev$
    \end{tabular}} & \\ \hline


  \raisebox{-2cm}{SM} 
& \raisebox{-2cm}{$g_{t},\,\gb,\,g_1,\,g_2,\,g_3$}
& \raisebox{-37mm}{\parbox{48mm}{\begin{picture}(48,48)\epsfig{file=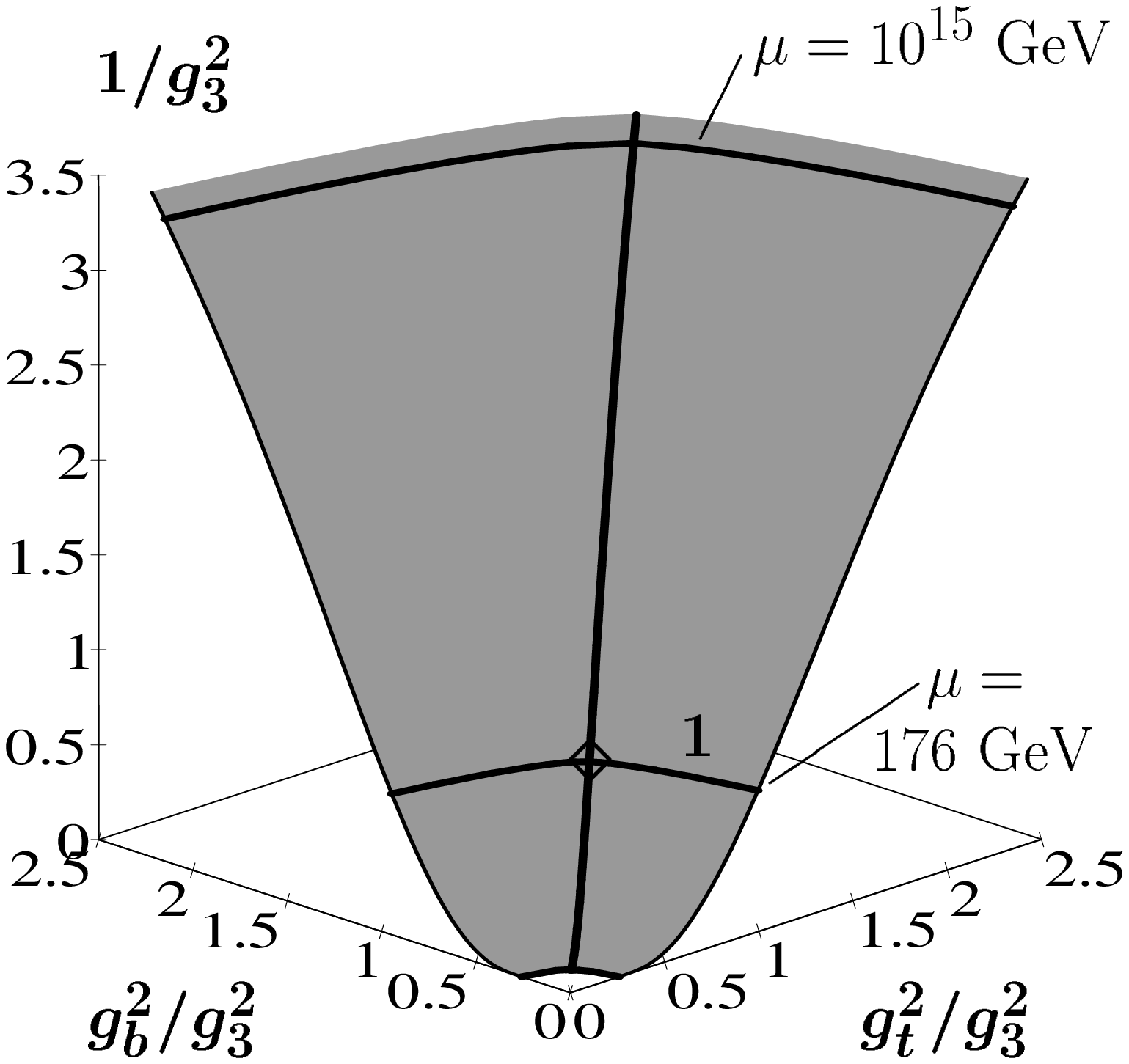,bbllx=120pt,
  bblly=270pt,bburx=561pt,
  bbury=685pt,height=4.8cm,width=4.8cm}\end{picture}}}
& \raisebox{-2cm}{fat line {\bf 1} in figure}
& \raisebox{-2cm}{symbol $\Diamond$ in figure}
&\\ \hline

\end{tabular}
\end{sideways}
}
\end{table}
 

\begin{table}
\parbox{171mm}{
\begin{sideways}
\begin{tabular}[b]{|c|c||c|c|c|} \hline

  \raisebox{-6mm}{model} & \raisebox{-6mm}{\begin{tabular}{c}considered\\
      couplings\end{tabular}} & \raisebox{-4mm}{\begin{tabular}{c} IR fixed surface for given\\
initial values of \gee, \gz, \gd\end{tabular}} & \raisebox{-4mm}{\begin{tabular}{c} IR fixed line\\
for $\mu=176\gev$\end{tabular}}&\raisebox{-4mm}{\begin{tabular}{c} IR fixed point\\
for $\mu=176\gev$\end{tabular}}
  \\[8mm] \hline


  \raisebox{-2.3cm}{MSSM} 
& \raisebox{-2.3cm}{$\htt,\,\hb,\,g_1,\,g_2,\,g_3$}
& \raisebox{-4cm}{\parbox{51mm}{\begin{picture}(50,50)\epsfig{file=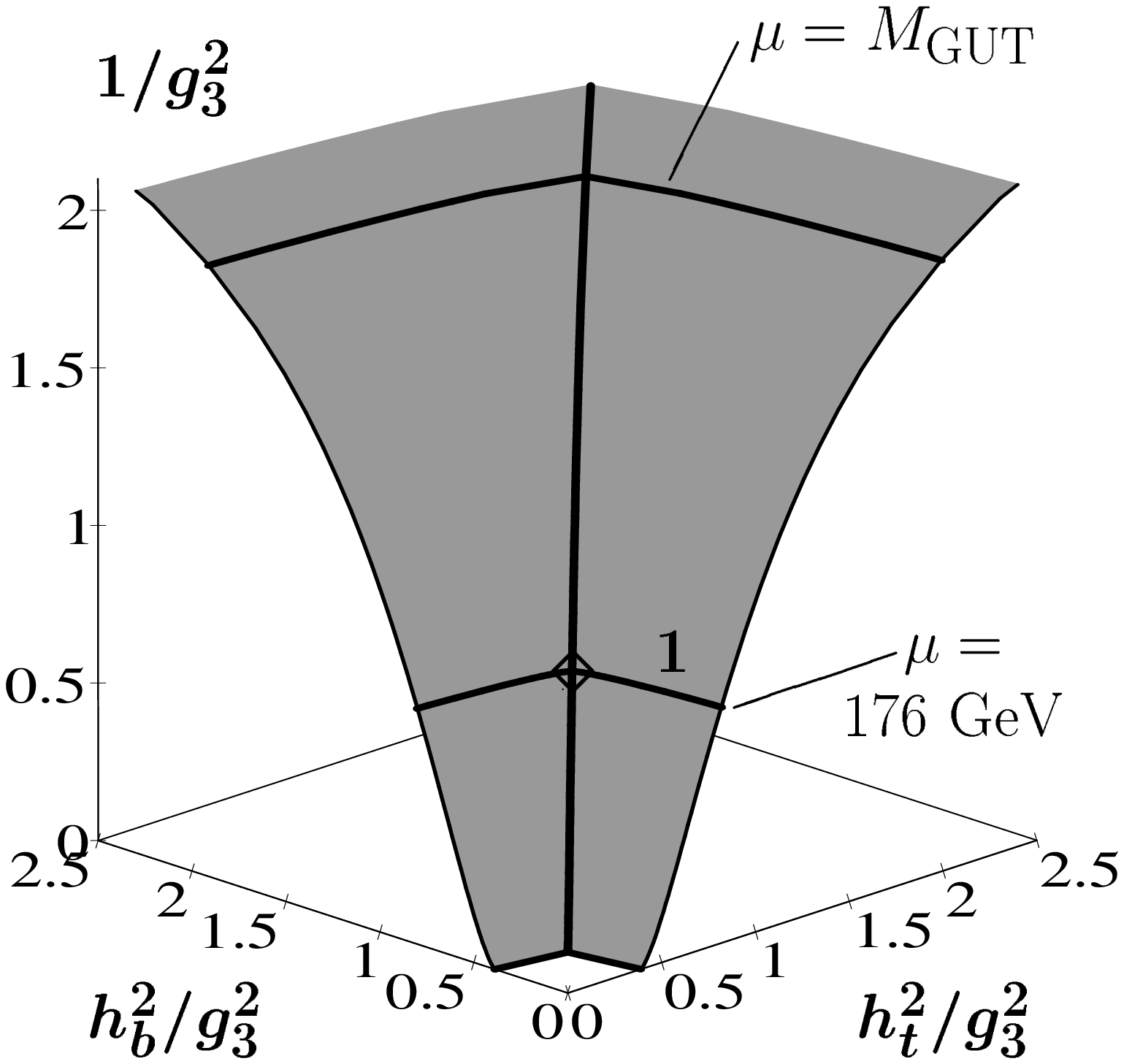,
bbllx=122pt,bblly=269pt,bburx=561pt,bbury=682pt,
height=5cm,width=5cm}\end{picture}}}
& \raisebox{-2.3cm}{fat line {\bf 1} in figure}
& \raisebox{-2.3cm}{symbol $\Diamond$ in figure}
\\ \hline



& 
& \raisebox{-4mm}{\begin{tabular}{c} IR fixed surface \\
for $\mu=176\gev$\end{tabular}} & & 
  \\[2mm] \hline


  \raisebox{-2.3cm}{SM}
& \raisebox{-2.4cm}{$\la,\,\gt,\,\gb,\,g_1,\,g_2,\,g_3$}
&
\raisebox{-42mm}{\parbox{51mm}{\begin{picture}(50,50)\epsfig{file=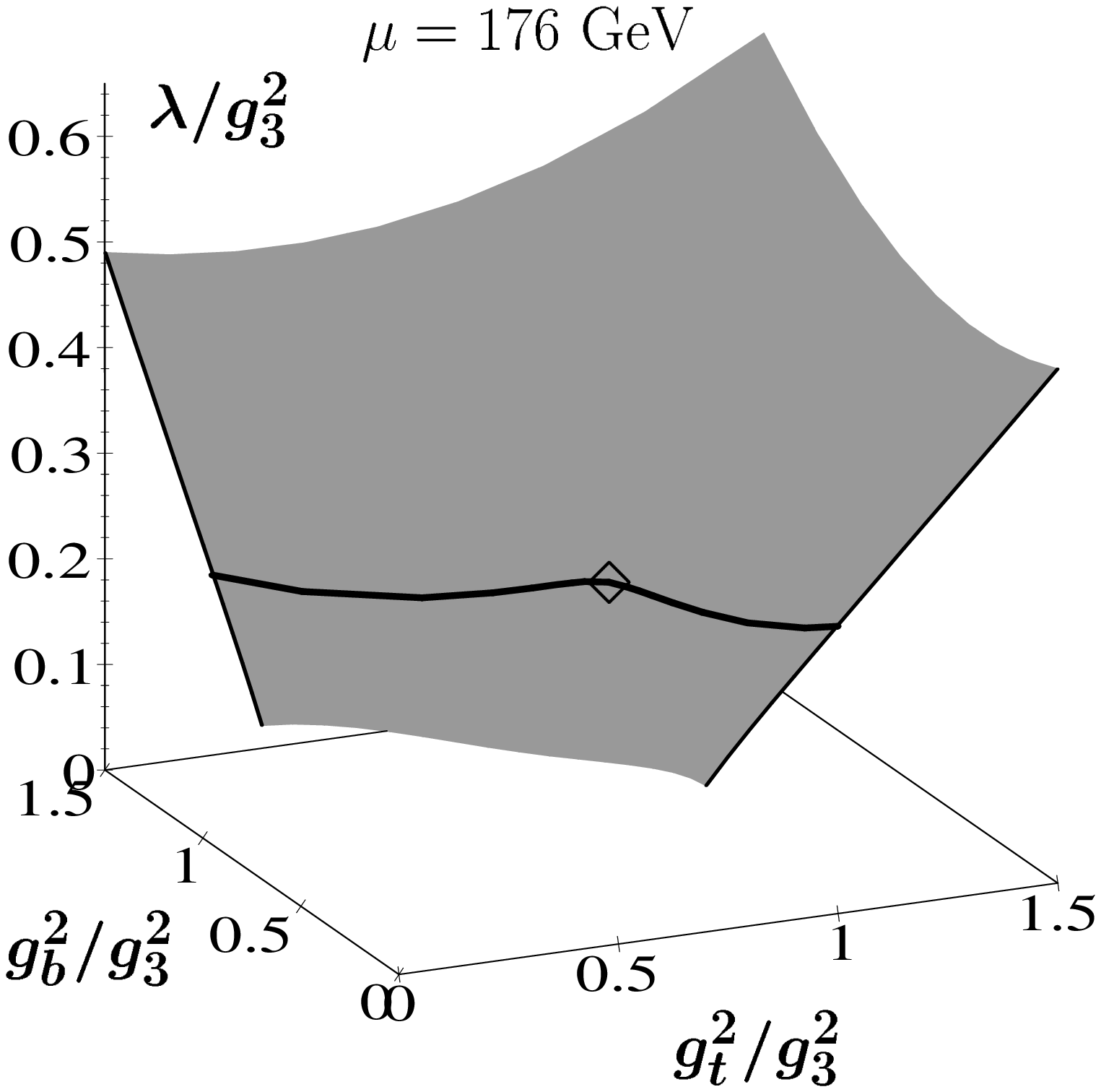,
bbllx=124pt,bblly=252pt,bburx=561pt,bbury=686pt,
height=5cm,width=5cm}\end{picture}}}
& \raisebox{-2.3cm}{fat line in figure}
& \raisebox{-2.3cm}{symbol $\Diamond$ in figure}
\\[2mm] \hline 
\end{tabular}
\end{sideways}
}
\caption[dum]{The IR fixed points, lines and surfaces, first in absence of
  the electroweak gauge couplings, then in presence of all gauge couplings.
  In both cases the number of couplings taken into account in the RGE
  increases entry by entry. With the exception of the Higgs selfcoupling
  \la, which is considered in the SM only, the SM and the MSSM are
  treated strictly in parallel. The table runs over four pages. Its
  detailed description is contained in the text of Sect. \ref{preview}.}
\label{fixman}
\end{table}

Table \ref{fixman} summarizes in four pages the RGE used (SM or MSSM), the sets of
nonzero couplings considered, the IR fixed points of the set of couplings
or of their ratios (characterized by a diamond in the figures), the IR
fixed lines (characterized by fat lines in the figures), the most
attractive IR fixed surface and, finally, bounds for couplings in
dependence on the UV cut-off scale \lam\ (characterized by thin lines in
the figures). As representative values $\lam
=10^4\gev,\;10^6\gev,\;10^{10}\gev$ and $10^{15}\gev$ have been chosen for
the SM and $\lam=\mgut\approx 2\cdot 10^{16}\gev$ for the MSSM.  Each horizontal entry line needs a few
commentaries which are given next.
\begin{itemize}
\item Starting with a single parameter, the Higgs self coupling \la\ of the
  SM, leads to the well studied four component $\Phi^4$ theory. The
  corresponding RGE for \la\ exhibits the much discussed ``trivial'' IR
  fixed point $\la=0$, attracting the (perturbatively admissible) RG flow,
  and leads to the well-known \lam\ dependent upper bounds which are well
  confirmed also in the framework of non-perturbative lattice calculations.
  The details are spelt out in Section \ref{higgs}.
\item The enlarged system of RGE for two parameters, the Higgs self
  coupling \la\ as well as the top Yukawa coupling \gt, has still a common
  ``trivial'' IR fixed point at $\la=\gt=0$, but shows already the
  remarkable feature of an IR attractive (linear) line, which is more
  strongly attractive than the IR fixed point; mathematically it is the
  special solution of the RGE which fulfills the boundary condition of a
  finite ratio $\la/\gts$ in the limit $\la,\gt\to\infty$. The RG flow from
  arbitrary (perturbatively allowed) UV initial values is then first
  towards the fixed line and subsequently close to or along this line
  towards the fixed point. Representative upper and lower bounds are
  displayed. The details are worked out in Section \ref{higgst}.
\item A next step includes the strong gauge coupling \gd. It is instructive
  to first discuss the set of two couplings, \gt\ and \gd\ in the SM, resp.
  \gd\ and \htt\ in the MSSM. It exhibits an IR attractive straight fixed
  line through the origin in the \gts-\gds\ plane resp. the \hts-\gds
  plane. More economically, this line (and similar ones in future
  variables) will be viewed as IR fixed points in the ratio variables
  \gts/\gds\ resp. \hts/\gds\ detailed in the table. The details are
  furnished in Sect. \ref{tgd}.
\item The set of three couplings \la, \gt\ and \gd\ is most economically
  reduced to the set of two ratios of couplings \la/\gds\ and \gts/\gds.
  This leads in the \la/\gds-\gts/\gds\ plane to a similar picture as in
  the \la-\gts\ plane in absence of \gd. However, the IR fixed point is not
  trivial any more. Again there is an IR fixed line, which is now
  nonlinear. It is characterized by an analogous boundary condition
  (finiteness for the ratio $\la/\gts$ in the limit $\la/\gds,\,\gts/\gds
  \to\infty$) and has the same slope in this limit; this fixed line is
  more strongly IR attractive than the fixed point.  Again the
  representative upper and lower bounds are displayed. The detailed
  discussion are found in Section \ref{Htgd}.
\item The set of non-zero variables \gt, \gb\ and \gd\ within the SM may be
  reduced to a discussion in the two ratios of variables \gts/\gds\ and
  \gbs/\gds; similarly the ratios \hts/\gds\ and \hbs/\gds\ become relevant
  in the MSSM. In both cases one finds qualitatively the same result: two
  IR fixed lines in the plane of the two ratios, the more attractive
  quarter-circle shaped one being marked by {\bf 1} the less attractive one
  by {\bf 2}. The IR attractive fixed point is at the intersection of the
  two lines. The RG flow is first towards the more attractive line and then
  close to or along this line towards the fixed point. The less attractive
  line as well as the fixed point imply exact top-bottom Yukawa coupling
  unification (in this approximation with vanishing electroweak gauge
  couplings) which is of eminent interest in the MSSM. The corresponding
  upper and lower bounds are also shown. The details are given in Sect.
  4.5.
\item The set of four parameters \la, \gt, \gb\ and \gd\ in the SM is
  discussed in the three-dimensional space of ratios of couplings \la/\gds,
  \gts/\gds\ and \gbs/\gds. There is a strongly IR attractive surface,
  containing all the fixed lines and fixed points which appeared in the
  discussion of the subspaces \gts/\gds\ versus \gbs/\gds, \la/\gds\ versus
  \gts/\gds\ and the corresponding one \la/\gds\ versus \gbs/\gds, which is
  trivially obtained by exchanging \gts\ for \gbs.  The RG flow is first
  towards the attractive surface, then close to or along the surface
  towards the more attractive fixed line {\bf 1} and finally close to this
  fixed line towards the top-bottom unifying IR fixed point. For details
  see Sect. 4.6.
\end{itemize}

So far, the electroweak gauge couplings \gee\ and \gz\ have been ignored.
They will be taken into account by enlarging the parameter space of ratios
of couplings \la/\gds, \gts/\gds\ and \gbs/\gds\ by adding the ratios
\ges/\gds\ and \gzs/\gds\ and treating them first as {\it free} variables. The common IR fixed point in this parameter space
is the so far determined fixed point in the variables \la/\gds, \gts/\gds\ 
and \gbs/\gds, supplemented by $\ges/\gds=\gzs/\gds=0$. This is an a
posteriori justification for ignoring \gee\ and \gz\, in a crude
approximation. Since, however $\gee(\mu=\mt)$ and $\gz(\mu=\mt)$ are
unequal to zero and thus the fixed point lies in a physically inaccessible region, the approximation is by far not good enough. In principle,
one has to look into the IR attractive four-dimensional surfaces in the
five-dimensional space of ratios of couplings, pick out the most strongly
attractive one and then feed in the known values for $\gee(\mu=\mt)$ and
$\gz(\mu=\mt)$ and discuss the resulting lower-dimensional manifold. These
become the IR attractive fixed manifolds which attract the RG flow for
increasing UV scale \lam\ and {\it fixed} IR scale, chosen to be
$\mu=\mt=176\gev$. This procedure is followed
in the table and the description below, again after
increasing the number of considered parameters in steps.
\begin{itemize}
\item The proposed procedure becomes most transparent, if first only the
  four parameters \gee, \gz, \gd\ and \gt\ resp. \htt\ are considered which
  allows to isolate and demonstrate by means of a figure the most
  attractive IR attractive surface in the three-dimensional space of the
  ratios \ges/\gds, \gzs/\gds\ and \gts/\gds\ resp. \hts/\gds. Within the
  surface the values of \gee/\gd\ and \gz/\gd\ are unconstrained. If one
  feeds in the experimentally determined values for \gee/\gd\ and \gz/\gd\ 
  at $\mu=m_Z$, the evolution of the two ratios from a high UV scale \lam\ 
  to this IR scale \mt=176\gev (with $\lam =10^{15}\gev$ in the SM and
  $\lam=\mgut \approx 2\cdot 10^{16}\gev$ in the MSSM) traces a line of
  finite length within the surface which is denoted by a fat dashed-dotted
  line in the figure. The variable \gts/\gds, resp. \hts/\gds, along this
  IR fixed line is displayed in a two-dimensional figure as a fat line,
  plotted conveniently as function of the variable 1/\gds. Also shown is
  the continuation of this RGE solution (small crosses) beyond the UV and
  IR scales. As expected it ends in the expected IR fixed point for
  $\ges/\gds=\gzs/\gds=0$. Note that at $\mu=\mt$ the value for \gts/\gds\ 
  resp. \hts/\gds\ is much higher than the IR fixed point value for
  $\gee=\gz=0$, a measure of the significant influence of \gee\ and \gz.
  The RG flow of \gts/\gds\, resp. \hts/\gds\ is then first very strongly
  attracted towards the IR fixed line (fat line) and then close to or along
  this line much more weakly towards the IR fixed point. For details see
  Sect. 5.1.
\item The set of parameters \gee, \gz, \gd, \gt\ and \gb\ leads to a
  three-dimensional IR fixed surface in the four-dimensional space of
  ratios \gts/\gds, \gbs/\gds, \ges/\gds\ and \gzs/\gds. Proceeding as
  above leads to an IR attractive two-dimensional surface for \gts/\gds\ 
  and \gbs/\gds\ versus 1/\gds. The relevant curve in the
  \gts/\gds-\gbs/\gds\ plane which replaces the fat IR attractive line of
  the case $\gee=\gz=0$, is read off for $\mu=\mt$. The corresponding
  figures for the MSSM are also shown. The details are spelt out in Sect.
  5.2.
\item Finally, including all parameters to be considered in this review,
  \la, \gt, \gb, \gee, \gz\ and \gd\ within the SM, lead to a
  four-dimensional IR attractive surface in the corresponding
  five-dimensional space of ratios \la/\gds, \gts/\gds, \gbs/\gds,
  \ges/\gds\ and \gzs/\gds. Feeding in as above the physical couplings
  \gee\ and \gz\ and evaluating them at $\mu=\mt$ leads to the
  two-dimensional surface in the \la/\gds-\gts/\gds-\gbs/\gds-space. For
  details see Sect. 5.3.
\end{itemize}   
 
Boundary conditions singling out the various IR manifolds, as far as not
mentioned above, will be discussed in Sects. \ref{noelweak} and
\ref{allgauge}.


\section {Infrared Fixed Points, Lines, Surfaces and Mass \newline Bounds 
in Absence of Electroweak Gauge Couplings\label{noelweak}}

This section fills in the information into Table \ref{fixman} in absence
of the electroweak gauge couplings, i.e. throughout this section 
\begin{equation}
\gee=\gz=0
\end{equation}
for all scales $\mu$; the couplings considered at the end of this section
will be \la, \gt, \gb\ (and marginally \gta) and \gds. This
section 
\begin{itemize}
\item may be considered as a first warm-up exercise with exact IR
  attractive fixed manifolds in the (one-loop) RGE and their physical
  implications, leading already to a reasonable approximation to physical
  reality.
\item It provides an excellent semi-quantitative insight into 
the dynamical origin of the triviality and vacuum stability bounds in the
Higgs-top mass plane, which become the tighter the larger is the UV
cut-off scale \lam.
\item Also it allows direct comparison with non-perturbative calculations
on the lattice which have been performed in the pure Higgs and the
Higgs-top sector of the SM in absence of all gauge couplings.
\end{itemize}
As advocated in Sect. \ref{preview} the procedure of gradual increase of
parameter space is followed, leading to less and less trivial
IR structures and furnishing increasingly improving approximations to
the SM resp. the MSSM . The inclusion of the electroweak couplings is
deferred to Sect. \ref{allgauge}. The final analysis, including two-loop RGE
and radiative corrections to the relations between couplings
and masses, is presented in Sect. \ref{masses}.

The concept of an IR attractive fixed point, line, surface,... will be
introduced step by step in conjunction with the applications. For
mathematical background reading we refer to Ref. \cite{guc}. For
completeness let us also add that the notions IR (UV) attractive and
repulsive used in this review are equivalent to the notions of IR (UV)
stable and unstable, respectively.  Furthermore, what physicists
prefer to call {\it fixed} lines, surfaces,..., is called by
mathematicians \cite{guc}, in fact more appropriately, {\it invariant}
lines, surfaces,... .

Let us remind the reader of the definition of a fixed point which will
allow most conveniently a generalization to fixed lines, surfaces,... .
The differential equation for the function $y(x)$ 
\begin{equation}
\frac{{\rm d}\,y}{{\rm d}\,x}=f(y)
\end{equation}
has a fixed point solution $y=c$ for constant $c$, if it stays at
$y(x)=c$ for {\it all} values of $x$ once its initial value
$y_0=y(x_0)$ is chosen equal to c. In this case of a single differential
equation (with a single dependent variable $y$) the fixed points are
identical with the zeroes of f(y). 

A system of $n$ coupled differential equations for $n$ dependent functions
$y_i(x)$
\begin{equation}
\frac{{\rm d\,}y_{i}}{{\rm d\,}x}=f_{i}(y_{1},y_{2},...,y_{n}),\ \ \
i=1,2,..,n,
\end{equation}
of the independent variable $x$ has a fixed point
\begin{equation}
y_{i}=c_{i}\ \ \  {\rm for}\ \ \  c_{i}={\rm const.},\ \ \ i=1,2,..,n
\end{equation}
if {\it all} $f_{i}$ vanish for $y_{i}=c_{i}$.

For future applications it is important to make the following point with
the aid of a simple example. Consider the set of two coupled differential
equations
\begin{eqnarray}
\frac{{\rm d\,}y_{1}}{{\rm d\,}x}&=&a y_1^2,\label{diff1}\\
\displaystyle{\frac{{\rm d\,}y_{2}}{{\rm d\,}x}}&=&b y_2(y_2-cy_1).
\label{diff2} 
\end{eqnarray} 
The right hand side of Eq. (\ref{diff2}) has a zero
at 
\begin{equation}
y_2=c y_1.
\label{flii}
\end{equation}
This equality may hold at some value $x=x_0$, signalling according to
the differential equation (\ref{diff2}) a vanishing of the derivative
of $y_2$ at $x=x_0$.  The relation (\ref{flii}), however, is not a
solution of the set (\ref{diff1},\ref{diff2}) of differential
equations for {\it all} values of x, i.e. it is {not} a fixed point of
the ratio $y_2/y_1$, or in other words not a fixed line in the
$y_2$-$y_1$-plane, (except at the point $y_1=0$): Eq. (\ref{diff1})
implies that $y_1$ is a non-constant function of $x$; correspondingly
$y_2$ in Eq. (\ref{flii}) is a non-constant function of $x$; this in
turn leads to a {\it nonvanishing left hand side} $ {\rm
d\,}y_{2}/{\rm d\,}x$ which does not match the {\it vanishing right
hand side}. A correct procedure is e.g. to rewrite the two equations
(\ref{diff1},\ref{diff2}) as
\begin{eqnarray}
\frac{{\rm d\,}y_{1}}{{\rm d\,}x}&=&a y_1^2,\nonumber\\
\displaystyle{\frac{{\rm d\,}\frac{y_2}{y_1}}{{\rm d\,}x}}&=&b
y_2\left(\frac{y_2}{y_1}-(c+\frac{a}{b})\right).\label{diffi}
\end{eqnarray}
Now, 
\begin{equation}
\frac{y_2}{y_1}=c+\frac{a}{b}
\end{equation} 
is indeed a fixed point of the of Eq. (\ref{diffi}) for the ratio $y_2/y_1$ , or
equivalently,
\begin{equation}
y_2=(c+\frac{a}{b})y_1
\label{flin}
\end{equation}
a fixed line in the $y_1$-$y_2$-plane. Obviously, the fake solution (\ref{flii}) is only
a good approximation to the correct solution (\ref{flin}), if $a/b\ll
c$. 

A fixed point is IR attractive or - equivalently IR stable - if it is
approached (asymptotically) by the ``top-down'' RG flow, i.e. by any
solution when evolved from the UV to the IR. IR fixed lines, surfaces,...
will be introduced most easily within the applications to follow. Like the
fixed points they will turn out to be special solutions, not determined by
initial value conditions but by boundary conditions. These boundary
conditions as a rule require a certain behaviour in a limit which is {\it
  outside} of the region of validity of perturbation theory. Nevertheless
the effect of IR attraction on the RG flow persists within the physical
perturbative region $\mt\lwig\mu\lwig\lam$.

In order to keep the discussion simple, let us fix the scales,
\begin{eqnarray}
{\rm the\ IR\ scale\ } \mu&=&\mt=176\gev,\\
\msusy&=&\mt=176\gev.
\end{eqnarray}
For the numerical calculations we use the experimental value \cite{par}
\begin{equation}
\alpha_3(m_Z)=0.117\pm 0.005\ \ {\rm leading\ to}\ \ \gds(\mt=176\gev)=
1.34,
\label{gdrei}
\end{equation}
if the three-loop QCD evolution \cite{marci} from $\mu=m_Z$ to
$\mu=\mt=176\gev$ is used.

The discussion in this section will be exclusively within the framework of
the one-loop RGE. In the absence of the electroweak gauge couplings the
one-loop contribution to the RGE may be read off from the general
expressions (\ref{RG1})-(\ref{RGtas}).

A partial decoupling of this coupled system of differential equations
may be obtained by introducing the following set of ratios of 
variables, the Higgs self coupling as well as the squares of the
Yukawa couplings divided by the square \gds\ of the strong gauge coupling,
\begin{eqnarray}
\rh&=&\frac{\la}{\gds}\ \ {\rm in\ the\ SM},\nonumber\\
\rt&=&\frac{\gts}{\gds}\ \ {\rm in\ the\ SM}\ \ {\rm  resp.}\ \
\rt=\frac{\hts}{\gds}\ \ {\rm in\ the\ MSSM},\nonumber\\ 
\rb&=&\frac{\gbs}{\gds}\ \ {\rm in\ the\ SM}\ \ {\rm  resp.}\ \
\rb=\frac{\hbs}{\gds}\ \ {\rm in\ the\ MSSM},\nonumber\\ 
\rta&=&\frac{\gtas}{\gds}\ \ {\rm in\ the\ SM}\ \ {\rm  resp.}\
\rta=\frac{\htas}{\gds}\ \ {\rm in\ the\ MSSM},
\label{ratios} 
\end{eqnarray}
and treating the ratio variables as functions of the independent
variable \gds. The resulting set of differential equations is
\begin{equation}
\begin{tabular}{c|c}
${\rm SM}$ & ${\rm MSSM}$\\ \\
${\displaystyle \frac{{\rm d}\,\gds}{{\rm d}\, t}
=-7\frac{g_3^4}{8\pi^2}}$ & ${\displaystyle \frac{{\rm d}\,\gds}{{\rm
d}\, t} =-3\frac{g_3^4}{8\pi^2}}$. \\ \\ 
$\ \ {\displaystyle -7\gds\frac{{\rm d}\,\rt}
{{\rm d}\,
\gds}=\rt(\frac{9}{2}\rt+\frac{3}{2}\rb+\rta-1)}\
\ $ &
$\ \ {\displaystyle -3\gds\frac{{\rm d}\,\rt}
{{\rm d}\, \gds}=\rt(6\rt+\rb-\frac{7}{3})}\ \ $\\ \\
$\ \ {\displaystyle -7\gds\frac{{\rm d}\,\rb}
{{\rm d}\,
\gds}=\rb(\frac{3}{2}\rt+\frac{9}{2}\rb+\rta-1)}\
\ $ &
$\ \ {\displaystyle -3\gds\frac{{\rm d}\,\rb}
{{\rm d}\, \gds}=\rb(\rt+6\rb+\rta-\frac{7}{3})}\ \
$\\ \\
$\ \ {\displaystyle -7\gds\frac{{\rm d}\,\rta}
{{\rm d}\,
\gds}=\rta(3\rt+3\rb+\frac{5}{2}\rta+7)}\
\ $ &
$\ \ {\displaystyle -3\gds\frac{{\rm d}\,\rta}
{{\rm d}\, \gds}=\rta(3\rb+4\rta+3)}\ \ $,\\ \\
\end{tabular}
\label{RGtbta}
\end{equation}
supplemented within the SM by
\begin{equation}
  {\displaystyle -7\gds\frac{{\rm d}\,\rh} {{\rm d}\, \gds}}={\displaystyle
    12\rh^2+6\rh\rt+6\rh\rb+2\rh\rta+7\rh-3\rt^2-3\rb^2-\rta^2}.
\label{RGh}
\end{equation}
It will be the basis of the Subsects. \ref{tgd}-\ref{Htbgd}.


\subsection{The Pure Higgs Sector of the SM \newline --
Triviality and an Upper Bound on the Higgs Mass\label{higgs}}

The first, though trivial, IR fixed point is met in the scalar sector of
the SM, i.e. the pure four-component $\phi^4$ theory in terms of a single
coupling, the Higgs selfcoupling \la. All the other SM couplings are
considered to be zero throughout this subsection. This model also allows
already a semi-quantative insight into the origin of the so-called triviality
bound \cite{cab}-\cite{lintriv} for the SM Higgs mass. Moreover, it allows comparison with
a large body of lattice calculations for the four component $\phi^4$ theory
which complement the perturbative results in the non-perturbative region.

The RGE in the pure Higgs sector is known even to three-loop order. In the
$\overline{{\rm MS}}$ scheme it is \cite{ladrei}
\begin{equation}
\frac{{\rm d}\,\la}{{\rm d}\,t}=\frac{3}{2\pi^2}\la^2-\frac{39}{32\pi^4}\la^3+\frac{7176+4032\zeta(3)}{(16\pi^2)^3}\la^4.
\label{rglam}
\end{equation}
The coefficient of the three-loop term is scheme dependent.

The key observation is that the RGE exhibits an IR attractive fixed
point\footnote{The two-loop differential equation has formally a further
  fixed point at $\la/(4\pi)=4\pi/13\simeq 0.97$ which lies, however,
  outside the range of validity of perturbation theory and is therefore
  physically meaningless.} at \la=0. Let us first list the relations which
will shed light on different facets of perturbative ``triviality'' and its
implications for the SM. For pedagogical purposes this is done in one-loop
order. The general solution of the RGE (\ref{rglam}) in one loop order is
\begin{equation}
\la(\mu)=\frac{1}{\frac{\displaystyle
1}{\la(\lam)}+\frac{\displaystyle 3}{\displaystyle 2\pi^2}
ln\frac{\lam}{\displaystyle \mu}}
\label{solla1}
\end{equation}
in terms of the unknown UV initial value \la(\lam). For $\mu$ increasing
beyond \lam, $\la(\mu)$ increases towards its Landau pole. We are, however
interested in the evolution towards the IR, i.e. in $\mu<\lam$. Since
$\la(\lam)>0$, $\la(\mu)$ is bounded from above by
\begin{equation}
\la(\mu)<\frac{2\pi^2}{3ln\frac{\lam}{{\displaystyle \mu}}}. 
\label{solla2}
\end{equation}

\begin{figure}
\begin{center}
\epsfig{file=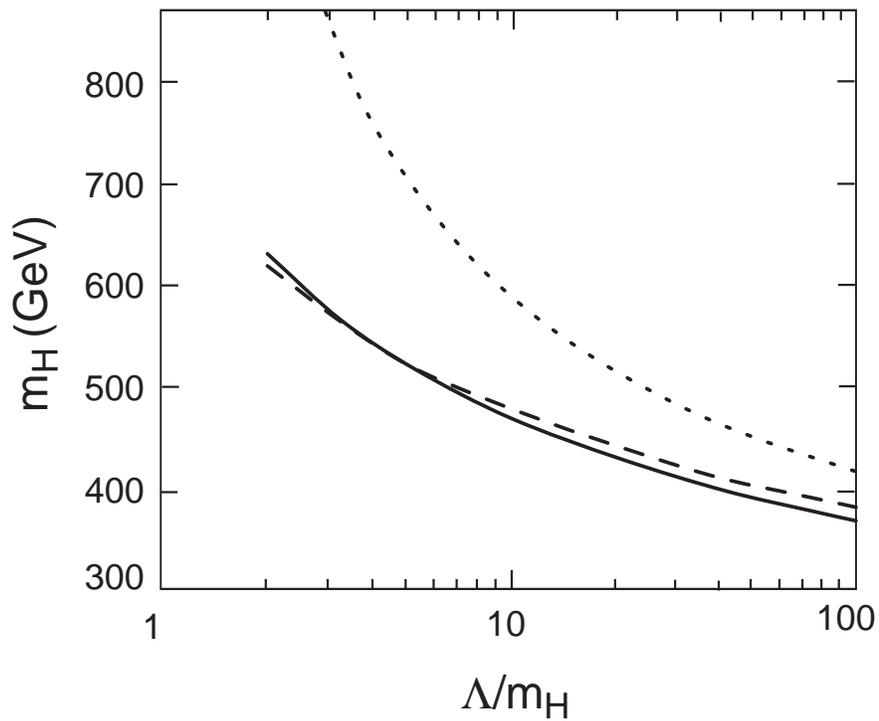,width=11.5cm}
\end{center}
\caption[dum]{Upper bound on the Higgs boson mass as a function of \lam/\mh,
  where \lam\ is the scale of new physics. The dotted curve is obtained by
  identifying \lam\ with the Landau pole of \la\ and is given by Eq.
  (\ref{tribo}). The solid curve \cite{marc} is the renormalization group improved
  unitarity bound (\ref{marval}). The dashed curve is the result of a
  lattice calculation of Ref. \cite{lues}. The figure was taken from Ref.
  \cite{marc}.}
\label{triv}
\end{figure}

Inserting the appropriate lowest order mass relation
$\mh=\sqrt{2\la(\mu=\mh)}\,v\,$ into the upper bound (\ref{solla2})
leads to the implicitely defined perturbative lowest order triviality
bound for the Higgs mass
\begin{equation}
\mh<\frac{2\pi/\sqrt{3}}{\sqrt{\ln\frac{\lam}{\mh}}}\;v
\label{tribo}
\end{equation}
with $v=(\sqrt{2}G_F)^{-1/2}$. It is exhibited in Fig. \ref{triv} as
dotted curve.

Eqs. (\ref{solla1})-(\ref{tribo}) may be interpreted as follows.
\begin{itemize}
\item The perturbative ``top down'' RG flow from $\mu=\lam$ to the IR
value at $\mu=\mh$, comprising all solutions of Eq.(\ref{triv})
starting from arbitrary perturbatively allowed initial values,
$\la(\lam)/(4\pi)<1$, is attracted {\it towards} the IR attractive
fixed point \la=0.
\item The IR value \la($\mu$) becomes the more independent of the
initial value \la(\lam) and the closer to the upper bound
(\ref{solla2}), the larger the UV initial value \la(\lam) is chosen
(within the framework of perturbation theory). The bound may be
interpreted to approximately collect the IR images of all
sufficiently large UV initial values.
\item The IR coupling $\la(\mu)$ as well as its upper bound
(\ref{solla2}) and the triviality bound (\ref{tribo}) for the Higgs
mass decrease with increasing value of the UV cut-off \lam.
\end{itemize} 

In the fictitious case of the pure $\phi^4$ theory, which as a
mathematical toy model need not be subject to any physical UV cut-off
\lam, the limit $\lam\rightarrow\infty$ can be performed and indeed
the full perturbative RG flow is drawn into the IR fixed point value
\la=0\, for the renormalized coupling, leading to a trivial
non-interacting theory (within the framework of perturbation theory,
discussed so far).

In the SM we do, however, expect a physical UV cut-off \lam\ to play the
role of the scale above which new physics enters, as expanded on in the
Introduction. So, as far as these simple-minded perturbative arguments go,
we expect a \lam\ dependent upper bound (\ref{tribo}) for the SM Higgs mass
which decreases for increasing UV cut-off \lam. As was first pointed out in
Ref. \cite{das}, this allows to determine an approximate {\it absolute}
(perturbative) upper bound for the Higgs mass: on the one hand the
triviality bound for the Higgs mass increases with decreasing \lam; on the
other hand the Higgs mass is a physical quantity, which the SM is supposed
to describe; so, for consistency, one has to require $\mh<\lam$. This
implies an absolute upper bound for the Higgs mass close to the region
where \mh\ and \lam\ meet.

A more subtle issue is to determine an absolute (perturbative) upper
bound for the Higgs mass \mh\ sufficiently much smaller than \lam,
such that the SM physics continues to hold even somewhat above \mh\
\begin{itemize}
\item without being too close to violating unitarity \cite{lee},\cite{marc},
\item without running out of the region of validity of perturbation
theory and 
\item without being significantly influenced by the nearby cutoff effects,
i.e. by the new physics becoming relevant at energies $O(\lam)$.
\end{itemize}
These three issues are of course not unrelated. In the following we shall
summarize efforts in the recent literature to determine such an absolute
bound for \mh\ on a quantitative level for each of these itemized issues
within the framework of perturbation theory. All of them apply to the full
SM and not only to its Higgs sector, which does, however, not make a
significant difference to the issue of an upper Higgs mass bound.  They end
up with very similar results. As we shall see, these results are also
supported in a mellowed form by non-perturbative lattice results in the
pure Higgs sector.  Altogether rather convincing conclusions can be drawn.

In order to implement the constraint of the unitarity bound, an interesting
improvement on the perturbative triviality bound (\ref{tribo}) was
introduced in Ref. \cite{marc}. The UV cut-off \lam\ is identified with the
momentum scale where perturbative unitarity is violated, an ansatz which goes
back to Ref. \cite{lee}. The tightest condition is obtained from the
(upper) unitarity bound on $|a_0(I=0)|$, where $a_0$ is the zeroth
partial wave amplidude of the isospin $I=0$ channel in $W_L W_L\rightarrow
W_L W_L$ scattering. The well known unitarity bound $|Re\,a_0(I=0)|\leq 1$
has been tightened to
\begin{equation}
|Re\,a_0(I=0)|\leq 1/2
\end{equation}
in Ref. \cite{lues}. In the limit of center of mass energy $\gg \mh$
the tree level expression for $a_0(I=0)$ is
\begin{equation}
a_0(I=0)=-5\la/(16\pi);
\end{equation}
thus, one may conclude that the maximally allowed value for \la(\lam)
is
\begin{equation}
\la(\lam)\le\frac{8\pi}{5}.
\end{equation}
Feeding this inequality into the one-loop relation (\ref{solla1}) leads
to the improved perturbative one-loop triviality bound
\begin{equation}
\mh\le\frac{2\pi/\sqrt{3}}{\sqrt{\ln\frac{\lam}{\mh}+\frac{{
\displaystyle 5\pi}}{{\displaystyle 12}}}}\;v
\label{marval}
\end{equation}
with $v=(\sqrt{2}G_F)^{-1/2}$. This so-called renormalization group
improved unitarity bound is clearly tighter than the bound (\ref{tribo});
it is exhibited as solid curve in Fig. \ref{triv}. One can read off that an
absolute upper Higgs mass bound, required to be a factor two below the
cut-off \lam, is reached for $\mh\simeq 600\gev$; applicability up to
$\lam=2\,{\rm TeV}$ requires a bound $\mh\lwig 530\gev$.

A recent quantitative analysis of the breakdown of perturbation theory for
a large Higgs mass was performed in Ref. \cite{ries}. The $\mu$ dependence
within different renormalization schemes (the $\overline{{\rm MS}}$ scheme
and the on mass shell scheme) is investigated in three physical observables
which are known at two-loop level. The criterion for validity of
perturbation theory is that the dependence on the renormalization scale
$\mu$ as well as on the scheme should diminish order by order in \la. The
conclusion \cite{ries} is that perturbation theory breaks down for
\mh=O(700\gev) and \mh\ must be $\mh\lwig{\rm O}(400\gev)$ for
perturbatively calculated cross sections to be trustworthy up to center of
mass energies of O(2\,TeV).

The effects of a nearby cut-off \lam\ in case of a large Higgs mass in
terms of corrections of the order of \mh/\lam\ have been
studied in Ref. \cite{cort}. The starting point is the contribution
of a virtual Higgs particle in the one-loop vacuum polarization diagrams of
weak gauge bosons
\begin{equation}
\Pi_{ij}^{\mu\nu}=-ig^{\mu\nu}(A_{ij}+q^2F_{ij}(q^2))+q^{\mu}q^{\nu}\ {\rm
  terms}, 
\end{equation}
where $i$ and $j$ stand for $W^{\pm},\ Z,\,\gamma$. At $q^2\leq m_Z^2$ only
$A_{ij}$ and the $F_{ij}(0)$ are retained, moreover
$A_{\gamma\gamma}=A_{\gamma Z}=0$ as required from the Ward identities, so
that the loop corrections are contained in the six quantities $A_{WW},\
A_{ZZ},\ F_{WW},\ F_{ZZ},\ F_{\gamma
  Z},\,F_{\gamma\gamma}$. Through renormalization three independent
combinations enter the definition of $\alpha,\ G_F,\ m_Z$ and there remain
three finite and scheme independent parameters, two of which show a
logarithmic dependence on the Higgs mass
\begin{eqnarray}
\epsilon_1&=&\frac{A_{ZZ}}{m_Z^2}-\frac{A_{WW}}{m_W^2}=\frac{3}{4}\frac{G_F\,\mt^2}{2\sqrt{2}\pi^2}+\frac{3}{4}\frac{G_F\,m_W^2}{2\sqrt{2}\pi^2}\tan^2\theta_W\ln(\frac{m_W^2}{\mh^2}),\nonumber\\
\epsilon_3&=&\cot\theta_W
F_{30}=\frac{1}{6}\frac{G_F\,m_W^2}{2\sqrt{2}\pi^2}\log(\frac{m_W^2}{\mt^2})-\frac{1}{12}\frac{G_F\,m_W^2}{2\sqrt{2}\pi^2}\ln(\frac{m_W^2}{\mh^2}),
\label{eps}
\end{eqnarray} 
where indices 3 and 0 stand for $W_3=\cos\theta_W\,Z+\sin\theta_W\,\gamma,\ 
W_0=\cos\theta_W\,\gamma-\sin\theta_W\,Z$ and $\theta_W$ is the Weinberg
angle. These two quantities are chosen as probes for the sensitivity to a
cut-off \lam.

\begin{sidewaysfigure}
\begin{tabular}{cc}
\boldmath $\epsilon_1$ & \boldmath $\epsilon_3$\\
\epsfig{file=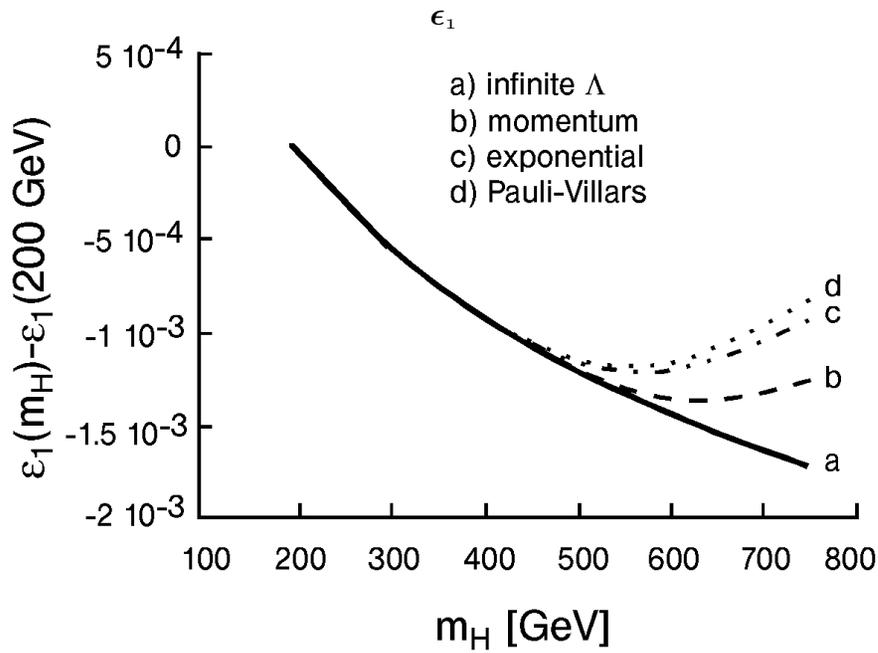,angle=-90,width=11.5cm}&
\epsfig{file=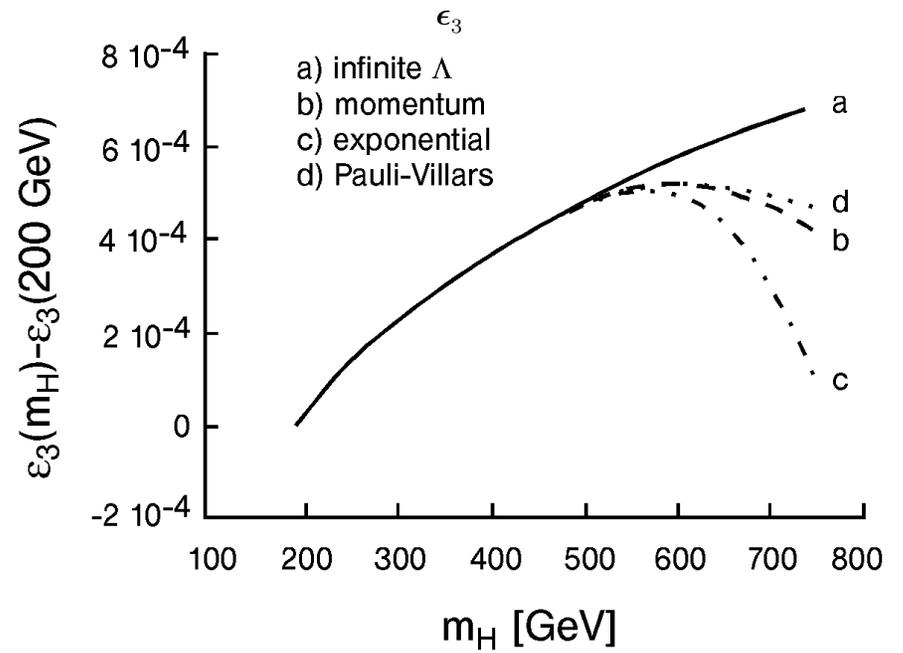,angle=-90,width=11.5cm}
\end{tabular}
\caption[dum]{Higgs boson mass dependence of the two quantities $\epsilon_1$ and
  $\epsilon_3$. The four curves represent the value at infinite cut-off and
  with finite cut-off corrections in the three regularization schemes
  analyzed. The figure was taken from Ref. \cite{cort}.}
\label{corte}
\end{sidewaysfigure}

The \mh\ dependent cut-off scale \lam, at which new physics is expected, is
extracted \cite{cort} from the inequality (\ref{tribo}) by replacing \mh\ 
in the logarithm on the right hand side by $v/\sqrt{2}$ and assuming \mh\ on the
left hand side to reach its maximal value
\begin{equation}
\lam=\frac{v}{\sqrt{2}}\,\exp\left(\frac{8\pi^2(v/\sqrt{2})^2}{3\mh^2}\right).
\label{cutoff}
\end{equation}
When a theory becomes sensitive to cut-off effects, it also becomes
regularization scheme dependent. The idea \cite{cort} is now to simulate
the effects of new physics at the \mh\ dependent cut-off \lam\ by
calculating the quantities $\epsilon_1$ and $\epsilon_3$ in three different
regularization schemes, i) by introducing a cut-off as an upper limit for
the momentum integration, ii) by employing the exponential representation
of euclidean propagators and ii) the Pauli-Villars regularization. In all
cases the relevant dimensionful cut-off parameter is replaced by Eq.
(\ref{cutoff}).  Results for the \mh\ dependence of $\epsilon_1$ and
$\epsilon_3$ in the different schemes, together with the result
(\ref{eps}), valid for $\lam\rightarrow\infty$, are shown in Fig.
\ref{corte} \cite{cort}.  In both cases the four curves split at a Higgs
mass higher than O(500\gev) due  to regularization scheme dependent
contributions of order \mh/\lam\ and one cannot trust the predictions of
the SM any more.

In summary, different sources within the framework of perturbation theory
come to the conclusion that the SM calculations as  functions of increasing
Higgs mass are trustworthy up to the upper bound
\begin{equation}
\mh\simeq {\rm O}(500\gev).
\end{equation}

The harder part of the proof of triviality and its implications for a
triviality bound for the SM Higgs mass is to continue it into the region of
large couplings, $\la(\lam)/(4\pi )\gwig 1 $, where perturbation theory
breaks down. In the early days close to rigorous proofs of triviality of
the one-component $\phi^4$ theory \cite{fro} and of the large $N$ limit of
the N-component theory \cite{barde} were given.  Meanwhile, the dominant
tool is lattice $\phi^4$ theory with the inverse lattice spacing playing
the role of the cut-off \lam, i.e. of the scale of new physics. Starting
with Refs. \cite{freedm}, \cite{das}, \cite{cal}, the triviality of the
scalar sector of the SM and therefore the need to retain a finite UV cutoff
\lam\ has meanwhile been established for some time by analytical and
numerical lattice calculations \cite{lue}, \cite{lues}, \cite{kut}. The
calculations are based on a representative class of lattice actions, all of
which respect the property of reflection positivity, the property in this
Euclidean formulation which corresponds to unitarity in Minkowski space.  A
lattice triviality bound \cite{lues} is shown in Fig. \ref{triv}
\cite{marc}. It is in surprisingly good agreement with the renormalization
group improved unitarity bound. A conservative conclusion \cite{hel} on an
absolute triviality bound for the Higgs mass as obtained from a
representative class of lattice actions is the following: the SM will
describe physics to an accuracy of a few percent up to energies of the
order 2 to 4 times the Higgs mass \mh\ only if
\begin{equation}
\mh\le 710\pm 60\,\gev.
\label{latt}
\end{equation}

Altogether, the non-perturbative lattice calculations confirm the results
obtained within the perturbative framework and even relax the absolute
upper bound on \mh\ somewhat.

The Higgs mass triviality bound is a weak function of the top Yukawa coupling,
resp. the top mass. This dependence will be explicitely discussed in
Sects. \ref{higgst}, \ref{Htgd} and \ref{secSM}.


\subsection{The Higgs-Top Sector of the SM\newline -- a first
IR Fixed Line and a First Vacuum Stability Bound\label{higgst}}

In the next step towards the SM a reduction to two coupling
parameters, the Higgs selfcoupling \la\ and the top Yukawa coupling \gt\ is
considered,
\begin{equation}
\la\neq 0,\;\gt\neq 0 \hspace{2cm}{\rm with\ all\ the\ other\ couplings}\,=0.
\end{equation}
This reduced system is of interest since even though it is still
``trivial'' with respect to both couplings, \la\ and \gt, it exhibits - as
we shall see - a non-trivial IR fixed line in the plane of the two
couplings which is more strongly IR attractive than the fixed point and it
can still be solved analytically. This fixed line turns out to be the
dynamical origin i) for the triviality bound, the upper bound of the Higgs
mass, which in this two coupling framework becomes a weak function of the
top mass as well as ii) for the lowest order vacuum stability bound, the
lower bound on the Higgs mass again being a function of the top mass.
Furthermore important features find again support by non-perturbative
lattice calculations in this \la-\gt-framework.

The corresponding RG equations (\ref{RGt}) and (\ref{RGla}) reduce in
one-loop order to
\begin{eqnarray}
\frac{{\rm d}\,\gts}{{\rm d}\,t}&=&\frac{9}{16\pi^2}g_t^4\label{gtred}\\
\frac{{\rm d}\,\la}{{\rm d}\,t}&=&\frac{6}{16\pi^2}(4\la^2+2\la\gts
-\gt^4)\label{lared}.
\end{eqnarray}
This is a coupled system of differential equations. A first
observation is that it exhibits a common 
\begin{equation}
{\rm IR\ attractive\ trivial\ fixed\ point\ at}\hspace{1cm}\la=0,\,\gt=0.
\label{fp}
\end{equation}
If the variable
\begin{equation}
R=\frac{\la}{\gts}
\end{equation}
is introduced, the system of RG equations (\ref{gtred}), (\ref{lared}) may be
rewritten in a {\it decoupled} form with {\it nested} solutions
$\gt=\gt(\mu)$ and $R=R(\gt(\mu))$
\begin{eqnarray}
\frac{{\rm d}\,\gts}{{\rm d}\,t}&=&\frac{9}{16\pi^2}g_t^4\\
\gts\frac{{\rm d}\,R}{{\rm d}\,\gts}&=&\frac{1}{3}(8 R^2+R-2).
\label{RGER}
\end{eqnarray}
The right hand side of the differential equation (\ref{RGER}) has 
two zeroes
\begin{equation}
R=\overline{R}=\frac{1}{16}(\sqrt{65}-1),\hspace{5mm}
R=\overline{\overline{R}}=\frac{1}{16}(-\sqrt{65}-1).
\end{equation}
The positive zero $\overline{R}$ is \cite{wet} an  
\begin{equation}
{\rm IR\ attractive\ fixed\ point\ in\ the\ variable}\,\;R{\rm\,\;at}
\hspace{5mm}R=\overline{R}=\frac{1}{16}(\sqrt{65}-1)\simeq 0.441.
\label{fl}
\end{equation}
An analogous IR fixed point had already been pointed out in an earlier
publication \cite{wet2} in application to a fourth heavy fermion
generation.  Here we meet for the first time a fixed point in the
ratio of couplings $R=\la/\gts$, which has also been termed (in a
context to be discussed in the next subsection) a ``quasi-fixed point''
\cite{pen}. In the $\la$-$\gt$-plane it corresponds to a fixed line,
which is linear and goes through the origin
\begin{equation}
\la=\frac{1}{16}(\sqrt{65}-1)\,\gts.
\label{fli}
\end{equation}
It has the property that the solution stays on the fixed line for {\it all}
values of $\mu$, once its initial value is chosen on it.  Its analytical
form is independent of initial values \lam, $\gt_{0}$ and $\la_0$. It is
the solution of the RGE which is defined by the boundary condition that the
ratio \la/\gts\ has a {\it finite} non-zero value in the limit
$\la,\,\gts\rightarrow\infty$; as announced earlier, this boundary
condition refers to a limit which leads outside the region of validity of
perturbation theory; nevertheless the IR attraction of this IR fixed line
applies to the RG flow in the perturbative region.

Going beyond Ref. \cite{wet}, one can determine analytically the way in which the
``top-down'' RG flow approaches the IR fixed line and finally the IR fixed
point from the analytical solution\footnote{This solution was first written
  down by F. Schrempp (unpublished).} to the system of differential equations
(\ref{RGER})
\begin{eqnarray}
\gts(t)&=&
\frac{g_{t0}^2}{1+\frac{{\displaystyle 9}}{{\displaystyle 16\pi^2}}g_{t0}^2\ln
\frac{\lam}{{\displaystyle \mu}}}\\ \nonumber\\
R(\gts)&=&\frac{\la(\gts)}{\gts}=\frac{\overline{R}-{\displaystyle \frac{R_0-
\overline{R}}{R_0-\overline{\overline{R}}}}\;\overline{\overline{R}}\;
({\displaystyle{\frac{\gts}{g_{t0}^2}}})^{\sqrt{65}/3}}{1-{\displaystyle
\frac{R_0-\overline{R}}
{R_0-\overline{\overline{R}}}}\;({\displaystyle
\frac{\gts}{g_{t0}^2}})^{\sqrt{65}/3}}. \label{geso}
\end{eqnarray}
for arbitrary (perturbatively allowed) initial values 
\begin{equation}
g_{t0}=\gt(\lam),\;\;\;R_0=\frac{\lambda_0}{g_{t0}^2}=\frac{\la(\lam)}
{\gts(\lam)}.
\end{equation}
 
Evolving from the UV to the IR, $\gt(\mu)$ indeed approaches 0 and
correspondingly the RG flow approaches the value $R=\overline{R}$,
i.e. the IR fixed line. A measure for the strength of IR attraction
towards the fixed line is the exponent in
$(\gts/g_{t0}^2)^{\sqrt{65}/3}$ which is large, $\sqrt{65}/3\simeq
2.69$. So, the RG flow is roughly as follows: first towards the IR
fixed line and then close to it or along it towards
the IR fixed point (\ref{fp}). 

The IR attractive line $\la=((\sqrt{65}-1)/16)\,\gts$ is the dynamical
origin for the triviality bound as well as for the vacuum stability bound.
The first important observation is that {\it no} solution of the considered
RGE can {\it cross} the IR fixed line. The line is the lower bound for all
solutions starting the top-down evolution from an initial value above it
and the upper bound for all solutions starting the evolution from an
initial value below it. The solutions starting from initial values
sufficiently closely above or below the line will end up on the line. The
IR images of the solutions starting from the largest possible initial value
of the Higgs self coupling \la\ allowed by perturbation theory will
constitute the upper, i.e. triviality bound for the IR values; according to
Sect. \ref{vac} the IR images of the lowest possible initial values of \la,
which are $\la_0$=0, constitute the lower, i.e. vacuum stability bound in
these lowest order considerations. Given a finite evolution path from some
initial scale $\mu=\lam$ to the IR scale, these bounds will be at some
finite difference from the IR fixed line, their position strongly depending
on \lam; they will be the closer to the line the larger the value of \lam.
For $\lam\rightarrow\infty$, which is of mathematical interest only, the RG
flow will first contract towards the IR fixed line and then towards the IR
fixed point \la=0, \gts=0. These lowest order bounds may be translated into
lowest order Higgs mass bounds in terms of the top mass in the Higgs-top
mass plane, using the lowest order relations $\mh=\sqrt{2\la(\mu=\mh)}v$
and $\mt=(v/\sqrt{2})\gt(\mu=\mt)$. The analytical formula for the \gts\ 
dependence of the triviality bound may be obtained from the
$\la_0\rightarrow\infty$, i.e. $R_0\rightarrow\infty$ limit of the general
solution (\ref{geso})
\begin{equation}
R(\gts)_{\rm triv}=\frac{\la(\gts)}{\gts}_{\rm triv}=\frac{\overline{R}-\overline{\overline{R}}
({\displaystyle{\frac{\gts}{g_{t0}^2}}})^{\sqrt{65}/3}}{1-({\displaystyle
\frac{\gts}{g_{t0}^2}})^{\sqrt{65}/3}},
\end{equation}
the one for the lowest order vacuum stability bound from the solution
(\ref{geso}) with initial value $\la_0=0$, i.e. $R_0=0$,
\begin{equation}
R(\gts)_{\rm vac. stab.}=\frac{\la(\gts)}{\gts}_{\rm
vac. stab.}=\frac{\overline{R}\left(1-
({\displaystyle{\frac{\gts}{g_{t0}^2}}})^{\sqrt{65}/3}\right)}{1-\frac{\overline{R}}{\overline{\overline{R}}}({\displaystyle
\frac{\gts}{g_{t0}^2}})^{\sqrt{65}/3}}.
\end{equation}

The corresponding figure in Table \ref{fixman} shows the \la-\gts-plane
with the trivial IR attractive fixed point (\ref{fp}) indicated as diamond,
the IR attractive fixed line (\ref{fli}) drawn as fat line and the
triviality upper and the vacuum stability lower bounds for four values of
\lam, $\lam=10^4,\;10^6,\;10^{10},$ $10^{15}\gev$, drawn as thin lines,
which cut out wedgeformed allowed regions which decrease for increasing
\lam. A thorough discussion including figures, more quantitative information about the degree of IR attraction as
well as proper reference to the literature will be given in Subsect.
\ref{Htgd}, where the inclusion of the strong gauge coupling leads to a more
realistic situation.

\begin{sidewaysfigure}
\begin{tabular}{cc} 
\epsfig{file=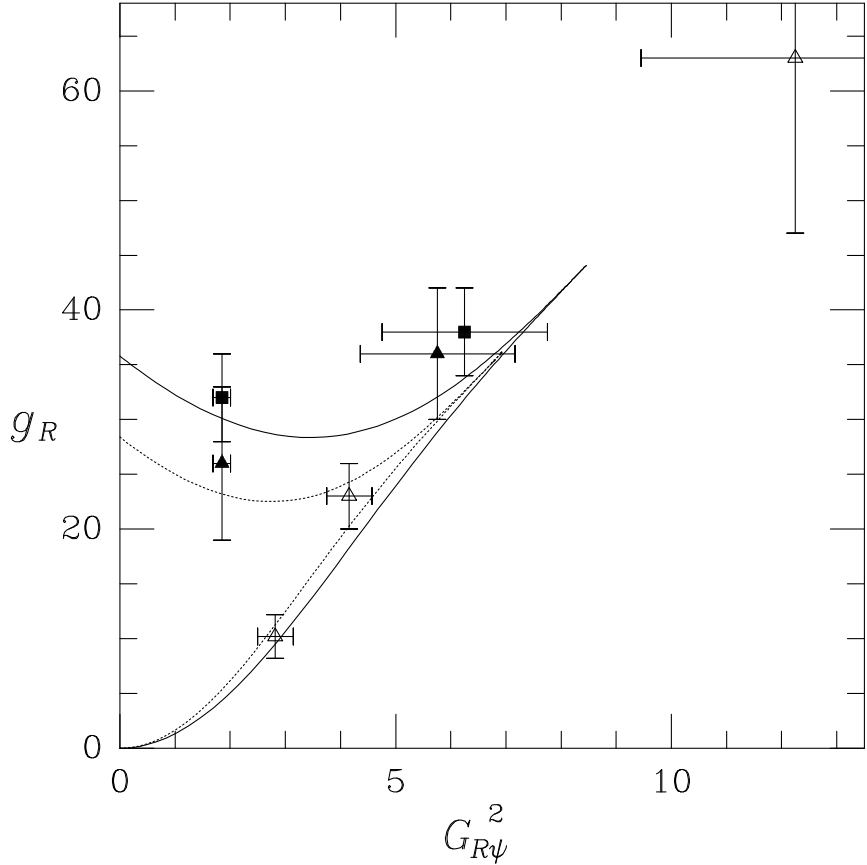,bbllx=106pt,bblly=186pt,bburx=496pt,
bbury=579pt,width=11.5cm}&
\raisebox{5mm}{\epsfig{file=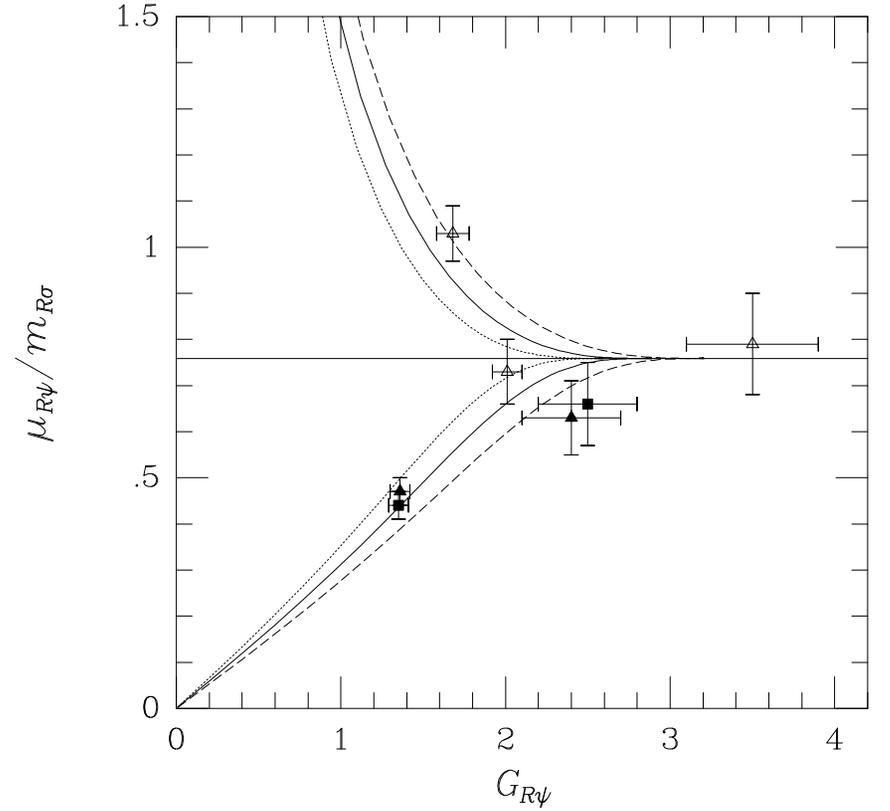,bbllx=76pt,bblly=189pt,bburx=496pt,
bbury=579pt,width=11.5cm}}\\
{\bf a)} & {\bf b)}
\end{tabular}
\caption[dum]{a) lattice results \cite{montv2} for the upper and lower bounds on $g_R$
  (corresponding to 6\la) as a function of $G^2_{R\psi}$ (corresponding to
  $g_t^2/2$) together with the one-loop estimates for scale ratios
  $\lam/\mh=3$ (solid curve) and $\lam/\mh=4$ (dotted curve). The open
  symbols denote the lower bound, the full symbols the upper bound; the
  triangles represent the $6^3\cdot 12$ lattice results, the squares the
  $8^3\cdot 16$ lattice results.  b) The mass ratio
  $\mu_{R\psi}/m_{R\sigma}$ (corresponding to $\mh/m_{t}$) as a function of
  $G_{R\psi}$ (corresponding to $g_t/\sqrt{2}$) in comparison with one-loop
  perturbative estimates for $m_{R\sigma}=0.75$ (dotted curve),
  $m_{R\sigma}=1$ (full curve) and $m_{R\sigma}=1.25$ (dashed curve). The
  figures were taken from Ref. \cite{montv2}.}
\label{latt1}
\end{sidewaysfigure}

Again it is of interest to go beyond perturbation theory and look for
confirmation to non-perturbative lattice calculations in the
Higgs-fermion sector. In Refs. \cite{montv1},\cite{montv2}
non- perturbative results for the triviality bound as well as the
vacuum stability bound have been obtained. The investigations are
based on a lattice action which is the sum of an $O(4)$ ($\simeq
SU(2)_L \times SU(2)_R$) scalar part and a fermionic part with chiral
$SU(2)_L \times SU(2)_R$ symmetry involving a heavy fermion doublet without
mass splitting, supplemented by a corresponding mirror fermion doublet
in order to overcome the notorious fermion doubler problem
accompanying regularization on the lattice.  Mass mixing between the
doublets is arranged such as to exactly decouple \cite{bore} the
mirror fermions from the physical spectrum. The calculations are
performed in the physically relevant broken symmetry phase. The
triviality bound results from bare Higgs selfcoupling $\la=\infty$,
the vacuum stability bound from $\la=10^{-6}$. 

The results are shown in Fig. \ref{latt1}a) \cite{montv2}, where the
renormalized Higgs selfcoupling $g_R$ corresponds to the IR coupling 6\la\ 
in our nomenclature and the renormalized Yukawa coupling $G_{R\psi}$ is to be
identified with the IR top Yukawa coupling $\gt/\sqrt{2}$. For comparison
the perturbative one-loop results for \lam =3\mh\ and \lam=4\mh\ are shown.
As in the pure Higgs case, the lattice calculations confirm the one-loop
perturbative bounds; the lattice bounds are ``absolute bounds'' in the same
sense as the $\Phi^4$ lattice bound (\ref{latt}) for the Higgs mass: they
allow the physics of the Higgs fermion sector to continue to be valid up to
scales \lam\ of the order of three to four times the Higgs mass.

An interesting result is the right-most point in Fig. \ref{latt1}a) \cite{montv2} with
large error bars at largest values of the two couplings. It
implies a lattice ``absolute'' upper bound (\ref{latt}) for the fermion
mass
\begin{equation}
m_f\lwig O(600\pm 80\gev), 
\end{equation}
which however leaves only a safety margin for the model up to a cut-off
$\lam\simeq 1.9\,\rm m_f$. The upper limit is provided by the requirement
of reflection positivity, the Euclidean analogue of unitarity in Minkowski
space. The rightmost point in Fig. \ref{latt1}a) \cite{montv2} is of
further interest, since it allows to trace the IR fixed line in the
\la-\gts\ plane, resp. the IR fixed point in the ratio R=\la/\gts, Eq.
(\ref{fl}), in the lattice data. In Fig. \ref{latt1}b) \cite{montv2} the
lattice bounds tend towards a ratio for the fermion mass divided by the
Higgs mass $\mu_{R\psi}/m_{R\sigma}=0.76$ which corresponds to the IR fixed
point position $R=\la/\gts=(\sqrt{65}-1)/16\simeq 0.441$, determined in Eq.
(\ref{fl}) .Thus all features of one-loop perturbation theory in the
Higgs-top sector are supported by non-perturbative lattice calculations.


\subsection{The Top-\protect\boldmath $\gd$ Sector of the SM and MSSM\newline -- a
Non-Trivial IR Fixed Point\label{tgd}}

As a first step towards including the strong gauge coupling \gd\ the
subset of the two couplings
\begin{equation}
\gt,\ \gd\neq 0,\ {\rm in\ the\ SM}\ {\rm resp.}
\ \htt,\ \gd\neq 0,\ {\rm in\ the\ MSSM}\ \hspace{8mm}{\rm with\ all\ the\ other\ couplings}=0
\end{equation}
is treated.  A {\it first non-trivial IR fixed point} appears in the
variable \rt=\gts/\gds\ in the SM and in the corresponding variable
\rt=\hts/\gds\ in the MSSM. Let us point out already here that this IR
fixed point will turn out to be at the heart of the quantum effects, as
encoded in the RGE, which will finally lead to an IR attractive top mass
value
\begin{eqnarray}
\mtp &\approx & 215\gev\ {\rm in\ the\ SM}\\
\mtp &\approx & O(190-200)\gev\sin\beta\ {\rm in\ the\ MSSM}
\label{anti}
\end{eqnarray}
after inclusion of further couplings, in particular of the electroweak
gauge couplings, of the two-loop contributions in the RGE and of the
radiative corrections leading to the pole mass.   

In case of the SM the fixed point was first pointed out in this form to be
discussed next by Pendleton and Ross \cite{pen}; their paper may justly be
considered to be the primer for all following investigations into IR fixed
points, lines, surfaces in the space of ratios of variables.  Earlier
references to an IR fixed point were made in Refs.  \cite{frogatt}.
Subsequent important results concerning the IR fixed point were obtained in
Ref. \cite{hill}, also to be discussed below, and in Refs.
\cite{wet2}-\cite{leung}, concentrating mainly on the issues of new heavy
fermion generations or of two Higgs doublets. The Pendleton-Ross IR fixed
point was rediscovered and expanded on in Refs.  \cite{kusizie} and
\cite{kusiziz} as a renormalization group invariant solution in the
framework of reduction of parameters, to be reviewed in Sect. \ref{parred}.
Later developments were made in Refs.  \cite{wet}-\cite{sch1}, \cite{schwi}.

The corresponding IR fixed point in the MSSM was discussed in Refs. \cite{alv}-\cite{schwi}. An explosive
development
\cite{taubuni}-\cite{ross1}
focused on the interrelated issues of tau-bottom(-top) Yukawa coupling
unification in supersymmetric grand unification and of an IR fixed top mass
value, Eq. (\ref{anti}), to be reviewed in Sect.
\ref{yukuni}.

The introduction of the ratio of coupling squares 
\begin{equation}
\rt=\frac{\gts}{\gds}{\rm \ in\ the\ SM}\;\;\;\;{\rm and}\;\;\;\;\rt=\frac{\hts}{\gds}{\rm \ in\ the\ MSSM},
\;\;\;\;\;\;{\rm respectively,}
\label{rt}
\end{equation}
in Refs. \cite{pen}-\cite{sch1} and \cite{alv}-\cite{schwi} leads to the
respective RG equations in \rt\ 
\begin{equation}
\begin{tabular}{c|c}
${\rm SM}$ & ${\rm MSSM}$\\ \\ 
$\;\;\;{\displaystyle -\frac{14}{9}\gds\frac{{\rm d}\,\rt}
{{\rm d}\, \gds}=\rt\left(\rt-\frac{2}{9}\right)}\;\;\;$ &
$\;\;\;{\displaystyle -\frac{1}{2}\gds\frac{{\rm d}\,\rt}
{{\rm d}\, \gds}=\rt\left(\rt-\frac{7}{18}\right)}\;\;\;$\\ \\
\end{tabular} 
\label{rgrt}
\end{equation}
in terms of the variable \gds. Following Refs. \cite{hill}-\cite{kusiziz},\cite{sch1}
for the SM and Ref. \cite{sch2} for the MSSM, the general
solutions of the RGE (\ref{rgrt}) for \rt\ and that for \gd, are
\begin{equation}
\begin{tabular}{c|c}
${\rm SM}$ & ${\rm MSSM}$\\ \\ $\;\;\;{\displaystyle
\rt(\gds)=\frac{2/9}{1-\left( 1-\frac{\displaystyle 2}{\displaystyle
9\rtz}\right)\left(\frac{\displaystyle \gds}{\displaystyle
\gdsz}\right)^{-1/7}}}\;\;\;$ & $\;\;\;{\displaystyle
\rt(\gds)=\frac{7/18}{1-\left( 1-\frac{\displaystyle 7}{\displaystyle
18\rtz}\right)\left(\frac{\displaystyle \gds}{\displaystyle \gdsz}
\right)^{-7/9}}}\;\;\;$\\ \\
${\displaystyle \gds(\mu)=\frac{\gdsz}{1-\frac{\displaystyle 7}
{\displaystyle 8\pi^2}\gdsz\ln\frac{\lam}{\displaystyle \mu}}}$ &
${\displaystyle \gds(\mu)=\frac{\gdsz}{1-\frac{\displaystyle 3}
{\displaystyle 8\pi^2}\gdsz\ln\frac{\lam}{\displaystyle \mu}}}$, 
\end{tabular} 
\label{solfp}
\end{equation}
where 
\begin{eqnarray}
\gdsz&=&\gds(\lam),\\
\rtz&=&\rt(\gds(\lam)). 
\end{eqnarray}

From Eqs. (\ref{rgrt}) and (\ref{solfp}) the following important analytical
results i) about an exact IR fixed point, first discovered by
Ref.\cite{pen} within the SM and ii) an effective IR fixed point, first
discovered by Ref. \cite{hill} within the SM, may be read off.  Further
references were given already in the introduction of this subsection.

\begin{itemize}
\item The differential equation for \rt(\gds) has an exact fixed point at 
\begin{equation}
\begin{tabular}{c|c}
${\rm SM}$ & ${\rm MSSM}$\\ \\ 
$\;\;\;\rt=\frac{2}{9}\;\;\;$ & $\;\;\;\rt=\frac{7}{18}.\;\;\;$\\ \\
\end{tabular}
\label{exfp}
\end{equation}
This fixed point corresponds to the special solution of the RGE defined by
the boundary condition that \rt\ approaches a finite value ($\not=0$) in
the limit $\gds\rightarrow 0$. This special solution remains in the fixed
point, once it has started in this fixed point value, irrespective of the
evolution of \gds\, i.e. in particular of the choice of initial values
\lam, \gdsz\, and of \gtsz.  In the \gts-\gds\ plane, resp. in the
\hts-\gds\ plane, this fixed point in \rt\ translates into a fixed line
(physicists' terminology) or invariant line (mathematicians' terminology
\cite{guc}), $\gts=2/9\gds$ and $\hts=7/18\gds$, respectively. This implies
that once the initial value is chosen on this line, the solution evolves
along this line; again the location of this line is independent of the
initial values \lam, \gdsz\ and \gtsz. Since, however, as anticipated in
the introduction of this Sect.  \ref{noelweak}, all further IR fixed
points, lines and surfaces will appear in spaces of {\it ratios} of
variables it is more economical to stick to the ratio \rt\ as new variable
in the following.
\item The IR fixed point (\ref{exfp}) in \rt\ is IR attractive, i.e. it
  attracts all solutions starting from arbitrary (perturbatively
  allowed) initial values \gdsz, $\rtz>0$ at $\mu=\lam$ for arbitrary
  \lam, in short the (perturbative) ``top-down'' RG flow, towards
  it. However, as has been pointed out in Ref. \cite{hill}, in case of
  the SM this attraction is exceedingly weak, like
  $(\gds/\gdsz)^{-1/7}$, where the logarithmic evolution of \gds\ is
  damped by the prohibitively small exponent, -1/7.  This is, however,
  not the case \cite{sch2} in the MSSM, where the corresponding
  exponent is -7/9, i.e. reasonably large. In any case, however, the
  full RG flow reaches the fixed point (\ref{exfp}) only in the
  limit $\gd\rightarrow\infty$ which is outside of the range of
  validity of perturbation theory and of mathematical interest only.
\item Of high interest is the upper bound \cite{cab}, \cite{mai},
\cite{hill}, \cite{wet2}, \cite{lintriv}, \cite{wet} for the variable
\rt. As has been worked out analytically in Ref. \cite{hill}, an
almost \rtz\ independent upper bound for \rt\ is approached for
  sufficiently {\it large} UV initial values \rtz.
  Neglecting\footnote{Following Ref. \cite{hill} it suffices in fact
  to neglect
  $\frac{9}{2}\rtz\left[\left(\frac{\gds}{\gdsz}\right)^{1/7}-1\right]$,
  resp.
  $\frac{18}{7}\rtz\left[\left(\frac{\gds}{\gdsz}\right)^{7/9}-1\right]$
  with respect to 1.}in Eqs. (\ref{solfp}) 1/\rtz\ with respect to 2/9
  and 7/18, respectively, allows to read off the {\it \rtz\ independent}
  upper bound, which could also be termed a triviality bound for the
  variable \rt
\begin{equation}
\begin{tabular}{c|c}
${\rm SM}$ & ${\rm MSSM}$\\ \\ 
$\;\;\;{\displaystyle \rt(\gds)<\frac{2/9}{1-\left(
\frac{\displaystyle \gds}{\displaystyle \gdsz}\right)^{-1/7}}}\;\;\;$ &
$\;\;\;{\displaystyle \rt(\gds)<\frac{7/18}{1-\left(
\frac{\displaystyle \gds}{\displaystyle \gdsz}\right)^{-7/9}}}\;\;\;$.\\
\end{tabular}
\label{bt} 
\end{equation}
Evaluated at the IR scale, this upper bound is roughly the IR image of {\it
all sufficiently large} UV initial values \rtz\ for
\rt. In the limits $\lam\rightarrow\infty$ or $\gd\rightarrow\infty$,
which of course both are physically not accessible, the upper bound
tends towards the Pendleton-Ross fixed point. Thus, as in the case of
the triviality bound for the Higgs selfcoupling \la, it is the {\it
finite} length of the RG evolution path from the UV to the IR which is
responsible for the finite gap between the upper bound (\ref{bt}) and
the IR fixed point (\ref{exfp}). The upper bound decreases for
increasing value of the UV cut-off \lam. 

Ref. \cite{hill} went even further and very intuitively reinterpreted the
upper bound (\ref{bt}) in case of the SM as an {\it effective
intermediate IR fixed point for solutions with large UV initial values
\rtz\ or $g_{t0}$} as follows. On the one hand, considering
$\frac{1}{7}\ln (\gds/\gdsz)$ as small expansion parameter, the
bound (\ref{bt}) may be rewritten approximately as
\begin{eqnarray}
\rt(\gds)_{\rm upper\
bound}&\approx&\frac{2}{9}\frac{7}{\ln\frac{\gds}{\gdsz}}\left(1+O(\frac{1}{7}\ln\frac{\gds}{\gdsz})\right),\
\ {\rm i.e.}\nonumber\\
\gts(\gds)_{\rm upper\ bound}&\approx&\frac{14}{9}\frac{\gds}{\ln\frac{\gds}{\gdsz}}. 
\label{appro}
\end{eqnarray}
On the other hand, in going back to the RGE for \gts,
\begin{equation}
 \frac{{\rm d}\,\gts}{{\rm d}\,
t}=\frac{\gts}{8\pi^2}\,\left(\frac{9}{2}\gts-8\gds\right),
\label{dgts}
\end{equation} 
Hill points out \cite{hill} that for {\it large} initial values
$\gtsz\gg\gdsz$ in the vicinity of the UV scale the running of \gts\ is
driven by the term $\frac{9}{2}\gts$ in the bracket, which justifies
to replace $8\gds$ in the bracket by some constant average value
$8\overline{\gds}$. In running with $\mu$ towards the \rtz\
independent upper bound in the IR region, a
{\it transient} slowing down in the running of \gts\ is expected in the
vicinity of the 
value of $\mu$ where the bracket and consequently ${\rm d}\,\gts/{\rm d}\,
t$ become zero for
\begin{equation}
\frac{9}{2}\gts\approx 8\overline{\gds}.
\label{zer} 
\end{equation}
Since this slowing down is expected to happen in the vicinity of the
upper bound (\ref{bt}), one may identify \cite{hill} 
\begin{equation}
\overline{\gds}\approx\frac{7}{8}\frac{\gds(\mu)}{\ln\frac{\gds(\mu)}{\gdsz}}.
\end{equation}
In the supersymmetric case, the analogon of the transient slowing down
condition (\ref{zer}) is
\begin{equation}
6\hts\approx\frac{16}{3}\overline{\gds},
\end{equation}
which was pointed out in Ref. \cite{bard}.
As was also stressed in Ref. \cite{hill}, the effective intermediate IR
fixed point is not a genuine fixed point, since its position depends
on the UV cut-off \lam\ as well as on the initial value \gdsz\ and it
is not attractive for smaller initial values \rtz. 

As the Hill intermediate effective fixed point appears as an
approximate interpretation of the upper bound (\ref{bt}), the
notions of an upper bound and of an effective fixed point have been
frequently treated as synonymous in the literature. In the following
we also shall repeatedly prefer the notion of  ``Hill effective
fixed point'' over the notion ``upper bound'', since it reflects more
directly the very strong IR attraction of the bound for solutions with
large \rtz\ as well as its independence of the (large) initial value \rtz.

\item A very important point to realize is that this strong attraction
  towards the upper bound, resp. the Hill effective fixed point, implies a
  strong effect of effacing the memory of the details of the UV physics in
  the IR physics: {\it any theory at some high UV scale \lam, which
    supplies for some dynamical reason sufficiently large UV initial values
    for the top Yukawa coupling, will end up in the upper bound, resp. in
    the Hill effective fixed point in the IR region}; the only memory
  retained is the scale \lam\ at which new physics is present. A very good
  example is (as has also been worked out by the authors) the appealing
  scheme running under the name of top condensation \cite{lin1}, e.g.
  reviewed in Ref. \cite{lin2}. At some high scale \lam\ a new physics
  sector provides the force for the condensation of top-antitop pairs thus
  providing a dynamical mechanism for the spontaneous electroweak symmetry
  breaking and a composite Higgs boson. The RG evolution of the naturally
  large UV initial values for the top Yukawa couplings largely effaces the
  details of the UV theory and ends up in the IR region in the \lam\ 
  dependent top Yukawa values of the IR upper bound, i.e. the Hill
  effective fixed point.
\end{itemize}

The final aim is of course to translate the fixed point position in the
variable \rt\ into the corresponding values for the top mass, which is in
lowest order
\begin{eqnarray}
  \mt&=&\sqrt{2/9}\gd(\mu=\mt)v/\sqrt{2}\approx 95\gev\ \ {\rm in\ the\ 
  SM},\\
  \mt&=&\sqrt{7/18}\gd(\mu=\mt)v/\sqrt{2}\sin\beta\approx
  126\gev\sin\beta\ \ {\rm in\ the\ MSSM}
\label{penmass}
\end{eqnarray}
In the approximation considered so far, already a comparatively large top
mass results. Also the upper bounds on the top mass corresponding to
the Hill effective fixed point may be estimated that way 
\begin{eqnarray}
\mt_{{\rm max}}&\approx& 570,\,460,\,300,\,230\gev\ {\rm resp.\ for}\ \lam=10^4,\,10^6,\,10^{10},\,10^{15}\gev\ {\rm in\ the\ SM},\\
\mt_{{\rm max}}&\approx& 175\sin\beta\gev\ {\rm for}\ \lam=\mgut\approx
2\cdot 10^{16}\gev\ {\rm in\ the\ MSSM}
\label{hillmass}
\end{eqnarray}

Given the approximation, this may be considered as a very convincing step
towards physical reality.  It shows that at the level of the quantum
effects, as encoded in the RGE equations, resides the intrinsic possibility
of a heavy top quark, i.e. a top quark much heavier than the other quarks
and leptons. In order to get quantitatively reliable results, one has i) to
switch back on the electroweak couplings \gee, \gz\ as well as the Higgs,
bottom and tau couplings, ii) use the full two-loop RGE and iii) take into
account the radiative corrections allowing to determine the pole mass \mtp.
These ``corrections'' (in particular the introduction of the electroweak
couplings) will turn out to significantly increase the top mass to the
values anticipated already in Eq. (\ref{anti}). The
corrections will be introduced step by step in the following sections.


\subsection{The Higgs-Top-\protect\boldmath $\gd$ Sector of the SM\newline -- a First Non-Trivial
Approximation\label{Htgd}}

Now the necessary information has been accumulated to discuss the
first approximation to the SM which is non-trivial from the
mathematical point of view and qualitatively already informative from the
physical point of view. A dynamical source for a heavy Higgs
boson arises which is intimately related to the one for the a heavy top quark
discussed in the last section.

The three couplings 
\begin{equation}
\la,\ \gt,\ \gd\ \not=0
\end{equation}   
are considered to be the only non-vanishing ones in the RGE of the SM
in this subsection.  

As above it is more economical to consider the ratios of couplings
\begin{equation}
\rt=\frac{\gts}{\gds}\ \ {\rm and}\ \ \rh=\frac{\la}{\gds}
\end{equation}
or even
\begin{equation}
\rt=\frac{\gts}{\gds}\ \ {\rm and}\ \ R=\frac{\rh}{\rt}=\frac{\la}{\gts}.
\end{equation}
By rewriting as in the last subsection \gds\ as function of t, \rt\
as function of \gds\ and finally \rh\ or alternatively R as a function
of \rt, one ends up with the following system of {\it decoupled}  
differential equations \cite{sch1} 
\begin{eqnarray}
{\displaystyle \frac{{\rm d}\,\gds}{{\rm d}\, t}}
&=&{\displaystyle -7\frac{g_3^4}{8\pi^2}}\\
{\displaystyle -14\gds\frac{{\rm d}\,\rt}
{{\rm d}\, \gds}}&=&{\displaystyle \rt\left(9\rt-2\right)}\\
{\displaystyle \rt\left(9\rt-2\right)\frac{{\rm d}\,\rh}
{{\rm d}\, \rt}}&=&{\displaystyle 24\rh^2+(12\rt +14)\rh-6\rt^2}  
\end{eqnarray}
or alternatively \cite{sch1}
\begin{eqnarray}
{\displaystyle \frac{{\rm d}\,\gds}{{\rm d}\, t}}
&=&{\displaystyle -7\frac{g_3^4}{8\pi^2}}\\
{\displaystyle -14\gds\frac{{\rm d}\,\rt}
{{\rm d}\, \gds}}&=&{\displaystyle \rt\left(9\rt-2\right)}\\
{\displaystyle (9\rt-2)\frac{{\rm d}\,R}
{{\rm d}\, \rt}}&=&{\displaystyle 24R^2+(3+\frac{16}{\rt})R-6}.\label{R} 
\end{eqnarray} 
The system of differential equations for \rt\ and \rh\, resp. R, has a
common IR fixed point first pointed out in Ref. \cite{pen} and
further analyzed in Refs. \cite{hill}, \cite{baggz}-\cite{kusiziz}, \cite{froge}, \cite{sch1}.
\begin{equation}
\rt=\frac{2}{9},\hspace{2cm}\rh={\displaystyle
\frac{\sqrt{689}-25}{72}},\ \ {\rm resp.}\ \ R={\displaystyle
\frac{\sqrt{689}-25}{16}}. 
\label{rtR}
\end{equation}
Thus, in presence of the gauge coupling \gd\ the nontrivial top fixed
point, discussed in the previous subsection, is supplemented by a
nontrivial Higgs fixed point. This common fixed point replaces the trivial
IR fixed point at $\la=\gts=0$ in absence of \gd. It is also interesting as
a crude approximation to physical reality, since it corresponds within this
lowest order perturbation theory to
\begin{eqnarray}
\mt&=&\sqrt{2/9}\gd(\mu=\mt)v/\sqrt{2}\approx 95\gev \\
\mh&=&\sqrt{(\sqrt{689}-25)/72}\gd(\mu=\mt)\sqrt{2}v\approx 53\gev.
\end{eqnarray}
Again, the common IR fixed point can be viewed in the presently discussed
approximation as a possible dynamical origin for large top quark and Higgs
masses! It will eventually lead to mass values
\begin{eqnarray}
\mtp &\simeq & 214\gev\ {\rm in\ the\ SM}\\
\mhp &\simeq & 210\gev,
\end{eqnarray}
when all the appropriate corrections are applied in the next sections.

As in the \la-\gts\ plane in absence of \gd, there is again an IR
attractive fixed line $\overline{r}_H(\rt)$ in the \rt-\rh-plane or
alternatively $\overline{R}(\rt)$ in the \rt-R-plane \cite{sch1}, with
prior indications already contained in Refs.
\cite{baggz}-\cite{kusiziz},\cite{froge}.
This fixed line is nonlinear in contradistinction to the one in absence of
\gd. It is the solution of the RGE singled out by the boundary condition of
approaching a {\it finite} ratio $R=\rh/\rt=\la/\gts$ in the limit
$\rt,\,\rh\rightarrow\infty$ or $\gts,\,\la\rightarrow\infty$. In fact
the limit is
\begin{equation}
\overline{\rho}_{H}(\rt)\rightarrow\frac{\sqrt{65}-1}{16}\,\rt,\ \ \ {\rm resp.}\ \ \ \overline{R}(\rt)\rightarrow\frac{\sqrt{65}-1}{16}\ \ {\rm for}\ \ \rt\rightarrow\infty.
\label{infi}
\end{equation}
Since in this limit the couplings \la\ and \gts\ are much larger than
\gds, we are back to the discussion led in Refs. \cite{wet2},
\cite{wet} and reported on in Subsect. 4.2. Indeed,
this limit is consistently identical with the slope (\ref{fli}) of the IR
attractive line in the \la-\gts-plane in absence of the strong gauge
coupling \gd. Now the fixed line is the solution of the RGE
interpolating also the trivial fixed
point at R=0 at \rt=0, which is now IR {\it repulsive} (i.e. UV
attractive), the IR attractive fixed point (\ref{rtR}) at \rt=2/9 and shows the
asymptotic behaviour (\ref{infi}) for $\rt\rightarrow\infty$.

The fixed line $\overline{R}(\rt)$ can be represented \cite{sch1} by infinite power
series expansions in powers of \rt\ around \rt=0, in powers of
$\rt-\frac{2}{9}$ around $\rt=\frac{2}{9}$ and in powers of 1/\rt\
in the limit $\rt\rightarrow\infty$
\begin{eqnarray}
\rt=0:\hspace{8mm}\overline{R}(\rt)&=&\frac{1}{3}\,\rt+\frac{1}{10}\,\rt^2-\frac{7}{132}\,\rt^3-\frac{79}{660}\,\rt^4+...,\label{zero}\\
\rt=\frac{2}{9}:\hspace{7mm}\overline{R}(\rt)&=&\frac{\sqrt{689}-25}{16}+27\frac{307-11\sqrt{689}}{1360}\,(\rt-\frac{2}{9})+...,\\
\rt\rightarrow\infty;\hspace{6mm}\overline{R}(\rt)&=&\frac{\sqrt{65}-1}{16}-\frac{17-\sqrt{65}}{42}\,\frac{1}{\rt}+\frac{485+11\sqrt{65}}{12789}\,\frac{1}{\rt^2}+...
.
\end{eqnarray}
A precursor of the expansion of the fixed line solution around \rt=0 is
found \cite{kusizie},\cite{kusiziz} in a one-loop solution of the program
of reduction of parameters to be discussed in Sect. \ref{parred}; a
precursor of the expansion around $\rt\rightarrow\infty$ had been put
forward in a four generation model \cite{baggz}.
\begin{figure}
\begin{center}
\epsfig{file=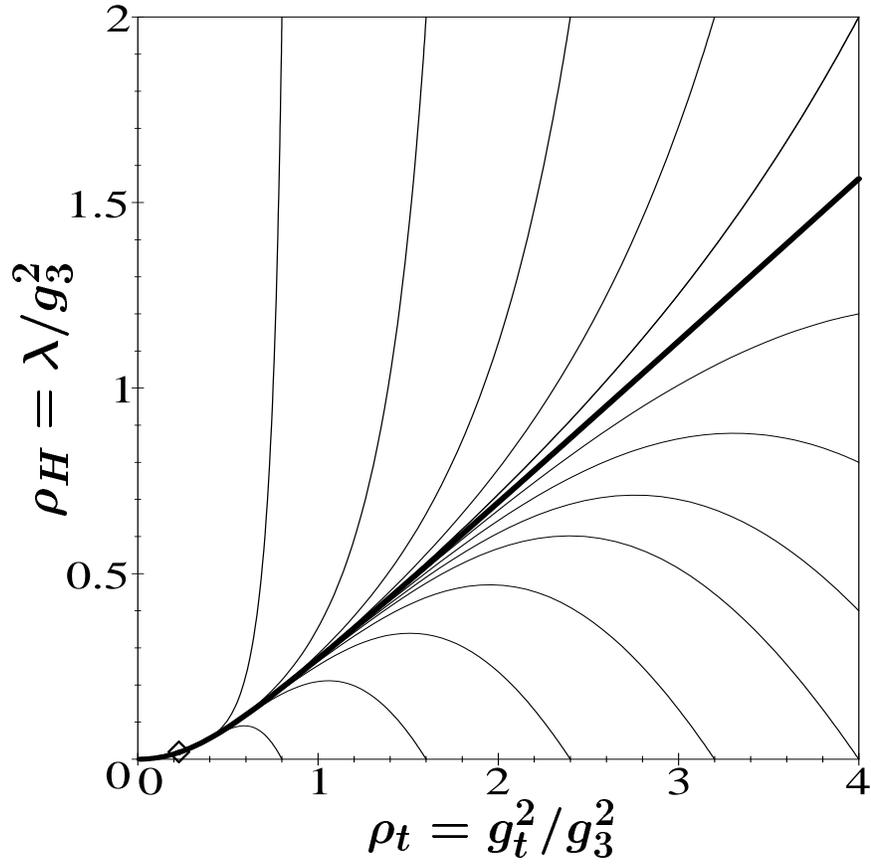,bbllx=114pt,bblly=247pt,bburx=558pt,
bbury=691pt,width=11.5cm}
\end{center}
\caption[dum]{The IR attractive fixed line (fat line) and the IR attractive fixed
  point (symbol $\Diamond$) in the \rh-\rt-plane; solutions (thin lines)
  representative for the ``top-down'' RG flow are shown, which demonstrate
  the strong IR attraction of the fixed line. This figure is an update of a
  figure in Ref. \cite{sch1}.}
\label{fixline}
\end{figure}

Before entering further analytical investigations let us present the main
numerical results in several illuminating figures.  Fig. \ref{fixline}
contains the IR fixed line in the \rt-\rh-plane (fat line); it is an update
of the corresponding figure in the R-\rt-plane in Ref. \cite{sch1}. As
expected it interpolates the IR repulsive fixed point at \rt=0, the IR
attractive fixed point at \rt=2/9 (diamond) and approaches a straight line
with slope $(\sqrt{65}-1)/16$ for $\rt\rightarrow\infty$. The RG flow from
the UV towards the IR is indicated by a set of solutions of the (one-loop)
RGE starting at representative UV initial values at \rt\ values above and
below the IR fixed point (thin lines). Clearly the solutions are much more
strongly attracted by the IR fixed line than by the IR fixed point.  They
first move towards the fixed line and then proceed close to or along the
line towards the fixed point. 

An important observation is that no solution starting from above the line
can end up below and vice versa. This is a general feature: any fixed line
divides a plane into two disjoint sectors.  If one were to follow the
solutions towards large values of \rt, all solutions above the fixed line
tend towards infinity and
all solutions below tend towards negative values. The fixed line solution,
being singled out by the boundary condition that \rh/\rt\ be finite in this
limit, is quasi the ``watershed'' between the two classes of solutions.
This is characteristic for a fixed line.
\begin{sidewaysfigure}
\begin{tabular}{cc}
{\bf UV} & {\bf IR}\\
\epsfig{file=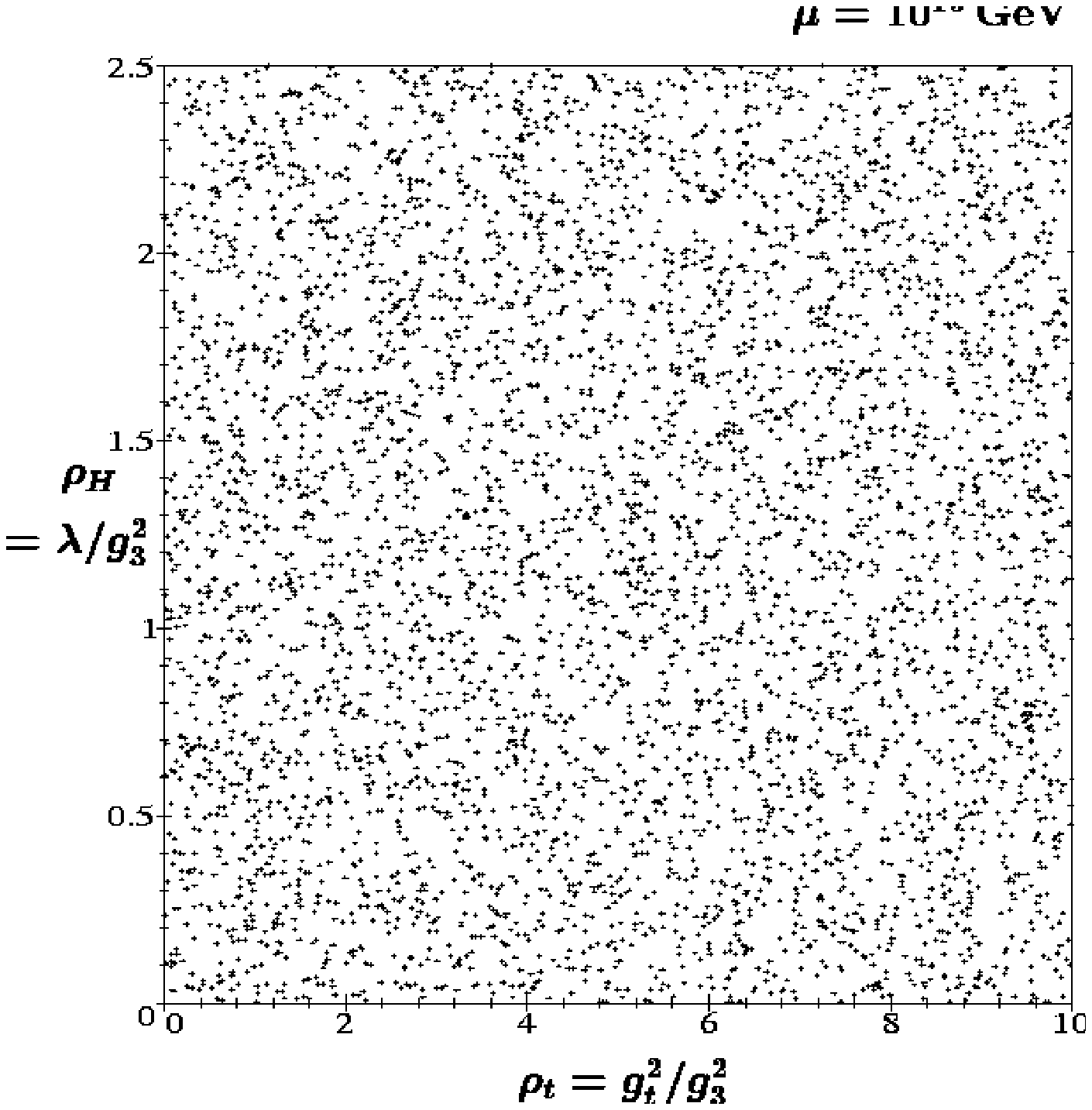,bbllx=32pt,bblly=342pt,bburx=472pt,
bbury=802pt,width=11.5cm}
&
\epsfig{file=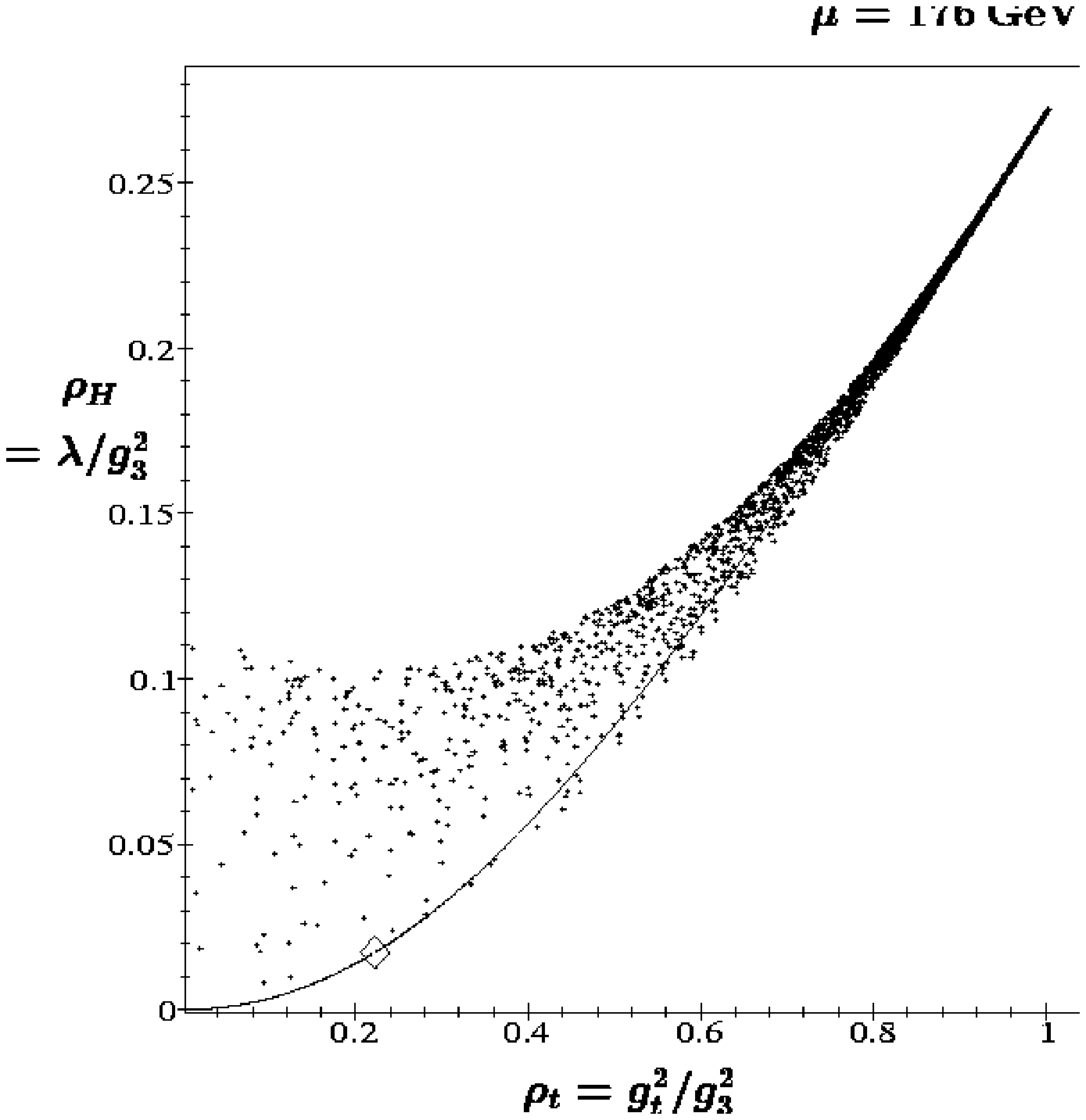,bbllx=33pt,bblly=345pt,bburx=469pt,
bbury=799pt,width=11.5cm}\\
\end{tabular}
\caption[dum]{Randomly chosen UV initial values in the \rh-\rt-plane at
  the UV scale $\mu=\lam=10^{15}\gev$ are subject \cite{sch3} to the
  RG evolution down to the IR scale $\mu=176\gev$; notice that the UV
  plane (figure on the left hand side) has scaled down by a factor of
  10 to the IR plane (figure on the right hand side). The strong IR
  attraction of the IR fixed line (fat line with symbol $\Diamond$ for
  the IR fixed point) is demonstrated. The tip of the line is the
  absolute upper bound on \rt\ and \rh.}
\label{uvir}
\end{sidewaysfigure}

A further important question is: if one starts with UV initial values
at $\mu=\lam=10^{15}\gev$, say, randomly distributed within the
perturbatively accessible large square $0\leq\rt\leq 10$,
$0\leq\rh\leq2.5$, as presented on the left hand side in
Fig. \ref{uvir}, how closely do the corresponding IR values at
$\mu=\mt=176\gev$ cluster around the IR fixed line? The result is
presented on the right hand side in Fig. \ref{uvir} \cite{sch3}:
notice first that the total range for IR values has shrunk to within a
square of one tenth of the length of the UV square, i.e.  within a
square $0\leq\rt\leq 1$, $0\leq\rh\leq 0.25$. Within this square the
points with larger values of \rt\ indeed cluster on or very closely to
the fixed line (fat line); for lower values of \rt\ the points above
the line fail to be drawn fully onto the line; their upper boundary is
the pendant of the ``triviality'' bound for $\lam=10^{15}\gev$.  (Even
though the allusion to a trivial IR fixed point does not apply any
more and the notion of an upper bound would be more appropriate we
follow the usage in the literature and maintain the expression
triviality bound). In comparison with the triviality bound for the
Higgs mass, discussed in the framework of the $\phi^4$ theory in
Subssect. 4.1, the inclusion of the top Yukawa coupling into the
discussion has turned the upper Higgs mass bound into a top mass
dependent bound \cite{mai}, \cite{lintriv}. The lower bound, the
one-loop vacuum stability bound (see Subsect. 2.6) \cite{mai},
\cite{linvac}-\cite{esp}, is the IR image of all points starting from
the UV initial values $R_0=0$, i.e. $\lambda_0=0$; they {\it all} end
up very closely to the fixed line (from below!). Thus for this large
value $\lam=10^{15}\gev$ for the UV scale the lowest order vacuum
stability bound is very close to the IR fixed line; the IR fixed line
is clearly an upper bound for the (lowest order) vacuum stability
bound. It becomes again clear from Fig. \ref{uvir} that it is IR
attraction of the fixed line rather than of the fixed point (diamond)
which determines the ``top-down'' RG flow.

The outermost tip of the clustering IR points is the IR image of {\it
all} the UV points with UV initial values $\rtz\geq 1$, i.e. of $90\%$
of the randomly chosen initial values \cite{cab}, \cite{mai},
\cite{hill}, \cite{wet2},
\cite{lintriv}, \cite{leung}, \cite{wet}, \cite{sch1}. This is the absolute
upper bound as well in \rt\ (which had been interpreted as Hill effective
fixed point) as well as in \rh, both calculated for $\lam=10^{15}\gev$; this
figure clearly demonstrates the attraction power of the upper bound and
also that it lies on the IR fixed line \cite{wet2}, \cite{leung},
\cite{wet}, \cite{sch1}.
\begin{figure}
\begin{center}
\epsfig{file=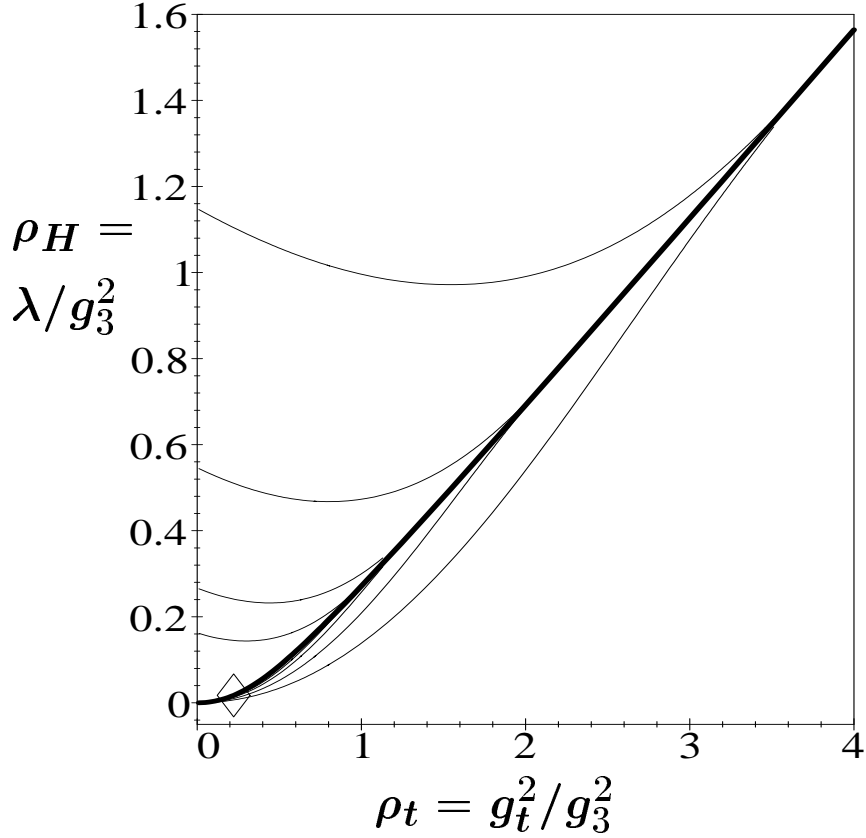,bbllx=8pt,bblly=333pt,bburx=464pt,
bbury=776pt,width=11.5cm}
\end{center}
\caption[dum]{The IR fixed line (fat line) and fixed point (symbol
$\Diamond$) in relation to the \lam\ dependent triviality bounds and vacuum
stability bounds (thin lines) for UV scale $\lam=10^4,\;10^6,\;10^{10},\;10^{15}\gev$,
bounding wedge-formed allowed regions. 
The IR attraction of the bounds towards the IR fixed
line is demonstrated. The tip of the wedge, the absolute upper bound
on \rt\ and \rh\
slides down the IR fixed line towards the IR fixed point for increasing
values of \lam.}
\label{wedges}
\end{figure}

So far we have discussed a {\it large} UV scale, $\lam=10^{15}\gev$.  Fig.
\ref{wedges} shows the IR fixed line (fat line) \cite{sch1} and fixed point (symbol
$\Diamond$) \cite{pen} in relation to the \lam\ dependent triviality
bounds and vacuum stability bounds (thin lines) \cite{cab},
\cite{mai}, \cite{lintriv}-\cite{esp}, \cite{sch1} for the UV scale
$\lam=10^4,\;10^6,\;10^{10},\;10^{15}\gev$ within these restricted
one-loop considerations. Clearly the \lam\ dependent tip of the wedge
slides down the IR line \cite{leung}, \cite{sch1} and the triviality
and vacuum stability bounds are drawn closer to the IR fixed line for
increasing values of \lam. In the physically inaccessible limit
$\lam\rightarrow\infty$ the upper bound, i.e.  the tip of the wedge,
is drawn into the Pendleton-Ross fixed point (symbol $\Diamond$).

Obviously, from Figs. \ref{uvir} and \ref{wedges}, we may conclude that the
IR attraction of the fixed line is the dynamical origin for the triviality
bound as well as for the vacuum stability bound. They are the boundaries of
the points which fail to reach the IR fixed line after evolution with the
{\it finite} evolution path from the UV scale \lam\ down to the IR scale
176\gev.

The discussion for couplings or rather for ratios of couplings may be
translated (within the lowest order treated in this section) into
results for the top and Higgs masses. The IR fixed line corresponds to
an IR attractive RG invariant top-Higgs mass relation. The triviality
and vacuum stability bounds, formulated here for the ratios
$\rh=\la/\gds$ or $R=\la/\gts$ of couplings, may be translated into
corresponding bounds for the Higgs mass as functions of the top mass
by means of the lowest order relations $\mt=\gt(\mt)v/\sqrt{2}$ and
$\mh=\sqrt{2\la(\mh)}v$.

The final results for the IR fixed line in the \mh-\mt-plane and the
corresponding triviality and vacuum stability bounds including all known
higher order corrections will be presented in Sect. \ref{masses}.

Analytical insight into all these features may be obtained \cite{sch1} from the
general solution of the one-loop Riccati differential equation (\ref{R})
in terms of the IR fixed line solution $\overline{R}$
\begin{equation}
R(\rt)=\overline{R}(\rt)+\frac{(R_{0}-\overline{R}_{0})\,{\rm
exp}[-F(\rt,\rtz)]}{1+\frac{8}{3}\,(R_{0}-\overline{R}_{0}){\displaystyle
\int_{\rt}^{\rtz}}{\rm
d}\rt'\,{\displaystyle \frac{{\rm
exp}[-F(\rt',\rtz)]}{\rt'-\frac{2}{9}}}}
\label{gensol}
\end{equation}
with $R_{0}=R(\rtz),\ \overline{R}_{0}=\overline{R}(\rtz)$ and
\begin{equation}
F(\rt,\rtz)=\frac{1}{3}\int_{\rt}^{\rtz}{\displaystyle\frac{{\rm
d}\rt'}{\rt'-\frac{2}{9}}}(16\overline{R}(\rt')+1+{\displaystyle
\frac{16}{3}}{\displaystyle \frac{1}{\rt'}}).
\end{equation}
The difference of any general solution and the fixed line solution,
$R(\rt)-\overline{R}(\rt)$, is given by the second term on the right hand
side of Eq. (\ref{gensol}), which in turn is controlled by the exponential
${\rm exp}[-F(\rt,\rtz)]$ in the numerator and the denominator. If the
initial value $R_{0}$ happens on the fixed line, i.e.
$R_{0}=\overline{R}_{0}$, then clearly $R(\rt)=\overline{R}(\rt)$, i.e. the
solution remains on the fixed line for all values of \rt. For values of \rt\
above and below the IR fixed point $\rt=\frac{2}{9}$ $F(\rt,\rtz)$ is
positive and increases with increasing evolution path; this reflects the IR
attractiveness of the fixed line. For the discussion of the rate of
attraction one has to distinguish two cases:
\begin{itemize}
\item $R_{0}-\overline{R}_{0}$ sufficiently small, such that the
denominator of Eq. (\ref{gensol}) remains close to one; the rate of
approach to the fixed line $\overline{R}(\rt)$ is given by ${\rm
exp}[F(\rt,\rtz)]$ in the numerator with a behaviour 
\begin{itemize}
\item for \rt, \rtz\ close to 2/9:
\begin{eqnarray}
{\rm exp}[-F(\rt,\rtz)]\approx 
\left(\frac{\rt-\frac{2}{9}}{\rtz-\frac{2}{9}}\;\frac{\rtz}{\rt}\right)^{\sqrt{689}/3}&=&\left({\displaystyle
\frac{\gds}{\gdsz}}\right)^{-\frac{1}{7}\sqrt{689}/3}\nonumber\\ &=&
\left(1-{\displaystyle \frac{7}{8\pi^2}}
\gdsz\ln({\displaystyle\frac{\lam}{\mu}})\right)^{\sqrt{689}/21},
\end{eqnarray}
with the high power $\sqrt{689}/3\approx 8.75$ beating the forbiddingly
low power 1/7 which monitores the exceedingly slow approach of \rt\
towards the IR fixed point 2/9. The resulting fairly large power with respect to
$\frac{\gdsz}{\gds}$, $\sqrt{689}/21\approx 1.25$, is the appropriate
measure for the strength of IR attraction of the fixed line $R(\rt)$.
\item for $\rt,\ \rtz\geq {\rm O}(1)$:
\begin{equation}
{\rm exp}[-F(\rt,\rtz)]\approx {\displaystyle (\frac{\rt-\frac{2}{9}}
{\rtz-\frac{2}{9}})}^{\sqrt{65}/3}\approx{\displaystyle (\frac{\gdsz}{\gds})}^{\frac{1}{7}\sqrt{65}/3}{\displaystyle
\frac{1}{\rtz^{\sqrt{65}/3}}}\rightarrow 0\ \ {\rm for}\ \ \rtz\rightarrow\infty
\end{equation}
with a strong rate of IR attraction for sufficiently large \rtz. This
is the rate controlling the approach to the upper bound, i.e. the Hill effective fixed
point(\ref{bt}) on the fixed line $\overline{R}(\rt)$. 
\end{itemize}
\item for large $R_{0}-\overline{R}_{0}$, the IR attraction towards
the fixed line $\overline{R}(\rt)$ is {\it strongly} enhanced, due to
the substantial increase of the denominator in solution
(\ref{gensol}). Even for $R_0\rightarrow\infty$ the general solution
$R(\rt)$ is attracted towards the fixed line to within a finite
difference 
\begin{equation}
R(\rt)\rightarrow \overline{R}(\rt)+{\displaystyle \frac{\frac{3}{8}{\rm
exp}[-F(\rt,\rtz)]}{\int_{\rt}^{\rtz}{\displaystyle \frac{{\rm
d}\rt'}{\rt'-\frac{2}{9}}}{\rm exp}[-F(\rt',\rtz)]}}\ \ {\rm for}\ \ R_0\rightarrow\infty,
\end{equation}
which shrinks for increasing length of evolution path (i.e. for
increasing value of \lam). This is the analytical form for the \lam\
dependent triviality bound.
\item Likewise, for $R_{0}=0$, the analytical form of the lowest order vacuum
stability bound is obtained from Eq. (\ref{gensol})
\begin{equation}
R(\rt)=\overline{R}(\rt)+\frac{-\overline{R}_{0}\,{\rm
exp}[-F(\rt,\rtz)]}{1-\frac{8}{3}\,\overline{R}_{0}{\displaystyle
\int_{\rt}^{\rtz}}{\rm
d}\rt'\,{\displaystyle \frac{{\rm
exp}[-F(\rt',\rtz)]}{\rt'-\frac{2}{9}}}} \ \ {\rm for}\ \ R_0=0.
\end{equation}
\end{itemize}

A final point concerns the hierarchy of IR attraction. The fixed point
(\ref{rtR}) is in fact the intersection point of {\it two} IR attractive
lines in the \rh-\rt-plane: the trivial one, \rt=2/9, attracting exceedingly
weakly like $(\gds/\gdsz)^{1/7}$, and the nontrivial one,
$\overline{\rh}(\rt)$, discussed in this section. This follows a general
rule \cite{guc}: A fixed point in a plane of two varables is the
intersection point of two fixed lines in the plane. The strength of
attraction of the two lines is regulated by the coefficients in the coupled
differential equations for the two variables. The general case is that one
of the fixed lines is more attractive for the RG flow than the other one;
the degenerate case in which both are roughly equally attractive, such that the
fixed point attracts the RG flow ``radially'', is, however, also
possible. Coming back to the \rh-\rt-plane: the point to make is that it is
gratifying that the {\it physically non-trivial} fixed line
$\overline{\rh}(\rt)$ is more strongly attractive than the more trivial
\rt=2/9 fixed line.


\subsection{The Top-Bottom-\protect\boldmath $\gd$ Sector of the SM and MSSM\newline --
Top-Bottom Yukawa Coupling Unification as an IR fixed Property\label{tbgd}}

The experimental top mass is much larger than the experimental bottom
mass. It is therefore interesting to see, which are the IR attractive
fixed manifolds in the top-bottom sector. 

Early analyses Refs. \cite{hill}-\cite{baggd} revealed already much of
the basic structures, though most of the conclusions were applied to
fictitious heavy higher fermion generations. Implicitely the analyses
\cite{taubuni}-\cite{ross1} of the consequences of tau-bottom Yukawa
coupling unification in supersymmetric unification for the IR physics
single out a narrow band of allowed values in the \tanb-\mt-plane which
turns out to lie in the vicinity of the IR fixed line in the top-bottom
system, to be discussed next. Why this is so and how the results look like
will be discussed in Sect. \ref{yukuni}. More recent analyses
\cite{sch2},\cite{schwi} meet the issue by confining the discussion to the
IR structure of the RGE in the SM and the MSSM.

Let us begin by a very brief discussion of a setting where the top and
bottom Yukawa couplings, \gt\ and \gb\ in the SM and \htt\ and \hb\ in
the MSSM, are the only non-zero couplings considered
(switching off for the moment also the strong gauge coupling). It is
easy to see from the resulting one-loop RG equations that there is a
trivial common fixed point, $\gt,\ \gb=0$, resp. $\htt,\ \hb=0$,
accompanied by an IR fixed line, the solution of the RGE distinguished
by the boundary condition that the ratio \gt/\gb, resp. \htt/\hb\ be
finite for $\gt,\ \gb\rightarrow\infty$. It is the line \gt=\gb\ in the
\gt-\gb-plane, resp. the line \htt=\hb\ in the \htt=\hb-plane. This
fixed line corresponds in this one-loop approximation to {\it exact
top-bottom Yukawa coupling unification at all scales $\mu$}. This is as it
turns out an
{\it IR attractive property} and a very intriguing result in the light of the
recently renewed interest in top-bottom Yukawa unification {\it at the UV
scale} in the MSSM as motivated from some GUT models.

Such an IR unification is of no interest in the SM, since it would imply in
this approximation \mt=\mb\ which is at strong variance with the
experimental situation. The subsequent discussion will show some kind of an
``escape route'' out of this dilemma.

The top-bottom Yukawa unification is, however, a fascinating and viable
option in the MSSM. The additional free parameter \tanb\ allows according to
the lowest order Eqs. (\ref{susymasses}) to have equal Yukawa couplings and any
disparity in the masses since in this order 
\begin{equation}
\frac{\mt}{\mb}=\frac{\htt}{\hb}\tanb
\end{equation}
with $\tan\beta$ being largely a free parameter.

After these introductory remarks, let us discuss the much more
non-trivial case of three non-zero couplings
\begin{equation}
\gt,\ \gb,\ ,\gd\neq0,\ \ {\rm resp.}\ \ \htt,\ \hb,\ \gd\neq0,
\end{equation}
while all the other couplings are put to zero in the RGE. The discussion
follows closely the analysis \cite{sch2}, which was performed in the MSSM. 

Again it is
economical to consider the following ratios of couplings 
\begin{equation}
\begin{tabular}{c|c}
${\rm SM}$ & ${\rm MSSM}$\\ \\ 
$\ \ \rt={\displaystyle \frac{\gts}{\gds}},\ \ \rb= {\displaystyle
\frac{\gbs}{\gds}},\ \ $ & 
$\ \ \rt={\displaystyle \frac{\hts}{\gds}},\ \ \rb= {\displaystyle
\frac{\hbs}{\gds}},\ \ $ \\ \\ 
\end{tabular}
\end{equation}
which lead to the following partially decoupled form of the RGE 
\begin{equation}
\begin{tabular}{c|c}
${\rm SM}$ & ${\rm MSSM}$\\ \\
$\;\;\;{\displaystyle -14\gds\frac{{\rm d}\,\rt}
{{\rm d}\, \gds}=\rt(9\rt+3\rb-2)}\;\;\;$ &
$\;\;\;{\displaystyle -3\gds\frac{{\rm d}\,\rt}
{{\rm d}\, \gds}=\rt(6\rt+\rb-\frac{7}{3})}\;\;\;$\\ \\
$\;\;\;{\displaystyle -14\gds\frac{{\rm d}\,\rb}
{{\rm d}\, \gds}=\rb(9\rb+3\rt-2)}\;\;\;$ &
$\;\;\;{\displaystyle -3\gds\frac{{\rm d}\,\rb}
{{\rm d}\, \gds}=\rb(6\rb+\rt-\frac{7}{3})}\;\;\;$.\\ \\
\end{tabular} 
\label{rtrbRGE}
\end{equation}
The obvious symmetry of the set of equations with respect to the
exchange of \gt\ and \gb, resp. of \htt\ and \hb, will be reflected in
all following results.

Clearly, the system of the two coupled differential equations
(\ref{rtrbRGE}) has a
number of fixed points; the only IR attractive one is the
following\footnote{One of us (B.S.) is grateful to W. Zimmermann and F.
  Schrempp for a communication and discussion on this point.} \cite{sch2};
\begin{equation}
\begin{tabular}{c|c}
${\rm SM}$ & ${\rm MSSM}$\\ \\
$\ \ \rt=\rb=\frac{1}{6}\ \ $ & $\ \ \rt=\rb=\frac{1}{3}\ \ $
\end{tabular} 
\label{rtrbfp}
\end{equation}  
the other fixed points are
\begin{equation}
\begin{tabular}{c|c}
${\rm SM}$ & ${\rm MSSM}$\\ \\
$\rt=0\ ({\rm IR\ repulsive}),\ \rb=0\ ({\rm IR\ repulsive}),$ & $\rt=0\ ({\rm IR\ repulsive}),\ \rb=0\ ({\rm IR\ repulsive}),
$\\ \\
$\rt=\frac{2}{9}\ ({\rm IR\ attractive}),\ \rb=0\ ({\rm IR\ repulsive}),$ & $\rt=\frac{7}{18}\ ({\rm IR\ attractive}),\ \rb=0\ ({\rm IR\ repulsive}),$\\ \\
$\rt=0\ ({\rm IR\ repulsive}),\ \rb=\frac{2}{9}\ ({\rm IR\ attractive}),$ & $\rt=0\ ({\rm IR\ repulsive}),\ \rb=\frac{7}{18}\ ({\rm IR\ attractive}),$\\
\end{tabular} 
\end{equation} 
There are two IR fixed lines in the \rt-\rb-plane. The one \cite{sch2} which
will turn out to be the less strongly IR attractive one, does not come as a
surprise;
\begin{equation}
\rt=\rb\ \ \ {\rm in\ the\ SM\ and\ the\ MSSM}.
\end{equation}
It signifies again top-bottom Yukawa coupling unification for all
scales $\mu$ with all the implications already discussed above.

The other one is the solution distinguished by the boundary condition
that it be finite$\neq0$ in the limit $\rb\rightarrow 0$ as well as in the
limit $\rt\rightarrow 0$. It is the solution \cite{sch2} which interpolates the
three  fixed points
\begin{equation}
\begin{tabular}{c|c}
${\rm SM}$ & ${\rm MSSM}$\\ \\
$(\rt=\frac{2}{9},\ \rb=0),$\ \ \ &\ \ \ 
$(\rt=\frac{7}{18},\rb=0),$\\$(\rt=\rb=\frac{1}{6}),$\ \ \ &\ \ \ $(\rt=\rb=\frac{1}{3}),$\\
$(\rt=0,\rb=\frac{2}{9}),$ \ \ \ &\ \ \ 
$(\rt=0,\rb=\frac{7}{18}).$\\
\end{tabular}
\end{equation}
This solution has roughly the shape of a quarter circle. Its end points are
the Pendleton-Ross fixed point (\ref{exfp}) at \rb=0 and the corresponding
mirror symmetric point under the exchange of \rt\ with \rb\ at \rt=0. The
line can be considered as the generalization of the Pendleton-Ross fixed
point into the \rt-\rb-plane. In case of the SM this IR fixed line has even
an implicit analytical solution\footnote{This solution was first written
  down by F. Schrempp (unpublished).}
\begin{eqnarray}
{\rm For}\ \rb\le\rt\hspace*{1cm}
\rt&=&\frac{(\rb-\rt)^2}{24\rt^{1/2}\rb^{3/2}}\ln\frac{\sqrt{\rt}-\sqrt{\rb}}{\sqrt{\rt}+\sqrt{\rb}}+\frac{1}{12}\frac{\rb+\rt}{\rb},\\
{\rm For}\ \rb\ge\rt\hspace*{1cm}
\rt&=&\frac{(\rb-\rt)^2}{24\rt^{1/2}\rb^{3/2}}\ln\frac{\sqrt{\rb}-\sqrt{\rt}}{\sqrt{\rb}+\sqrt{\rt}}+\frac{1}{12}\frac{\rb+\rt}{\rb}.
\end{eqnarray}
The two IR attractive fixed lines intersect each other in their only
common point, the IR attractive fixed point (\ref{rtrbfp}).

Next we need some analytical insight into the respective strengths of
attraction of the two IR fixed lines as well as of the IR fixed point
(\ref{rtrbfp}). In absence of an analytical solution one proceeds by
linearizing the system (\ref{rtrbRGE}) of differential equations for
\rt\ and
\rb\ in the neighbourhood of the common IR fixed point (\ref{rtrbfp}) and by
solving it analytically for UV initial values
\begin{equation}
\rtz=\rt(\mu=\lam),\ \rbz=\rb(\mu=\lam).
\end{equation}
The solutions are \cite{sch2}
\begin{equation}
\begin{tabular}{c|c}
${\rm SM}$ & ${\rm MSSM}$\\
$\ \
\rt(\gds)=\frac{1}{6}+\frac{1}{2}(\rtz+\rbz-\frac{1}{3})\left(
\frac{\gds}{\gdsz}\right)^{-\frac{1}{7}}\ \ $ &
$\ \ \rt(\gds)=\frac{1}{3}+\frac{1}{2}(\rtz+\rbz-\frac{2}{3})\left(
\frac{\gds}{\gdsz}\right)^{-\frac{7}{9}}\ \ $\\[4mm]
$\phantom{\ \ \rt(\gds)=\frac{1}{6}}+\frac{1}{2}(\rtz-\rbz)\left(
\frac{\gds}{\gdsz}\right)^{-\frac{1}{14}}\ \ $ &
$\phantom{\ \ \rt(\gds)=\frac{1}{3}}+\frac{1}{2}(\rtz-\rbz)\left(
\frac{\gds}{\gdsz}\right)^{-\frac{5}{9}}\ \ $\\ \\
$\ \ \rb(\gds)=\frac{1}{6}+\frac{1}{2}(\rtz+\rbz-\frac{1}{3})\left(
\frac{\gds}{\gdsz}\right)^{-\frac{1}{7}}\ \ $ &
$\ \ \rb(\gds)=\frac{1}{3}+\frac{1}{2}(\rtz+\rbz-\frac{2}{3})\left(\frac{\gds}{\gdsz}\right)^{-\frac{7}{9}}\
\ $\\[4mm]
$\phantom{\ \ \rb(\gds)=\frac{1}{6}}-\frac{1}{2}(\rtz-\rbz)\left(
\frac{\gds}{\gdsz}\right)^{-\frac{1}{14}}\ \ $ &
$\phantom{\ \ \rb(\gds)=\frac{1}{3}}-\frac{1}{2}(\rtz-\rbz)\left(
\frac{\gds}{\gdsz}\right)^{-\frac{5}{9}}.\ \ $\\ \\
\end{tabular}
\label{rtrbappr}
\end{equation}
In this approximation the second IR attractive line has the form
$\rt+\rb=\frac{1}{3}, {\rm resp.} \frac{2}{3}$. One can immediately read
off from the approximate analytical solution (\ref{rtrbappr})
\begin{itemize}
\item As well in the SM as in the MSSM, the quarter-circle IR
attractive line is more attractive, the relevant negative power of \gds/\gdsz\
being 1/7 to be compared to 1/14 in the SM and 7/9 as compared to 5/9
in the MSSM. Thus the ``top-down'' RG flow from all UV initial
values above and below the fixed line proceeds roughly first
towards the quarter-circle fixed line and then close to or along this
line towards the IR attractive fixed point (\ref{rtrbfp}). Only if the
solution happens to start on the \rt=\rb\  fixed line it remains on it
for the whole evolution path, i.e. for all values $\mu$ between \lam\
and \mt, leading to top-bottom Yukawa unification at all scales. 
\item In the SM both exponents, 1/7 as well as 1/14, are exceedingly
small, in particular the exponent 1/14 is so ridiculously small (given
the evolution from \lam\ to \mt) that it does not seem to be a
serious dilemma that the physical Yukawa couplings are so far from
equality. This is the basis of the ``escape route'' mentioned earlier
in this section. This is a practical argument which is, however, not
satisfactory from the mathematical point of view. 
\item The corresponding negative exponents in the MSSM are much larger
than in the SM; the quarter circle fixed line attracts with a strength
7/9, the \rt=\rb\  fixed line with a strength 5/9. 
\end{itemize}

\begin{sidewaysfigure}
\begin{tabular}{cc}
\raisebox{3mm}{\epsfig{file=rtrbSUSY.ps,bbllx=21pt,bblly=339pt,bburx=471pt,
bbury=770pt,width=11.5cm}}&
\epsfig{file=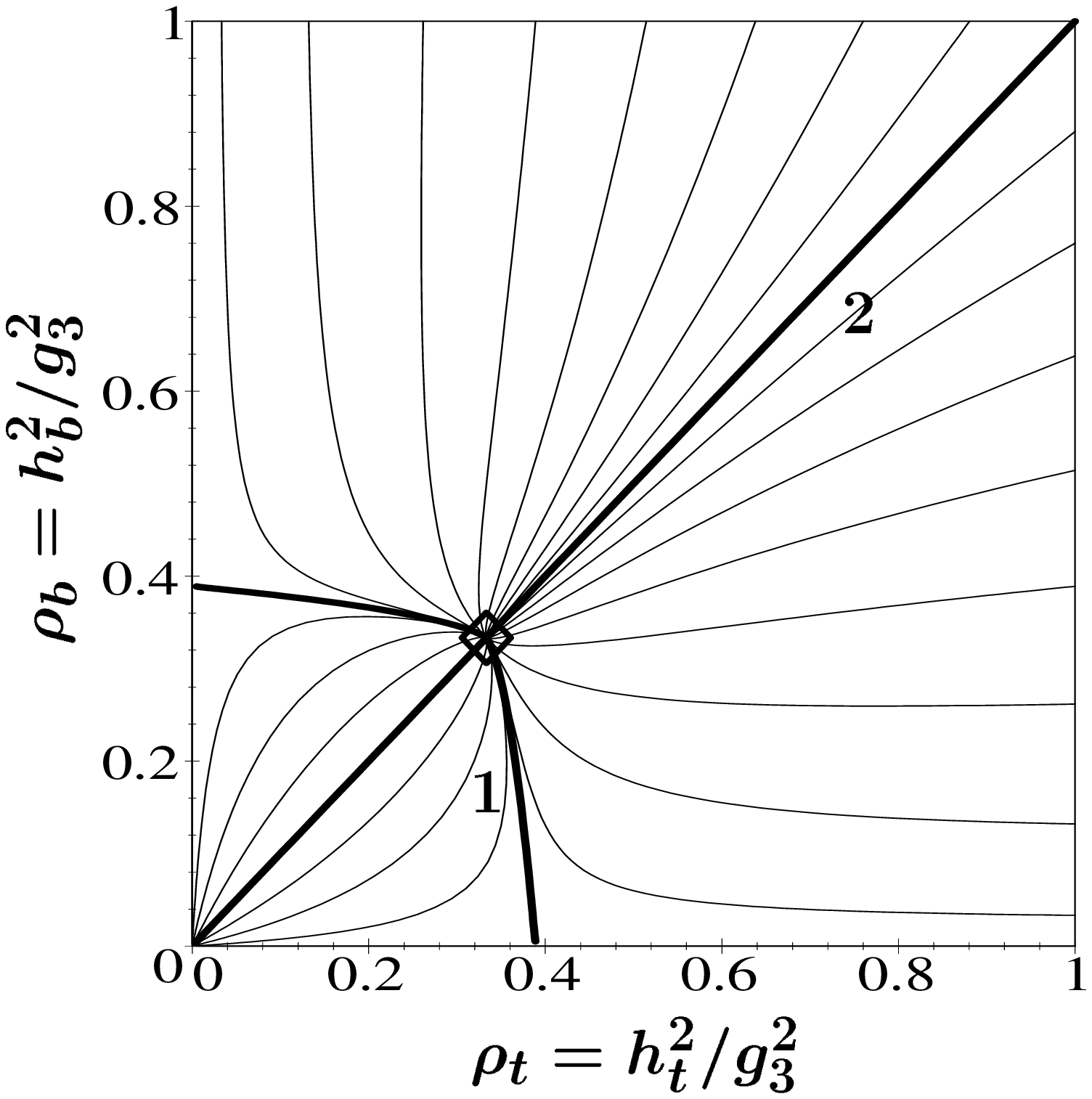,bbllx=73pt,bblly=247pt,bburx=507pt,
bbury=685pt,width=11.5cm}\\
{\bf a)} & {\bf b)}
\end{tabular}
\caption[dum]{The more attractive IR fixed line (fat line {\bf 1}),
  the less attractive IR fixed line (fat line {\bf 2}) and the IR
  attractive fixed point at their intersection (symbol $\Diamond$) in the
  \rb-\rt-plane of the MSSM. In addition a) The upper bound for
\lam=\mgut (thin line), playing the role of a Hill type effective fixed line, is shown. b) The ``top-down'' RG flow, indicated by
  representative solutions (thin lines), is shown to be first drawn
  towards the more strongly attractive fixed line line {\bf 1} and
  then close to it or along it towards the IR attractive fixed
  point. Line {\bf 2} implements top-bottom Yukawa unification at all
  scales $\mu$.}
\label{rtrbflow}
\end{sidewaysfigure}

There is again an upper boundary in the \rt-\rb-plane which collects
the IR images of solutions with sufficiently {\it large} UV initial value
\rtz\ or \rbz. Since it is essentially independent of sufficiently
large UV initial values \rtz\ or \rbz\ and very strongly attractive,
we shall also call it a Hill type {\it effective} IR fixed line. The
Hill effective fixed point, discussed in Subsect. \ref{tgd}, is one
point on this line, the point for \rbz=0. Due to the symmetry under
exchange of the variables \rt\ and \rb, there is a mirror symmetric Hill
effective fixed point for \rtz=0. This upper boundary is an IR effective
fixed line in the sense that it strongly attracts all solutions of the RGE
with sufficiently {\it large} initial values \rtz\ or \rbz\ and that it is
independent of those initial values. It is {\it not} a genuine IR fixed
line since i) it depends on the UV scale \lam, in fact it shrinks with
increasing \lam\ towards the genuine quartercircle shaped fixed line, and
ii) it does not attract RGE solutions with smaller initial
values \rtz\ or \rbz. The Hill type effective
fixed line is again dependent on the length of the evolution path,
i.e. on \lam\ (in the limit $\lam\rightarrow\infty$, which is of
mathematical interest only, the upper boundary, i.e. the Hill type
effective line tends towards the more attractive IR fixed line).

Let us add two illustrative figures for the MSSM. In Figs. \ref{rtrbflow}a)
and \ref{rtrbflow}b) we show the two IR fixed lines, the one denoted by
{\bf 1} is the more attractive one, the one marked by {\bf 2} is the less
attractive one, which incorporates top-bottom Yukawa unification at all
scales $\mu$; their intersection is the IR attractive fixed point, marked
by a symbol $\Diamond$. For comparison in Fig. \ref{rtrbflow}a) the
upper boundary, playing the role of a Hill type effective
fixed line, as calculated for an UV scale $\lam=\mgut\approx 2\cdot
10^{16}\gev$ is shown. In Fig. \ref{rtrbflow}b) the RG flow, first towards
the IR fixed line {\bf 1} and then towards the IR fixed point is
demonstrated with a selection of general solutions (thin lines).  Again the
more attractive fixed line appears as the ``watershed'' between solutions
which tend towards infinity for increasing scale $\mu$ and solutions which
tend towards zero in the same limit. Fig.  \ref{rHtb} in Subsect.
\ref{Htbgd} demonstrates the RG flow in case of the SM. Finally, Fig. \ref{net} in
Sect. \ref{masses} will show the contraction of UV initial values at
\lam=\mgut, subject to the MSSM ``top-down'' two-loop RG evolution, into
the close vicinity of the IR fixed lines for the more realistic case where
all gauge couplings are included and two-loop RGE equations are used.

An inclusion of the $\tau$ Yukawa coupling into the RGE of the MSSM has
been taken into account in Ref. \cite{sch2}. The fixed point (\ref{rtrbfp})
is then supplemented by the trivial fixed point value \rta=0, which turns
out to be rather strongly attractive, thus justfying a posteriori its
omission in the present discussion.


\subsection{The Higgs-Top-Bottom-\protect\boldmath $\gd$ Sector of the SM\newline
  -- a First IR Fixed Surface\label{Htbgd}}

\begin{figure}
\begin{center}
\epsfig{file=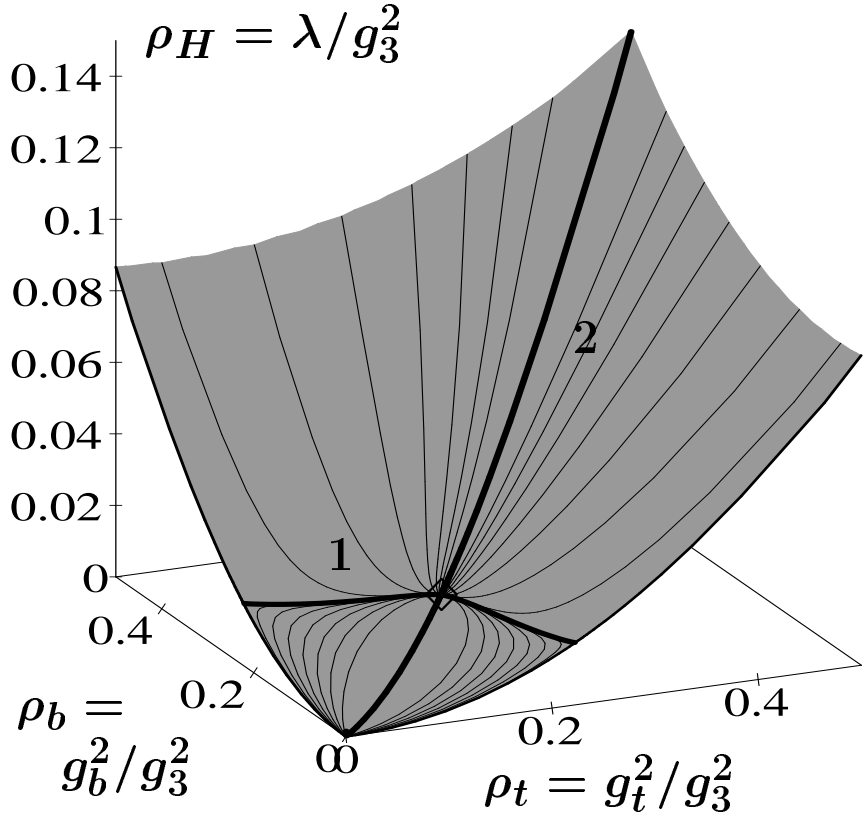,bbllx=119pt,bblly=259pt,bburx=560pt,
bbury=677pt,width=11.5cm}
\end{center}
\caption[dum]{The IR attractive surface in the \rh-\rb-\rb-space containing the more
  attractive IR fixed line (fat line {\bf 1}), the less attractive one (fat
  line {\bf 2}) and the IR fixed point (symbol $\Diamond$) at their
  intersection. The ``top-down'' RG flow is first drawn towards the surface
  (not shown), then within the surface towards the IR fixed line {\bf 1},
  then along or close to this line towards the IR fixed point, demonstrated
  by representative solutions (thin lines). The figure
  was taken from Ref. \cite{schwi}.}
\label{rHtb}
\end{figure}
Finally, the discussion of
\begin{equation}
\la,\ \gt,\ \gb,\ \gd\ \neq\ 0
\end{equation}
leads us to a reasonable approximation of the SM and to the first IR
attractive fixed surface \cite{schwi}.  The one-loop RGE for the three ratio variables
\rh=\la/\gds, \rt=\gts/\gds\ and \rb=\gbs/\gds\ are given by the Eqs.
(\ref{rtrbRGE}), supplemented by
\begin{equation}
{\displaystyle -7\gds\frac{{\rm d}\,\rh}
{{\rm d}\,
\gds}}=12\rh^2+6\rh\rt+6\rh\rb+7\rh-3\rt^2-3\rb^2.
\end{equation}
The common fixed point \cite{schwi} of the set of three differential equations is
\begin{equation}
\rt=\rb=\frac{1}{6},\ \ \rh=\frac{\sqrt{89}-9}{24}.
\label{fpHtb}
\end{equation}

In the space of the three ratio variables \rh, \rt\ and \rb\ there exists
a non-trivial, very strongly IR attractive surface \cite{schwi}, shown in
Fig. \ref{rHtb}. It is bounded for \rb=0 by the IR attractive fixed line
$\overline{\rh}$(\rt) in the \rh-\rt-plane as discussed in Subsect. \ref{Htgd},
and -- due to the mirror symmetry of the RGE under the exchange of \rt\ and
\rb\ -- for \rt=0 by the corresponding equally IR attractive line
$\overline{\rh}$(\rb)  in the \rh-\rb-plane. For large values of \rt=\rb=$\rho$ it has the
expansion
\begin{equation}
\rh(\rho)=\frac{1}{2}\rho-\frac{1}{6}+\frac{1}{36}\frac{1}{\rho}+...\ .
\end{equation} 
The IR attractive fixed surface fulfils the boundary condition of finite
ratios \rh/\rt\ and \rh/\rb\ for \rt\ or \rb\ increasing towards infinity. It
is the ``watershed'' between solutions above the surface, for which the
ratios \rh/\rt\ or \rh/\rb\ tend towards infinity, and solutions below the
surface which tend towards negative values. Again the IR fixed surface
separates the \rh-\rt-\rb-space in two distinct regions: no solution of the
RGE can penetrate from above to below the surface or vice versa. 

Solutions starting their evolution on the fixed surface, evolve within the
surface, independent of the UV initial values $\rh_0$, \rtz, \gdsz\ and
\lam.  The full ``top-down'' RG flow (not shown in Fig. \ref{rHtb}) is
first strongly attracted towards the IR fixed surface, from above and from
below, and then proceeds on it or close to it as displayed in Fig.
\ref{rHtb}. Within the surface we rediscover the generalizations of the two
IR attractive fixed lines, the more attractive quarter circle line (fat
line {\bf 1}) and the less attractive line \rt=\rb\ (fat line {\bf 2}),
which were discussed in Subsect. \ref{tbgd}. So the RG flow within the
fixed surface (indicated by the thin lines for representative solutions in
Fig. \ref{rHtb}) is towards the more attractive fixed line, from above and
from below, and then along or close to it towards the common fixed point
(\ref{fpHtb}) denoted by the symbol $\Diamond$.

A last comment concerns again the hierarchy of IR attraction. There are in
fact three IR attractive surfaces in the \rh-\rt-\rb-space. The by far most
strongly IR attractive non-trivial surface, discussed above, and the two
more trivial ones, rising vertically over the quarter circle line and the
\rt=\rb\ line, respectively, in the \rt-\rb-plane. In fact the fixed lines
{\bf 1} and {\bf 2} in Fig. \ref{rHtb} are each an intersection between the
non-trivial and the corresponding more trivial ones. The criterion for
which of the three IR attractive surfaces is the most attractive one lies
buried in the specific coefficients of the RGE. Again it is gratifying to
find the {\it physically most non-trivial} surface to be the most strongly
IR attractive one.


\section{Infrared Fixed Points, Lines, Surfaces in
the Presence of All Gauge Couplings\label{allgauge}}

In this section the analyses are extended to include the electroweak gauge
couplings \gee\ and \gz. Before entering the analytical discussion, let us
summarize in words the most important development of the last years
\cite{alv}-\cite{ross1}, \cite{mond}, \cite{zoup}  concerning the RG flow of the top Yukawa coupling
in the MSSM. 
\begin{itemize}
\item The top down RGE flow of the top Yukawa coupling $\htt(\mu)$ from
  large UV initial values $\htt_0\gg 1$ at the UV scale \lam=\mgut\ is
  focused essentially into one point at the IR scale $\mu=\mt=176\gev$,
  $\htt(176\gev)\approx 1.1$. This is not all too surprising, since one
  expects the phenomenon of an upper bound, acting as a Hill
 effective fixed point \cite{hill}, as
  discussed in Subsect. \ref{tgd}, also to be present if the electroweak
  couplings are switched on; it was precisely the power of the Hill
  effective fixed point to collect the IR images of large UV initial
  values.
\item More surprising is the fact that also smallish UV initial values for
  the ratio $\rtz=\htt_0^2/\gdsz$ are RG evolved {\it upwards} towards a
  {\it larger} IR value. A corresponding figure will be supplied later in
  this section. Thus, e.g. an initial value as small as, say,
  $\htt_0\approx 0.4$ is evolved towards $\htt(176\gev)\approx 0.8$.
\item Thus, it is perfectly safe to conclude \cite{alv}-\cite{sch2},
  \cite{taubuni}-\cite{ross1}, \cite{mond}, \cite{zoup} that in the top Yukawa
  coupling
  there is a very strongly  
\begin{equation}
{\rm IR\ attractive\ fixed\ point\ at}\ \ \  \htt=O(1)
\end{equation} which directly leads, with all radiative
  corrections applied, to the 
\begin{equation}
{\rm IR\ attractive\ top\ mass}\ \ \ \mt=O(190-200)\gev\sin\beta.
\end{equation}  
The near-at-hand qualitative interpretation, implied to a certain extent
also in references quoted above, is that switching on the electroweak gauge
couplings effects the genuine Pendleton-Ross type fixed point (\ref{exfp})
and the effective Hill fixed point (\ref{bt}) both to move up and to move
more closely together, combining their focusing power for the ``top-down''
RG flow. As we shall see from the following analytical discussion, this is
roughly what happens.
\item The effect is qualitatively also present in the SM, but the focusing
  effect for the RG flow is much less pronounced.
\end{itemize}

The analysis \cite{schwi} described in the following traces the dynamical
origin of these phenomena quasi with a mathematical magnifying glass back
to IR attractive fixed manifolds in the spaces of ratios of couplings. In
case of the MSSM the analytical insight is much improved, but the practical
consequences beyond those known already are somewhat limited.  However, in
the case of the SM the improved insight allows the exact determination of an
IR attractive top mass, Higgs mass and top-Higgs mass relation in presence
of the electroweak gauge couplings.

The presentation still sticks to the one-loop RGE, since in that framework
all discussed IR fixed manifolds are exact. The appropriate two-loop
results as well as their translation into
results for the Higgs and fermion masses, mass relations and mass bounds
including radiative corrections are
reviewed in Sect. \ref{masses}.

In the following it is again consistently economical to consider the
ratios of couplings \rh, \rt, \rb\ and \rta\ introduced in
Eqs. (\ref{ratios}), supplemented by the ratios
\begin{equation}
\re=\frac{\ges}{\gds},\ \ \ {\rm and}\ \ \ 
\rz=\frac{\gzs}{\gds}.
\end{equation}
The one-loop RGE, rewritten for these variables, are
\begin{equation}
\begin{tabular}{c|c}
${\rm SM}$ & ${\rm MSSM}$\\ \\
${\displaystyle \frac{{\rm d}\,\gds}{{\rm d}\, t}
=-7\frac{g_3^4}{8\pi^2}}$ & ${\displaystyle \frac{{\rm d}\,\gds}{{\rm
d}\, t} =-3\frac{g_3^4}{8\pi^2}}$,\\  
\end{tabular} 
\end{equation}
where \gd\ is treated as before as a function of $t$, and
\begin{equation}
\begin{tabular}{c|c}
${\rm SM}$ & ${\rm MSSM}$\\ \\
${\displaystyle \gds\frac{{\rm d}\,\re}
{{\rm d}\,
\gds}=-\re-\frac{41}{70}\re^2}$&${\displaystyle 
\gds\frac{{\rm d}\,\re}{{\rm d}\,
\gds}=-\re-\frac{11}{5}\re^2}$\\ \\
${\displaystyle \gds\frac{{\rm d}\,\rz}
{{\rm d}\,
\gds}=-\rz+\frac{19}{42}\rz^2}$&${\displaystyle 
\gds\frac{{\rm d}\,\rz}{{\rm d}\,
\gds}=-\rz-\frac{1}{3}\rz^2}$\\ \\
${\displaystyle -7\gds\frac{{\rm d}\,\rt}
{{\rm d}\,
\gds}}=$ & ${\displaystyle -3\gds\frac{{\rm d}\,\rt}
{{\rm d}\, \gds}}=$\\ \\ ${\displaystyle \rt(\frac{9}{2}\rt+\frac{3}{2}\rb+\rta-\frac{17}{20}\re-
\frac{9}{4}\rz-1)}$ &
${\displaystyle \rt(6\rt+\rb-\frac{13}{15}\re-3\rz-\frac{7}{3})}
$\\ \\
${\displaystyle -7\gds\frac{{\rm d}\,\rb}{{\rm d}\,
\gds}}=$ & ${\displaystyle -3\gds\frac{{\rm d}\,\rb}
{{\rm d}\, \gds}}=$\\ \\ ${\displaystyle \rb(\frac{3}{2}\rt+\frac{9}{2}\rb+\rta-\frac{1}{4}\re-
\frac{9}{4}\rz-1)}$ &
${\displaystyle \rb(\rt+6\rb+\rt+\rta-\frac{17}{15}\re-3\rz-
\frac{7}{3})}$\\ \\
${\displaystyle -7\gds\frac{{\rm d}\,\rta}
{{\rm d}\,
\gds}}=$ & ${\displaystyle -3\gds\frac{{\rm d}\,\rta}
{{\rm d}\, \gds}}=$\\ \\ ${\displaystyle \rta(3\rt+3\rb+\frac{5}{2}\rta-\frac{9}{4}\re-
\frac{9}{4}\rz+7)}$ &
${\displaystyle \rta(3\rb+4\rta-\frac{9}{5}\re-3\rz+3)}$,\\
\end{tabular}
\label{rtrbrta12} 
\end{equation}
supplemented within the SM
\begin{eqnarray}
{\displaystyle -7\gds\frac{{\rm d}\,\rh}
{{\rm d}\,
\gds}}&=&{\displaystyle 12\rh^2+6\rh\rt+6\rh\rb+2\rh\rta-\frac{9}{10}
\rh\re-\frac{9}{2}\rh\rz+7\rh}\nonumber\\
& &{\displaystyle -3\rt^2-3\rb^2-\rta^2+\frac{27}{400}\re^2+\frac{9}{40}
\re\rz+\frac{9}{16}\rz^2}.
\label{rh12}
\end{eqnarray}
In Eqs. (\ref{rtrbrta12}) and (\ref{rh12}) the ratios are treated as
functions of the independent variable \gds.

The general one-loop solutions for \re\ and \rz\ are
\begin{equation}
\begin{tabular}{c|c}
${\rm SM}$ & ${\rm MSSM}$\\ \\
$\re(\gds)={\displaystyle
  \frac{\re_0}{\frac{\gds}{\gdsz}\left(\frac{41}{70}\re_0
    +1\right)-\frac{41}{70}\re_0}}$ & $\re(\gds)={\displaystyle
  \frac{\re_0}{\frac{\gds}{\gdsz}\left(\frac{11}{5}\re_0
    +1\right)-\frac{11}{5}\re_0}}$ \\ \\ 
$\rz(\gds)={\displaystyle
  \frac{\rz_0}{\frac{\gds}{\gdsz}\left(-\frac{19}{42}\rz_0
    +1\right)+\frac{19}{42}\rz_0}}$ & $\rz(\gds)={\displaystyle
  \frac{\rz_0}{\frac{\gds}{\gdsz}\left(\frac{1}{3}\rz_0
    +1\right)-\frac{1}{3}\rz_0}}$.\\ \\  
\end{tabular} 
\end{equation}
Let us, for the purpose of mathematical insight, treat for a moment the
limit $\gds\rightarrow\infty$, which is of course unphysical, since it
leads outside the framework of applicability of the perturbative RGE
(\ref{rtrbrta12}) and (\ref{rh12}). In this limit, the variables \re\ and
\rz\ tend towards zero.  Thus, the coupled system of differential equations
(\ref{rtrbrta12}), supplemented by Eq. (\ref{rh12}) in case of the SM, has
a single common IR attractive fixed point (in the unphysical region of the RGE)
\begin{equation}
\begin{tabular}{c|c}
${\rm SM}$ & ${\rm MSSM}$\\ \\
$\rt=\rb={\displaystyle \frac{1}{6}},\ \ \rh={\displaystyle \frac{\sqrt{89}-9}{24}},\ \ \rta=0,\ \re=\rz=0$ &
$\rt=\rb={\displaystyle \frac{1}{3}},\ \ \rta=0,\ \re=\rz=0$ \\ \\ 
\end{tabular}
\label{fp12}
\end{equation}
which is identical to the one in absence of the electroweak gauge
couplings, supplemented by \re, \rz=0. This feature allows in retrospect
the consistent discussion of the RGE in {\it absence} of the electroweak
gauge couplings (and in absence of the $\tau$ Yukawa coupling) performed in
the last section. This circumstance, combined with the knowledge that the
electroweak gauge couplings are numerically small in the IR region, might
lead one to expect that the systematic inclusion of the electroweak
couplings amounts only to a small correction. This is not quite true.
Already when including the electroweak couplings by their constant, i.e.
non-running, averages \cite{pen} or by a more elaborate averaging procedure
\cite{sch1},\cite{sch2}, the position of the IR attractive fixed points is
shifted considerably, leading to substantially higher IR attractive top and
Higgs mass values. The many analyses
\cite{hill}-\cite{leung},\cite{froge},\cite{alv}-\cite{ross1}, \cite{mond},
\cite{zoup} quoted in
the introduction of this section, corroborate this statement in more or less
exact frameworks.

There is an intricate reason \cite{schwi} behind the strong effect of the
electroweak gauge couplings: there exist very strongly attractive IR fixed
manifolds defined by boundary conditions for {\it large} values of \re\ and \rz, which persist to be
strongly attractive even in the IR region where \re\ and \rz\ are small as
compared to one. This is most transparently studied in the case of the top
sector in presence of all gauge couplings in the next subsection. 

The procedure leading to success starts by treating \re\ and \rz\ as free
variables in parallel to \rh, \rt, \rb\ (and \rta). This increases the
space of ratio parameters by two dimensions. Such a procedure was put
forward for the first time in Refs. \cite{kusiziz} in the
framework of reduction of parameters to be discussed in Sect.
\ref{parred}. Within this enlarged space the IR attractive fixed subspaces
are determined and the full RG flow from the UV to the IR towards these IR
manifolds is considered (disregarding for the moment experimentally known
initial values for \re\ and \rz). On the one-loop level the resulting IR
fixed manifolds are exact. Of course all the IR fixed manifolds terminate
in the (unphysical) IR fixed point (\ref{fp12}).

The next step will consist in feeding in
the experimentally measured initial conditions \cite{par}
$\alpha(m_Z)=1/127.9$ and $sin^2\hat{\theta}_W(m_Z)=0.2319\pm 0.0005$
($\overline{\rm MS}$); ignoring the errors and evolving up to
$\mu=\mt=176\gev$, we find
\begin{eqnarray}
\ges(\mt=176\gev)&=&0.215,\ \gzs(\mt=176\gev)=0.418,\ \ {\rm leading\ to}\nonumber\\
\re(\mt=176\gev)&=&0.160,\ \rz(\mt=176\gev)=0.312,
\label{gez}
\end{eqnarray}
if the initial value (\ref{gdrei}), $\gds(\mt=176\gev)=1.34$, is taken into
account. This relates $\re(\mu)$ to $\rz(\mu)$ which will be called henceforth
``experimental relation between \re\ and \rz''. Both functions are then known
functions of the scale $\mu$, or, more
conveniently for our purposes, they become {\it known}
functions of 1/\gds, which in turn is a known function of the scale $\mu$. 
Finally, for calculating IR attractive top and Higgs mass values, one is
interested in evaluating \re\ and \rz\ at the IR scale, chosen as
$\mu=\mt=176\gev$ in the following, feeding in the values in
Eq. (\ref{gez}). 

So, finally the $n$-dimensional IR fixed manifold for free variables \re\ 
and \rz\ shrinks to an $n$-2 dimensional manifold if the IR values for \re\ 
and \rz\ are introduced. It is very important to realize that these
submanifolds become IR fixed points, lines, surfaces,... in the following
sense: the ``top-down'' RG flow tends more and more closely towards them
from above and from below, if the UV scale \lam\ increases, while the IR
scale (in our choice) $\mu=\mt=176\gev$) remains constant. In the limit
$\lam\rightarrow\infty$ while keeping the IR scale fixed, which is of
mathematical interest only, the full RG flow is drawn onto them. 

Let us also point out that the IR scale, chosen as 176\gev in the
following, is not really a free parameter. In determining the top mass
\mtp\ or the Higgs mass \mhp\ from an IR fixed point or fixed line, the
appropriate IR scale $\mu$ is determined implicitely from the conditions
\begin{eqnarray}
\frac{v}{\sqrt{2}}\gt(\mu=\mtp)&=&\mtp(1+\delta_t(\mu=\mtp)),\label{IRtSM}\\
v\sqrt{2}\sqrt{\la(\mu=\mhp)}&=&\mhp(1+\delta_H(\mu=\mhp))\ \ {\rm in\ the\ SM},\label{IRhSM}\\
\frac{v}{\sqrt{2}}\htt(\mu=\mtp)\sin\beta&=&\mtp(1+\delta_t(\mu=\mtp))\ \
{\rm in\ the\ MSSM}.\label{IRt}
\end{eqnarray}
Since, however, the dependence on $\mu$ in \gt, \la and \htt\ is only
logarithmic, the correction to the result will turn out to be negligible.

Again, for simplicity, \msusy=\mt=176\gev\ is chosen throughout this section.


\subsection{The Top Sector of the SM and MSSM\label{t123}}

The minimal and most instructive subsystem of ratios of couplings in the
presence of all gauge couplings is \rt, \re\ and \rz.  It provides also the
basis for the IR fixed point in the top mass of the MSSM,
$\mt=O(190-200)\gev\sin\beta$ \cite{baggv}-\cite{ross1}.

To start with, there are several IR fixed surfaces to be found in the \rt-\re-\rz\ space.
The first surface to be put forward in the literature \cite{kusiziz} turns
out to be not the most strongly IR attractive one. It was found in the
framework of parameter reduction, where IR attraction is not a criterion;
it will be discussed in Sect.  \ref{parred}.

By far the most strongly IR attractive surface \cite{schwi} is
characterized by its boundary conditions for {\it large} values of \re\ and
\rz. These limits
again lie outside of the region of validity of perturbation theory;
however, as we had also experienced in other cases, the surface defined by
these boundary conditions determines the properties in the perturbative
region.

The origin of this surface is most easily exposed, if first two limiting
cases \rz=0, $\rz\neq 0$ and \rz=0, $\re\neq 0$ are discussed.

For \rz=0, $\re\neq 0$ a strongly IR attractive line in the \rt-\re-plane
appears with the following properties \cite{schwi}
\begin{equation}
\begin{tabular}{c|c}
${\rm SM}$ & ${\rm MSSM}$\\ \\
\multicolumn{2}{c}{${\rm for}\ \re\rightarrow\infty {\rm \ it\
                        behaves\ asymptotically\ as}$}
\\ \\
\ \ $\rt\rightarrow {\displaystyle
\frac{11}{10}}\re $\ \ &\ \ $ \rt\rightarrow {\displaystyle 
\frac{56}{45}}\re $\ \ \\ \\
\multicolumn{2}{c}{${\rm with\ the\ general\ solution\ approaching\ 
it}\ {\it in\ this\ limit\ as}$}\\ \\
\ \ ${\displaystyle \frac{\rt}{\re}}\simeq {\displaystyle \frac{{\displaystyle
\frac{11}{10}}}{1-\left(1-{\displaystyle \frac{11}{10}\frac{\rez}{\rtz}}\right)\left(\frac{\re}{\rez}\right)^
{99/82}}}$\ \
&\ \ ${\displaystyle \frac{\rt}{\re}}\simeq {\displaystyle \frac{{\displaystyle
\frac{56}{45}}}{
1-\left(1-{\displaystyle \frac{56}{45}}{\displaystyle \frac{\rez}{\rtz}}\right)\left(\frac{\re}{\rez}\right)^
{112/99}}}.$\ \ \\ \\
\end{tabular}
\label{linert1} 
\end{equation}
Clearly, this fixed line is strongly IR attractive, since \re\ decreases in
its evolution towards the IR; the attraction towards the fixed line is
controlled by the large exponent 99/82, resp. 112/99 of \re. Incidentally
the fixed line corresponds to an IR fixed point in the variable
\begin{equation}
\rt/\re=\gts/\ges,\ {\rm resp.}\ \hts/\ges.
\end{equation}

Similarly instructive is the limit \cite{schwi}, where \rt\ is considered as a
function of \rz\ only with \re=0, which is even qualitatively different
for the SM and the MSSM.
\begin{equation}
\begin{tabular}{c|c}
${\rm SM}$ & ${\rm MSSM}$\\ \\ 
\ \ ${\rm there\ is\ an\ IR\ attractive\ fixed\ point}$\ \
&\ \ ${\rm there\ is\ an\ IR\ attractive\ fixed\ line}$\
\ \\
\multicolumn{2}{c}{$\hspace{33mm}{\rm in\ the\ \rt-\rz\ plane}$}\\ \\
\ \ $\rz={\displaystyle \frac{42}{19}},\ \rt={\displaystyle
\frac{227}{171}}$\ \ &\ \ $\rt={\displaystyle \frac{2}{3}}\rz\ {\rm for}\ \rz\rightarrow
\infty$\ \ \\ \\
\multicolumn{2}{c}{$\hspace{33mm}{\rm The\ general\ solution}$}\ \ \\
\ \ ${\rm for}\ \rz,\ \rzz\ {\rm near}\ {\displaystyle
      \frac{42}{19}}\ {\rm is}$\ \ &\ \ for large values of \rz\ is\\ \\
\ \ $\rt\simeq {\displaystyle \frac{{\displaystyle
    \frac{227}{171}}}{{\displaystyle
    1-\left(1-\frac{227}{171}\frac{1}{\rtz}\right)\left(\frac{\gds}{\gdsz}\right)^
{-227/266}}}}$\ \ &\ \ $\rt\simeq {\displaystyle 
\frac{{\displaystyle \frac{2}{3}\rz}}{{\displaystyle
    1-\left(1-\frac{2}{3}\frac{\rzz}{\rtz}\right)\left(\frac{\rz}{\rzz}\right)^4}}}$\
  \ \\ \\
\end{tabular}
\label{linert2} 
\end{equation}
Clearly, the fixed point resp. the fixed line is strongly IR
attractive: \rz\ decreases in its evolution towards the IR and the exponents,
227/266 resp. 4, controlling the attraction, are large.

The resulting two-dimensional IR attractive surface in the
\rt-\re-\rz\ plane is bounded by the above mentioned IR fixed lines
and fixed point. 

\begin{sidewaysfigure}
\begin{tabular}{cc}
{\bf SM} & {\bf MSSM}\\
\epsfig{file=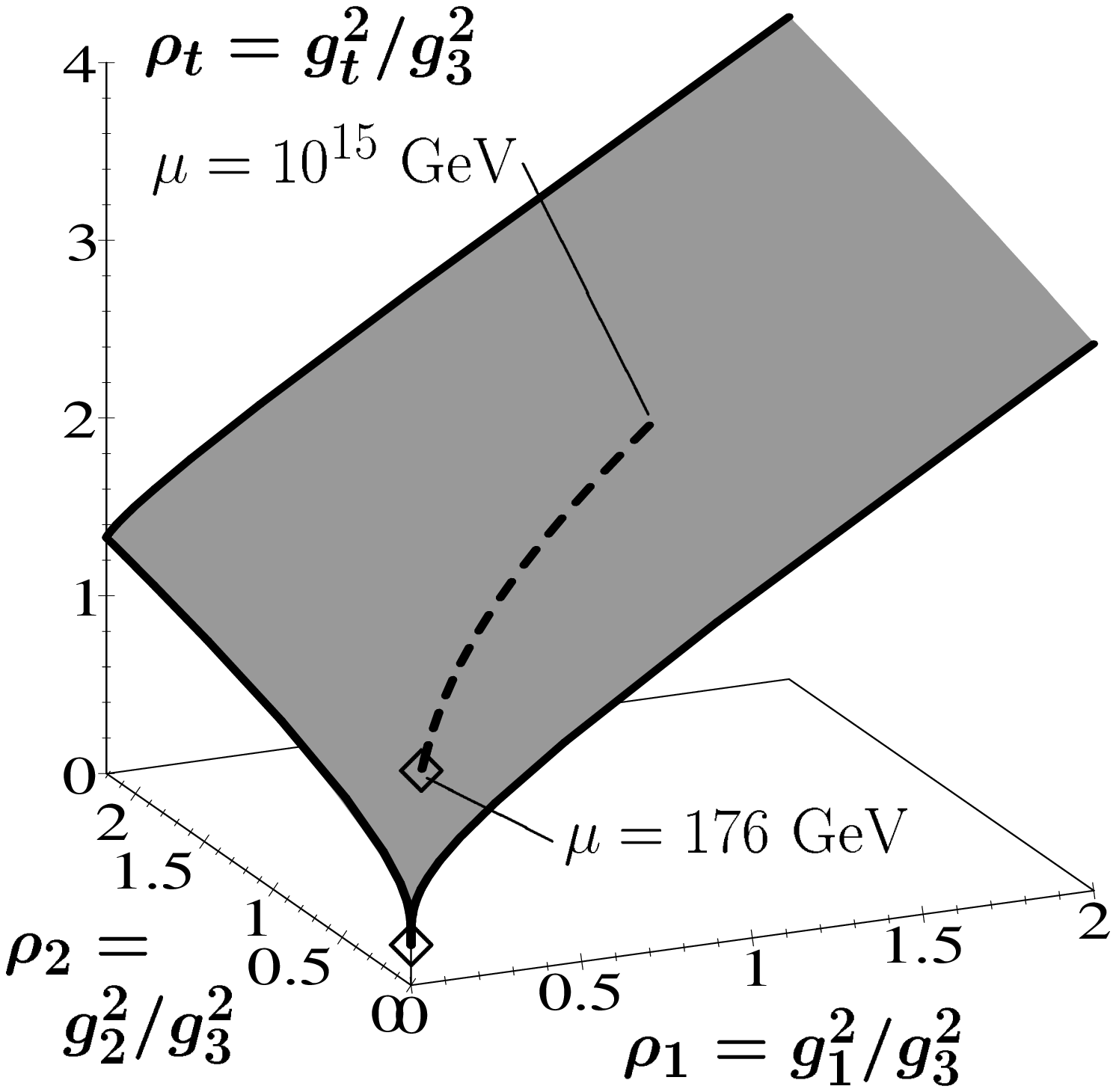,bbllx=123pt,bblly=258pt,bburx=552pt,
bbury=679pt,width=11.5cm}&
\epsfig{file=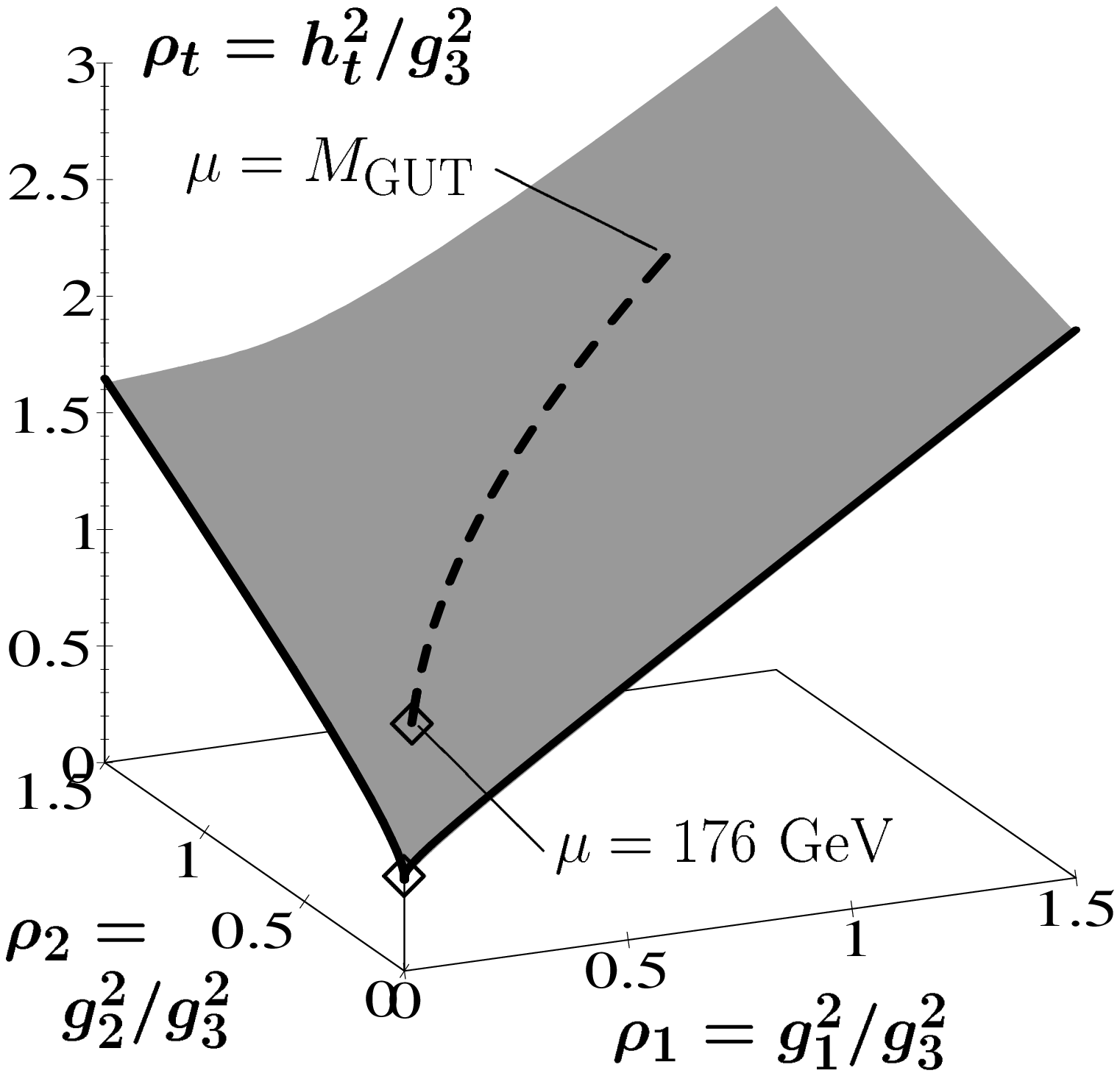,bbllx=123pt,bblly=258pt,bburx=562pt,
bbury=679pt,width=11.5cm}\\
\end{tabular}
\caption[dum]{The strongly attractive IR fixed surfaces in the
  \rt-\re-\rz-space with \re\ and \rz\ considered to be free variables.
  With the input of the experimental initial values for \re\ and \rz, the
  evolution of \re\ and \rz\ from a high UV scale to the IR scale
  $\mu=176\gev$ traces an IR fixed line (fat broken line) on the surface.
  At its IR tip is the IR fixed point (symbol $\Diamond$) towards which the
  RG flow is drawn. The figure was taken from Ref. \cite{schwi}.}
\label{rt12}
\end{sidewaysfigure}

The numerical determination of the IR fixed surface in the
\rt-\re-\rz-space is displayed in Fig. \ref{rt12} for the SM and for the
MSSM. The surface merges into the IR fixed point \rt=2/9, \re=\rz=0 for the
SM and into \rt=7/18,\re=\rz=0 , as all ``top-down'' RG solutions do. The
important point, however, is that the IR fixed surface is much more
strongly IR attractive than the IR fixed point. Thus the ``top-down'' RG
flow with free UV initial values for \rt, \re\ and \rz\ is first attracted
towards the surface and then close to it or along it towards the fixed
point.
\begin{sidewaysfigure}
\begin{tabular}{cc}
{\bf SM} & {\bf MSSM}\\
\epsfig{file=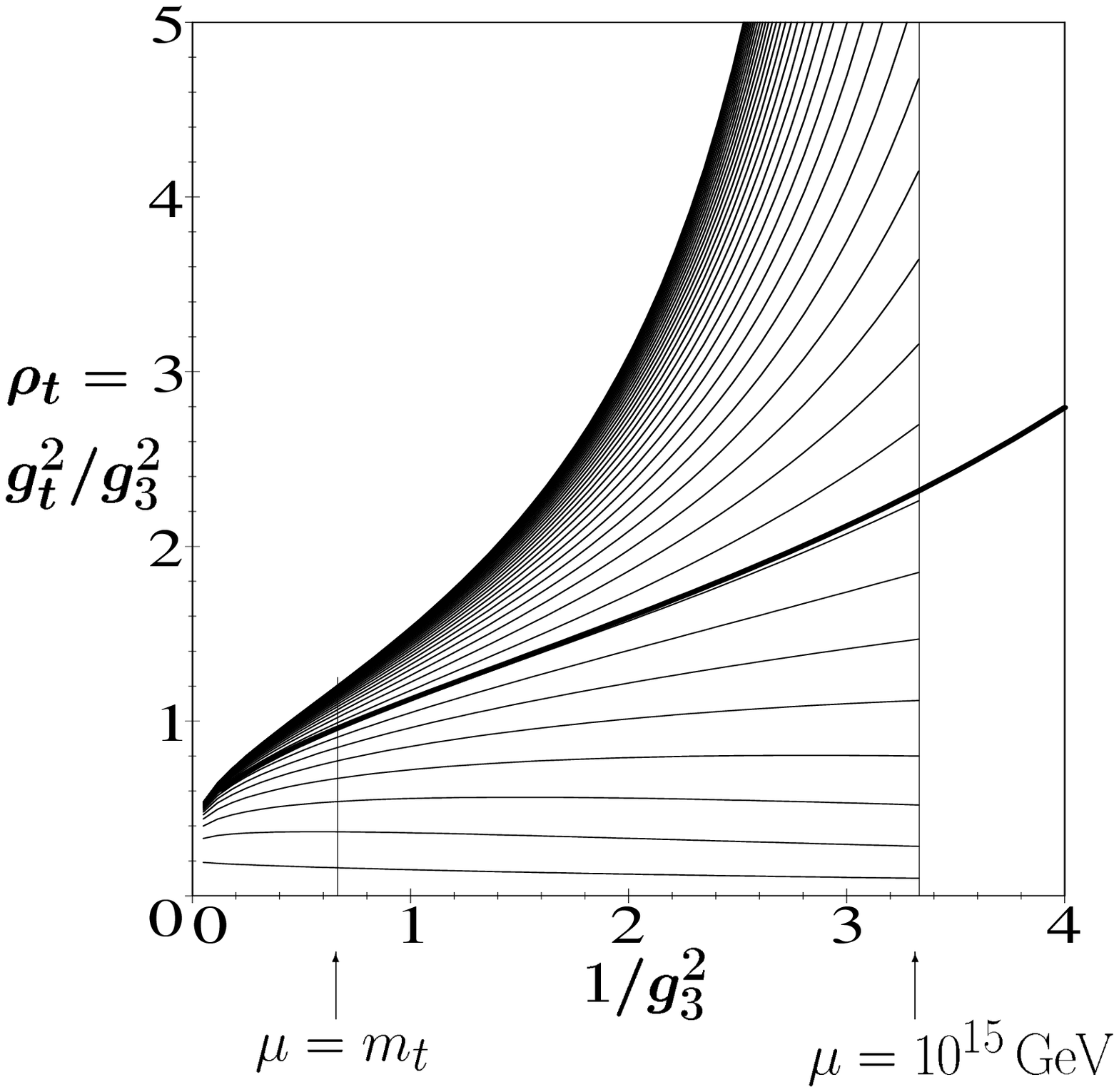,bbllx=14pt,bblly=311pt,bburx=492pt,
  bbury=770pt,width=11.5cm}&
\epsfig{file=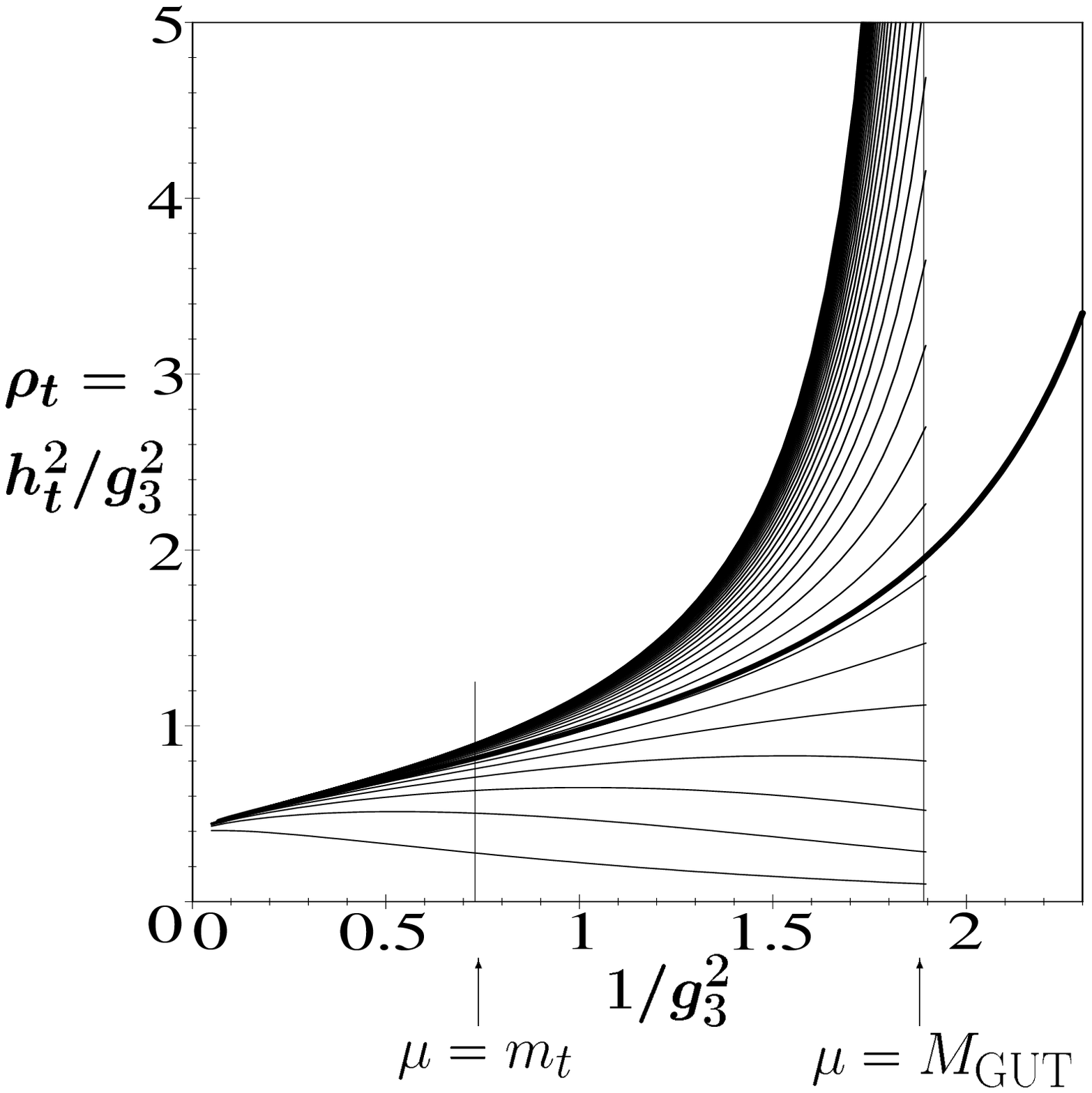,bbllx=14pt,bblly=311pt,bburx=492pt,
  bbury=770pt,width=11.5cm}\\
\end{tabular}
\caption[dum]{\rt\ as a function of $\mu$ or, more conveniently, as a
function of 1/\gds. The IR attractive fixed line (fat line) in presence
of all gauge couplings, identical
with the fat broken in Fig. \ref{rt12}, is shown to attract the RG
flow, represented by selected solutions (thin lines). The figure was
taken from Ref. \cite{schwi}.}
\label{rtr1r2flow}
\end{sidewaysfigure}

The surface can be represented analytically in various regions by
double power series expan- sions\footnote{Incidentally in the MSSM even an IR attractive fixed
line exists in the \re-\rz\ plane, \re=(5/33)\rz,
again characterized by a boundary condition in the unphysical region,
which even is not far from the line representing
the ``experimental relation'' between \re\ and \rz, but not
sufficiently near to make a point. So, this fixed line will be ignored
subsequently.} . For the SM they are \cite{schwi}
\begin{eqnarray}
 & & {\rm for}\ \re\rightarrow\infty,\ {\rm around}\ \rz=0\ {\rm in\
   powers\ of}\ 1/\re\ {\rm and}\ \rz\nonumber\\
 \nonumber\\
\rt&=&{\displaystyle \frac{11}{10}\re+\frac{176}{181}-\frac{2150720}
{8616143}\frac{1}{\re}+\frac{99}{362}\rz+\frac{7656563200}{35869003309}\frac{1}{\re^2}
+\frac{1298880}{8616143}\frac{1}{\re}\rz}
\nonumber\\ &
&-\frac{384780}{8616143}\frac{1}{\re}\rz^2+...\label{xyz1}\\ 
& &{\rm for}\ \re\rightarrow\infty,\ {\rm around}\ \rz= 42/19\ {\rm in\ powers\ of
}1/\re\ {\rm and}\ \frac{42}{19}-\rz\nonumber\\ \nonumber\\
\rt&=&\frac{11}{10}\re+\frac{5423}{3439}-\frac{418655600}{3110427623}
\frac{1}{\re}-\frac{99}{362}(\frac{42}{19}-\rz)+\frac{1270201090400}
{10696760595497}\frac{1}{\re^2}\nonumber \\
& &+\frac{7642800}{163706717}\frac{1}{\re}
(\frac{42}{19}-\rz)-\frac{384780}{8616143}\frac{1}{\re}(\frac{42}{19}-\rz)^2+...\label{xyz2} 
\end{eqnarray}
\begin{sidewaysfigure}
\begin{tabular}{cc}
{\bf SM} & {\bf MSSM}\\
\epsfig{file=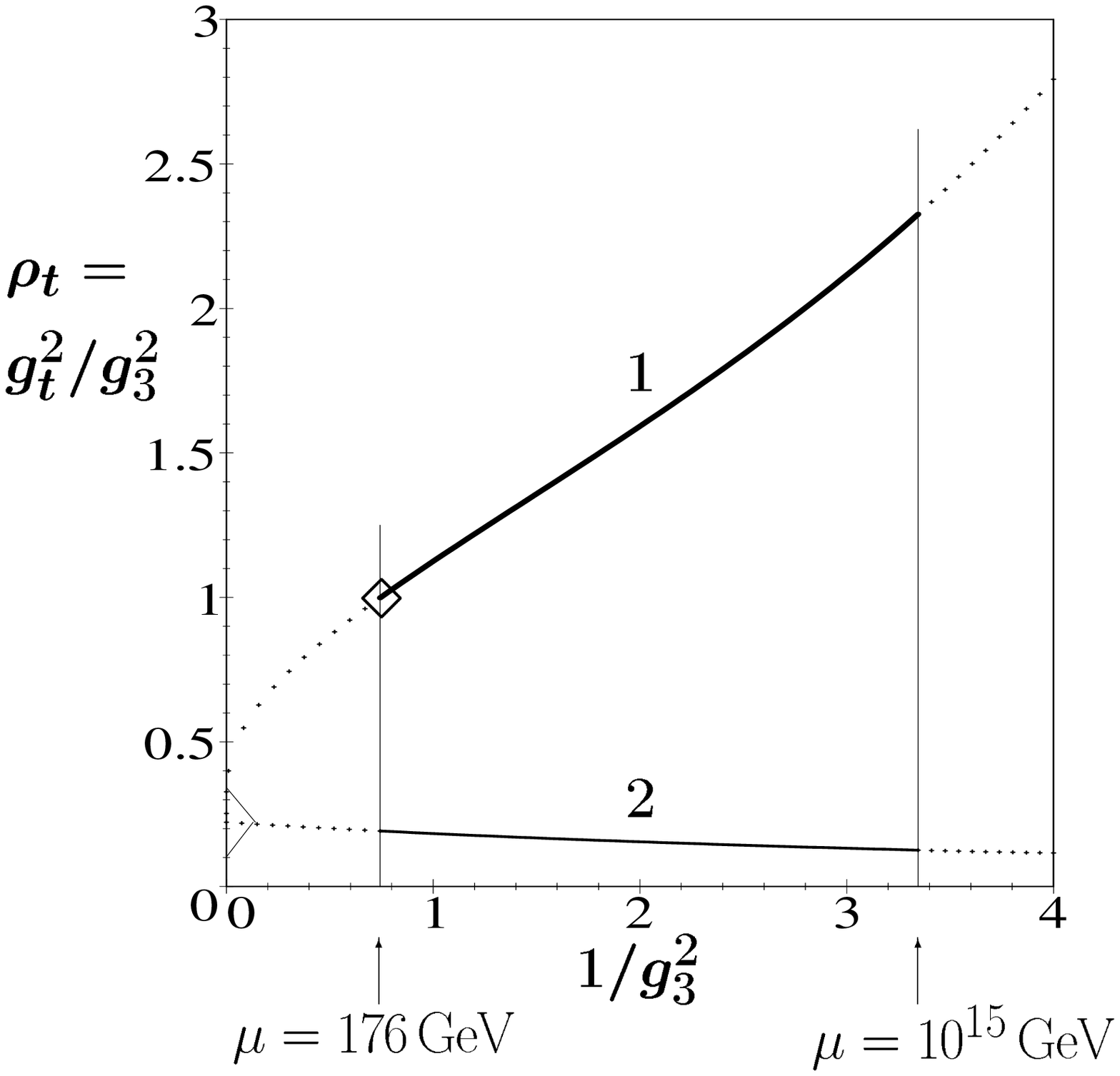,bbllx=14pt,bblly=316pt,bburx=491pt,
  bbury=769pt,width=11.5cm}&
\epsfig{file=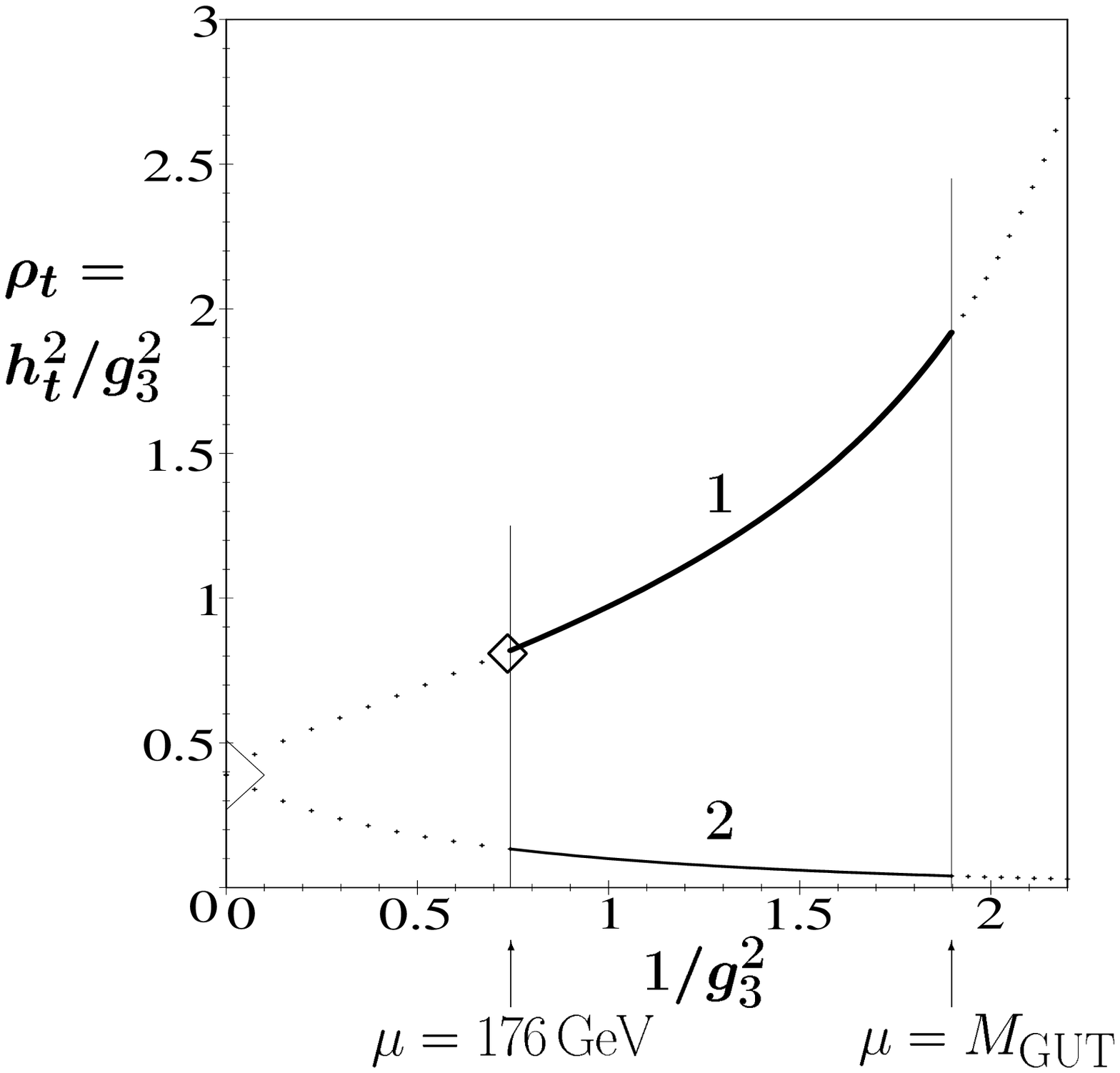,bbllx=14pt,bblly=314pt,bburx=492pt,
  bbury=770pt,width=11.5cm}\\
\end{tabular}
\caption[dum]{The IR attractive fixed line (fat line {\bf 1}) in presence
  of all gauge couplings as in Fig.  \ref{rtr1r2flow} with the IR
  attractive fixed point (symbol $\Diamond$). In comparison a much less IR
  attractive line (thin line {\bf 2}) is shown, which is favoured within
  the program of reduction of parameters to be discussed in Sect.
  \ref{parred}. The figure was taken from Ref. \cite{schwi}.}
\label{rtr1r2fili}
\end{sidewaysfigure}

Feeding in the initial values for \re, \rz\ and \gds, the resulting
experimental relation between \re\ and \rz\ singles out a line in the IR
attractive surface in the \re-\rz-\rt\ space, which has been depicted by a
fat line in Fig. \ref{rt12}; the line has finite length, running from an UV
scale \lam\ down to the IR scale \mt=176\gev. Plotting the \rt\ values of
the surface along this line as a function of 1/\gds\ results in the fat line
in Fig. \ref{rtr1r2flow}. Shown are also general solutions representing the
RG flow (thin lines).  

Here we find first of all the quantitative support for the
qualitative discussion led at the beginning of Sect. \ref{allgauge}: the
``top-down'' RG flow {\it from above and from below} is focused into a
very narrow band of IR values in case of the MSSM. There are many similar
figures for the RG flow in the literature
\cite{baggv}-\cite{frogz},
\cite{ara}-\cite{ross1},
mostly for the Yukawa coupling \htt. In case of the SM the focusing
effect, though qualitatively similar, is seen to be much less pronounced.

Next let us discuss the effect of the line cut out of the IR attractive
fixed surface, (fat broken line in Fig. \ref{rt12}, fat line in Fig.
\ref{rtr1r2flow}) displayed\footnote{The second line {\bf 2}
    will be discussed in Sect. \ref{parred}.}  again as the isolated line {\bf 1} Fig.
\ref{rtr1r2fili}. It clearly acts like an IR attractive line on this RG
flow for all solutions starting from above and from below the line. (This
becomes even more evident when persuing the solutions towards large values
of $\mu$, where all the solutions above the line are drawn towards it from
infinity, while those below the line are drawn towards it from zero). The
IR attraction is stronger in the MSSM, as expected from the analytical
discussion, and somewhat less strong in the SM.

At the IR scale, $\mu=\mt=176\gev$, the following conclusions can be drawn. 
\begin{itemize}
\item The IR point (symbol $\Diamond$) on the fixed line {\bf 1} in Fig.
  \ref{rtr1r2fili} plays the role of an IR attractive fixed point in
  presence of the electroweak gauge couplings as well in the SM as in the
  MSSM. It is a fixed point in the sense that for increasing UV scale \lam
  (which is of mathematical interest only), while keeping the IR scale
  $\mu=\mt=176\gev$ fixed, the RG flow contracts towards this point from
  below and from above. This fixed point replaces in the presence of the
  electroweak couplings the Pendleton-Ross fixed point (\ref{exfp}).
\item The IR image of all solutions starting from a high initial value of
  \rt\ or in other words the effective Hill fixed point in presence of the
  electroweak gauge couplings, represents the upper bound of
  all IR points.
\item This Hill fixed point is {\it distinctly above} the IR fixed point,
  however, in case of the MSSM very close to it since the IR fixed point is
  very strongly IR attractive. This was anticipated already in the
  introduction to Sect. \ref{allgauge}. For the SM the distinction has a
  sizeable effect to be discussed below.
\item In the perturbatively inaccessible region below $\mu=\mt$ the IR
  fixed line tends towards the genuine IR fixed point at \re=\rz=0 and
  \rt=2/9 for the SM resp. \rt= 17/18 for the MSSM, as expected. This part
  of the line is indicated by small crosses in Fig.  \ref{rtr1r2fili}. The
  fact that the fixed line rises so strongly between \re=\rz=0 and the
  still rather small \re\ and \rz\ values (\ref{gez}) at the IR scale is
  responsible for the strong increase of the IR fixed \rt\ value and
  correspondingly for the IR fixed top mass when switching on the
  electroweak gauge couplings.
  \end{itemize}
  It is worthwhile to translate the positions of the fixed point (symbol
  $\Diamond$) and of the Hill upper bound in the variable \rt\ into top
  mass values already at this level. One finds from
  $\mt(\mu=\mt)=\sqrt{\rt(\mu=\mt)}\gd (\mu=\mt) (v/\sqrt{2})\sin\beta$ the
  values of approximately $180\gev\sin\beta$ and $190\gev\sin\beta$,
  respectively. These values will be lifted by radiative corrections to
  $\mtp\approx 190\gev\sin\beta$ and $\mtp\approx 200\gev\sin\beta$,
  respectively. This shows that  the range of values
  $(190-200)\gev\sin\beta$ quoted in the literatur covers the range between
  the fixed point and the upper bound. It is, however gratifying to see
  that the IR fixed value lies at the lower end of that range admitting a top
  mass compatible with the experimental mass within errors even at
  larger values of $\sin\beta$. (This fact had been pointed out already in an
  approximate treatment of the inclusion of the electroweak gauge couplings
  in Ref. \cite{sch2}).

  The enhanced mathematical insight is very advantageous for the SM. There
  the IR attraction is much weaker than in the supersymmetric case.
  Therefore the IR fixed point and the Hill effective fixed point are
  distincly different. It is most appropriate to use the IR fixed point to
  assign an IR attractive value for top mass in the SM, which will be done
  in Sect.  \ref{masses}.

\subsection{The Top-Bottom Sector of the SM and MSSM\label{tb123}}

Next the analysis is extended to the top-bottom sector where we expect an
IR attractive fixed line at the IR scale. Let us repeat that early results
were obtained in Refs. \cite{hill}-\cite{baggd}, followed by the
analyses\cite{taubuni}-\cite{ross1} in the framework of supersymmetric
grand unification with tau-bottom Yukawa unification to be reviewed in
Sect. \ref{yukuni}. This framework allowed to trace
an allowed region in the \tanb-\mt plane which turns out to lie
in the vicinity of the IR fixed line. In this section we
restrict the discussion again exclusively to the search for the IR manifolds
in the RGE of the SM and MSSM; an approximate treatment  was given in
Ref. \cite{sch2} for the MSSM, the exact treatment for the SM as well as
the MSSM is found in Ref. \cite{schwi}.

The extension of this search to the \rt-\rb\ sector leads \cite{schwi} to a
strongly IR attractive three-dimensional subspace in the four dimensional
\rt-\rb-\re-\rz\ space (the analogue of Fig. \ref{rt12} in the \rt-\re\rz\ 
case). The analytical treatment includes e.g. the boundaries for
$\re\rightarrow\infty$, \rz=0
\begin{equation}
\begin{tabular}{c|c}
SM\ \ \ &\ \ \ MSSM\nonumber\\ &  \nonumber\\
$\rt=\frac{11}{10}\re$ for \rb=0\ \ \ &\ \ \  $\rt=\frac{56}{45}\re$ for \rb=0\\ & \nonumber\\
$\rb=\frac{29}{30}\re$ for \rt=0\ \ \  \ \ \ & $\rb=\frac{53}{45}\re$ for \rt=0\\ & \nonumber\\
$\rt=\frac{7}{8}\re,\ \rb=\frac{27}{40}\re$\ \ \  \ \ \ & $\rt=\frac{566}{525}\re,\
\rb=\frac{524}{525}\re.$\\ & \nonumber\\
\end{tabular}
\end{equation}
\begin{sidewaysfigure}
\begin{tabular}{cc}
{\bf SM} & {\bf MSSM}\\
\epsfig{file=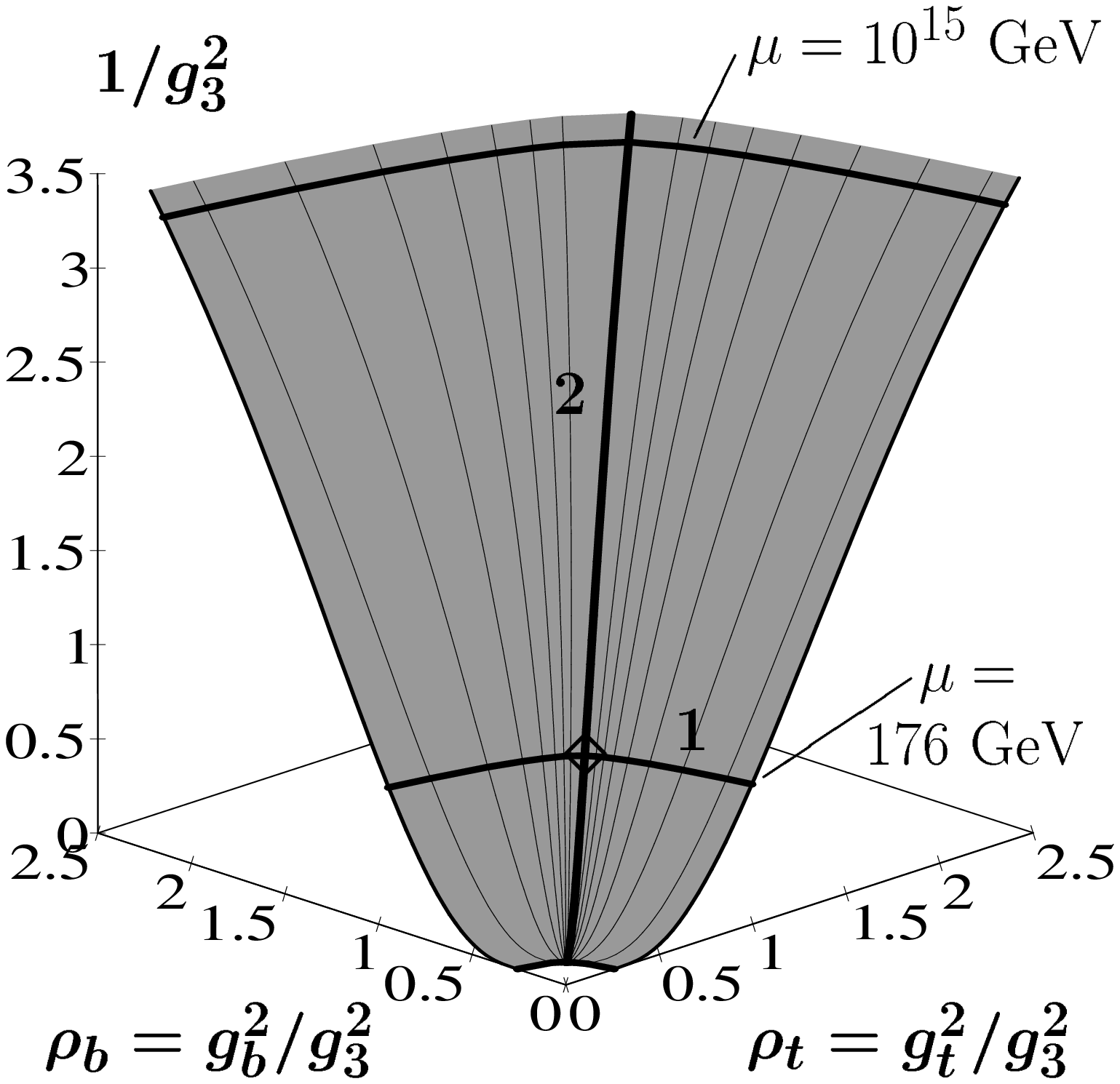,bbllx=122pt,bblly=258pt,bburx=562pt,
bbury=687pt,width=11.5cm}&
\epsfig{file=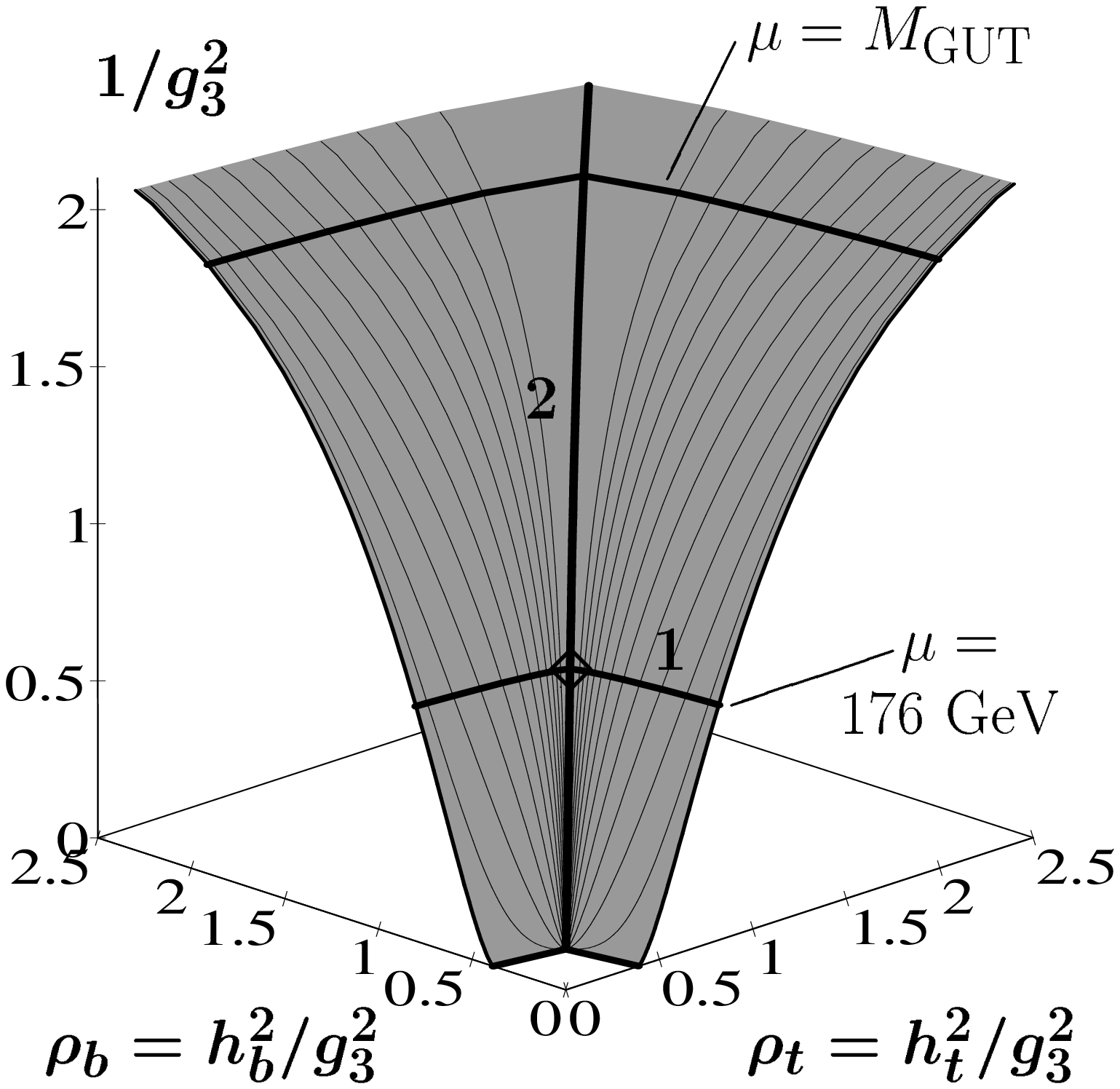,bbllx=118pt,bblly=258pt,bburx=562pt,
bbury=682pt,width=11.5cm}\\
\end{tabular}
\caption[dum]{IR attractive fixed surface in the presence of all gauge
  couplings in the space \rt,\rb\ as functions of $\mu$ or 1/\gds. Shown
  are the more attractive IR fixed line (fat line {\bf 1}), the less
  attractive fixed line (fat line {\bf 2}), the latter implementing
  approximate top-bottom Yukawa unification at all scales $\mu$, the IR fixed point
  (symbol $\Diamond$) at their intersection. The RG flow within the
  surface is represented by selected solutions (thin lines). The figure was
  taken from Ref.  \cite{schwi}.}
\label{tbgdsurf}
\end{sidewaysfigure}
\begin{sidewaysfigure}
\begin{tabular}{cc}
{\bf UV} & {\bf IR}\\
\epsfig{file=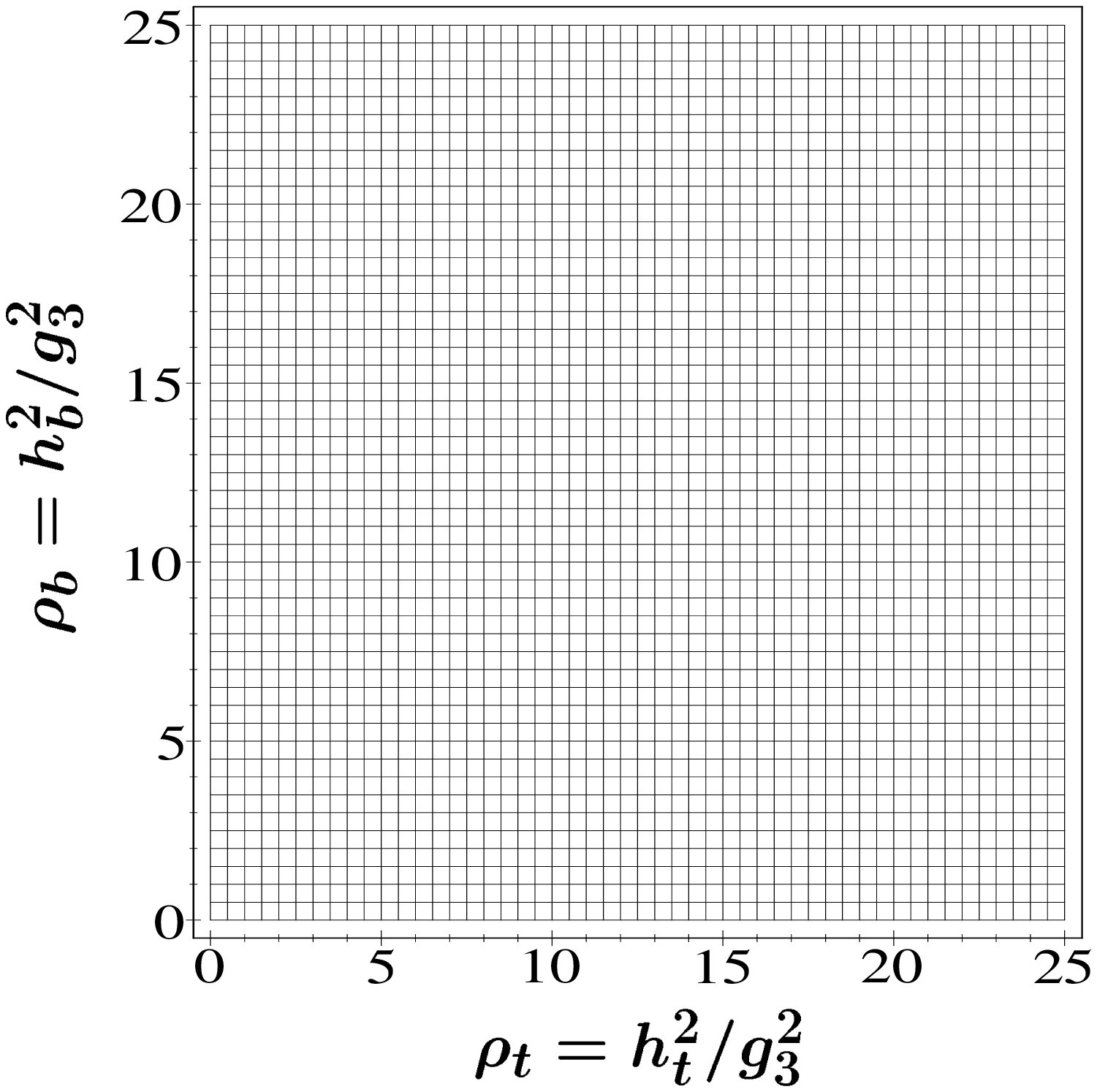,bbllx=72pt,bblly=247pt,bburx=506pt,
bbury=679pt,width=11.5cm}
&
\epsfig{file=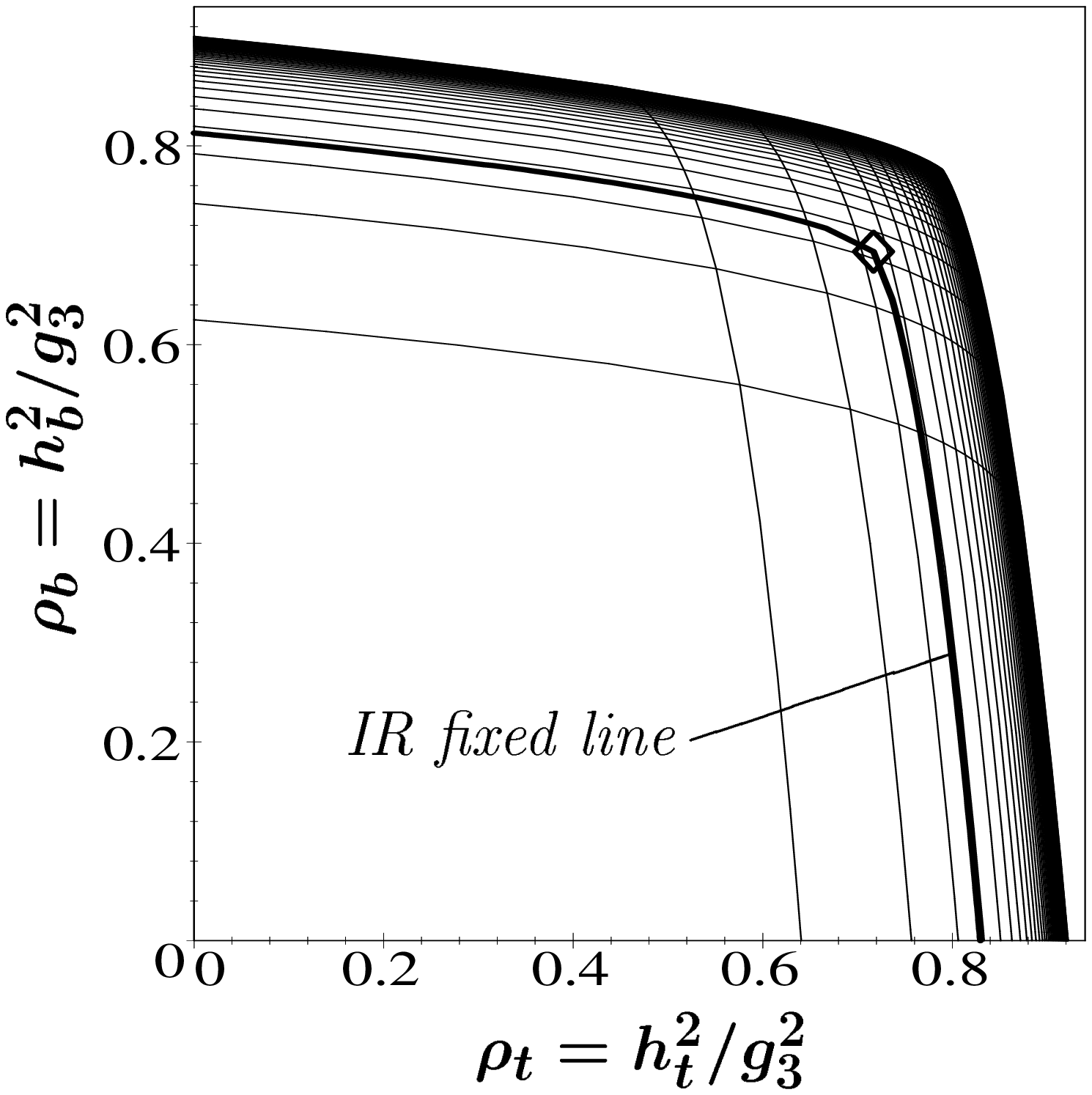,bbllx=72pt,bblly=247pt,bburx=509pt,
bbury=679pt,width=11.5cm}\\
\end{tabular}
\caption[dum]{The lattice points are chosen as UV initial values in the
\rt-\rb-plane at
$\lam=\mgut\approx 2\cdot 10^{16}\gev$ and subject to the RG evolution
of the MSSM in presence of all gauge couplings to the IR scale $\mu=176\gev$; notice that the UV plane
(figure on the left hand side) has scaled down by a factor of 25 to
the IR plane (figure on the right hand side). The very strong IR
attraction of the IR fixed line (fat line) and the IR fixed point
(symbol $\Diamond$) on it becomes apparent. The upper bound of all
lines constitute the Hill effective fixed line. The figure was taken from
Ref. \cite{schwi}.}
\label{net}
\end{sidewaysfigure}
The three-dimensional fixed subspace cannot be represented graphically any
more.  Injecting the experimental relation between \re\ and \rz\ as a
function of 1/\gds, leads to the projection into a two-dimensional IR
attractive fixed surface in the \rt-\rb-1/\gds\ space, shown in Fig.
\ref{tbgdsurf} which attracts the RG flow from above and below. Also
displayed are the RG flow within the surface (thin lines), which is
followed by solutions which start on the surface or have been attracted
onto it. Running down the IR surface to scale $\mu=\mt$ singles out an IR
attractive fixed line in the $\rt(\mu=\mt)$-$\rb(\mu=\mt)$ plane (the fat
line {\bf 1} in Fig. \ref{tbgdsurf}. This line substitutes in the presence
of the electroweak gauge couplings the role of the quarter circle fixed
line, determined for \re=\rz=0 in Subsect \ref{tbgd}. It has a similar
shape, but lies at substantially higher values. Again the line and the IR
fixed point on it (Diamond) is distingished by the fact that for increasing
UV scale \lam, while keeping the IR scale $\mu=\mt=176\gev$ fixed, the RG
flow concentrates more and more closely towards the line and shrinks along
the line more and more closely towards the IR fixed point (except for UV
initial values \rt=0 or \rb=0). The IR fixed point in \rt\, determined for
\rb=0 in the last subsection, lies on this IR fixed line at \rb=0.
(Evolution into the perturbatively unaccessible region below $\mu=\mt$,
draws all solutions running in the IR attractive surface towards \re=\rz=0
and the genuine IR fixed lines and finally the genuine IR fixed point in
the \rt-\rb\ plane discussed in Sect. \ref{tbgd}).

The resulting IR fixed line (fat line {\bf 1} in Fig. \ref{tbgdsurf}) is
plotted in Fig. \ref{net} as fat line, the IR fixed point is denoted by the
symbol $\Diamond$. To be more realistic, the figure contains
already the result from a two-loop RGE evolution. Fig. \ref{net} also
illustrates the dramatic IR attraction of this fixed line and of the fixed
point on it. On the left hand side a dense lattice in the large plane
$0\lwig\rt,\rb\leq 25$ is shown. The lattice points are taken as UV initial
values at the scale $\mgut\approx 2\cdot 10^{16}\gev$. The right hand side
shows the IR image of the lattice at the scale $\mu=\mt=176\gev$. Notice
first of all that the lattice has shrunk by a factor of 25 in each
dimension to within a square $0\lwig\rt,\rb\leq 1$. Secondly one sees, how
the large initial values \rt\ or \rb\ have shrunk towards the boundary, the
Hill effective fixed line, which is close to the IR fixed line but distinct
from it. The IR fixed line is independent of the UV scale \lam=\mgut. The
Hill effective line, however, reflects the choice $\lam=\mgut\approx 2\cdot
10^{16}\gev$; in the mathematical limit $\lam\rightarrow\infty$, while
keeping the IR scale fixed, the upper boundary tends towards the IR
fixed line. The whole RG flow first is attracted by the IR fixed line from
above and below and then proceeds along or close to the IR fixed line
towards the IR fixed point (with the exception of the UV initial values
\htt=0 and \hb=0). The IR fixed line and the upper boundary have moved much
closer together in the presence of the electroweak gauge couplings.

Let us also come back to the important issue of the top-bottom Yukawa
unification at all scales. The fat line {\bf 2} in Fig. \ref{tbgdsurf}
signalizes that top-bottom Yukawa unification at all scales $\mu$ survives
the inclusion of the electroweak gauge couplings as an approximate
property. It is weakly IR attractive. Also the IR fixed point ($\Diamond$)
in Figs. \ref{tbgdsurf} and \ref{net} implies approximate top-bottom Yukawa
unification at the IR scale $\mu=\mt=176\gev$.

The translation of the results in the \rt-\rb-plane into the
\tanb-\mtp-plane of the MSSM will be performed in Sect. \ref{masses}.

\subsection{The Higgs-Top-Bottom Sector of the SM\label{Htb123}}

\begin{figure}
\begin{center}
\epsfig{file=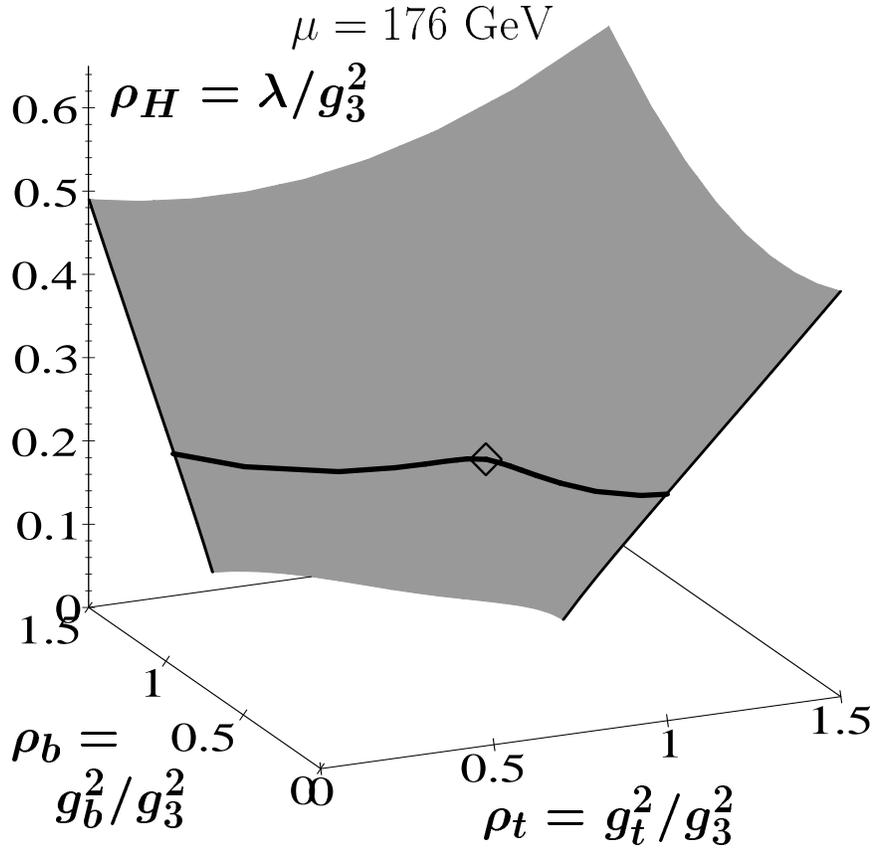,bbllx=123pt,bblly=259pt,bburx=560pt,
bbury=687pt,width=11.5cm}
\end{center}
\caption[dum]{The strongly IR attractive surface in the \rh-\rt-\rb-space at the
  IR scale $\mu=176\gev$ in the presence of all gauge couplings, the IR
  attractive line (fat line) and the IR attractive fixed point (symbol
  $\Diamond$). The IR attraction of the RG flow is first towards the fixed
  surface, then along or close to the surface towards the fixed line and
  finally along or close to the line towards the fixed point. The figure
  was taken from Ref. \cite{schwi}.}
\label{rHtbmu}
\end{figure}

The two-dimensional IR attractive surface, discussed in the absence of the
electroweak gauge couplings, presumably turns into a four-dimensional IR
attractive subspace in the five dimensional \rh-\rt-\rb-\re-\rz\ space.
Injecting the experimental relation between \re\ and \rz\ and evolving down
to $\mu=\mt$ leads to the two-dimensional IR attractive fixed surface in
the $\rh(\mu=\mt)$-$\rt(\mu=\mt)$-$\rb(\mu=\mt)$-space shown in Fig.
\ref{rHtbmu}, replacing in the presence of the electroweak gauge couplings
the surface Fig. \ref{rHtb}. The surface in Fig.  \ref{rHtbmu} \cite{schwi}
is again distinguished by the fact that in the (unphysical) limit
$\lam\rightarrow\infty$, while keeping the IR scale fixed, the RG flow is
drawn first towards it, then within the surface towards the IR fixed line
(fat line) and finally along or close to the line towards the fixed point
($\Diamond$). (Again the IR fixed surface, Fig. \ref{rHtb}, is approached
in the perturbatively unaccessible region for $\mu$ below \mt).


\section{Infrared Attractive Top and Higgs Masses, Mass
Relations and Mass Bounds\label{masses}}

The previous sections served to develop the subject step by step
pedagogically, still keeping to the one-loop level, since it allows to
determine the IR fixed manifolds exactly.  Next we are going to review the
results on IR attractive top and Higgs mass values and IR attractive mass
relations and on Higgs and top mass bounds on the professional level. The
present state of the art in most publications is the following.
\begin{itemize}
\item Two-loop RGE for the couplings involved are used. At the two-loop level
  the IR fixed manifolds are mathematically not exact any more, but
  numerically well-determined within the perturbatively allowed region
  $\mt\lwig\mu\lwig\lam$; they turn out to shifted by at most 10$\%$ with
  respect to the one-loop ones. 
\item In order to determine masses from couplings, the two-loop running
  couplings are first related to the running masses in the $\overline{\rm
    MS}$ scheme according to the relations (\ref{mfr}) for the fermions and
  (\ref{mhr}) for the Higgs boson.  Then in most publications the radiative
  corrections, relating the running masses to the physical pole masses (or
  at least the most important ones, the radiative QCD corrections to the
  quark masses), as detailed in Sect. \ref{rad} are applied. For
  convenience let us collect the relevant formulae again, including the
  matching conditions (\ref{match}) for Yukawa couplings in the MSSM for
  the transition from the RGE of the SM to the RGE of the MSSM at the scale
  \msusy,
  \begin{eqnarray}
 m_f(\mu)&=&\mfp(1+\delta_f(\mu)) \;\;\;{\rm with}\nonumber\\ 
  m_f(\mu)&=&\frac{v}{\sqrt{2}}
  g_f(\mu)=\frac{1}{\sqrt{2\sqrt{2}G_F}}g_f(\mu);\;\;{\rm for}\; f=t,b,\tau\nonumber\\
\gt(\msusy^{-}) & = & \htt(\msusy^{+})\sin\beta,\ \ {\rm in\ case\ of\ the\ MSSM}\nonumber\\ \gb(\msusy^{-}) & = &
  \hb(\msusy^{+})\cos\beta,\ \ {\rm in\ case\ of\ the\ MSSM}\nonumber\\ \gta(\msusy^{-}) & = &
  \hta(\msusy^{+})\cos\beta\ \ {\rm in\ case\ of\ the\ MSSM}\nonumber\\
  \mh(\mu)&=&\mhp(1+\delta_H(\mu)) \;\;\;{\rm with}\nonumber\\ 
  \mh(\mu)&=&\sqrt{2\lambda(\mu)}v=\sqrt{\frac{\sqrt{2}}{G_F}\la(\mu)}.
\end{eqnarray}
The choice of the IR scale varies from $\mu=m_Z$ to $\mu=\mt$ in the
literature, which can give rise to slight deviations e.g. in the precise
determination of mass bounds. Typically the supersymmetry breaking scale
\msusy\ is varied between \mt\ and 1\,TeV; for the determination of an
upper bound of the light Higgs mass in the MSSM values up to 10\,TeV are
considered.
\item The known bottom and tau masses, Eqs. (\ref{massb},\ref{masstau}),
  have to be evolved from $\mu=\mb$, resp.  from $\mu=\mta$ up to the IR
  scale. For this purpose, one makes use of the relations between the
  running masses and the $\mu$ independent pole masses, collected in Sect.
  \ref{rad}. (See. e.g. Ref. \cite{bar} for a detailed presentation
  including also a figure of the rather sizeable dependence of
  $\mb(\mu=\mt)$ on the initial value for $\alpha_s(m_Z)$, where
  $\alpha_s=\gds/(4\pi)$). Notice that the bottom mass quoted in Eq.
  (\ref{massb}) is the running mass $\mb(\mu=\mb)$ in the $\overline{\rm
    MS}$ scheme.
\end{itemize}  

\subsection{Top Mass and \protect\boldmath $\tan\beta$ in the MSSM\label{secMSSM}}

Let us start with the more conspicous and well known results in the MSSM,
the much-quoted strongly IR attractive fixed point \cite{alv}-\cite{ross1},
\cite{mond}, \cite{zoup}
\begin{equation}
\mtp=O(190-200)\gev\sin\beta
\label{every}
\end{equation}
and a tendency for IR values of \tanb\ to settle around $\tanb=O(60)$ 
and the analytical dissection of these results.

\begin{figure}
\begin{center}
\epsfig{file=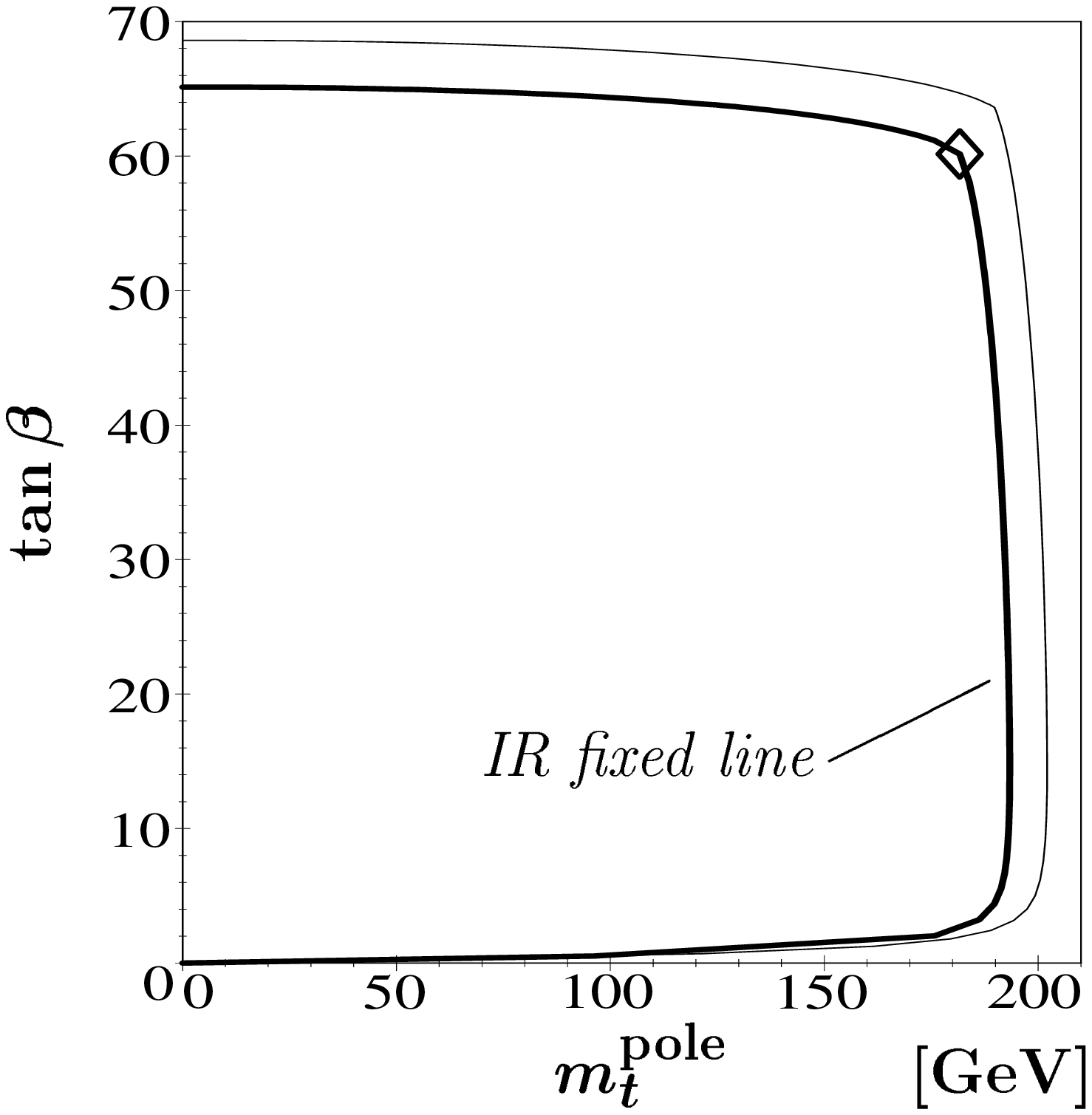,bbllx=123pt,bblly=247pt,bburx=548pt,
bbury=684pt,width=11.5cm}
\end{center}
\caption[dum]{The very strongly IR attractive fixed line (fat line) in the
  \tanb-\mtp-plane of the MSSM and the IR attractive fixed point (symbol
  $\Diamond$) are shown; the fixed point implements approximate top-bottom
  Yukawa unification. The upper boundary with the interpretation of a Hill
  type effective fixed line, i.e. the IR image of all large initial
  values for \htt\ or \hb\ at \lam=\mgut, is shown as thin line. The figure
  was taken from Ref. \cite{schwi}.}
\label{tanbmt}
\end{figure}

In the top-bottom-tau sector at the IR scale to be set for the purpose of
the argument at $\mu=\mt=176\gev$, there are four unknown parameters to be
considered, the top, bottom and tau Yukawa couplings at $\mu=\mt=176\gev$
and \tanb. The input of the experimental tau and bottom masses, leads to
two unknown parameters at $\mu=\mt=176\gev$, which have been chosen in the
literature most sensibly to be \mtp\ and \tanb. The IR fixed line in the
\rt-\rb-plane at $\mu=\mt$, Fig.  \ref{net}, can then be translated into an
IR fixed line in the \tanb-\mtp-plane, Fig. \ref{tanbmt} \cite{schwi}, by
remembering the definitions \rt=\hts/\gds\ and \rb=\hbs/\gds\ and inserting
$\gds(\mu=\mt)=1.34$ , Eq. (\ref{gdrei}). Two-loop RGE are used and all
radiative corrections are applied (for \hta=0). An approximate treatment
had also been given in Ref. \cite{sch2}. The resulting IR fixed line
implies a strongly IR attractive relation between the top mass and the MSSM
parameter \tanb.

Let us emphasize again that the IR fixed line (fat line) is independent of
the scale \lam, while the Hill type effective fixed line (thin line)
represents the upper IR bound for $\lam=\mgut\approx 2\cdot 10^{16}\gev$: in
the mathematical limit $\lam\rightarrow\infty$, while keeping the IR scale
fixed, the upper bound along with all other solutions tends towards the
fixed line.  The IR fixed point, implementing approximate top-bottom Yukawa
unification, is denoted by a symbol $\Diamond$ in Fig. \ref{tanbmt}. The
whole RG flow is attracted from above and below very strongly first towards
the IR fixed line and then along or close to it towards the IR fixed point
(with the exception of the solutions starting from initial values \htt=0 or
\hb =0).

The results to be read off Fig. \ref{tanbmt} are the following \cite{schwi}
\begin{itemize}
\item In the large \tanb\ interval
\begin{equation}
1\lwig\tanb\lwig62
\end{equation}
the IR fixed line corresponds to top mass values 
\begin{equation}
150\gev\lwig\mtp\lwig 190\gev
\end{equation}
well compatible with the experimental top mass (\ref{CDF}) within the
experimental bounds. A good approximation for the IR fixed line for not too
large values of \tanb\ is
\begin{equation}
\mtp\approx 192\gev\sin\beta.
\end{equation}
\item The IR fixed point (for \hta=0) lies at
\begin{equation}
\mtp\approx 182\gev,\ \ \ \ \tanb\approx 60,
\label{tanbmtfp}
\end{equation}
implementing approximate top-bottom unification.  This top mass value
agrees amazingly well with the experimental value.
\item The upper bound, i.e. the Hill type effective fixed line, is very
  nearby and may be approximately parametrized by
\begin{equation}
\mtp\approx 202\gev\sin\beta.
\end{equation}
\end{itemize}

Obviously the result (\ref{every})
quoted in the literature \cite{alv}-\cite{ross1}, \cite{mond}, \cite{zoup}
covers the band of \mtp\ values between the IR fixed line and the upper
bound; so, Fig. \ref{tanbmt} contains quasi a dissection of this band into
a genuine IR fixed line, attracting all solutions, and an upper bound. It
is gratifying to notice that the genuine IR fixed line lies at the lower
end of \mtp\ values and thus allows a more favourable comparison with the
experimental top mass value.

Let us anticipate already that the shape of the band between the IR fixed
line and the upper bound in the \mtp-\tanb-plane roughly coincides with
the band of allowed IR values resulting from the requirement of tau-bottom
Yukawa unification \cite{ara}-\cite{ross1} in the framework of
supersymmetric grand unification. To this issue we devote Sect. 7, where we
review why tau-bottom Yukawa unification focuses the IR physics onto or
close to the IR fixed line in the \mtp-\tanb-plane.

Here we come back to the knowledge that in determining the top mass \mtp\ 
from an IR fixed line or point, the IR endpoint of the RG evolution, so far
chosen to be 176\gev\, is in fact {\it not} a free parameter: rather the IR
scale has to be identical to the top mass \mtp\ which in turn is determined
implicitely from the condition (\ref{IRt}).  Since, however, the dependence
on $\mu$ in \htt\ is only logarithmic, the correction to the result is
negligible.  

This analysis had been performed with \msusy=\mt=176\gev and
$\alpha_s(m_Z)=0.117$. Of course the result varies if these two parameters
are varied. Such a variation will be reviewed in Sect. \ref{yukuni} as
well.

Altogether the IR attractive top mass value emerging within the
framework of the MSSM is very close to the experimental value and it is
gratifying that its IR attraction is so strong. Still, one has to await an
experimental confirmation of the supersymmetric scenario and a measurement
of \tanb\ before drawing any further conclusions.

\subsection{Top and Higgs Masses and Top-Higgs Mass relation in the 
SM\label{secSM}}

In the SM the input of the known bottom mass leads to a determination of
the Yukawa couplings \gb\ at the IR scale which we shall set, in order to
be definite, at $\mu=\mt=176\gev$. With the known input of $\gds(\mu=\mt)$,
this can be turned into a knowledge of the ratio variable \rb=\gbs/\gds\ at
$\mu=\mt$. Now we turn to the two-loop analogue of the IR fixed surface at
$\mu=\mt$ in the Higgs-top-bottom sector, Fig. \ref{rHtb}, i.e. in the
\rh=\la/\gds-\rt=\gts/\gds-\rb=\gbs/\gds-space. The insertion of \rb\ at
$\mu=\mt$ cuts out of the surface an IR fixed line in the \rh-\rt-plane
with a fixed point in it, the fixed line being much more strongly IR
attractive than the fixed point. Translating \rh\ and \rt\ at $\mu=\mt$
into \la\ and \gt\ at $\mu=\mt$ and these after inclusion of all radiative
corrections into \mhp and \mtp, one ends up with the following result in
the \mhp-\mtp-plane.
\begin{itemize}
\item A strongly IR attractive fixed line in the \mhp-\mtp-plane,
implying a strongly IR attractive top-Higgs mass relation
\item and on the fixed line a weakly IR attractive fixed point in the
  \mhp-\mtp-plane, corresponding to IR fixed point top and Higgs  mass
  values. This fixed point plays the role of the Pendleton-Ross fixed point
  (\ref{rtR}) in the presence of the electroweak gauge couplings.
\end{itemize} 
\begin{figure}
\begin{center}
\epsfig{file=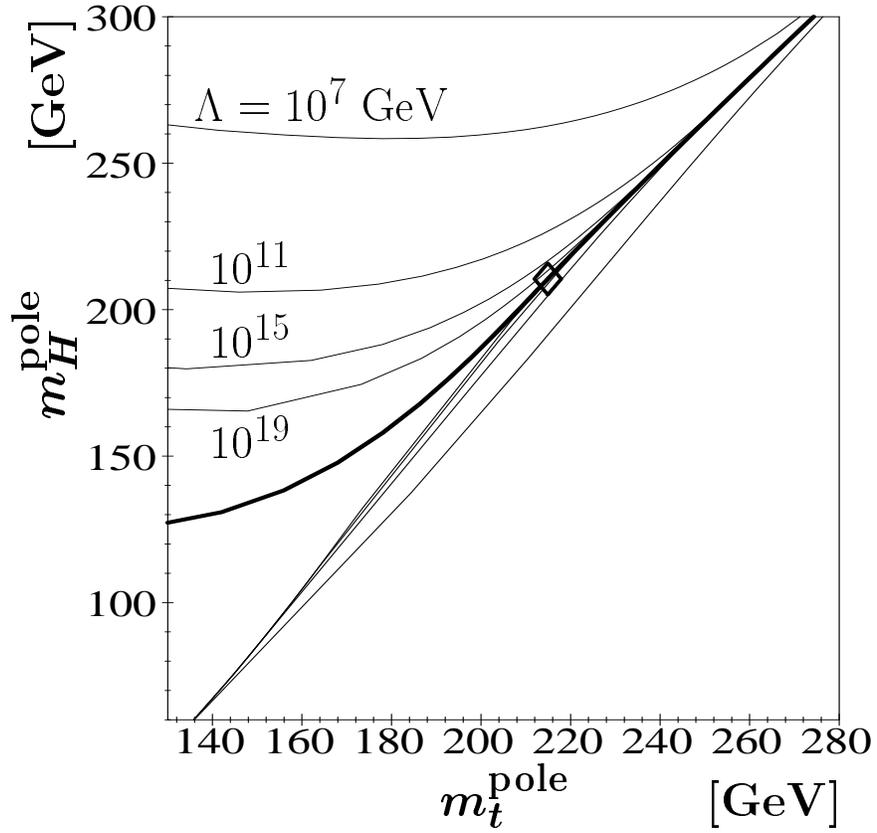,bbllx=112pt,bblly=250pt,bburx=563pt,
bbury=683pt,width=11.5cm}
\end{center}
\caption[dum]{The strongly IR attractive fixed line (fat line) in the
  \mhp-\mtp-plane of the SM and the weakly IR attractive fixed point
  (symbol $\Diamond$), as well as the triviality and vacuum stability
  bounds (thin lines) for $\lam=10^7,\;10^{11},\;10^{15},\;10^{19}\gev$ are
  shown. The figure was taken from Ref. \cite{schwi}.}
\label{mHmt}
\end{figure}
The strongly IR attractive top-Higgs mass relation \cite{schwi} is shown in
Fig. \ref{mHmt} (fat line) above \mtp=150\gev.
(Below a top mass of 150\gev the theoretical determination of the
fixed line starts to become unreliable). It may be considered as the
non-trivial update of the corresponding figure in Ref. \cite{sch1}; see
also Refs. \cite{wet2}, \cite{wet}, \cite{leung}.
The weakly attractive IR fixed point, characterized by a the symbol $\Diamond$ in
Fig. \ref{mHmt} is at \cite{schwi}
\begin{eqnarray}
{\rm IR\ attractive\ top\ fixed\ point\ mass}\hspace*{1cm}\mtp&\approx& 214\gev,\nonumber\\
{\rm IR\ attractive\ Higgs\ fixed\ point\ mass}\hspace*{1cm}\mhp&\approx& 210\gev.
\end{eqnarray}
The IR attractive top mass value is clearly outside the combined one
standard deviation errors of
the experimental top mass; still it is impressive that the SM, which is not
endorsed with an additional free parameter like \tanb\ in the MSSM, has an IR
attractive top mass value so relatively close to the experimental value.
It has to be stressed, however that this IR fixed point is very weakly
attractive, in contradistinction to the top mass fixed point in the MSSM.

The IR fixed line is much more strongly IR attractive than the IR fixed point. So it seems justified to attach more significance
to it. Evaluating the line at the experimental top mass value,
\mtp=176\gev, say, leads to the corresponding Higgs mass value \cite{schwi}
\begin{equation}
\mhp\approx 156\gev,\ \ {\rm IR\ attractive\ top-Higgs\ mass\ relation,\
  evaluated\ at\ \mtp=176\gev}.
\end{equation} 
This is a very interesting Higgs mass value, but again, experimental
confirmation is needed, before any further conclusions can be drawn.

As we had pointed out already in previous sections, the RG flow is first
drawn towards the IR fixed line and then close to it or along it towards
the IR fixed point. The attraction is stronger above the fixed point than
below. This is again reflected in the combined triviality and vacuum
stability bounds (thin lines in Fig. \ref{mHmt}) for four representative
values of $\lam=10^7,\ 10^{11},\,10^{15},\ 10^{19}\gev$. The allowed region
is within the ``wedges''. (For the theoretical discussion and the quotation
of the relevant literature we refer to Subsects. 2.6, 4.1-4.3). The tip of
the wedge is the upper bound, which slides down the IR fixed line for
increasing value of the UV scale \lam\ (towards the IR fixed point in the
mathematical limit $\lam\rightarrow\infty$, while keeping the IR scale
fixed). (For a thorough discussion of the
lower bounds see Subsect. \ref{lowsm}).

Notice again that the IR scale, set throughout this analysis to
$\mu=176\gev$, is not a free parameter. Rather the IR scale has to be
identical to \mtp, resp. \mhp, which in turn is determined implicitely from
the condition (\ref{IRtSM}), resp. (\ref{IRhSM}).  Since, however, the
dependence on $\mu$ in the couplings is only logarithmic, the corrections
to the results are very small.

\subsection{Lower Bound on the Higgs Mass in the SM\label{lowsm}}

In order to be prepared for future searches for the SM Higgs boson at the
LEP200 upgrade of LEP and at the LHC collider, it is important to have a
precise analysis of the lower bound, i.e. the vacuum stability bound
\cite{mai}, \cite{linvac}-\cite{esp}, on the SM Higgs mass as a function of
the top mass and the UV scale \lam. There are two
recent professional analyses in Refs. \cite{isi} and \cite{casa} (see also
Refs. \cite{cas} for providing the basis and Ref. \cite{esp} for a
refinement) which provide such a lower bound and which agree to within a
few GeV well within their quoted errors of 3-5\gev.

\begin{sidewaysfigure}
\begin{tabular}{cc}
{\bf a)}&{\bf b)}\\
\epsfig{file=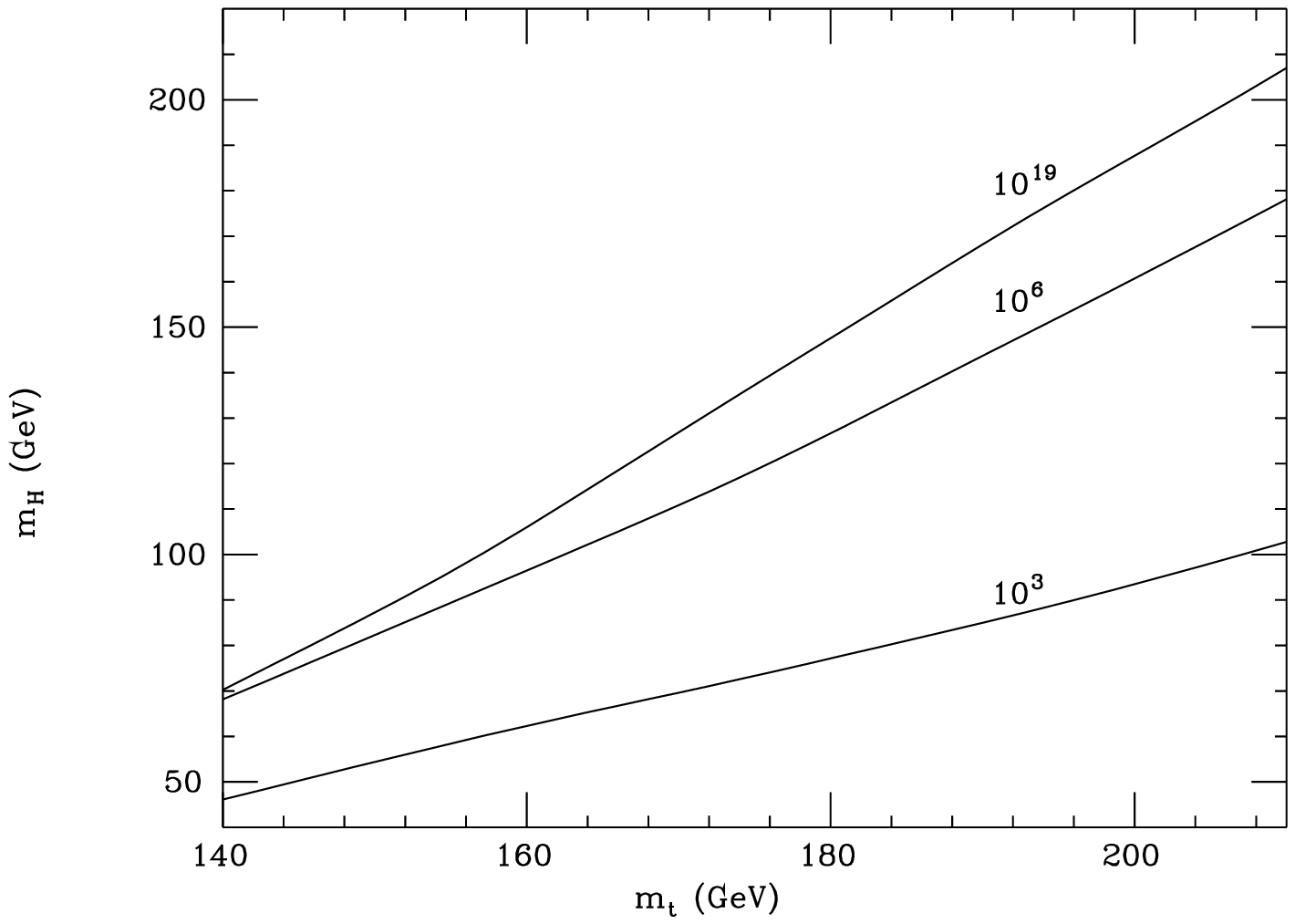,bbllx=25pt,bblly=155pt,bburx=433pt,
bbury=446pt,width=11.5cm}&
\epsfig{file=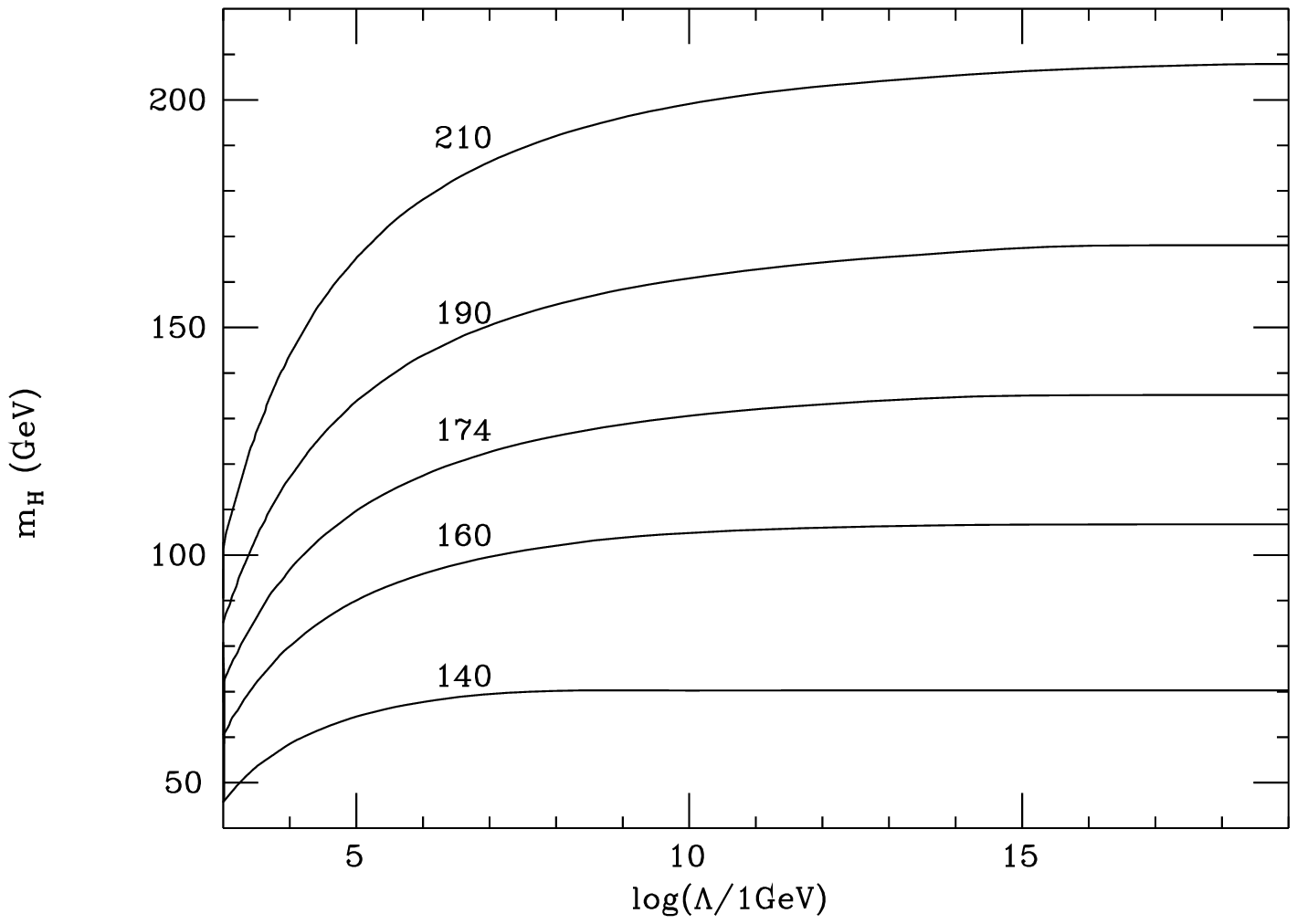,bbllx=25pt,bblly=155pt,bburx=433pt,
bbury=445pt,width=11.5cm}\\
{\bf a)} & {\bf b)}
\end{tabular}
\caption[dum]{a) Vacuum stability bounds for the SM Higgs mass \mhp\ as a
  function of the top mass \mtp\ for different values of the UV scale \lam.
  b) Limits on the Higgs mass \mhp\ as a function of the UV scale \lam\ for
  various values of the top mass \mtp. The figures were taken from Ref.
  \cite{isi}. }
\label{isid}
\end{sidewaysfigure}
\begin{figure}
\begin{center}
\epsfig{file=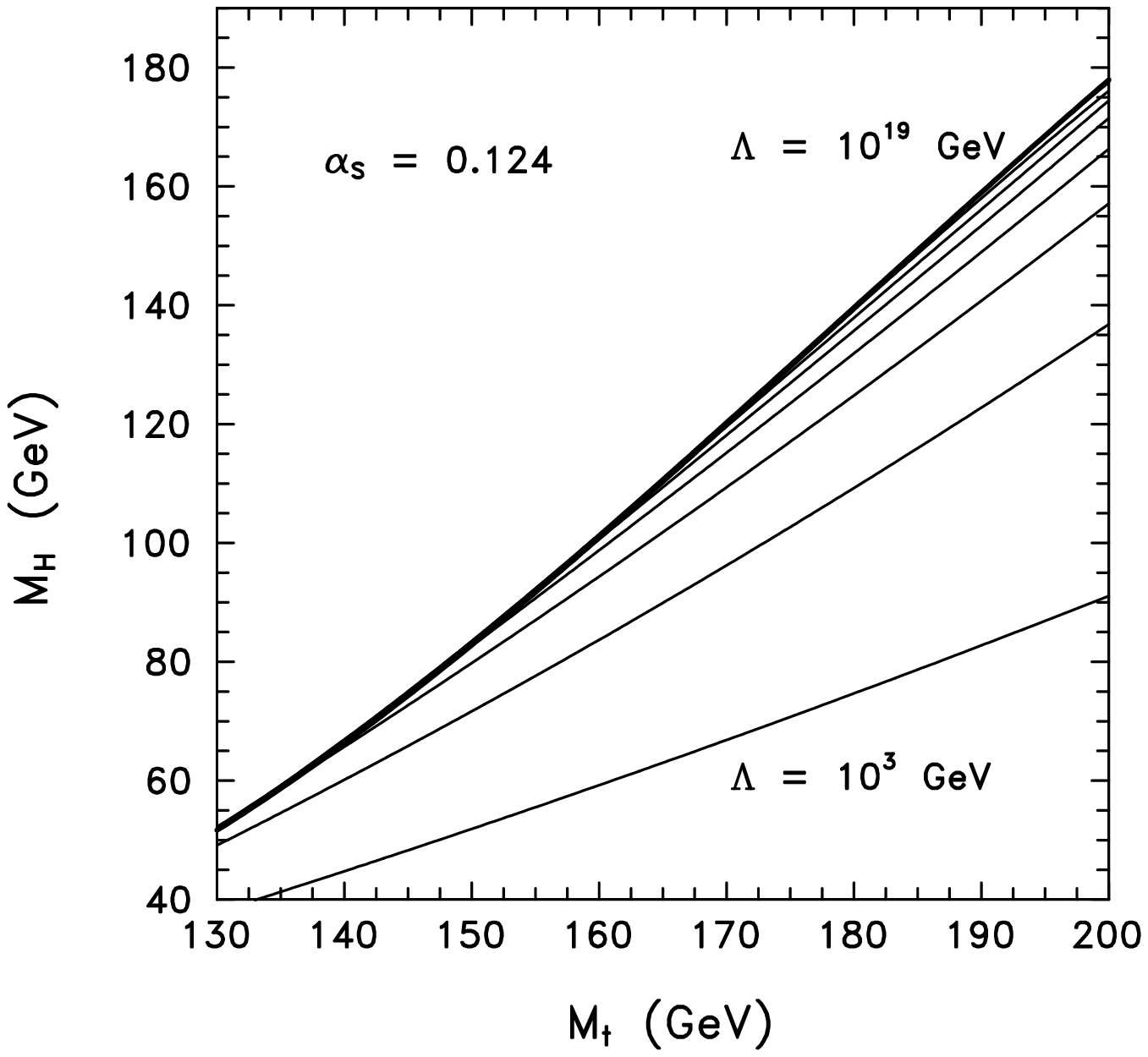,angle=-90,bbllx=133pt,bblly=147pt,bburx=495pt,
bbury=537pt,width=11.5cm}
\end{center}
\caption[dum]{Vacuum stability bounds for the SM Higgs mass \mhp\ as a
  function of the top mass \mtp\ for different values of the UV scale \lam. The figure was taken from Ref. \cite{casa}.}
\label{espinosa}
\end{figure} 

In both analyses the vacuum stability bound is determined for various UV
scales \lam; both assume the minimal UV scale \lam\, at which new physics
could possibly appear without having significant influence at the
electroweak scale, to be $\lam=10^3\gev$.  In this sense the vacuum
stability bound for $\lam=10^3\gev$ provides the ``absolute'' lower
bound on the Higgs mass. In Ref. \cite{isi} the
bound was determined from the requirement that
the two-loop running Higgs selfcoupling $\la(\mu)$ reach zero at
$\mu=\lam$, as justified in Sect. \ref{vac}. 
Ref. \cite{casa} used their refined method, leading to a nearly scale
independent one-loop effective potential which is RG improved to the
next-to-leading order, mentioned at the end of Sect. \ref{vac}. Both
include the radiative corrections in relating the running $\overline{\rm MS}$
couplings 
to the physical Higgs and top pole masses (see the two footnotes in
Sect. \ref{rad} in this context). In Ref. \cite{isi} $\alpha_s(m_Z)=0.118$
is used, in Ref. \cite{casa} $\alpha_s(m_Z)=0.124$. In Ref. \cite{isi} an
analytical approximation for the lower Higgs mass bound is given for the
two extremal values for \lam\, which allows to vary \mtp and $\alpha_s(m_Z)$ 
\begin{eqnarray}
\mhp&>&135+2.1(\mtp-174)-4.5\left(\frac{\alpha_s(m_Z)-0.118}{0.006}\right)\
  \ {\rm for} \lam=10^{19}\gev,\\
\mhp&>&72+0.9(\mtp-174)-1.0\left(\frac{\alpha_s(m_Z)-0.118}{0.006}\right)\
  \ {\rm for} \lam=10^3\gev,
\end{eqnarray}
in Ref. \cite{casa} for comparison
\begin{equation}
\mhp>127.9+1.92(\mtp-174)-4.25\left(\frac{\alpha_s(m_Z)-0.124}{0.006}\right)\
  \ {\rm for}\ \lam=10^{19}\gev.
\end{equation}
The results of the two
analyses are exhibited in
Figs. \ref{isid} and \ref{espinosa}, respectively.      

One important conclusion is that for a top mass larger than 150\gev\ the
discovery of a SM Higgs boson at LEP200 would imply that the SM breaks down
at a scale \lam\ much smaller than a grand unifying scale of
$O(10^{15})\gev$. Actually, as can be inferred with the additional
information about the triviality (upper) bound for the Higgs mass in Fig.
\ref{mHmt}, for \mtp=176\gev and $\lam\gwig 10^{15}\gev$, only a mass range
\begin{equation}
130\gev\lwig\mhp\lwig190\gev
\end{equation}
is allowed.

\subsection{Upper Bound on the Lightest Higgs Mass in the MSSM\label{upmssm}}

The conclusions for the lightest Higgs boson in the MSSM are completely
different. This is due to the fact that in the MSSM the tree level mass of
the lightest Higgs boson is given in terms of the electroweak gauge
couplings, Eq. (\ref{lowmh}), which reaches maximally the value of the Z
boson mass for $\cos^2\beta=1$. The remaining task is to calculate the
higher order radiative corrections. 

\begin{figure}
\begin{tabular}{cc}
\epsfig{file=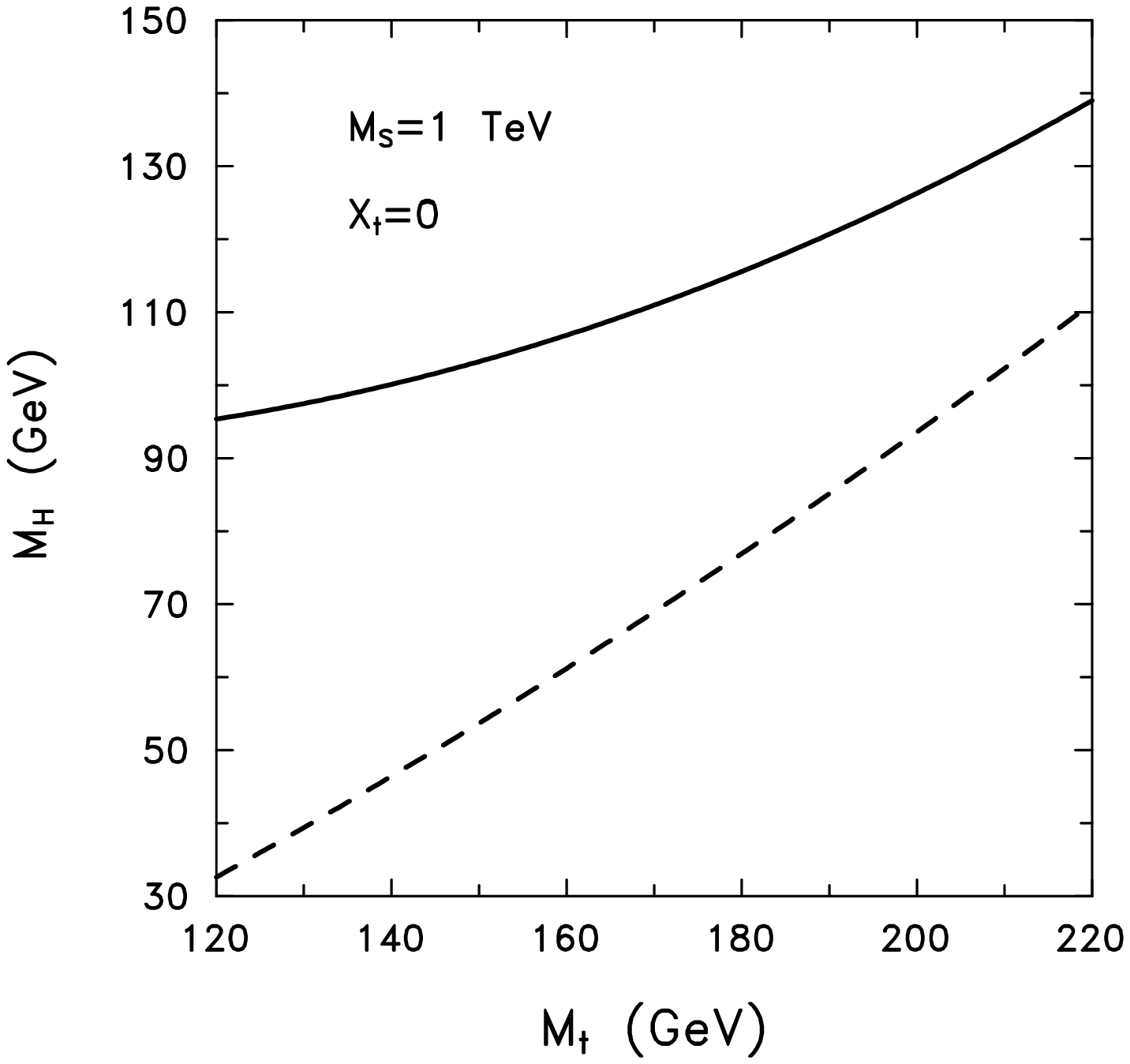,bbllx=95pt,bblly=141pt,bburx=487pt,
bbury=554pt,angle=-90,width=8cm}&\epsfig{file=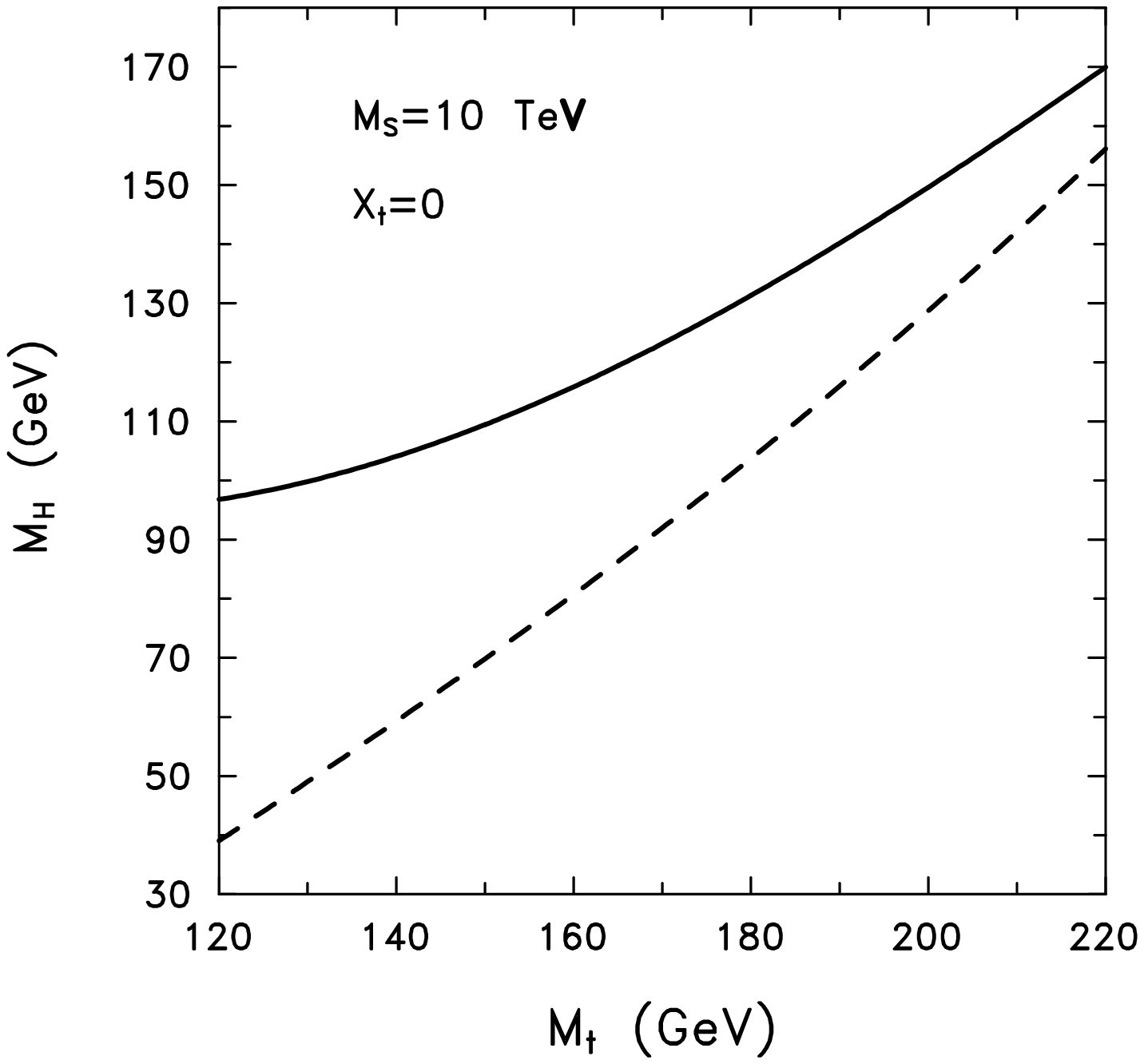,bbllx=101pt,bblly=141pt,bburx=486pt,
bbury=554pt,angle=-90,width=8cm}\\[3mm]
{\bf a)}&{\bf b)}\\
\epsfig{file=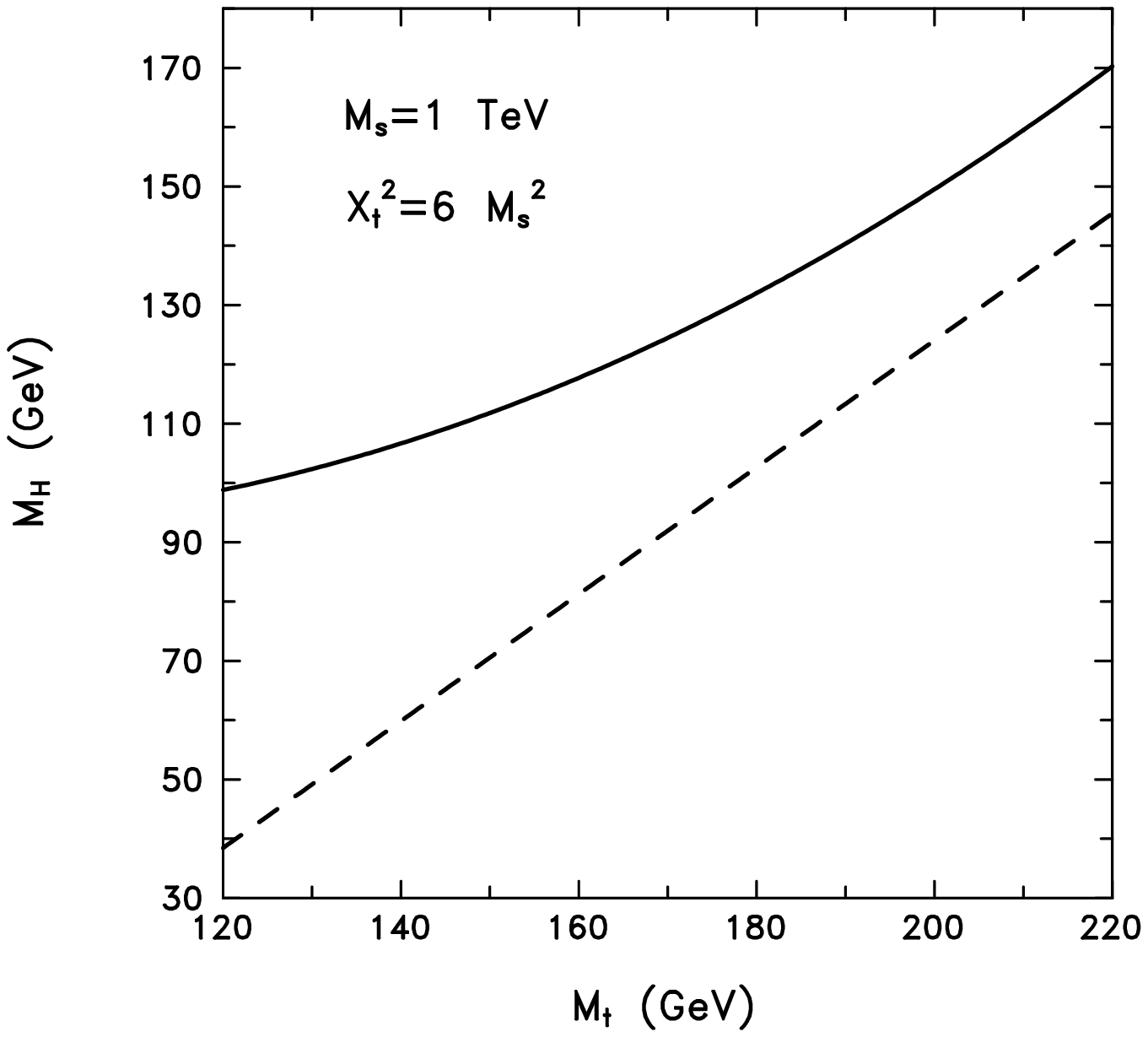,bbllx=69pt,bblly=75pt,bburx=463pt,
bbury=429pt,width=8cm}&\epsfig{file=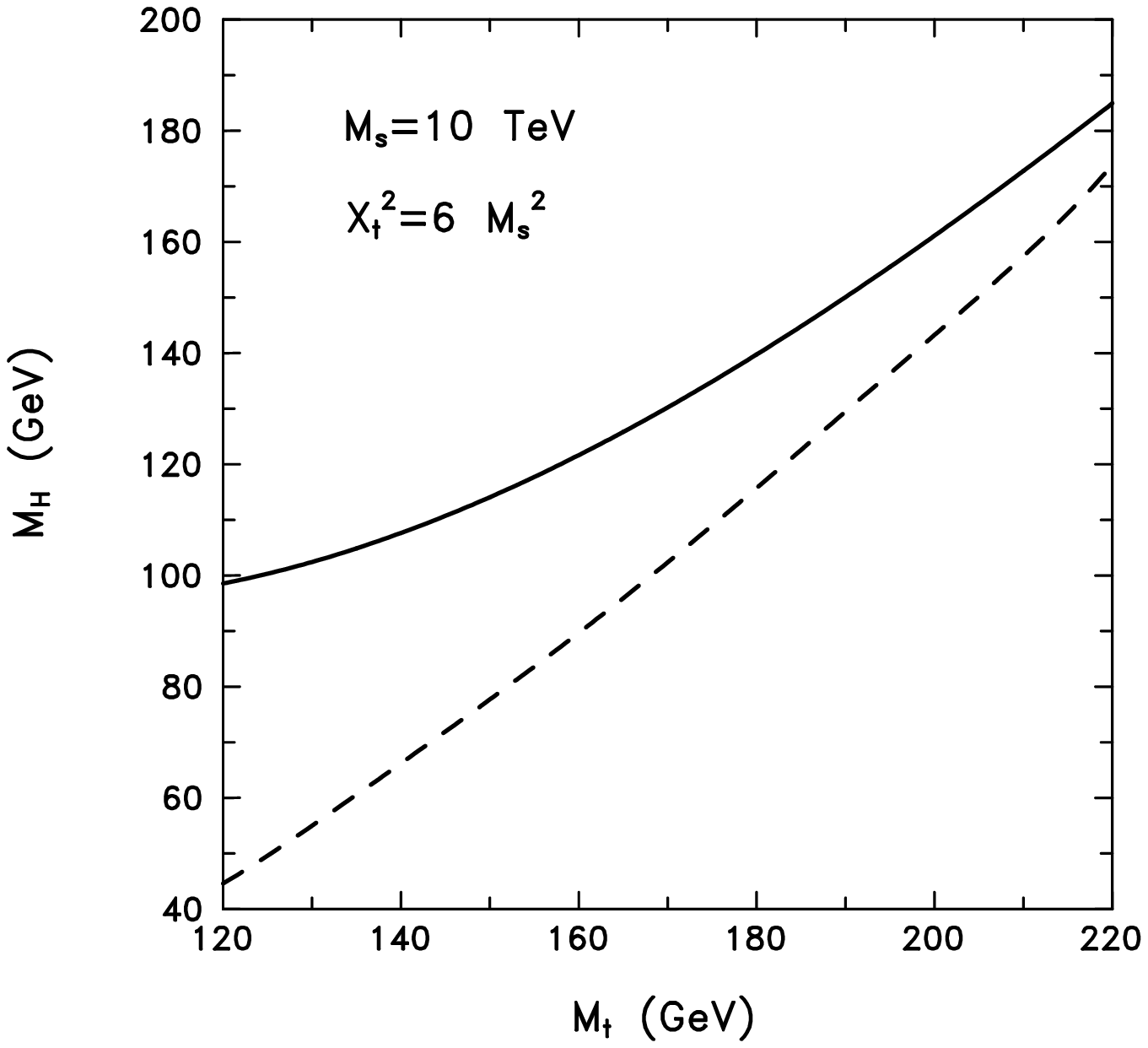,bbllx=72pt,bblly=75pt,bburx=464pt,
bbury=433pt,width=8cm}\\
{\bf c)}&{\bf d)}
\end{tabular}
\caption[dum]{The upper bound on the mass  \mhp\ of the lightest Higgs mass
  boson in the MSSM as a function of the top mass \mtp\ for
  $\cos^2\beta=1$ (solid line) and $\cos^2 2\beta=0$ (dashed line) and values
  of the supersymmetry scale \msusy\ and the mixing parameter $X_t$: a)
  $\msusy=1\,{\rm TeV}$, $X_t=0$, b)
  $\msusy=10\,{\rm TeV}$, $X_t=0$, c)
  $\msusy=1\,{\rm TeV}$, $X_t^2=6\msusy^2$, d)
  $\msusy=10\,{\rm TeV}$, $X_t=6\msusy^2$. The figure was taken from Ref. \cite{cas}.}
\label{xxx}
\end{figure}

Let us start with a qualitative discussion. The radiative corrections are
mainly due to the running of the Higgs selfcoupling according to the
(two-loop) RGE of the SM from the scale $\mu=\msusy$, where its initial
value, Eq. (\ref{lowmh}), is fixed, to the scale $\mu=\mh$, in dependence
on the size of the top mass. Since the initial value at $\mu=\msusy$ is
well {\it below} the IR attractive Higgs-top mass relation, displayed in
Fig. \ref{mHmt}, one expects the evolution to {\it increase} the Higgs
selfcoupling and correspondingly the Higgs mass. The increase will be the
stronger, the longer is the evolution path, i.e. the larger is the value
for \msusy. The upper bound for the Higgs mass, as a function of the top
mass, will have to settle below or at best on the IR attractive top-Higgs
mass relation.

There is a vast literature on the determination of the higher order
radiative corrections \cite{kun}-\cite{wagne}
which cannot all be reviewed here.  We shall concentrate on the results of
a recent, very careful analysis \cite{cas}. In Ref. \cite{cas} the supersymmetry breaking
scale is assumed to be well above the lightest Higgs mass; $\msusy=1\,{\rm
  TEV}$ and $\msusy=10\,{\rm TEV}$ are the values for which results are
presented. The authors \cite{cas} include the one-loop threshold
contribution due to a possible stop mixing at \msusy in the following form,
\begin{equation}
\Delta\la=\frac{3\htt^4}{16\pi^2}\frac{X_t^2}{\msusy^2}\left(1-\frac{X_t^2}{12\msusy^2}\right),
\label{corr}
\end{equation}
This term has to be added to the initial value for the tree level Higgs
selfcoupling at $\mu=\msusy$ on the right hand side of Eq. (\ref{lowmh}).
$X_t$ is a measure of the stop mixing. The authors consider two extreme
values for $X_t$, $X_t^2=6\msusy^2$ for maximal threshold effects and
$X_t=0$, for absence of mixing.

They use the framework \cite{cas} of an almost scale independent one-loop
effective potential, RG improved at the two-loop level, which has been
quoted already in Sects. \ref{vac} and \ref{lowsm}, and include the full
radiative corrections relating the $\overline{\rm MS}$ masses to the
physical pole masses \mhp\ and \mtp. Since the tree level Higgs mass lies
well below the IR attractive fixed line in the \mhp-\mtp-plane,
Fig. \ref{mHmt}, the RG evolution from $\mu=\msusy$ down to $\mu=\mh$
gives positive corrections.
Fig. \ref{xxx} shows the result of the analysis. The bound depends on 
the parameters \msusy, $X_t$ and $\cos^2 2\beta$ and of course on \mtp.

An analytic approximation for the most conservative bound is
\cite{cas}
\begin{equation}
\mhp<126.1+0.75(\mtp-174)-0.85\left(\frac{\alpha_s(m_Z)-0.124}{0.006}\right)
\end{equation}
which can be compared with the lower bounds on the SM Higgs mass discussed in
Sect. \ref{lowsm}. For a top mass of order of 176\gev\ the upper bound on the
Higgs mass is of the order of 130\gev.

\section{Supersymmetric Grand Unification Including Yukawa 
Unification\label{yukuni}}


As has been developed in Sect. \ref{gut}, an interesting tool to constrain
the ``top-bottom'' RG flow by {\it symmetry relations} between initial
values at the UV scale is provided in the framework of minimal
supersymmetric grand unification implying in addition tau-bottom or even
tau-bottom-top Yukawa unification. Recent analyses have been performed in
the gauge sector 
\cite{ssgut},\cite{bar},\cite{lang1},\cite{langa},\cite{lang2} and
including Yukawa coupling unification, more recently combined with the
issue of an IR attractive top mass,
\cite{taubuni}-\cite{ross1}.
Most of them go beyond the framework considered in this review and cover in
addition issues like i) inclusion of all fermion masses and mixing angles
into the analysis and investigation of the quantitative implications of
{\it ans\"atze} (textures) for the fermion mass matrix at the grand
unification scale \mgut\ with tau-bottom Yukawa unification implemented,
ii) exploration of the Higgs sector, iii) implementation of a realistic
supersymmetric particle spectrum, iv) radiative electroweak symmetry
breaking, v) search for IR fixed points in the supersymmetry breaking mass
terms or vi) investigation of the fixed point structure of the underlying
gauge theory. We are going to lead the discussion for the heaviest fermion
generation only within the framework outlined in Sect. \ref{thfr}.

These analyses have two global aspects.
\begin{itemize}
\item They represent a crucial test of supersymmetric grand unification
  including tau-bottom (-top) Yukawa unification at the quantitative level.
\item More interestingly in the context of this review, they demonstrate
  that tau-bottom unification requires the IR values for the top mass to be
  close to its IR fixed point or, more generally, the IR values in the
  \tanb-\mtp-plane to be close to the IR fixed line discussed in Subsect.
  6.1.
\end{itemize}
The program involves several steps.  A first step is the exploration
\cite{ssgut}, \cite{bar},\cite{lang1},\cite{langa},\cite{lang2} of gauge
coupling constant unification (\ref{gegzgd}). The preferred way to perform
this analysis is to assume gauge coupling constant unification
(\ref{gegzgd}) at a scale \mgut\ which at this stage is not specified; so
there are two unknown parameters, the unified gauge coupling at $\mu=\mgut$
and the unification scale \mgut\ itself; in addititon there is a mild
dependence on the supersymmetry breaking scale \msusy\ which regulates the
transition from the RGEs of the MSSM to those of the SM. These parameters
are adjusted such that the two-loop RG evolution of the MSSM gauge
couplings, including among others the two-loop contribution of the large
top Yukawa coupling, leads to the experimental values for $\alpha(m_Z)$ and
$\sin^2\theta_W$; this procedure results in a value for \mgut\ and for
$\alpha_s(m_Z)$, both depending mildly on \msusy. As has been stressed in
Ref.  \cite{lang1},\cite{langa},\cite{lang2} it is important to estimate
the theoretical error for $\alpha_s(m_Z)$, allowing for a variation of
\msusy\ within reasonable bounds and for threshold corrections and
nonrenormalizable operator corrections at the low and high scales. A most
recent analysis \cite{lang2} along these lines determines \mgut\ to be
$\mgut\simeq 3\cdot 10^{16}\gev$ and $\alpha_s(m_Z)$ with the appropriate
theoretical errors to be
\begin{equation}
\alpha_s(m_Z)\simeq 0.129 \pm 0.010,
\end{equation}
which is larger than the experimental value, however, still within the
experimental and theoretical errors.

The analysis so far fixes the gauge sector and \mgut, having used the
experimental value of the top quark mass already for the two-loop
contribution of the top Yukawa couplings to the RGE of the gauge couplings.

Next, the two-loop RGEs for the heavy fermion generation, the tau, bottom
and top quark are considered. There are to start with four completely free
parameters, the initial values for the top, bottom and tau Yukawa couplings
at the unification scale \mgut, say, and the parameter \tanb,
characterizing the ratio of the two different Higgs vacuum expectation
values in the supersymmetric theory; a further rather constrained parameter
is the value of \msusy.

Now, the tau-bottom Yukawa unification (\ref{bta})
\begin{equation}
\hta(\mu=\mgut)=\hb(\mu=\mgut),
\end{equation} 
is introduced as well as the known tau and bottom masses within their
errors, Eqs.  (\ref{massb},\ref{masstau}).

What happens then is rather subtle (see e.g. Ref. \cite{bar} for a
comprehensive presentation) and requires some preparatory remarks. In
Subsect. \ref{rad} and the introduction to Sect. \ref{masses} we have seen, how to translate the
given input of the top and bottom masses into the MSSM bottom and
$\tau$ Yukawa couplings at some higher scale $\mu$. For definiteness,
let us fix \msusy=\mt=176\gev. Then the MSSM bottom and $\tau$ Yukawa
couplings at this scale are given as follows
\begin{eqnarray}
\mb(\mu=\mt)&=&\frac{v}{\sqrt{2}}\,\hb(\mu=\mt)\cos\beta,\\
\mta(\mu=\mt)&=&\frac{v}{\sqrt{2}}\,\hta(\mu=\mt)\cos\beta.
\end{eqnarray}
in terms of the unknown parameter $\cos\beta$. Now, one realizes that
the input of tau-bottom Yukawa unification on the one hand and of the tau and bottom
masses on the other hand leads to fixing the ratio 
\begin{equation}
R_{b/\tau}(\mu)=\frac{\hb(\mu)}{\hta(\mu)} 
\end{equation}
at {\it two} scales
\begin{eqnarray}
R_{b/\tau}(\mu=\mgut)&=&1\hspace{3cm}{\rm and}\nonumber\\
R_{b/\tau}(\mu=\mt)&=&\frac{\mb(\mu=\mt)}{\mta(\mu=\mt)}.
\label{init}
\end{eqnarray}
The input of {\it two initial values for the same quantity $R_{b/\tau}(\mu)$ at two different
scales $\mu$} can only be accommodated, by tuning the contribution of the
top Yukawa coupling $\htt(\mu)$, which enters the RGE of \hb\ on the
one-loop level and is absent from the one-loop RGE of \hta. This can be
made most transparent analytically \cite{ander},\cite{bar} by looking at the one-loop
RGE for $R_{b/\tau}$
\begin{equation}
\frac{{\rm d}\,R_{b/\tau}}{{\rm
    d}\,t}=\frac{R_{b/\tau}}{16\pi^2}\left(\frac{4}{3}\ges-\frac{16}{3}\gds+\hts+3\hbs-3\htas\right),
\label{diffR}
\end{equation}  
the solution of which has to accommodate the two initial values (\ref{init}).
For small \tanb\ the bottom and tau Yukawa couplings do not play a
significant role in this RGE. As the running gauge couplings are fixed at
$\mu=\mt$ and $\mu=\mgut$ and thus not
available for the fine tuning of the initial values (\ref{init}), clearly 
the contribution \cite{raby},\cite{dimop},\cite{zral},\cite{bar}
\begin{equation}
  \exp\left(\frac{1}{16\pi^2}\int\limits_{\mt}^{\mgut}{\rm
    d}\ln\mu\;\hts(\mu)\right)
\label{integ}
\end{equation}
is crucial in the fine tuning of the ratios
$R_{b/\tau}(\mgut)$ and $R_{b/\tau}(\mt)$ to their values prescribed by tau-bottom
unification and the bottom and tau masses, respectively. Now, the running of the top
Yukawa coupling $\htt(\mu)$ is in principle not free to choose, it has to
satisfy its own RGE in the system of coupled first order differential
equations and furthermore the product $(v/\sqrt{2})\htt(\mu=\mt)\sin\beta$
involving its initial value $\htt(\mu=\mt)$, adorned with the appropriate
radiative corrections, is required to match the experimental top mass.

Now comes the key observation which provides a way out: the RGE for
$\htt(\mu)$ exhibits the strongly attractive IR fixed point, discussed at
length in Sects.  \ref{preview}-\ref{masses} and this fixed point is, as we
know, well compatible with the experimental value for \mt. Setting the IR
value $\htt(\mu=\mt)$ onto or close to the fixed point value, admits
practically {\it any} sufficiently large UV starting value $\htt(\mgut)$.
This circumstance allows to adjust the UV value $\htt(\mu=\mgut)$ such as to
accommodate the 
ratios $R_{b/\tau}(\mgut)$ and $R_{b/\tau}(\mt)$ as prescribed by tau-bottom
unification and the bottom and tau masses, respectively, without altering
the IR value $\htt(\mu=\mt)$ and without fixing the free parameter \tanb.

As it turns out, the accommodation of the two conditions (\ref{init})
implies rather large values for the UV initial value \htt(\mgut). There is
a tight corner in the analysis: the necessary value for \htt(\mgut) depends
sensitively on the value of $\mb(\mt)$ which in turn depends on $\alpha_s
(m_Z)$ (via the evolution of \mb\ from $\mu=\mb$ to $\mu=\mt$): the
required value for $\htt(\mgut)$ increases with $\alpha_s (m_Z)$ and for
$\alpha_s (m_Z)$ increasing beyond the value 0.12 the top Yukawa coupling
$\htt(\mgut)$ becomes so large that it leaves the safe region of
perturbation theory. This casts doubt on the perturbative unification.

For larger values of \tanb\ the bottom Yukawa coupling becomes relevant
besides the top Yukawa coupling on the right hand side of the differential
equation (\ref{diffR}). Fortunately, there is a whole IR fixed line
available in the \rt-\rb\ plane, viz. in the \htt-\hb\ plane, which may
take over for larger values of \tanb\ the role of the IR fixed point in the
argument led for small values of $\tan\beta$ above. This is exactly what
happens. From the discussion in Sects. \ref{tb123} and 6.1 one
would expect that for any value $\tanb\lwig50-60$ there should be an IR
starting point on or close to the half-circle type IR fixed line in the
\htt-\hb\ plane, resp. in its translation into the \tanb-\mtp-plane, which
similarly allows to accommodate the two initial values (\ref{init}) for the
ratio $R_{b/\tau}$. As announced in Sects.  \ref{tbgd}, \ref{tb123} and
6.1, the exploitation \cite{taubuni}-\cite{ross1} of the
consequences of tau-bottom Yukawa unification for the IR physics isolates
narrow allowed bands of IR values in the \tanb-\mtp-plane, which turn out
to be close to the IR attractive fixed line as translated into the
\tanb-\mtp-plane. So in this very implicit way the IR attractive fixed line
in the \tanb-\mtp-plane had been isolated without explicitely having been
recognized as such.

\begin{figure}
\begin{center}
\epsfig{file=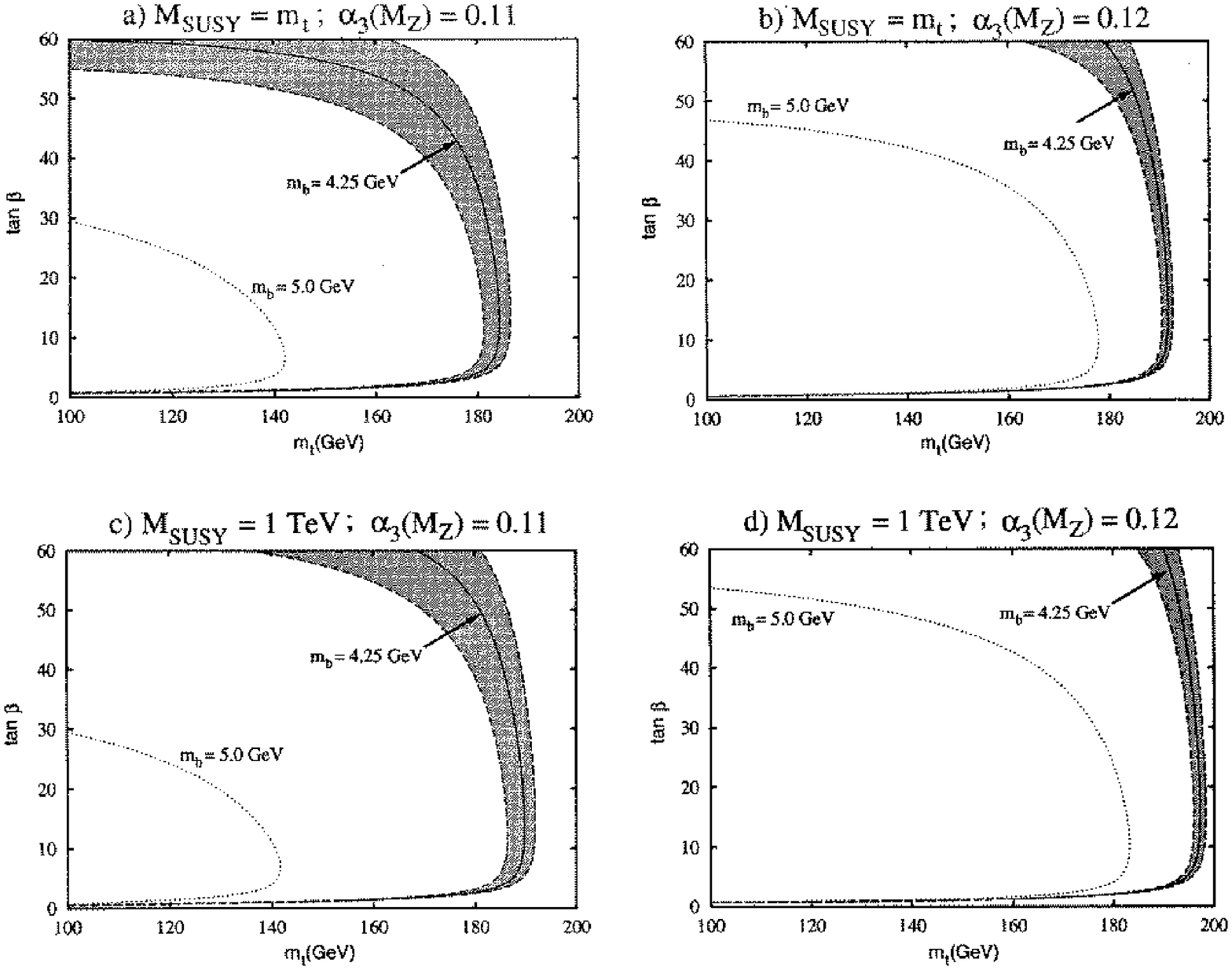,angle=-90,width=16.8cm}
\end{center}
\caption[dum]{Contours of constant \mb\ in the \tanb-\mt-plane obtained
  from the RG evolution of the MSSM subject to the tau-bottom Yukawa
  unification constraint at \mgut\ for the parameters a) \msusy=\mt,
  $\alpha_s(m_Z)=0.11$, b) \msusy=\mt, $\alpha_s(m_Z)=0.12$, c)
  $\msusy=1\,{\rm TeV}$, $\alpha_s(m_Z)=0.11$, d) $\msusy=1\,{\rm TeV}$,
  $\alpha_s(m_Z)=0.12$. The top mass shown is the $\overline{\rm MS}$ mass
  \mt; the curves experience a shift of $5\%$ towards higher values if
  plotted against \mtp. With this shift the contours appear to lie in the
  close vicinity of the IR fixed line shown in Fig. \ref{tanbmt}. The
  figure was taken from Ref. \cite{bar}.}
\label{barger1}
\end{figure}

\begin{figure}
\begin{center}
\epsfig{file=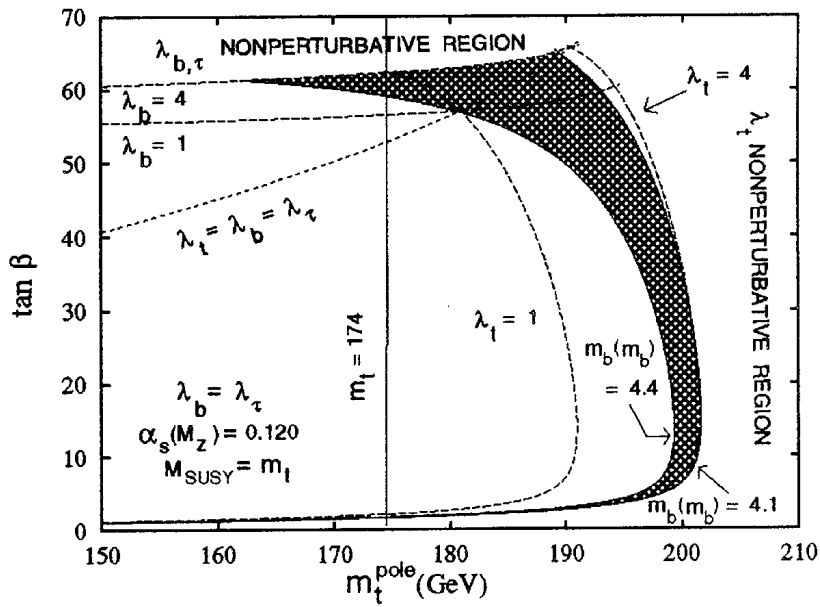,angle=-90,width=11.5cm}
\end{center}
\caption[dum]{An update of Fig. \ref{barger1} for \msusy=\mt\ and 
  $\alpha_s(m_Z)=0.12$ with \mt\ replaced by \mtp. Also shown is the line
  which results from tau-bottom-top Yukawa unification at \mgut. The
  contours appear to lie in the close vicinity of the IR fixed line, the IR
  result of tau-bottom-top Yukawa unification in the close vicinity of the
  fixed point shown in Fig. \ref{tanbmt}. The figure was taken from Ref.
  \cite{barmini}.}
\label{barger2}
\end{figure}

\begin{figure}
\begin{center}
\epsfig{file=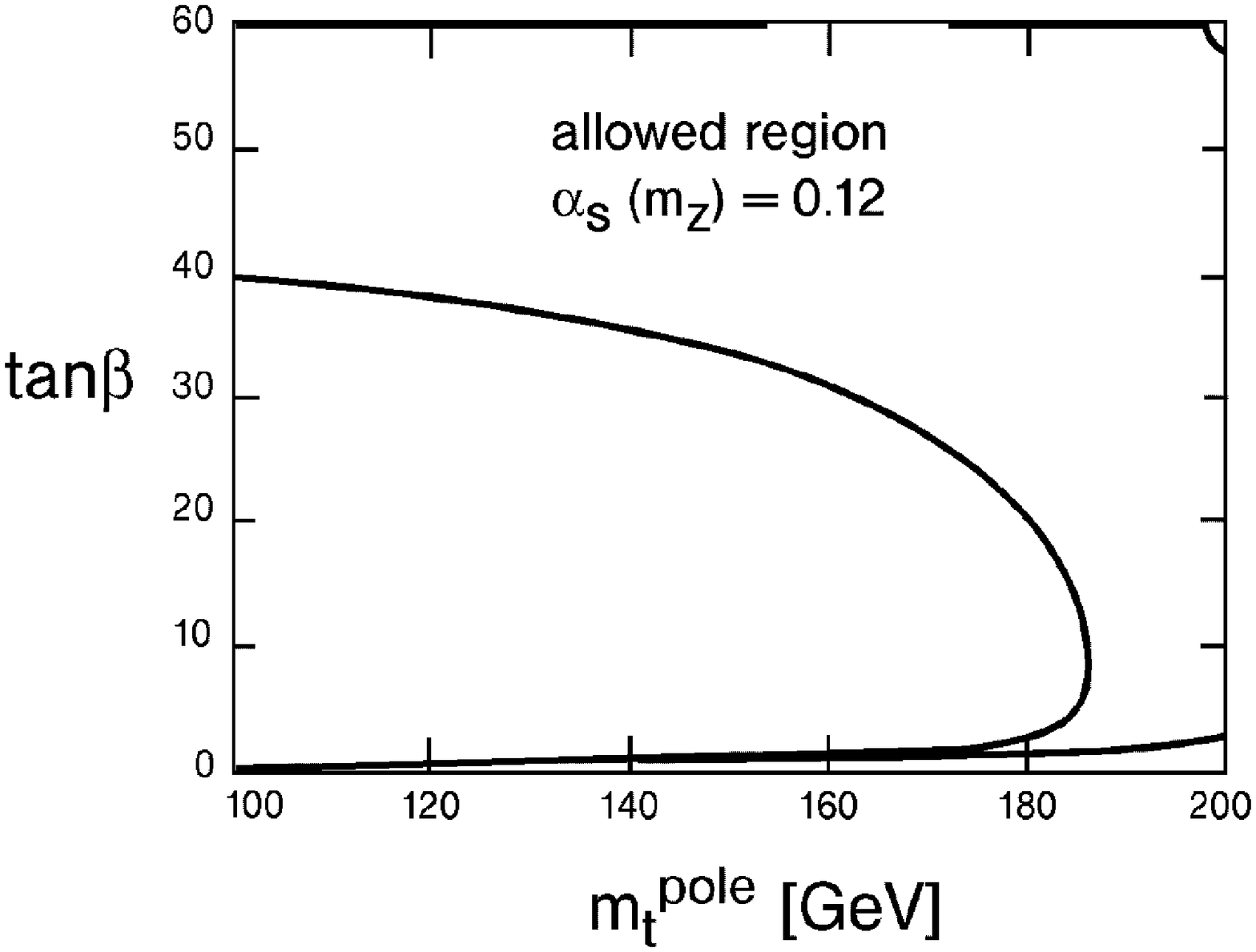,angle=-90,width=11.5cm}
\end{center}
\caption[dum]{The region in the \tanb-\mtp-plane which is consistent with
  tau-bottom Yukawa unification and with the constraint (\ref{langi}) on
  the bottom mass for $\alpha_s(m_Z)=0.12$. This figure is adapted from
  Ref. \cite{langa}.}
\label{lan}
\end{figure}

Let us present typical results from Refs.
\cite{bar},\cite{barmini},\cite{langa}.  They are represented in form of
allowed regions in the \tanb-\mtp-plane at a suitable IR scale. The
analyses are based on the two-loop RGE of the MSSM, the grand unification
scale is determined by the unification of the electroweak gauge couplings
and $\alpha(m_Z)$ is treated as a free parameter.  In Refs.  \cite{bar},
\cite{barmini} the IR scale was chosen to be \mt=150\gev and the analysis
was performed for two values of the supersymmetry breaking scale,
\msusy=\mt=150\gev and $\msusy=1\,{\rm TeV}$, and two values of
$\alpha(m_Z)$, $\alpha(m_Z)=0.11$ and $\alpha(m_Z)=0.12$. The input bottom
mass was chosen according to Eq.  (\ref{massb}) as $4.25\pm 0.15\gev$. For
this narrow band of bottom mass values the requirement of tau-bottom
unification admits a narrow allowed bands in the \tanb-\mt-plane shown as a
narrow dark region in Fig.  \ref{barger1} for a) \msusy=\mt,
$\alpha_s(m_Z)=0.11$, b) \msusy=\mt, $\alpha_s(m_Z)=0.12$, c)
$\msusy=1\,{\rm TeV}$, $\alpha_s(m_Z)=0.11$, d) $\msusy=1\,{\rm TeV}$,
$\alpha_s(m_Z)=0.12$. Obviously, the allowed region is shifted towards
larger top masses for increasing scale \msusy and for inreasing
$\alpha(m_Z)$. The top mass in these figures is not \mtp but rather
$\mt(\mu=\mt)$, so about $5\%$ have to be added in order translate to it
into \mtp. The corresponding updated figure \cite{barmini} in the
\tanb-\mtp-plane for \msusy=\mt=150\gev and $\alpha(m_Z)=0.12$ is shown in
Fig. \ref{barger2}. In Ref.  \cite{langa} the ansatz
\begin{equation}
\mb(5\gev)=\rho^{-1}\mb^0(5\gev)\leq4.45\gev,
\label{langi}
\end{equation}
was made, where $\mb^0$ is the prediction for \mb\ without the theoretical
corrections incorporated in the parameter $\rho^{-1}$. This correction
parameter is to incorporate all uncertainties in the running from $m_Z$ up
to \mgut; a theoretical estimate leads to $\rho^{-1}=0.85$ as reasonable
value. The resulting allowed region for $\alpha(m_Z)=0.12$ is shown in Fig.
\ref{lan}. 

A comparison e.g. of Fig. \ref{barger2} with Fig. \ref{tanbmt} shows
that indeed tau-bottom Yukawa unification requires the IR result of
the ``top-down'' RG flow to lie on or close to the IR fixed line. A
closer look reveals that it is rather the upper bound, the Hill
effective fixed line, which is determined from the analyses
implementing tau-bottom unification.

A final issue is the tau-bottom-top Yukawa unification (\ref{tbta}) within
supersymmetric grand unification. In this context it is a symmetry
constraint $\htt(\mu=\mgut)=\hb(\mu=\mgut)$ for the UV initial values of
\htt\ and \hb, required to hold in addition to the tau-bottom Yukawa
unification. In Sects. \ref{tbgd} and \ref{tb123} an IR attractive fixed
line was isolated which implements approximate top-bottom Yukawa
unification {\it at all scales $\mu$}; correspondingly we expect the
top-bottom Yukawa unification input at $\mu=\mgut$ to persist down to the
IR scale. Indeed, the analysis \cite{barmini}, Fig. \ref{barger2}, shows
that the IR image of combined tau-bottom-top Yukawa unification at the UV
scale is at large values of \tanb, i.e. in the vicinity of the IR fixed
point in Fig. \ref{tanbmt}.

Altogether the following features emerge from the analyses described and figures shown  
\begin{itemize}
\item Grand unification of gauge couplings requires the strong gauge
  coupling $\alpha_s(m_Z)$ to be rather large, but still within the
  experimental and theoretical errors. Taking the central value
  $\alpha_s(m_Z)=0.129$ emerging from the most recent analysis of this
  kind at face value, tau-bottom Yukawa unification requires UV top Yukawa
  couplings which are too large to lead to perturbatively reliable
  results. 
\item In a setting, where only unification of the electroweak gauge
  couplings is required and $\alpha_s(m_Z)$ is allowed to vary between the
  values 0.11 and 0.12, tau-bottom Yukawa unification may be implemented.
  As a consequence the allowed IR parameters become tightly constrained as
  a function of the bottom mass value and a weak function of the
  supersymmetry breaking scale \msusy\ and the value of $\alpha_s(m_Z)$.
  Taking Fig. \ref{barger2} as a guideline and the experimental top mass
  (\ref{CDF}), \mt=176\gev with errors, at face value, two narrow windows
  for solutions for \tanb\ remain, one at small values $1\lwig \tanb \lwig
  4$ and one at large values $42\lwig\tanb\lwig 66$.
\item Additional top-bottom Yukawa unification at the grand unification
  scale totally fixes the heavy quark generation sector, settling \tanb\ at
  $\tanb=O(60)$, i.e. close to the value of the IR fixed point
  (\ref{tanbmtfp}).
\end{itemize}

Several interesting lessons can be learnt from the whole discussion.
\begin{itemize}
\item The very existence of the IR fixed point in \htt\ and, more
  generally, of the IR fixed line in the \htt-\hb\ plane, which are
  inherently to a large extent independent of the UV initial values at
  $\mu=\mgut$ of \htt, resp. of \htt\ and \hb, provide the basis for being
  able to accommodate tau-bottom unification and the physical tau and
  bottom masses simultaneously.  Formulated more pointedly, their function
  is twofold: they implement highly acceptable top mass values and their
  independence of the UV initial value allows to accommodate the
  experimental bottom and top masses as well as tau-bottom Yukawa
  unification. Seen from the gauge and Yukawa unification point of view,
  the presence of the IR fixed point and line is a very fortunate
  circumstance.
\item As we have seen, the supersymmetric grand unification scenario with
  tau-bottom Yukawa unification enforces rather large initial values for
  \htt(\mgut) or for \hb(\mgut).
  These are exactly the initial values for solutions of the RGE which are
  drawn safely {\it onto} the IR fixed point for small \hb(\mgut),
  resp. the IR fixed line in the \tanb-\mtp-plane. It is for
  small initial values that the RGE solutions fail to end up in the IR
  fixed point or line, even given the long evolution path from \mgut\ to
  \mt. Thus, seen from the IR physics point of view, the tau-bottom Yukawa
  unification preselects by a symmetry principle at \mgut\ a portion of the
  ``top-down'' RG flow which focuses at the IR scale in the IR fixed
  manifolds. This is again a very
  gratifying circumstance.
\item The IR physics in the top-bottom sector of a supersymmetric grand
  unification scenario is to a rather high degree independent of the
  details of the theory around and above the unification scale \mgut. What
  matters for the IR physics is the specific form of the RGE, which had
  been chosen to be those of the MSSM, and the long evolution path for the
  RGE flow from a very high UV scale \lam=\mgut.
\end{itemize}
In summary, there is an amazing conspiracy between the ultraviolet physics
issues and IR physics issues at work. 

As mentioned already there is a large body of interesting papers \cite{ara}-\cite{ross1}in the
framework of supersymmetric grand unification, usually adorned with
tau-bottom Yukawa unification
\begin{itemize}
\item which explore larger grand unifying groups or
\item which include all quarks and leptons and their mixing parameters into
  the numerical analysis or
\item all physical Higgs bosons in the two-Higgs sector of the MSSM or 
\item a realistic spectrum of supersymmetric partners of the ordinary
  particles or
\item radiative electroweak symmetry breaking or
\item search for IR fixed points in the supersymmetry breaking mass terms or
\item investigation of the IR fixed point structure of the underlying grand
  unifying gauge theory
\end{itemize}
Their discussion goes beyond the scope of this review. The qualitative
conclusions about the IR fixed point for \mt, resp. the dependence of the
top mass on \tanb, persist. There are of course changes on the quantitative
level. 

A very interesting question is: which is the extent of the loss of memory
of UV physics at the IR scale. This can be ``measured'' by varying the
physics at the UV scale and evaluating the effect of this variation in the
top-bottom sector of the MSSM at the
IR scale as e.g. proposed and analysed in Ref. \cite{ross1}. Similar
investigations \cite{mond},\cite{zoup} were performed in the framework of
reduction of parameters to be discussed in the next section. The top mass
IR fixed point turns out to be remarkably stable towards such variations.

\section{Program of Reduction of Parameters\label{parred}}

The program of reduction of parameters was formulated quite generally
\cite{zioe}-\cite{oesizi} for renormalizable field theories with more than
one coupling. It goes beyond the framework of this review to summarize the
full content of the program \cite{zioe}-\cite{zoup} including all the
recent developments \cite{mond},\cite{zoup}; after some introduction we
shall mainly confine the discussion to its application to the SM and
occasionally to the MSSM. This restriction will, unfortunately, not allow
to do full justice to the scope and beauty of the approach. What will
emerge, however, is that the program, starting with the very first
publications \cite{zioe}-\cite{oesizi} already as a systematic search for
special solutions of the RGE, has had an effect as role model for the
search for IR attractive fixed manifolds, reported on in previous sections,
with which it is interrelated to a certain extent.

Starting point of the reduction program is a renormalizable field theory
with a set of $n+1$ couplings $\la_{i}(\mu)$ for $i=1,...,n$ and $g(\mu)$;
generically $g$ is an asymptotically free (gauge) coupling.  The running of
these couplings as functions of the scale parameter $\mu$ is controlled by
the perturbative renormalization group equations, $n+1$ coupled non-linear
differential equations for these couplings.

The complete reduction program amounts to \cite{zioe}-\cite{oesizi},
\cite{zoup} a systematic search for special solutions of these RGE
\begin{itemize}
\item which establish {\it renormalization group invariant relations
    between all couplings} in the form $\la_{i}=\la_{i}(g)$; this
  requirement is shown to imply 
    the relations between the $\beta$ functions $\beta_{\la_{i}}={\rm
    d\,}\la_{i}/{\rm d\,}\ln(\mu)$ and $\beta_{g}={\rm d\,}g/{\rm
    d\,}\ln(\mu)$
\begin{equation}
\beta_{g}\frac{{\rm d\,}\la_{i}}{{\rm d\,}g}=\beta_{\la_{i}}\ \ \ {\rm
  for}\ \ \ i=1,...,n
\end{equation} 
\item and which are {\it not} determined by initial value conditions but 
\item rather by the theoretically appealing boundary condition that
  with the coupling $g$ also the couplings $\la_{i}$ vanish, i.e. become
  {\it asymptotically free} in the UV limit $\mu\rightarrow\infty$
\begin{equation}
\la_{i}\rightarrow 0\hspace{1cm}{\rm for}\hspace{1cm}g\rightarrow 0.
\label{asfree}
\end{equation}  
\end{itemize}
In the most restricted form, the dependent couplings $\la_{i}(g)$ emerge as
expansions in powers of $g$, typically in the form 
\begin{equation}
\la_{i}^2(g)=\rho_{1i}\,g^2 + \rho_{2i}\,g^4 + \rho_{3i}\,g^6 + ...
\label{expan}
\end{equation}
with constants $\rho_{1i},\ \rho_{2i},...$; less restricted forms allow the
expansions also to contain fractional powers or logarithms of $g$. The
reduction is called {\it non-trivial}, if all the first order coefficients
$\rho_{1i}$ in the expansion (\ref{expan}) are positive, {\it trivial} if
they are zero or, more generally, if $\la_{i}^2/g^2\rightarrow 0$ for
$g\rightarrow 0$; of course there are also mixed modes of partly nontrivial and
partly trivial expansions possible; for a comprehensive recent discussion see
Ref. \cite{zoup}.

In any case, complete reduction amounts to reducing the number $n+1$ of
independent couplings to a single one, $g$. Provided a reduction is
complete, it allows in principle a reduction {\it to all orders in
perturbation theory} and in practice to the order to which the RGE have
been calculated. 

It is instructive to relate the result of this program to a general search
for IR fixed points, lines, surfaces of the RGE as advocated in earlier
sections. A complete non-trivial reduction of the type (\ref{expan}) with
positive constants $\rho_{1i}$ amounts {\it at the one-loop level} to
determining \cite{oe} an {\it IR attractive i.e. IR stable fixed
point} $\la_{i}^2/g^2=\rho_{1i}$ for $i=1,...,n$ of the one-loop RGE in the
space of the {\it ratios} of couplings $\la_{i}/g$. In the framework of
reduction of parameters, this point has the interpretation as an UV
repulsive i.e. UV unstable fixed point solution, it exists as such in
principle to all orders of perturbation theory, more precisely the
existence of a renormalization group trajectory that asymptotically
approaches the point in the UV limit is ensured.  Despite this vicinity to
the search for IR fixed manifolds at the one-loop level, the important
difference to this search lies in the selection criterion: instead of
selecting special solutions which are IR attractive for the whole RG flow,
special solutions are singled out which link all couplings in such a way
that they decrease to zero simultaneously in the UV limit $\mu\rightarrow\infty$.

In the application of the program of reduction of parameters to the SM
\cite{kusizie},\cite{kusiziz} the ``driving'' coupling $g$ is identified
with the strong gauge coupling \gd; the condition (\ref{asfree}) then
implies the physical condition that as many couplings of the SM as possible
are linked in a renormalization group invariant manner to \gd\ such as to
share its property of asymptotic freedom in the UV limit
$\mu\rightarrow\infty$. A complete reduction fails in the sector of gauge
couplings \gee, \gz\ and \gd. The $U(1)$ gauge coupling \gee\ being not
asymptotically free also prohibits the performance of the formal limit
$\mu\rightarrow\infty$, but rather limits the final analysis to within a finite
interval below a physical UV scale \lam, introduced already in the
introduction.

A partial reduction can be performed nevertheless. The analysis
\cite{kusizie},\cite{kusiziz} of the partial reduction program in the
matter sector of the SM then proceeds in a two-step process: in a first
step the electroweak gauge couplings are switched off, \gee=\gz=0; in a
second step the electroweak interactions are systematically and
consistently included as corrections in form of a double power series
expansion in the ratios \re=\ges/\gds and \rz=\gzs/\gds. As results at the
bottom line of the analyses \cite{kusizie},\cite{kusiziz} are two solutions, a non-trivial
one and a trivial one.

Let us start with the non-trivial solution. In the absence of electroweak
couplings a consistent and physically meaningful non-trivial solution is
found in the top-Higgs-\gd\ sector; at one loop it is identical \cite{oe}
with the IR attractive fixed point by Pendleton and Ross \cite{pen}, Eq.
(\ref{rtR}), in the variables \rt=\gts/\gds\ and \rh=\la/\gds
\begin{equation}
\gts=\frac{2}{9}\,\gds,\hspace{1cm}\la=\frac{\sqrt{689}-25}{72}\,\gds,
\label{kszol}
\end{equation}
  at the two-loop level it
  is given \cite{ku},\cite{ziwieku} by the expansion in powers of \gds
\begin{eqnarray}
\gts&=&\frac{2}{9}\,\gds+\frac{31359+41\sqrt{689}}{15552}\,\frac{\gd^4}{16\pi^2}+...,\nonumber\\
\la&=&\frac{\sqrt{689}-25}{72}\,\gds+\frac{147015115-535843\sqrt{689}}{3856896}
\,\frac{\gd^4}{16\pi^2}+... .\label{kutl}
\end{eqnarray}
At the one-loop level the electroweak couplings are included \cite{kusiziz}
as perturbations for small values of the ratios \re=\ges/\gds\ and
\rz=\gzs/\gds\ in form of a double power series expansion in powers of \re\
and \rz\ into the ansatz for a special solution of the RGE for sufficiently
small values for \re\ and \rz, subject to the boundary condition that for
$\re,\rz\rightarrow 0$ the unperturbed solution (\ref{kszol}) is recovered;
the result is
\begin{eqnarray}
\rt=\gts/\gds&=&\frac{2}{9}-\frac{17}{540}\,\re-\frac{1}{12}\,\rz+
\frac{799}{64800}\,\re^2+\frac{119}{9360}\,\re\rz-\frac{1}{288}\,\rz^2+...
\label{rtzi}\\
\rh=\la/\gds&=&\frac{\sqrt{689}-25}{72}-\frac{83\sqrt{689}-1295}{66960}\,\re-\frac{7\sqrt{689}-163}{1488}\,\rz+... .
\label{rhzi}
\end{eqnarray}
This expansion has a finite radius of convergence in \re\ and \rz\ around
\re=\rz=0, but the solution may be extended beyond it numerically.

It is interesting to translate the one-loop result (\ref{rtzi}) into the
language of IR attractive manifolds\footnote{One of us (B.S.) is grateful
  to W. Zimmermann for stimulating the following discussion and
  investigation \cite{baeun} by a letter.}: it constitutes an IR attractive
fixed surface in the three-dimensional \rt-\re-\rz- space which attracts
the ``top-down'' RG flow very weakly. This can be seen by solving
\cite{baeun} the linearized version of the RGE (\ref{rtrbrta12}) and
 around the common IR attractive fixed point
\rt=2/9 and \re=\rz=0, for \rb=\rta=0; it has the IR fixed solution
(\ref{rtzi}) with only the linear terms in \re\ and \rz\ present. The
general solution to this linearized problem can also be obtained
\cite{baeun}; it shows that the IR attractive fixed surface (\ref{rtzi})
attracts the ``top-down'' RG flow in the vicinity of the IR fixed point
\rt=2/9, \re=\rz=0 with the small power $(\gds)^{-1/7}$, i.e.  exceedingly
weakly.  This surface had not been discussed in Sect.  \ref{t123}, since it
loses the competition with the much stronger IR attractive fixed surface,
given in form of analytic expansions in Eqs.  (\ref{xyz1}), (\ref{xyz2}) and
numerically in Fig.  \ref{rt12}. This may be seen most transparently by
following Ref.  \cite{kusiziz} and Sect.  \ref{t123}, i.e. by introducing
the experimental initial values for $\gee(m_Z)$ and $\gz(m_Z)$. This allows
to express \re\ and \rz\ as functions of 1/\gds, and plot the resulting
projection of the two surfaces into the \rt-1/\gds\ plane. The resulting
two IR attractive fixed lines are displayed in Fig. \ref{rtr1r2fili}. The
``top-down'' RG flow may be read off from Fig. \ref{rtr1r2flow}; clearly
the fat line {\bf 1}, representing the more attractive fixed line discussed
in Sect. \ref{t123}, is much more strongly attractive than the thin line
{\bf 2}, representing the projection of the solution (\ref{rtzi}) emerging
from the one-loop reduction program.

The conclusions arrived at in the last paragraph are important for the
search for IR attractive manifolds where the rate of IR attraction
counts, while it carries little weight for the parameter reduction approach
which is dictated by the boundary condition of asymptotic freedom in the UV
limit. In any case the discussion has exhibited yet another interrelation
between IR motivated and UV motivated physics issues.

The inclusion of \re\ and \rz\ into the solutions (\ref{rtzi}) and
(\ref{rhzi}) in form of a double power series in powers of \re\ and \rz\
cannot be extended, even not in principle, to the UV limit $\mu\rightarrow
\infty$, since \re\ will eventually become large and run out of the region
of validity of perturbation theory. Thus, in a way, the parameter reduction
scheme faces in application to the SM a similar problem as the search for
IR attractive fixed manifolds: the increase of \re\ for increasing $\mu$
prevents the former from reaching the mathematical UV limit
$\mu\rightarrow\infty$, while the increase of \gds\ for decreasing $\mu$
prevents the latter from reaching the mathematical IR limit $\mu\rightarrow
0$ \cite{ziIR}. Both approaches have to make do with a {\it finite} interval
between the IR scale $\mu\simeq\mt$ and an UV scale $\lam\lwig10^{19}\gev$.

The top and Higgs mass values determined from the one-loop results
\cite{kusiziz} by evaluating \rt\ and \rh\ at $\mu=m_Z$ are of the order of $\mt\simeq 91.3\gev$ and $\mh\simeq
64.4\gev$. More precise values emerge from the two-loop approach to be
presented next. Even though the resulting top mass value is excluded
meanwhile by experiment, the discussion provides a welcome deeper insight
into the parameter reduction method on a higher loop level.

At the two-loop level \cite{ku},\cite{ziwieku} besides an expansion in
powers of \re\ and \rz\ also an expansion in powers of the small parameter
\rb=\gbs/\gds\ is included. Eqs. (\ref{rtzi}) and (\ref{rhzi}) are then 
replaced by

\begin{eqnarray}
\rt&=&\frac{2}{9}-\frac{17}{540}\,\re-\frac{1}{12}\,\rz-\frac{1}{5}\,\rb+
\frac{799}{64800}\,\re^2+\frac{119}{9360}\,\re\rz-\frac{1}{288}\,\rz^2+
\frac{9}{400}\,\re\rb-\frac{54}{175}\,\rb^2\nonumber\\
 & &-\frac{5593}{972000}\,\re^3-\frac{323}{62400}\,\re^2\rz+\frac{17}{56160}\,\re\rz^2-\frac{1}{1728}\,\rz^3-0.009\,\re^2\rb-0.001\,\re\rz\rb\nonumber\\
 & &+0.0029\,\re^4+0.0025\,\re^3\rz-0.00008\,\re^2\rz^2+0.00005\,\re\rz^3-0.00014\,\rz^4+...\nonumber\\
 &
 &+\frac{\gds}{4\pi^2}(\frac{31359+41\sqrt{689}}{62208}-0.2231\,\re-0.8262\,\rz+0.1690\,\re^2+0.1824\,\re\rz\nonumber\\ & &-0.0664\,\rz^2+...)+...\nonumber\\
\rh&=&\frac{\sqrt{689}-25}{72}-\frac{83\sqrt{689}-1295}{66960}\,\re-\frac{7\sqrt{689}-163}{1488}\,\rz-0.037161\,\rb\nonumber\\
 & &+.022843\,\re^2+0.0531783\,\re\rz+0.1092913\,\rz^2+0.0362\,\re\rb+0.024\,\rz\rb\nonumber\\
 &
 &+0.2725\,\rb^2-0.01722\,\re^3-0.032795\,\re^2\rz-0.02158\,\re\rz^2+0.00885\,\rz^3\nonumber\\
 & &-0.0160\,\re^2\rb+0.0215\,\re\rz\rb+0.09348\,\rz^2\rb\nonumber\\
 & &+0.0124\,\re^4+0.0226\,\re^3\rz+0.0152\,\re^2\rz^2+0.00588\,\re\rz^3+0.00923\,\rz^4+...\nonumber\\
 &
 &\frac{\gds}{4\pi^2}(\frac{14701515-535843\sqrt{689}}{15427584}-0.03088\,\re-0.1205\,\rz\nonumber\\
 & &+0.0835\,\re^2-0.0115\,\re\rz-0.00083\,\rz^2+...)+... 
\end{eqnarray} 
where the decimal numbers are numerical approximations.

Including a physical value for the bottom mass (5\gev), radiative corrections
relating the pole masses for the top quark and the Higgs boson to their
$\overline{{\rm MS}}$ masses at $\mu=m_Z$ the two-loop results lead to the
following pole masses \cite{ku}
\begin{equation}
\mtp=99.2\pm5.7\gev\hspace{1cm}{\rm and}\hspace{1cm}\mhp=64.6\pm0.9\gev.
\end{equation}
The predicted top mass is too low to be compatible with the experimental
value.

It is interesting to ask whether the analogous parameter reduction analysis
leads to a success in the framework of the MSSM. The answer is no
\cite{baeun}. The one-loop result in absence of the electroweak gauge
couplings is identical with the IR attractive fixed point solution
(\ref{fp12}) a la Pendleton and Ross, the one-loop expansion of \rt\ in
powers of \re\ and \rz\ leads to an IR attractive fixed surface in the
\rt-\re-\rz-space \cite{baeun} which is again less strongly IR attractive
as the one (Fig \ref{rt12}) discussed in Sect. \ref{t123}. Introducing
the experimental initial conditions for \re\ and \rz, leads to the solution
drawn by the thin line {\bf 2} in Fig. \ref{rtr1r2fili}, which is clearly
less strongly attractive than the one drawn by a fat line {\bf 1} representing the solution
discussed in Sect. \ref{t123}. 

The ``trivial'' solution of the parameter reduction program applied to the
SM leads to even smaller top and Higgs masses. There are, however, again
very interesting cross relations to the search for IR attractive manifolds
at the one-loop level. The one-loop trivial solution \cite{kusiziz} in
absence of electroweak couplings
\begin{equation}
\rh=\frac{1}{3}\,\rt^2+\frac{1}{10}\,\rt^3+...
\end{equation}
establishes a relation between the ratios of couplings \rh\ and \rt. It
turns out to be identical with the expansion (\ref{zero}) of the IR
attractive fixed line in the \rt-\rh-plane around \rt=\rh=0 (remembering
that $R=\rh/\rt$). This solution extends up to the IR fixed point at
\rt=2/9, i.e. is historically a precursor of the low \rt\ end of the fixed line shown in
Fig. \ref{fixline}. The corresponding one-loop expansion including the electroweak
corrections \cite{kusiziz} is
\begin{eqnarray}
\rh&=&\frac{1}{3}\,\rt^2+\frac{1}{10}\,\rt^3\nonumber\\
 &
 &+\frac{27}{2800}\,\re^2(1+\rt)+\frac{9}{280}\,\re\rz(1+\rt)+\frac{9}{112}\,\rz^2(1+\rt)\nonumber\\
 &
 &-\frac{351}{56000}\,\re^3-\frac{1143}{78400}\,\re^2\rz-\frac{129}{15680}\,\re\rz^2+\frac{33}{3136}\,\rz^3-\frac{2}{15}\,\re\rt^2+... .
\end{eqnarray}
Again this is the expansion of the corresponding three-dimensional IR
attractive fixed manifold in the \rt-\rh-\re-\rz-space, discussed in
Sect. \ref{Htbgd} (for \gb=0), around \rt=\rh=\re=\rz=0. 

In a number of recent interesting publications \cite{mond},\cite{zoup} the
program of reduction of parameters is applied to supersymmetric grand
unified theories for scales $\mu\ge\mgut$ {\it above} the grand unification
scale \mgut. The advantage is that theories can be singled out which allow
a complete reduction. The details of the applications go clearly beyond this
review, however, some of the conclusions are very pertinent.

The resulting RG invariant relations between the gauge coupling(s) and the
Yukawa couplings for $\mu\ge\mgut$ are summarized under the headline
``gauge-Yukawa unification''. There are typically several physically
appealing solutions to each discussed theory, each one providing RG
invariant relations between all dimensionless coupling parameters.
Generically the UV initial value for the top Yukawa coupling is
sufficiently large; in many cases the bottom Yukawa coupling is of the
order of the top coupling. The authors \cite{mond},\cite{zoup} then explore
the corresponding strongly constrained ``top-down'' RG flows from \mgut\ to
the IR scale \mt\ according to the RGE of the MSSM with \msusy\ varying
within reasonable bounds. Here their various solutions for various theories
are caught to a certain extent in the trap of the strongly IR attractive
fixed manifolds and and in particular of the IR fixed point for the top
mass discussed in Sects.  \ref{allgauge} and \ref{masses}. Thus, in zeroth
approximation one expects the IR results to be rather independent of the UV
input. The authors work out to which extent the results vary for the
different theories and for the different solutions within a given theory.
Their top masses lie around 190\gev, as expected, with a variation of at
most 10\gev\ from one model or solution to another one; the values for \tanb\
are around 63, signalling top-bottom unification, see Sects. \ref{tb123}
and \ref{masses}.

These gauge-Yukawa reduction scenarios have two features in common with
supersymmetric grand unification with tau-bottom Yukawa unification
discussed in Sect. \ref{yukuni}.
\begin{itemize}
\item The top-bottom sector of the IR region is a rather insensitive
  testing ground for the details of the theory around and above
  $\mu=\mgut$.
\item The dynamics at the UV scale \mgut\ singles out largish values for the
  UV initial values for the top and generically also for the bottom Yukawa
  coupling. This in turn selects that portion of the ``top-down'' RG flow
  which is contracted fully onto the IR fixed point in \htt, resp. on the
  IR fixed line in the \htt-\hb\ plane. This enhances strongly the
  significance of these IR fixed manifolds.
\end{itemize}

\section{Conclusions}

The efforts to trace a possible dynamical origin for the top and Higgs
masses at the level of the quantum effects, as encoded in the RGE, have
been reviewed. The SM and the MSSM have been considered in parallel.

The most important answers to this question lie in IR attractive fixed
lines and fixed points which are approached by the ``top-down'' RG flow
from an UV scale \lam\ to the IR scale O$(v)$
\begin{itemize}
\item in the \tanb-\mtp-plane of the MSSM:
\begin{itemize}
\item for small bottom Yukawa couplings the much quoted very strongly IR
  attractive fixed point
  for the top mass 
\begin{equation}
\mtp=O(190-200)\gev\sin\beta
\label{aaa}
\end{equation}
which is resolved into 
\begin{eqnarray}
{\rm a\ genuine\ IR\ fixed\ point\ at}\hspace*{5mm}\mtp&\approx& 190\gev\sin\beta,\\
{\rm and\ an\ upper\ bound,\ the\ IR\ image\ of\ large}\hspace*{7mm}\ \ \ \ \ & &\nonumber\\{\rm UV\ initial\ values\ for\ the\ top\ Yukawa\ coupling}\hspace*{5mm}\mtp&\approx& 200\gev\sin\beta.
\end{eqnarray}
\item for unconstrained bottom Yukawa couplings a very strongly IR attractive fixed line in the
  \tanb-\mtp-plane, Fig. \ref{tanbmt}, with an IR attractive fixed point at
\begin{equation}
\mtp\approx 182\gev,\hspace{1cm}\tanb\approx 60,
\end{equation}
implying approximate bottom-top Yukawa unification.  
\end{itemize}
\item in the \mhp-\mtp-plane of the SM:
\begin{itemize}
\item a weakly IR attractive fixed point at 
\begin{equation}
\mtp\approx 214\gev,\hspace{0.6cm}\mhp\approx 210\gev,
\end{equation}
\item lying on a strongly attractive IR fixed line in the \mhp-\mtp-plane,
Fig. \ref{mHmt}, 
\begin{equation}
{\rm which\ implies}\ \ \ \mhp\approx 156\gev\ \ \ {\rm for}\ \ \ \mtp=176\gev.
\label{bbb}
\end{equation}
\end{itemize}
\end{itemize}

As it turns out, the IR attractive top and Higgs mass values are roughly of
the order of the electroweak scale, $v/\sqrt{2}$, much larger than the
masses of all other matter particles. In particular in the MSSM the
resulting top mass $\mtp=O(190-200)\gev\sin\beta$ is well compatible with
the experimental value, but also the corresponding SM value $\mt\approx
215\gev$ is not very far from it. This is altogether a very striking result. Of
course, as we knew from the outset, in order to judge the significance of
these results for physical reality, one needs further experimental
information. The MSSM needs an experimental support for the
supersymmetry content and an experimental value for \tanb, the SM as well
as the MSSM await the experimental detection of the Higgs boson.

An intricate interrelation between UV physics and IR physics in the MSSM, which has received
much attention in the literature, may be summarized as follows.
\begin{itemize}
\item The UV symmetry constraint of tau-bottom Yukawa coupling unification
  in supersymmetric grand unification focuses the RG flow towards the IR
  scale much more strongly into the IR fixed point top mass value, more
  precisely onto the IR fixed line in the \tanb-\mtp-plane, than the
  unconstrained RG flow.
\item It appears to be the very presence of the IR fixed point in the top
  mass, resp. of the IR fixed line in the \tanb-\mtp-plane, which allows to
  implement tau-bottom Yukawa unification at the UV scale as well as the
  physical values of the tau and bottom masses.
\item Bottom-top Yukawa unification, a symmetry property at the unification
  scale in (supersymmetric) grand unification, is also an IR attractive
  property.
\end{itemize}

Much effort has been spent in this review to work out the underlying IR
fixed manifolds (fixed points, lines, surfaces,...) in the space of ratios
of parameters. When all gauge couplings are included, the relevant space is
the space of variables \gts/\gds, \gbs/\gds, \ges/\gds, \gzs/\gds, treated
absolutely in parallel for the SM and the MSSM; in case of the SM the space
is further increased by the variable \la/\gds. These higher dimensional
manifolds appear to be essential for an insight into the IR fixed
points and lines in presence of all gauge couplings. The results are
summarized in a compact form in table 2 in Sect. \ref{preview}. A second
important issue was to assess the respective rates of approach towards the
various IR attractive manifolds as far as possible analytically. From this
collective investigation the following hierarchies emerge.
\begin{itemize}
\item The SM and the MSSM have a very similar IR fixed manifold structure.
  However, the IR attraction for the ``top-down'' RG flow is systematically
  stronger in the MSSM than in the SM.
\item Generically {\it nontrivial} higher dimensional IR attractive
  manifolds are more strongly IR attractive than the lower
  dimensional ones.
\end{itemize}
This last statement needs qualification: In the high dimensional
multiparameter space an IR attractive fixed point generically lies at the
intersection of two IR attractive 
lines. Unless accidentally the strength of attraction is equal for both
lines, it is always one of the two lines which attracts more strongly. The
lines in turn are intersections of two IR attractive surfaces, one of which
will be more attractive, etc. The point is that most of these lines and
surfaces are trivial ones, e.g. in the sense that they correspond to one
variable held constant. It turns out that typically the most non-trivial
lines, surfaces, etc. turn out to be the most attractive ones, leading to
non-trivial IR attractive relations between the considered parameters.

The triviality and vacuum stability bounds on the SM Higgs mass
provide a measure for the distance of the boundaries of the
``top-down'' RG flow from the IR fixed point and fixed line in the
\mhp-\mtp-plane as a function of the UV scale \lam. These bounds are
also of relevance for future Higgs searches. Therefore, the most
recent precise determinations of the vacuum stability bound in the SM
as well as of the upper bound on the mass of the lightest Higgs boson
in the MSSM have been included in this review.

Let us close with an outlook from two complementary points of view.

One could entertain the ``conservative'' expectation that IR physics should
be largely independent of UV physics. This may be achieved in a
renormalizable theory by arranging the parameters to match the IR
attractive (i.e. IR stable) fixed points or fixed lines etc. of the RGEs.
In this case the UV scale \lam, at which new physics arises, the details of
the new physics as well as the rate of approach towards the IR fixed
manifolds, become largely irrelevant.  The only input which counts then is
the specific form of the RGEs in the vicinity of the IR scale. Seen under
this angle, the IR fixed point structures in Eqs.  (\ref{aaa})-(\ref{bbb})
are a promising starting point.  An investigation into IR fixed manifolds
in the remaining quark and lepton sector (even if they were to turn out to
be only weakly IR attractive) would be a necessary step towards further
support of this point of view.

Very interesting issues have recently been raised in the literature
\cite{ross1}, \cite{mond}, \cite{zoup}, which rely very much on an
interrelation between new physics at a high UV scale and the known IR
physics.
\begin{itemize} 
\item From our present knowledge it may appear that whichever new physics
  in supersymmetric grand unification one formulates: as long as the UV
  initial values of the top Yukawa couplings are fixed by the UV dynamics
  to be larger than O(1), the RGE evolution will lead into the ``trap'' of
  the IR fixed point for the top mass; if the bottom Yukawa coupling is
  large in addition, one ends up with \tanb=O(60). The issue is then
  \cite{ross1}, \cite{mond}, \cite{zoup} to investigate, how ``stable''
  this memory loss of UV physics in the top-bottom sector is with respect
  to ``variations'' in the new physics sector.  First answers \cite{ross1},
  \cite{mond}, \cite{zoup} indicate that it is remarkably stable.
\item A much furtherreaching issue \cite{ross1} is the question to which
  extent the quark and lepton masses and mixing angles as well as the soft
  supersymmetry breaking mass terms are determined in terms of the IR fixed
  point structure of the theory lying beyond the SM. First answers
  \cite{ross1} indicate that the knowledge of the symmetries and multiplet
  content of the theory beyond the SM is largely sufficient to determine
  the IR structure and to lead to very promising results in the low energy
  sector of the theory.
\end{itemize}

{\bf Acknowledgements:} One of us (B.S.) is grateful to J.R. Espinosa, G.
Isidori, F. Jegerlehner, B. Kniehl, I. Montvay, F.  Schrempp, F. Wagner, P.
Weisz, W. Zimmermann for valuable communications and/or enlightning
discussions. We also thank F. Schrempp for granting the inclusion of some
unpublished results, obtained partly by himself, partly in collaboration
with one of us (B.S). One of us (B.S.) is grateful to the DESY theory group
for the continuous hospitality.

\end{document}